\documentclass[acmtog, authorversion]{acmart}

\newcommand{\final}{1}
\newcommand{\isarxiv}{1}
\newcommand{\supp}{1}

\newcommand{\hili}{1}

\usepackage{color}
\usepackage{ifthen}
\usepackage{float}
\usepackage{alltt}
\usepackage{tikz}
\usepackage{fp}
\usepackage[scientific-notation=true]{siunitx}
\usepackage{amsmath}
\usepackage{amssymb}
\usepackage{amsthm}
\usepackage{newlfont} % for Box
\usepackage{floatflt}
\usepackage{wrapfig}
\usepackage{fixltx2e}
\usepackage{subfig} % for subfloat
\usepackage{multirow}
\usepackage{multicol}
\usepackage{cleveref}
\usepackage{slashbox}
\usepackage[export]{adjustbox}
\usepackage{CJKutf8} % Chinese
\usepackage{microtype}        % Improved Tracking and Kerning
\usepackage{ccicons}          % Cite your images correctly!
\usepackage{balance}       % to better equalize the last page

\usepackage[ruled]{algorithm2e} % For algorithms
\usepackage[noend]{algcompatible}

\usepackage{booktabs} %for toprule, midrule, bottomrule in table
\usepackage{slashbox} %for toprule, midrule, bottomrule in table
\usepackage{threeparttable}

\usepackage{tabularx}

\newcommand{\pcomment}[1]{\mbox{// {\it #1}}}  % Pseudo-code comments.
\renewcommand{\pcomment}[1]{\texttt{\small{// #1}}}
\newcommand{\tabtextsize}{\footnotesize}

\newcommand{\Caption}[2]{\caption[#1]{{\footnotesize #1} {\footnotesize #2}}}

\newcommand{\emacsquote}[1]{{``#1''}}

\definecolor{LiyiColor}{rgb}{0.7,0,0} % color for Sith
\definecolor{XiaoyinColor}{rgb}{0.56,0.34,0.62} % purple for Mace Windu
\definecolor{PatColor}{rgb}{0.25,0.5,0.4}
\definecolor{ScottColor}{rgb}{0,0.4,0} % color for the Jedi Consulars (e.g. Yoda)
\definecolor{MohamedColor}{rgb}{0,0,0.8} % color for the Jedi Guardians (e.g. Obiwan)
\definecolor{MuhammadColor}{rgb}{0,0.5,0}
\definecolor{AhmedColor}{rgb}{0.87,0,0.5}
\definecolor{DineshColor}{rgb}{0.3,0.6,0.6} 
\definecolor{ChonhyonColor}{rgb}{0.7,0.4,0.4} 
\definecolor{VijayColor}{rgb}{0.7,0.4,0.4} 
\definecolor{HarishColor}{rgb}{0.4,0.7,0.4} 
\definecolor{YusuColor}{rgb}{0.44,0.66,0.38} % opposite that of Xiaoyin
\definecolor{LauraColor}{rgb}{1.0,0.63,0.63}
\definecolor{AnjulColor}{rgb}{0.2, 0.0, 0.35}

\newcommand{\liyi}[1]{{\color{LiyiColor} Li-Yi: #1 $\qed$}}
\newcommand{\mohamed}[1]{{\color{MohamedColor} Mohamed: #1 $\qed$}}
\newcommand{\xiaoyin}[1]{{\color{XiaoyinColor} Xiaoyin: #1 $\qed$}}
\newcommand{\muhammad}[1]{{\color{MuhammadColor} Muhammad: #1 $\qed$}}
\newcommand{\ahmed}[1]{{\color{AhmedColor} Ahmed: #1 $\qed$}}
\newcommand{\scott}[1]{{\color{ScottColor} Scott: #1 $\qed$}}
\newcommand{\pat}[1]{{\color{PatColor} Pat: #1 $\qed$}}
\newcommand{\dmanocha}[1]{{\color{DineshColor} Dinesh: #1 $\qed$}} % Dinesh Manocha
\newcommand{\chpark}[1]{{\color{DineshColor} Chonhyon: #1 $\qed$}} % Chonhyon Park
\newcommand{\vijay}[1]{{\color{VijayColor} Vijay: #1 $\qed$}} % Vijay Natarajan
\newcommand{\harish}[1]{{\color{HarishColor} Harish: #1 $\qed$}} % Harish D <harishd10@gmail.com> 
\newcommand{\yusu}[1]{{\color{YusuColor} Yusu: #1 $\qed$}} % Yusu Wang
\newcommand{\laura}[1]{{\color{LauraColor} Laura: #1 $\qed$}} % Laura Swiler
\newcommand{\anjul}[1]{{\color{AnjulColor} Anjul: #1 $\qed$}} % Anjul Patney

\definecolor{ReviewerColor}{rgb}{0.1,0.56,0.5} %
\newcommand{\reviewer}[2][]{{\color{ReviewerColor} Reviewer\##1: #2 $\qed$}} % reviewers

\newcommand{\warning}[1]{{\it\color{red} #1}}
\newcommand{\todo}[1]{{\color{red}todo: #1}} % to do before submisstion
\newcommand{\note}[1]{{\it\color{blue} #1}}
\newcommand{\nothing}[1]{}

\definecolor{VideoColor}{rgb}{0.9,0.4,0}
\newcommand{\video}[1]{{\color{VideoColor} Video: #1 $\qed$}}
\definecolor{AudioColor}{rgb}{0.1,0.6,1.0}
\newcommand{\audio}[1]{{\color{AudioColor} Audio: #1 $\qed$}}
\definecolor{DeadlineColor}{rgb}{0.9,0.4,0}
\newcommand{\deadline}[1]{{\bf\color{DeadlineColor} ETA: #1}}

\definecolor{figred}{rgb}{1,0,0}
\definecolor{figgreen}{rgb}{0,0.6,0}
\definecolor{figblue}{rgb}{0,0,1}
\definecolor{figpink}{rgb}{1,0.63,0.63}

\ifthenelse{\equal{\final}{1}}
{
\renewcommand{\liyi}[1]{}
\renewcommand{\mohamed}[1]{}
\renewcommand{\xiaoyin}[1]{}
\renewcommand{\muhammad}[1]{}
\renewcommand{\ahmed}[1]{}
\renewcommand{\scott}[1]{}
\renewcommand{\pat}[1]{}
\renewcommand{\dmanocha}[1]{}
\renewcommand{\chpark}[1]{}
\renewcommand{\vijay}[1]{}
\renewcommand{\harish}[1]{}
\renewcommand{\yusu}[1]{}
\renewcommand{\laura}[1]{}
\renewcommand{\anjul}[1]{}
\renewcommand{\reviewer}[2][]{}

\renewcommand{\warning}[1]{}
\renewcommand{\note}[1]{}
\renewcommand{\todo}[1]{}
\renewcommand{\video}[1]{}
\renewcommand{\audio}[1]{}
\renewcommand{\deadline}[1]{}
}
{}

\newcommand{\pseudocode}{Algorithm}
\floatstyle{plain}
\newfloat{algorithm}{tbhp}{lop}
\floatname{algorithm}{\pseudocode}

\newcommand{\filename}[1]{\url{#1}}
\newcommand{\foldername}[1]{\url{#1}}

\ifdefined\email
\else
\newcommand{\email}[1]{\url{#1}}
\fi

\newcommand{\figref}[1]{\Cref{#1}}
\algblockdefx{FORP}{ENDFP}[1]%
  {\textbf{for }#1 \textbf{do in parallel}}%
  {\textbf{end for}}

\newcommand{\RETURN}{\STATE \textbf{return}}

\newcommand{\funct}[1]{\texttt{#1}}

\newlength{\tempheight}
\newlength{\tempwidth}

\newcommand{\rowname}[1]% #1 = text
{\rotatebox{90}{\makebox[\tempheight][c]{\textbf{#1}}}}

\newcommand{\columnname}[1]% #1 = text
{\makebox[\tempwidth][c]{\textbf{#1}}}

\newcommand{\domainsym}{\Omega}
\newcommand{\dimnumber}{d} % dimensionality of the sample space
\newcommand{\samplesym}{s}
\newcommand{\sample}{\samplesym{}}
\newcommand{\sampleprime}{\samplesym^{\prime}{}}
\newcommand{\sampleprimeprime}{\samplesym^{\prime\prime}{}}
\newcommand{\samplestar}{\samplesym^{*}{}}
\newcommand{\samplesetsym}{\mathcal{S}}
\newcommand{\sampleset}{\samplesetsym{}}
\newcommand{\samplenumber}{n}
\newcommand{\neighbors}{N} %number of them
\newcommand{\neighborset}{\mathcal{N}} %the set of them

\newcommand{\rnumber}{r}

\newcommand{\rconflict}{r_f}

\newcommand{\rcoverage}{r_c}

\newcommand{\rratio}{\beta} %\liyi{Don't use \rho, as it can be confused as the $\rhonumber$ below used since [Lagae and Dutre 2008]}

\newcommand{\lodmethod}{\emph{IntersectionSearch}}

\newcommand{\hidmethod}{\emph{CenterSearch}}
\newcommand{\Hidmethod}{\emph{CenterSearch}}

\newcommand{\addatpoint}{\emph{InsertAtPoint}} 
 
\newcommand{\addinball}{\emph{InsertInBall}} 
\newcommand{\addonray}{\emph{InsertOnRay}}

\newcommand{\bigbetachance}{\epsilon}
\newcommand{\dart}{\mbox{\ensuremath{k}-d~dart}} % use \darttext\ with trailing slash for a space
\newcommand{\darts}{\mbox{\ensuremath{k}-d~darts}} % use \darttext\ with trailing slash for a space
\newcommand{\Darts}{\mbox{\ensuremath{k}-d~darts}} %this should still be lower-case k, since k is a math variable.
\newcommand{\kddart}{\dart}
\newcommand{\kddarts}{\darts}

\newcommand{\disk}{D}

\newcommand{\spoke}{\ell} %cursive letter el, clearly different from the number 1 or capital I.
\newcommand{\maxspokeattempts}{m}%perhaps t

\newcommand{\rx}{\rconflict} % disk radius (not same as exclusion distance for spatially varing radii)
\newcommand{\rvor}{\rcoverage} % distance to power vertex
\newcommand{\mps}{MPS}

\newcommand{\void}{\textrm{void}}
\newcommand{\rvoid}{r_\textrm{void}}

\newcommand{\area}{\textrm{Area}} % use parenthesis

\newcommand{\expected}{\textrm{E}} % fancy double stroke E
\newcommand{\phiti}[1]{p_{#1}(\textrm{hit})} %number of hits
\newcommand{\pmissi}[1]{p_{#1}(\textrm{miss})} %probability of a miss
\newcommand{\pmissmi}[1]{p_{#1}^\hammerlimit(\textrm{miss})} %probability of $m$ misses

\newcommand{\vdarearatio}{R}  %area(void)/area(disk)

\newcommand{\sandia}{Sandia National Laboratories}

\newcommand{\unc}{UNC Chapel Hill}
\newcommand{\hku}{University of Hong Kong}
\newcommand{\ucd}{University of California at Davis}
\newcommand{\nvidia}{NVIDIA Research}
\newcommand{\adobe}{Adobe Research}

\setcitestyle{square}
\def\textrightarrow{$\rightarrow$}

\SetAlFnt{\small}
\SetAlCapFnt{\small}
\SetAlCapNameFnt{\small}
\SetAlCapHSkip{0pt}
\IncMargin{-\parindent}

\renewcommand{\Caption}[2]{\caption[#1]{{#1} {#2}}}

\LetLtxMacro\Oldparagraph\paragraph
\renewcommand{\paragraph}[1]{\Oldparagraph*{{#1}}}

\newcommand\pargrph[1]{\paragraph{#1}}

\LetLtxMacro\Oldsubsubsection\subsubsection
\renewcommand\subsubsection[1]{\paragraph{#1}}

\definecolor{NewColor}{rgb}{0.9,0.4,0}

\definecolor{DeleteColor}{rgb}{0.1,0.6,1.0}
\newcommand{\delete}[1]{{\color{DeleteColor} #1}}
\definecolor{MoveColor}{rgb}{0.5,0.1,0.5}

\ifthenelse{\isundefined{\hili}}
{

\renewcommand{\delete}[1]{}

}
{
\ifthenelse{\equal{\hili}{0}}
{

\renewcommand{\delete}[1]{}

}
{}
}

\algnewcommand\algorithmicforeach{\textbf{for each}}
\algdef{S}[FOR]{ForEach}[1]{\algorithmicforeach\ #1\ \algorithmicdo}

\newcommand{\HidMethod}{{Spoke-Darts}} % no reason to use this except ALL-CAPS section and paper titles
\renewcommand{\hidmethod}{{spoke-dart sampling}}
\renewcommand{\Hidmethod}{{Spoke-dart sampling}}
\newcommand{\spokedart}{{spoke-dart}}

\newcommand{\spokedarts}{{spoke-darts}}
\newcommand{\Spokedarts}{{Spoke-darts}}
\newcommand{\linespokesalg}{{line-spokes}}
\newcommand{\favoredspokesalg}{{favored-spokes}}
\newcommand{\twospokesalg}{{two-spokes}} % there is no reason to use this except ALL-CAPS section and paper titles
\newcommand{\Linespokesalg}{{Line-spokes}}
\newcommand{\Favoredspokesalg}{{Favored-spokes}}
\newcommand{\Twospokesalg}{{Two-spokes}} % there is no reason to use this except ALL-CAPS section and paper titles

\newcommand{\spoketext}{spoke}

\newcommand{\spokestext}{spokes}

\newcommand{\linespoketext}{line-spoke}

\newcommand{\linespokestext}{line-spokes}
\newcommand{\Linespokestext}{Line-spokes}

\newcommand{\bridsonmethod}{Point-Annulus}
\newcommand{\bridsonmethodours}{Point-Annulus*p}
\newcommand{\hammerlimit}{\maxspokeattempts} 

\newcommand{\anchor}{a}
\newcommand{\interval}{I}

\newcommand{\spokesym}{\spoke}
\newcommand{\linespoke}{\spoke_1}
\newcommand{\linespokeprime}{\spoke_1^\prime}

\newcommand{\activepool}{\mathcal{P}}
\newcommand{\dist}{d}

\newcommand{\cellradius}{\delta}

\newcommand{\dgraph}{\mathcal{D}} %Delaunay graph
\newcommand{\dgraphast}{\mathcal{D^\ast}} %approximate Delaunay graph
\newcommand{\witness}{\omega}
\newcommand{\neighborcandidates}{\mathcal{M}} %set f of vertices that might be a neighbor with the current vertex.

\newcommand{\approptoinn}[2]{\mathrel{\vcenter{
  \offinterlineskip\halign{\hfil$##$\cr
    #1\propto\cr\noalign{\kern2pt}#1\sim\cr\noalign{\kern-2pt}}}}}

\newcommand{\knr}{k(d)}

\newcommand{\knrlinep}{k_{\textrm{l,p}}(d)}
\newcommand{\knrlinea}{k_{\textrm{l,b}}(d)}
\newcommand{\knrtwo}{k_{\textrm{two}}(d)}
\newcommand{\knrfavored}{k_{\textrm{fav}}(d)}
\newcommand{\volunit}{V_d}

\graphicspath{
{.}
{figs/raster/}
{figs/handdrawn/}
{figs/concept/}
{figs/blog/}
{figs/mitsuba}
{figs/}
{figs/optimization/}
}
\acmJournal{TOG}
\acmYear{2018}
\acmVolume{37}
\acmNumber{2}
\acmArticle{22}
\acmMonth{5} 
\acmDOI{10.1145/3194657}
\keywords{line sampling, high dimension, blue noise, Delaunay graph, global optimization, motion planning}

\setcopyright{acmlicensed}

\begin{document}

%\title{Optimal Sizing Function for Point Sampling Applications}
%\title{Sparse Maximal Poisson Disk Sampling}
%\title{Generating A maximal Blue Noise Distribution without suffering from the Curse-Of-Dimensionality}
%\title{Maximal Blue Noise Sampling without Curse-of-Dimensionality}
%\title{Reducing Dimensionality Curse for Maximal Blue Noise Sampling} %\liyi{Trying to come up with a cuter title.}
%\title{Spoke Darts for Efficient High Dimensional Blue Noise Sampling} %\liyi{emphasize the key points: efficient + high dimensional, up front, without getting distracted by less important (and arguable) claim about maximality.}
%% \title{Spoke Darts for Better Blue Noise Sampling and High Dimensions} 
%\newcommand{\mytitle}{Spoke Darts for Efficient Step Blue Noise in any Dimension} 
%\newcommand{\mytitle}{Spoke Darts for Efficient Sampling in any Dimension} 
%\newcommand{\mytitle}{Spoke Darts for Efficient Blue Noise Sampling in any Dimension} 
%\newcommand{\mytitle}{Spoke Darts for Efficient Blue Noise in any Dimension} 
%\newcommand{\mytitle}{Step Blue Noise in any Dimension} 
%\newcommand{\mytitle}{Efficient Blue Noise Sampling in any Dimension} %\liyi{I prefer this over any specific mentioning of spoke dart or step blue noise, which are subsets of our project.}
\newcommand{\mytitle}{\HidMethod{} for High-Dimensional Blue-Noise Sampling} %\liyi{sounds better than the previous one} 

\def\emptyauthor{}

\title{\mytitle}

\author{Scott A. Mitchell}
\affiliation{\sandia{}} 
\author{Mohamed S. Ebeida}
\affiliation{\sandia{}} 
\author{Muhammad A. Awad}
\affiliation{\ucd{}} 
\author{Chonhyon Park}
\affiliation{\unc{}}
\author{Anjul Patney}
\affiliation{\nvidia{}} 
\author{Ahmad A. Rushdi}
\affiliation{\ucd{}}
\affiliation{\sandia{}}
\author{Laura P. Swiler}
\affiliation{\sandia{}} 
\author{Dinesh Manocha}
\affiliation{\unc{}}
% \liyi{hope nobody minds I being the last author?} % scott{that's fine} \liyi{thanks I like to be Yoda....}
\author{Li-Yi Wei}
\affiliation{\hku{}} 
\affiliation{\adobe{}}

\renewcommand\shortauthors{Mitchell, Ebeida, Awad, Park, Patney, Rushdi, Swiler, Manocha, Wei}

\begin{teaserfigure}
  \centering
  \subfloat[8D Delaunay graph]{
    \label{fig:teaser_delaunay_graph}
    \ifthenelse{\equal{\isarxiv}{1}}
    {\includegraphics[width=0.2\linewidth]{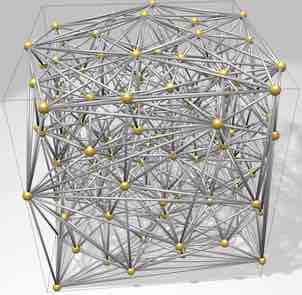}}
    {\includegraphics[width=0.2\linewidth]{figs/delaunay_3d.png}}
  }
  \subfloat[100D global optimization]{
    \label{fig:teaser_optimization}
    \ifthenelse{\equal{\isarxiv}{1}}
    {\includegraphics[width=0.25\linewidth]{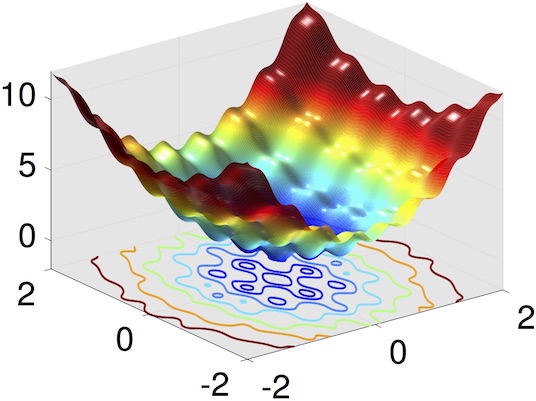}}
    {\includegraphics[width=0.25\linewidth]{figs/optimization/bohachevsky.pdf}}
  }
  \subfloat[8D rendering]{
    \label{fig:teaser_rendering}
    \ifthenelse{\equal{\isarxiv}{1}}
     {\includegraphics[height=0.2\linewidth]{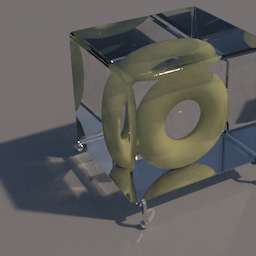}}
     {\includegraphics[height=0.2\linewidth]{figs/mitsuba/torus-in-glass/256spp/spoke_8D.png}}
  }%subfloat
  \subfloat[23D motion planning]{
    \label{fig:teaser_motion_planning}
    \ifthenelse{\equal{\isarxiv}{1}}
    {\includegraphics[height=0.2\linewidth]{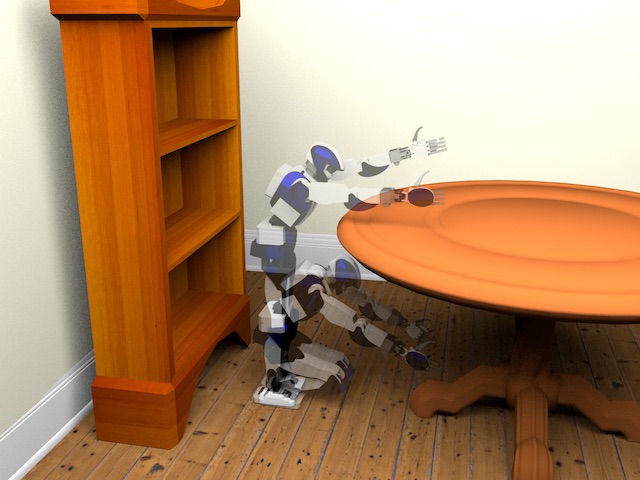}}
    {\includegraphics[height=0.2\linewidth]{figs/motionplanning/hrp4.png}}
  }
\Caption{\Hidmethod{} for high-dimensional applications:}
{%
Delaunay graph construction, optimization, rendering, and motion planning.}
\label{fig:teaser}
\end{teaserfigure}

%  \protect\subref{fig:teaser_delaunay_graph}
% \protect\subref{fig:teaser_optimization}
%  \protect\subref{fig:teaser_rendering}
% \protect\subref{fig:teaser_motion_planning}

\nothing{
\liyi{(December 24, 2013) make sure we can provide at least one good image for each application; otherwise it could be an indication that the application is not good for graphics. =D}
\mohamed{(December 27, 2013) I will do my best =D}
\liyi{(January 19, 2014) I added images to \subref{fig:teaser_delaunay_graph} and \subref{fig:teaser_optimization} but they are both low dimensional; better show high dimensional versions if we know how to visualize them.
In particular, \subref{fig:teaser_delaunay_graph} needs something other than surface mesh which seems quite misleading even more so than a 2D optimization function. Mohamed, do you have a Delaunay graph image of a 2D box domain? Maybe we can use that, less misleading than 3D objects.}
\liyi{Is there a version of \filename{delaunay_data.png} that has dark drawings over a white background? The current one, with the reverse color space, doesn't look good here.}
}%nothing
 % \liyi{(February 25, 2017) Better figure layout; we want to show this as early as possible.}

\begin{abstract} 
Blue noise sampling has proved useful for many graphics applications, but remains under-explored in high-dimensional spaces due to the difficulty of generating distributions and proving properties about them. 
We present a blue noise sampling method with good quality and performance across different dimensions.
The method, spoke-dart sampling, shoots rays from prior samples and selects samples from these rays.
It combines the advantages of two major high-dimensional sampling methods: the locality of advancing front with the dimensionality-reduction of hyperplanes, specifically line sampling.
We prove that the output sampling is saturated with high probability, with bounds on distances between pairs of samples, and between any domain point and its nearest sample.
% New sampling rules over these segments create step blue noise, with a flat frequency spectrum, which avoids high-frequency artifacts.
%
% Our algorithm creates blue noise directly and scales well to high dimensions.
%
We demonstrate spoke-dart applications for approximate Delaunay graph construction, global optimization,  and robotic motion planning. 
Both the blue-noise quality of the output distribution, and the adaptability of the intermediate processes of our method, are useful in these applications.
\end{abstract}

\nothing{\liyi{February 25, 2017) Looks asymmetric to put particular emphasis on global optimization.}%liyi
Our global optimization algorithm uses a variation of spoke-darts that adaptively creates non-uniform blue noise, demonstrating both the flexibility of our approach and the usefulness of its output.
\scott{(March 2, 2017) perhaps it is asymmetric wrt applications, but I think it is worth emphasizing that the method is useful beyond just outputting blue noise, because the reviewers are skeptical that the blue noise output is useful. And it is worth emphasizing that the blue noise output is actually used in one of our applications, optimization. That wasn't clear in prior drafts. Also, I note we list our applications twice, once in the first sentence and once in the last. Suggest we just do it in the last. The first sentence gets too complicated with it in there.}
\liyi{(March 3, 2017) Good point. Even I did not notice this fine point before. I also updated the corresponding part in the introduction, which was not very clear before.}%liyi
}%nothing

\begin{CCSXML}
<ccs2012>
<concept>
<concept_id>10010147.10010371.10010382.10010386</concept_id>
<concept_desc>Computing methodologies~Antialiasing</concept_desc>
<concept_significance>300</concept_significance>
</concept>
</ccs2012>
\end{CCSXML}

\ccsdesc[300]{Computing methodologies~Antialiasing}


\iffalse
%\category{I.3.5}%{Computing Methodologies}%
%{Computer Graphics}{Computational Geometry and Object Modeling}

\category{I.3.3}%{Computing Methodologies}%
{Computer Graphics}{Picture/Image Generation---\emph{Antialiasing}}
% I.3.3 Computing Methodologies, Computer Graphics, Picture/Image Generation---Antialiasing

\category{I.4.1}%{Computing Methodologies}%
{Image Processing and Computer Vision}
{Digitization and Image Capture---\emph{Sampling}}
% I.4.1 Computing Methodologies, Image Processing and Computer Vision, Digitization and Image Capture---Sampling

% see "General Terms" in http://www.acm.org/about/class/1998
\terms{Algorithms, Theory, Experimentation}
\fi

\maketitle

%%%%%%%%%%%%%%%%%%%%%%%%%%%%%%%%%%%%%%%%%%%%

\section{Introduction}
\label{sec:introduction}

%%%%%%%%%%%%%%%%%%%%%%%%%%%%%%%%%%%%%%%%%%%%

% paragraph topics: well-spaced, and runtime
Sampling is a core technique for various scientific and engineering applications. 
\nothing{
% samitch: the next few sentences set the stage for integration, specifically rendering, to be our main application, but it isn't. 
% samitch: it also sets the stage for blue-noise, and step blue-noise, to be paramount.
% Need to revise
%In many applications it is impossible to compute quantities exactly, as, for example, they would require casting an %infinite number of light rays.
%Fortunately, exact computations are rarely necessary, as variations below some resolution are imperceptible.
%There is no need to synthesize a texture down to molecular resolution; a pixel or two is adequate.
}%nothing
Sampling allows us to approximate continuous quantities in tractable space and time via discrete samples.
The samples should be well-spaced for efficiency, and yet random enough to avoid structural aliasing.
Low-discrepancy sequences \cite{niederreiter1992random,Keller:2012:AMC} are known for generating well-spaced samples, but their inherently deterministic nature is prone to produce regular \nothing{sample }patterns, which can cause aliasing.
Blue noise sampling \cite{Mitchell:1987:GAILS,ulichney1988dithering,Lagae:2007:CMPD,Ebeida:2012:SAM} can synthesize samples that are simultaneously random and well-spaced, but it is computationally more expensive, especially as the dimension increases.
Hyperplane sampling~\cite{kddarts_arxiv} is a random approach that scales to any dimension, but the output distribution is not guaranteed to be well-spaced.
%Thus, blue noise has been applied mainly to low dimensional (e.g. 2D) domains.
Thus, high-dimensional (e.g. $\geq$ 6D) blue-noise sampling remains elusive, even though high-dimensional spaces are common in geometry, optimization, rendering, robotics and other applications.

%\input{rdf_spectra_teaser_fig}

% paragraph topics: irregular domains motivates advancing front, but its runtime is a problem. 
% There are many techniques for sampling domains with different dimensions.
%Low-discrepancy sequences have the distinct advantage of linear scaling in both output size and dimension~\cite{niederreiter1992random}.
%Other approaches\nothing{ for square domains} scale less well; for example, the size of a background grid grows exponentially with dimension~\cite{Ebeida:2012:SAM}.
Advancing front techniques~\cite{firstspheremesh,Li99biting:advancing,fastspherepacking,Dunbar:2006:PDS,Bridson:2007:FPD} are able to efficiently sample from irregular domains; in contrast many other methods are tailored to domains that are hyperrectangles. The ability to handle irregular domains is an advantage in some contexts, but the complexity of computing and storing the geometry of fronts grows exponentially with dimension.
\nothing{\liyi{(February 25, 2017) Already did in the previous work section, which is more suitable as the introduction is supposed to sell our own work!}%liyi
% say something about Bridson
\bridsonmethod~\cite{Bridson:2007:FPD} is one popular advancing front technique that does scale well, by avoiding the need to create the front geometry explicitly.
}%nothing

% samitch: fill in more detail
We present a new algorithm that has the advantages of both advancing front and hyperplane sampling.
It scales to high dimensions by avoiding computing the front geometry. 
It uses line sampling, selecting the next sample from a line segment through a prior sample.

Its output has guaranteed blue noise properties.
We provide bounds on the spatial properties of our output, including saturation, that apply to any dimension.
We show algorithmic time and space complexities that avoid the curse of dimensionality.
We provide experiments that confirm these theoretical bounds and trends, and compare to related methods.

Traditional blue noise does well at avoiding low frequency artifacts.
To avoid artifacts in high frequency areas, 
the community has developed an interest in \emph{step blue noise} \cite{Heck:2013:BNS}, where the frequency spectrum resembles a step function without oscillations.  
%Here we also consider ``soft blue noise,'' a variation where the radial power and dual distances transition from zero to a constant value along a soft ramp, as opposed to an abrupt step or a step with oscillations. 
Until now, the only way to create these distributions was an expensive post-processing optimization of an initial distribution.
Variations of our algorithm can create \emph{soft blue noise}, potentially avoiding high and low frequency artifacts.

Beyond blue noise, 
the adaptability and efficiency of our methods facilitate diverse applications, as shown in \Cref{fig:teaser}.
For {\em approximate Delaunay graph construction} and {\em global optimization}, the advancing-front and radial exploration provide advantages when sampling from the irregular shape of the local domains, even when the global domain is a hyper-rectangle.
Moreover, global optimization benefits from the intermediate process of our sampling method, not just the final output sample sets. % in contrast to many prior graphics applications.
\nothing{\liyi{(February 25, 2017) This does not seem helpful.}%liyi
(We also demonstrate the feasibility of high-dimensional rendering using blue noise, but \hidmethod\ does not show an advantage there because the domain is regular, blue and soft-blue noise does not improve visual quality, and the spectrum's DC component is paramount~\cite{oztireli2016integration,pilleboue2015variance}.)
}%nothing
For {\em robotic motion planning}, the ability to do advancing front over irregular domains may prove useful for adaptively exploring narrow regions of the configuration space.

%In particular, we are able to effectively explore the configuration space of robots with 23 degrees of freedom for motion planning. We are able to approximate high-dimensional Delaunay graphs, the list of the most significant neighboring samples in all directions.  Using these graphs we are able to adaptively enrich a blue noise sample set to explore the space of an unknown function, supplanting the use of crude axis-aligned rectangles, in a common global optimization algorithm.

%\input{contribution}

%\noindent \textbf{Contribution Summary} 
%\subsubsection{Contribution} 
The contributions of this paper include:
\begin{itemize}
	\item The idea of \emph{\hidmethod,} which combines the advantages of the locality of advancing-front 
	%(\bridsonmethod) 
	with the dimension-mitigation of hyperplane sampling, specifically line-sampling;
	% (\kddarts);
	\item Direct algorithms for blue noise in high dimensions;
	\item Proven and demonstrated time, memory, and saturation bounds that scale well;
	%\item
	%Output that is locally saturated and well spaced: the distance between any two samples is at least $r$, and with high probability the distance between any domain point and its nearest sample is at most $2r$ for \linespokesalg; 
	\item Applications using spokes for high dimensional Delaunay graphs, global optimization, and motion planning;
	\item Open source software~\cite{spokedartspubliccode}.	
\end{itemize}

To our knowledge, we provide the first method for probabilistically-guaranteed locally-saturated blue noise in high dimensions,
and the first direct method for {\em soft blue noise} in $\dimnumber>2$.
By ``direct'' we mean that samples are placed once, when they are generated, and never moved.
We demonstrate blue noise in dimensions 2--30, and applications in dimensions up to 100.
% TwoRadii MPS from "Variable Radius..." paper did 2d

%%%%%%%%%%%%%%%%%%%%%%%%%%%%%%%%%%%%%%%%%%%%

\section{Related Work}
\label{sec:related}
\label{sec:previous_work}

\begin{figure}[!htb]
	\centering
	 \subfloat[Point set]
	{
		\includegraphics[width=0.1\textwidth]{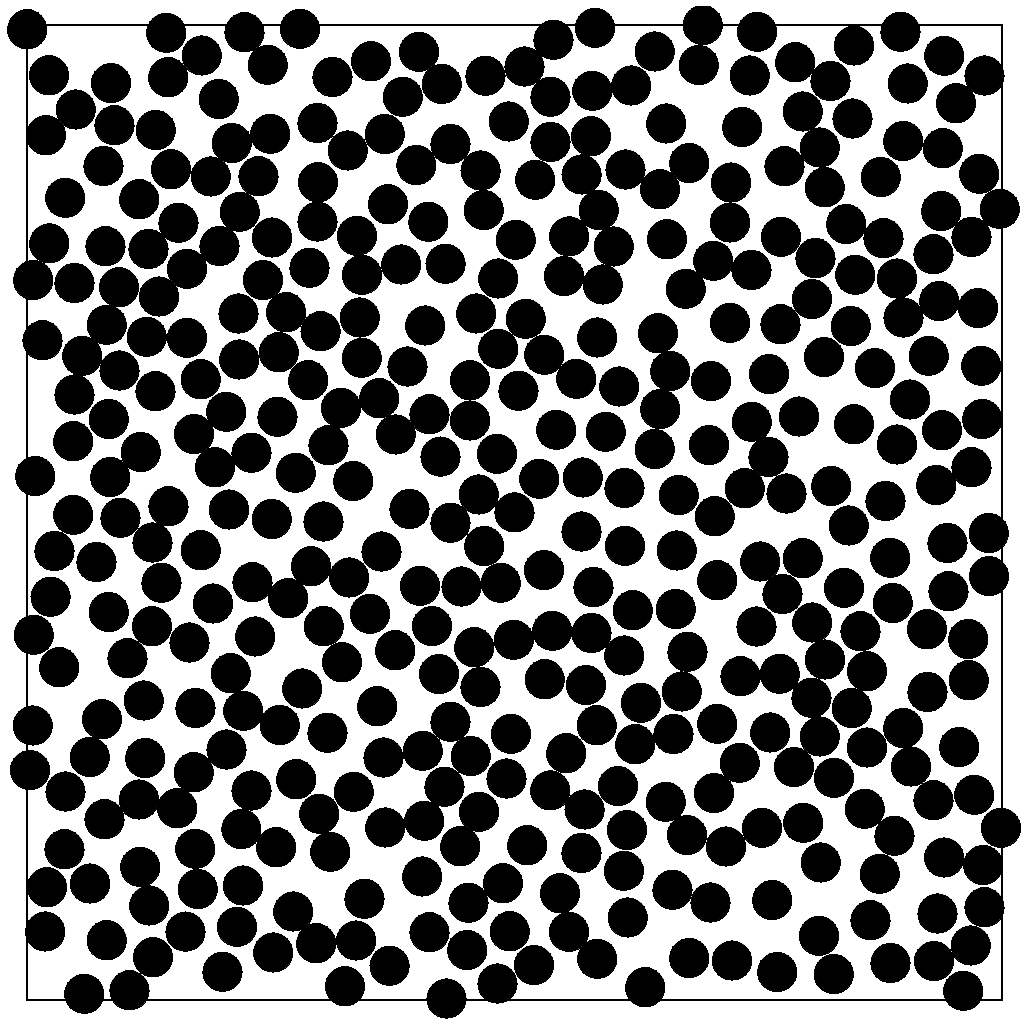}
	}
	\subfloat[Periodogram]
	{
		\includegraphics[width=0.1\textwidth]{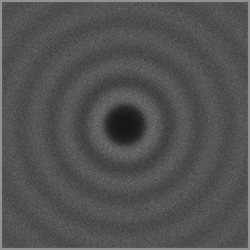}
	}
	\subfloat[RDF]
	{
		\input{figs/distributions/2_P_198_0spoke_411222_beta_points_rdf_nolabel}
	}
	\subfloat[RP]
	{
		%\documentclass{standalone}
%\usepackage{tikz}
%\begin{document}
\FPeval{\mpsk}{0.414595 /  0.002515625000000}%
\FPeval{\mpsrraw}{0.005031250000000}%
%\FPeval{\mpsr}{ \mpsrraw * \mpsk * 10. / 8.}%
\FPeval{\mpsr}{0.766211 * 10. / 8.}%
\FPeval{\mpsfinalr}{ 8.0 * \mpsr }%
\FPeval{\mpsoriginalmaxx}{10.}%
\FPeval{\mpsxscale}{ \mpsoriginalmaxx / \mpsfinalr }%
% 0.3 for 1/3 column width
% 0.8 for original constant
\FPeval{\mpsxscales}{ 0.35 * 0.640048 * \mpsxscale }%
\FPeval{\mpsyscales}{0.36}% 
\FPeval{\mpsframewidth}{0.1}%
%\FPeval{\mpsframewidth}{0.1 * \mpsxscale}
\FPeval{\framemax}{2.1 + \mpsframewidth}%
\FPeval{\framemin}{0.0 - \mpsframewidth}%
% original xscale of 
\begin{tikzpicture}[xscale=0.75*\mpsxscales, yscale=1.5*\mpsyscales]
  \draw[help lines,dashed] (0.000000,1.000000) -- (\mpsfinalr,1.000000);
  \begin{scope}
    \clip (0.000000,\framemin) rectangle (\mpsfinalr,\framemax);
    \draw
    (	0	,	4.2	) --
(    0.024388	,	0.071837) --
(	0.073165	,	0.066131625	) --
    (	0.146331	,	0.068626222	) --
(	0.219496	,	0.069006889	) --
(	0.292661	,	0.067993111	) --
(	0.365826	,	0.068048444	) --
(	0.438992	,	0.068576444	) --
(	0.512157	,	0.070710333	) --
(	0.585322	,	0.075069444	) --
(	0.658488	,	0.078665333	) --
(	0.731653	,	0.085415444	) --
(	0.804818	,	0.089895889	) --
(	0.877984	,	0.096056222	) --
(	0.951149	,	0.104387889	) --
(	1.024314	,	0.115392667	) --
(	1.097479	,	0.131197	) --
(	1.170645	,	0.149199111	) --
(	1.24381	,	0.175078333	) --
(	1.316975	,	0.215099	) --
(	1.390141	,	0.273506667	) --
(	1.463306	,	0.355910333	) --
(	1.536471	,	0.460469667	) --
(	1.609636	,	0.599413556	) --
(	1.682802	,	0.768080889	) --
(	1.755967	,	0.938843889	) --
(	1.829132	,	1.107747333	) --
(	1.902298	,	1.240158333	) --
(	1.975463	,	1.354227667	) --
(	2.048628	,	1.439029667	) --
(	2.121794	,	1.509006778	) --
(	2.194959	,	1.568655667	) --
(	2.268124	,	1.571232889	) --
(	2.341289	,	1.556915333	) --
(	2.414455	,	1.523826444	) --
(	2.48762	,	1.474027778	) --
(	2.560785	,	1.418365444	) --
(	2.63395	,	1.337733667	) --
(	2.707116	,	1.290422	) --
(	2.780281	,	1.214315222	) --
(	2.853446	,	1.136642556	) --
(	2.926612	,	1.035382556	) --
(	2.999777	,	0.942057222	) --
(	3.072942	,	0.865869333	) --
(	3.146107	,	0.802039778	) --
(	3.219273	,	0.760294	) --
(	3.292438	,	0.716620667	) --
(	3.365603	,	0.691641778	) --
(	3.438769	,	0.671490222	) --
(	3.511934	,	0.663522778	) --
(	3.585099	,	0.663263444	) --
(	3.658265	,	0.673095778	) --
(	3.73143	,	0.702298667	) --
(	3.804595	,	0.741702556	) --
(	3.87776	,	0.781533111	) --
(	3.950926	,	0.818233	) --
(	4.024091	,	0.865703333	) --
(	4.097256	,	0.921547889	) --
(	4.170422	,	0.989679778	) --
(	4.243587	,	1.038714667	) --
(	4.316752	,	1.088028556	) --
(	4.389917	,	1.131359444	) --
(	4.463083	,	1.180075333	) --
(	4.536248	,	1.220177667	) --
(	4.609414	,	1.255649	) --
(	4.682579	,	1.270168889	) --
(	4.755744	,	1.272912556	) --
(	4.828909	,	1.256502111	) --
(	4.902074	,	1.240939333	) --
(	4.97524	,	1.226869889	) --
(	5.048405	,	1.199859889	) --
(	5.121571	,	1.165881889	) --
(	5.194736	,	1.122189667	) --
(	5.267901	,	1.078908778	) --
(	5.341066	,	1.032557889	) --
(	5.414232	,	0.983630222	) --
(	5.487397	,	0.934292222	) --
(	5.560562	,	0.895576778	) --
(	5.633728	,	0.858732	) --
(	5.706893	,	0.828658667	) --
(	5.780058	,	0.804518222	) --
(	5.853223	,	0.792374	) --
(	5.926389	,	0.791715556	) --
(	5.999554	,	0.795139333	) --
(	6.072719	,	0.802531222	) --
(	6.145885	,	0.818488667	) --
(	6.21905	,	0.840267667	) --
(	6.292215	,	0.869243444	) --
(	6.36538	,	0.893167778	) --
(	6.438546	,	0.925602222	) --
(	6.511711	,	0.955135556	) --
(	6.584876	,	0.996767667	) --
(	6.658041	,	1.032469889	) --
(	6.731207	,	1.068633667	) --
(	6.804372	,	1.097873667	) --
(	6.877537	,	1.117977778	) --
(	6.950703	,	1.134994556	) --
(	7.023868	,	1.148662222	) --
(	7.097034	,	1.160013444	) --
(	7.170198	,	1.163316667	) --
(	7.243364	,	1.162868889	) --
(	7.316529	,	1.157285333	) --
(	7.389695	,	1.145740444	) --
(	7.46286	,	1.130616667	) --
(	7.536025	,	1.109389444	) --
(	7.60919	,	1.091632111	) --
(	7.682356	,	1.060104444	) --
(	7.755521	,	1.030825667	) --
(	7.828686	,	0.998424778	) --
(	7.901852	,	0.968535778	) --
(	7.975017	,	0.938336111	) --
(	8.048182	,	0.909424222	) --
(	8.121347	,	0.882584111	) --
(	8.194512	,	0.867169667	) --
(	8.267678	,	0.855885667	) --
(	8.340843	,	0.854245111	) --
(	8.414008	,	0.851781333	) --
(	8.487174	,	0.858030111	) --
(	8.560339	,	0.864012444	) --
(	8.633505	,	0.876813556	) --
(	8.70667	,	0.890106	) --
(	8.779835	,	0.911243222	) --
(	8.853001	,	0.935872	) --
(	8.926166	,	0.965392222	) --
(	8.999331	,	0.997746333	) --
(	9.072496	,	1.025491444	) --
(	9.145661	,	1.053875333	) --
(	9.218827	,	1.074037556	) --
(	9.291992	,	1.089020333	) --
(	9.365157	,	1.103338889	) --
(	9.438323	,	1.113729778	) --
(	9.511488	,	1.124305444	) --
(	9.584653	,	1.130528333	) --
(	9.657819	,	1.128448778	) --
(	9.730984	,	1.123267222	) --
(	9.804149	,	1.105092556	) --
(	9.877315	,	0.974675444	) ;
  \end{scope}
  \foreach \x in {0,2,4,6,8}
        \FPeval{\xx}{ \mpsr * \x }%
      \draw (\xx cm,-0.024000) -- (\xx cm,-0.3) node[below] {\x};
\FPeval{\xx}{ 0.3 * \mpsr}%
  \foreach \y in {0,1,2}
    \draw (\xx,\y cm) -- (0.000000,\y cm) node[left] {\y};
  \draw (0.000000,0.0) rectangle (\mpsfinalr,\framemax);
  \FPeval{\midx}{\mpsfinalr * 0.5}%
  %\node[below=0.4cm,text height=10pt,text depth=3pt] at (\midx,-0.100000) {RP};
\end{tikzpicture}
%\end{document}
	}
	\caption{Randomness metrics from PSA~\protect\cite{Schlomer:2011:ASA}.}
%	(b) The point set's periodogram: its Fourier spectrum. (c) Radial Distance Function (RDF): the histogram of pairwise distances between points. (d) Radial Power (RP): the radial average of the periodogram, the dual of the RDF.  % this is both well-known, and described already in the intro when.
 \label{fig:rdf_spectra_lineSpokes}
\end{figure}

%%%%%%%%%%%%%%%%%%%%%%%%%%%%%%%%%%%%%%%%%%%%

Blue noise has many graphics applications \cite{Chen:2013:BBN} in rendering \cite{Cook:1986:SSCG,Sun:2013:LSS}, texturing \cite{Lagae:2005:PODF}, stippling \cite{Balzer:2009:CCPD,Fattal:2011:BNPS,deGoes:2012:BNT}, geometry processing \cite{Alliez:2003:APR,Oztireli:2010:SSM}, animation \cite{Schechter:2012:GSA}, visualization \cite{Li:2010:ABNS}, and numerical computation \cite{kddarts_arxiv}.
Blue noise sampling can be achieved by various methods, such as dart throwing (also known as Poisson-disk sampling) \cite{Cook:1986:SSCG} and relaxation \cite{Lloyd:1983:RLA}.

\nothing{
\liyi{(July 25, 2017)
TODO: check this upcoming HPG 2017 paper \cite{Wang:2017:FMP} to see if it is related in any way.
}%liyi
}%nothing

Two main spatial properties are used to characterize blue noise distributions: (1) randomness and (2) well-spaced-ness.
Randomness avoids aliasing while well-spaced-ness reduces noise and improves efficiency.

\textit{Randomness} is typically characterized by the frequency spectrum of the sample distribution~\cite{ulichney1988dithering}, a feature of the output of some process rather than the randomness of the process itself.
The spectral properties can be measured by the radial power (RP) \cite{Lagae:2007:CMPD} in the frequency domain, or equivalently in the primary domain such as differential vectors (differential domain analysis) \cite{Wei:2011:DDA} and radial distance function (RDF) \cite{Oztireli:2012:ASP}.
See \Cref{fig:rdf_spectra_lineSpokes} for an example of these measures.
Many traditional algorithms for blue noise produce a step-like RDF, but a RP spectrum with oscillations, and these can produce visible artifacts in high frequency areas~\cite{Heck:2013:BNS}.
Step blue noise has a RP resembling a step function, without oscillations.
Stair blue noise \cite{Kailkhura:2016:SBN} provides additional degrees of freedom over step blue noise for tuning spectral characteristics.
By \emacsquote{soft blue noise} we loosely mean that both the RP \emph{and} RDF are steep but smooth ramps without oscillations.
This can be preferred because of lower aliasing, as demonstrated by recent results of using this type of noise for the classical zone-plate \nothing{test }pattern~\cite{Kopf:2006:RWT,Heck:2013:BNS,Subr:2013:FAS}. 
Our prior two-radii sampling directly produced soft blue noise in 2D, without post-processing~\cite{Mitchell:2012:VRPD}.

\textit{Well-spaced} samples, on the other hand, mean that samples are not too close to one another, yet no domain point is too far from a sample.
%We measure this by saturation.
One way to measure well-spaced-ness is discrepancy \cite{Shirley:1991:DQM,Keller:2012:AMC}.
\liyi{(October 22, 2017)
What is the relationship between discrepancy and saturation?
Can be an interesting to derive their math relations/bounds and compare.
}%liyi
Another measure is {\em saturation}, which depends on two radii:
coverage radius $\rcoverage$ for maximum domain to sample distance, and conflict radius $\rconflict$ for minimum inter-sample distance~\cite{Mitchell:2012:VRPD,Ebeida:2013:ISC}.
Saturation is then quantified using their ratio $\beta = {\rcoverage}/{\rconflict}$; the lower the $\beta$, the higher and better the saturation.
%Although there are practical random algorithms that achieve $\beta=1$ in low dimensions, none do so in high-dimensions. 
Saturation is desired in many contexts, as described in the extensive literature on maximal Poisson-disk sampling (MPS)~\cite{Jones:2005:EGPDS,Cline:2009:DRS,Gamito:2009:AMPD,Ebeida:2011:EMP,Ebeida:2012:SAM,Yan:2013:GPA} and low discrepancy sampling \cite{Ahmed:2016:LBN}.
For some applications, it is unclear how important saturation is as the dimension increases. 

% dreaded error appears here. First pass, this paragraph is mostly on the prior page. Second pass, parts of it are on the next page
% \pdfendlink ended up in different nesting level than \pdfstartlink
%
Despite the potential applications for high dimensional sampling, most sample-generation algorithms are low dimensional, in part because of the curse of dimensionality --- many blue noise algorithms do not scale well to high dimensions (e.g. tiling \cite{Kopf:2006:RWT,Ahmed:2016:LBN,Wang:2017:FMP,Ahmed:2017:APS}), especially when seeking high saturation.
The sampling methods that scale well with dimension do not provide a guarantee of local saturation, while those providing local saturation have exponential complexity.
The algorithms closest to obtaining both of these goals are based on advancing-front~\cite{firstspheremesh,Bridson:2007:FPD} or \kddarts~\cite{kddarts_arxiv}, as detailed below.

\subsection{Advancing front\nothing{ methods\liyi{(February 25, 2017) Be consistent; if we did not say ``methods'' for other parts, do not say it here.}}}

Advancing front methods were initially proposed for meshing \cite{firstspheremesh,Li99biting:advancing,fastspherepacking} and later adopted for sampling in graphics \cite{Jones:2005:EGPDS,Dunbar:2006:PDS,Bridson:2007:FPD}.
The basic idea is to draw new samples from regions around existing samples (the front) and expand towards the rest of the domain. 
Most methods build some form of the front boundaries explicitly, and some construct the union of spheres~\cite{Li99biting:advancing,fastspherepacking}.
These methods are intractable in high dimensions because the number of intermediate-dimensional faces grows factorially with dimension. 
In practice, implementing the geometric primitives for the constructions would be challenging as well. 

In~\bridsonmethod~\cite{Bridson:2007:FPD}, a key innovation is to represent the front boundary implicitly, by a list of sample disks touching the front.
\bridsonmethod\ does rejection sampling around a prior sample, selecting a point uniformly by volume from the $[\rconflict, \rcoverage]$ annulus around it.
The sample is removed from the front after a fixed number (30) of consecutive rejections.
Its advantage is locality, mitigating the effects of domain size. 
This enables tractable runtime in high dimensions.%
\footnote{As published, step 0 constructs a background grid. Replacing it with a $k$-d tree improves runtime from $2^{O(d)}O(n)$ to $O(d n^2),$ the same complexity as our \hidmethod.}
The single page sketch in Bridson~\shortcite{Bridson:2007:FPD} does not analyze saturation by dimension. We postulate that the method guarantees that a large fraction of the annulus volume is saturated, but does not bound the uncovered volume outside all annuli. More significantly, we have discovered that its output has an undesirable artifact, a sharp discontinuity in the density of points at the outer boundary of annuli, as further demonstrated in \Cref{sec:analysis}.

\subsection{Hyperplane sampling}

\Darts~\cite{kddarts_arxiv} uses hyperplanes for Poisson-disk sampling: select a random axis-aligned hyperplane, find its uncovered subset, and select a point from this subset. A rejection occurs only when the entire hyperplane is covered. 
Its advantage is that hyperplanes mitigate the effects of high dimensions.
%A hyperplane is especially efficient for dealing with arrangements of spheres, because its intersection with a sphere is a lower-dimensional analytic sphere.
Its disadvantage is that it does not guarantee \emph{local} saturation, because hyperplanes are selected \emph{globally} from the entire domain.
(Global dart throwing \cite{Dippe:1985:ATS,Cook:1986:SSCG} has similar issues.)

In principle, using hyperplanes of any dimension is possible, up to the dimension of the domain. However, the difficulty is actually performing and representing the necessary geometric primitives over this object. In \Darts, only 1D lines and 2D planes were demonstrated. In the present work, we merely use lines, 1D hyperplanes.
The method in Sun et al.~ \shortcite{Sun:2013:LSS} samples lines and line-segments for rendering applications, including 3D motion blur, 4D lens blur, and 5D temporal light fields.
For determining sample positions it relies on subroutines that do not scale well to high dimensions.

\subsection{Combining advancing front with line search}
\label{sec:building_blocks}

\begin{figure}[tbh]
  \centering
  {
    \includegraphics[height=1.5in]{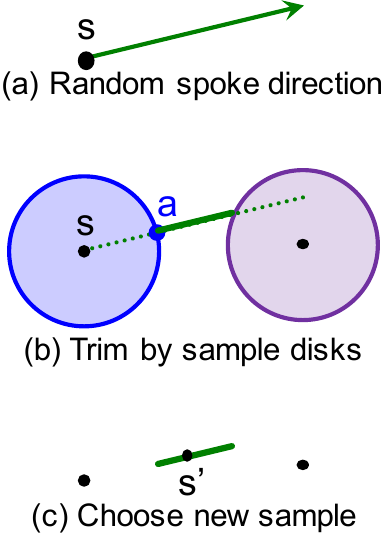}
    \quad \quad \quad
    \includegraphics[height=1.5in]{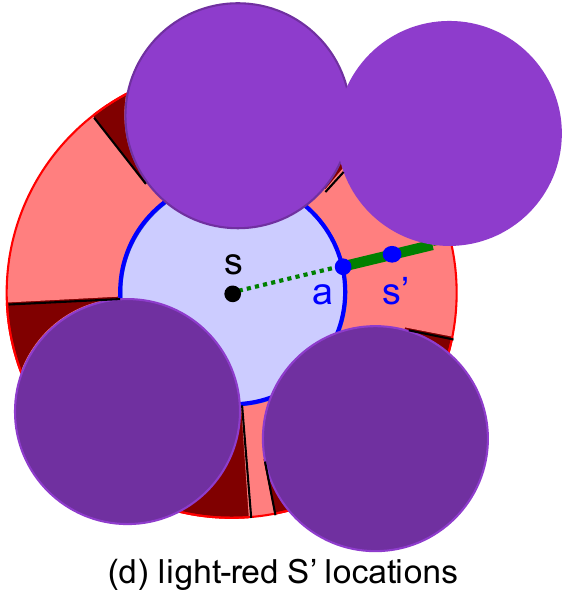}
   }   
  \Caption{\Linespokesalg{} in 2D.}
   {(a) A (green) spoke is a randomly-oriented line segment through a prior sample $\sample$.
(b) It is trimmed by sample disks, keeping the (solid) subsegment containing anchor point $\anchor$.
(c) The next sample $\sampleprime$ is chosen from the trimmed segment. 
(d) Because of the anchor point, the next sample will be in the subset of the annulus that is light-red, not the dark-red regions on the far side of other disks. 
}
   \label{fig:line_spokes}
   \label{fig:spokes} 
\end{figure}

Our key idea is to combine the advantages of advancing front and hyperplane sampling.
%the local saturation of \bridsonmethod\ and the dimension-mitigation of \kddarts.
Specifically, spoke-darts replaces the point-sampling of Bridson~\shortcite{Bridson:2007:FPD} with line-sampling. 
A \emph{spoke} is a line segment passing through a point, at a random orientation; see \Cref{fig:line_spokes}.
We employ a constant number (12) of consecutive rejections before advancing the front, retaining good run-time scalability across dimensions. However, 12 consecutive rejections provides a local saturation\nothing{ distance} guarantee that is \emph{the same in all dimensions}. 
%They also enable \emph{step} blue noise. Selecting a sample point uniformly from a segment results in traditional blue noise.
%We achieve step blue noise by selecting the sample point non-uniformly from a spoke.
%We achieve  \emph{soft step blue noise} by generating a second spoke through the first-spoke point, and selecting the final sample from it.  
We can generate different blue noise profiles. 
In particular, we can avoid the spike in the distribution at the sampling radius by non-uniform sampling from a spoke segment and by generating a second spoke through a point on the first spoke. These two spokes mimic the two radii in Mitchell et al.~\shortcite{Mitchell:2012:VRPD}, and produce a similar \emph{soft blue noise} profile.

%%%%%%%%%%%%%%%%%%%%%%%%%%%%%%%%%%%%%%%%%%%%

\section{Spoke-Dart Blue Noise Sampling}
\label{sec:algorithm}

%%%%%%%%%%%%%%%%%%%%%%%%%%%%%%%%%%%%%%%%%%%%

\Hidmethod{} generates new samples from the current sample set boundary and gradually expands towards the rest of the domain. The key features distinguishing our algorithm from prior methods are (1) how the front is described and advanced, and (2) how new samples are drawn. The front is described by the boundary of the union of disks around samples, but  its geometry is not explicitly constructed. New samples are selected by generating a random \spoketext\ (radial line) through an existing sample, trimming it by existing sample disks, and selecting an uncovered point from the remaining sub-spoke. 

\subsubsection{Algorithm summary}
Our top level algorithm follows.
We initialize the output set with one sample and put it into the active pool of front points.
When this pool becomes empty our algorithm terminates.
We remove a sample $\sample$ from the pool and try to generate new samples $\sampleprime$ from random \spokestext{} $\spokesym$ through $\sample$. 
Accepted samples are added to the pool.
We keep throwing \spokestext{} from the same sample until $\hammerlimit=12$ consecutive \spokestext{} failed to generate an acceptable sample.
Radius $\rnumber$ is the minimum allowed distance between any two samples.
Our method is summarized in \Cref{alg:spoke}.

\begin{algorithm}
  \Caption{\Spokedarts{} for blue noise sampling.}{}
  \label{alg:spoke}
  \begin{algorithmic} [1]
    \REQUIRE sample domain $\domainsym$\nothing{ with distance $\dist$ and conflict $\rnumber$ measure}
    \ENSURE output sample set $\sampleset$
    \STATE $\sample \leftarrow \funct{RandomSample}(\domainsym)$
    \STATE $\sampleset \leftarrow \left\{ \sample \right\}$ \pcomment{all samples}
    \STATE $\activepool \leftarrow \left\{ \sample \right\}$ \pcomment{active pool, FIFO queue}
    \WHILE{$\activepool$ not empty}
      \STATE $\sample \leftarrow \funct{PopFront}(\activepool)$
      \STATE $\neighbors \leftarrow \funct{CollectNeighbors}(\sample,\sampleset)$
      \STATE $reject \leftarrow 0$
      \WHILE{$reject < \hammerlimit$ (=12)}
        \STATE $\spoke, \anchor \leftarrow \funct{RandomSpoke}(\sample, \interval)$ 
		\STATE $\spoke \leftarrow \funct{TrimSpoke}(\spoke, \anchor, \neighbors)$ %\pcomment{\Cref{alg:line_spoke} or \ref{alg:plane_spoke}}
        \IF{$\spoke$ is empty}
            \STATE $reject \leftarrow reject + 1$
        \ELSE
					 \STATE $\sampleprime \leftarrow \funct{RandomSample}(\spoke)$		
           \IF{TwoSpokes}
               \STATE $\sampleprime \leftarrow $ \funct{SecondSpoke}($\sampleprime,\neighbors$)
             \ENDIF
            \STATE add $\sampleprime$ to $\neighbors$, $\sampleset$, and the end of $\activepool$
%            \STATE $\sampleset \leftarrow \sampleset \bigcup \left\{ \sampleprime \right\}$
%            \STATE $\activepool \leftarrow \activepool \bigcup \left\{ \sampleprime \right\}$
%            \STATE $\neighbors \leftarrow \neighbors  \bigcup \left\{ \sampleprime \right\}$
            \STATE $reject \leftarrow 0$
        \ENDIF
      \ENDWHILE
    \ENDWHILE
    \RETURN{} $\sampleset$
  \end{algorithmic}
\end{algorithm}

%
%Next, we present more details on different parts of the algorithm.

% --------------------------
\subsection{Spokes}
\label{sec:spoke_darts}
The line of a spoke passes through a sample $\sample$, and the spoke is the interval $\interval$ of distances from $\sample$.
A spoke has a distinguished \emph{anchor} point $\anchor \in \interval$ used to select which segment to retain during trimming; see \Cref{fig:spokes}.
For \linespokesalg, $\interval=[\rnumber,2\rnumber]$ and the anchor lies at $\rnumber$, because, as in \bridsonmethod, the uncovered region starts at $\rnumber$ and the extent of the local front we wish to consider is $2\rnumber$.
%e.g.\ $2 \rnumber$ as in \bridsonmethod.
% Other algorithms use different $\interval$.

%At one extreme, anchored spokes, we test whether a particular ``anchor'' point of this line is uncovered, then find the 
%At one extreme, we also consider an arc of a great circle around the $d$-dimensional Poisson-disk 
%\input{spokes_new_fig} %\liyi{(February 7, 2016) I moved \Cref{fig:spokes} to \Cref{sec:building_blocks} for better layout and easier reference.}

% --------------------------
\subsection{CollectNeighbors}
\label{sec:neighbor_collection}
For a sample on the front, for each spoke we trim it by iterating over the nearby samples.
For spoke-darts with spoke extent $2\rnumber$, a sample is a neighbor if its center distance is less than $3\rnumber$, because that is the farthest away a sample can lie and still have its disk overlap the spoke.
A key efficiency is to gather all neighbors once \emph{before} any trimming operations.

In our implementation, a $k$-d tree saves time over exhaustive search for small $\dimnumber$ and large $\samplenumber$.
\Cref{fig:crossover_body,fig:crossover} show a speedup for $\dimnumber < 7$ and $\samplenumber \ge 200,0000$.
We maintain a $k$-d tree of the entire point set.
We collect the subtree of neighbors, and update the tree and subtree as we successfully add new samples.

% --------------------------
\subsection{RandomSpoke}
\label{sec:randomspoke}
A \linespoketext{} is generated by selecting a line with random orientation, by choosing a point $p$ from the surface of the disk around $\sample$, uniformly by area.
% of the corresponding sample $\sample$, and extended with length $\spokeradius$ away from $\sample$.
%A random \planespoketext{} is generated by selecting the plane containing two such points.

To pick $p$ we use the classical method of Muller~\shortcite{Muller}, as follows.
Generate each of the vector's $\dimnumber$ coordinates independently from a normal (Gaussian) distribution. Then linearly scale the vector of coordinates to the disk radius.
The reason this works is because the level sets of a $d$-dimensional Gaussian distribution are $d$-spheres.

% --------------------------
\subsection{TrimSpoke}
\label{sec:trim}
Trimming subtracts out the portion of a segment that is covered by a neighbor disk, leaving just its uncovered subset.
For efficiency, we just keep the one subsegment that contains a distinguished \emph{anchor point}; this is considerably faster than finding all uncovered segments.
Further, we do a prepass and discard the entire spoke if the anchor is covered.
These primitives are efficient, linear in dimension, and the prepass avoids square roots.
%See \Cref{alg:all_segments} for an efficient $| \neighbors | \log | \neighbors |$ implementation, which gathers all disk-line intersections then computes the depth of each interval between intersections.
%This subtraction primitive is efficient in all dimensions.
In \Cref{fig:line_spokes}, these shortcuts mean that the next sample will be chosen from the light red part of the annulus only, and not the dark red portions.
This potentially affects the output distribution characteristics, but the saturation proof takes it into account.

%
%For overall efficiency, we quickly discard some spokes and spoke segments.
%It is acceptable to discard a spoke even if part of it is not covered.
%This retains the correctness of the algorithm, and affects the run-time and final distribution characteristics.
%The fastest primitive is to determine if a single \emph{anchor point} $\anchor$ is covered by a disk, since that just involves checking the squared distance between the point and each neighbor.
%%See Algorithms~\ref{alg:anchored_point}, \ref{alg:anchored_segment} and \ref{alg:trim_blue_noise}.
%
%Retaining just the uncovered segment containing anchor $\anchor$ is considerably faster than finding all uncovered segments.
%In \Cref{fig:line_spokes}, this means that the next sample will be chosen from the light red subset of the annulus only, and not the dark red portions. 
%(This still gives blue noise, but with a  higher peak distribution at distance $r$ than traditional MPS: 4 versus 3.)
%For efficiency we avoid the overhead of keeping track of all of the resultant segments, and only keep the one containing the anchor point $a,$ if any.
%This also avoids wasting time with further searches in the case that the spoke is completely covered, which is common as the algorithm nears termination.
%Many combinations of these heuristics are possible; 
%the ones we found most useful are summarized in 
%\Cref{alg:trim_blue_noise} and \Cref{alg:trim_wheel}.

%\input{line_spoke_alg}
%\input{all_segments_alg}
%\input{trims_alg}
%\input{anchor_covered}
%\input{plane_spoke_alg}

% --------------------------
\subsection{RandomSample}
\label{sec:randomsample}
For blue noise, it is sufficient to pick a sample \emph{uniformly by length} from an uncovered spoke segment.

One might assume that picking a point uniformly by the swept volume, dependent on the dimension, would generate better quality blue noise. 
However, we found this detrimental for our algorithms, and also for the prior work of \bridsonmethod~\cite{Bridson:2007:FPD}. It generates worse blue noise than traditional MPS algorithms; see \Cref{sec:bridsonexperiments}.
Exploring more sophisticated selection criteria led us to our \twospokesalg\ algorithm.
%; see \Cref{sec:twospokesalg} for a summary and 

\subsection{\Twospokesalg}
\label{sec:twospokesalg}

\begin{figure}[tb]
  \centering
  {
     \includegraphics[height=1.42in]{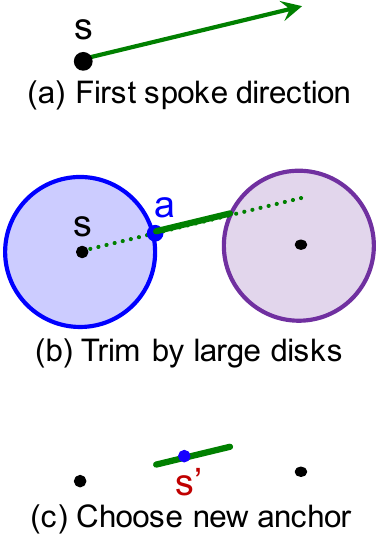}
     \quad
     \includegraphics[height=1.44in]{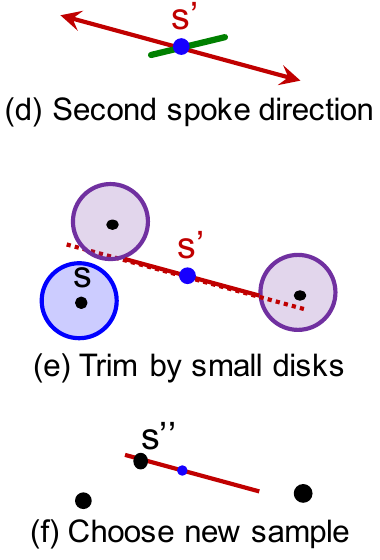}
     \quad
     \includegraphics[height=1.05in]{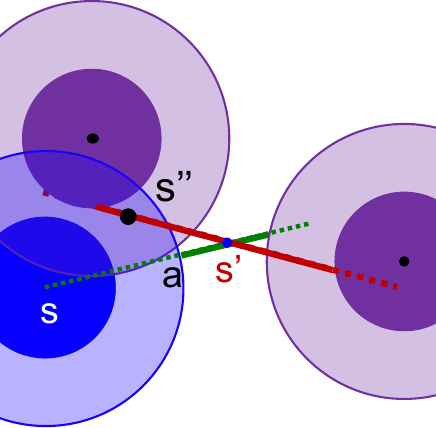}
  }
  \Caption{}
   {(a)--(c) \Twospokesalg\ starts much the same as \linespokesalg, only using a longer spoke and trimming by $2\rnumber$-disks.
(d)--(f) A second spoke is trimmed by $\rnumber$-disks, and the new sample $\sampleprimeprime$ is chosen from it.
}
    \label{fig:two_spokes}
% \label{fig:spokes} 
\end{figure}

\Twospokesalg\ is an algorithm variation that further randomizes the placement of samples; see \Cref{fig:two_spokes}.
Its output distribution avoids the traditional spike at the sampling radius, and mitigates other artifacts.
%It also mitigates the artifacts at twice the sampling radius of \bridsonmethod\ and \linespokesalg.
We make the first spoke longer, and shoot a second spoke from an uncovered point on the first spoke.
The first spoke has $\interval=[2\rnumber,4\rnumber]$ with anchor at $2\rnumber$ and is trimmed by radius-$2\rnumber$ sample disks.
The second spoke has $\interval=[-2\rnumber,2\rnumber]$ with anchor at $\sampleprime$ and is trimmed by radius-$\rnumber$ sample disks.
RandomSample is approximately uniform by swept volume from the nearer spoke end. 

\liyi{(October 25, 2017)
Here, it can help to provide some intuitions on why we designed the algorithm above and why it has the nice properties below.
}%liyi

\Twospokesalg\ shares the following properties with two-radii Poisson-disk sampling~\cite{Mitchell:2012:VRPD}.
Their spectra are similar, and the distance between samples is at least $\rnumber$. 
A new sample's large $2\rnumber$-disk covers $\sampleprime$, and no other large disk covers it so far, ensuring progress and algorithm termination.
%, with additional criteria to avoid any sharp transitions to zero probability of placement. % confusing
There is a simple parameterization of the two spoke lengths that starts at \linespokesalg, then trades away saturation to gain randomness; see \Cref{sec:alg_variants}.

%%%%%%%%%%%%%%%%%%%%%%%%%%%%%%%%%%

\section{Analysis and Guarantees}
\label{sec:analysis}

%%%%%%%%%%%%%%%%%%%%%%%%%%%%%%%%%%

%\subsection{Summary of saturation $\rratio$ by $\epsilon$, $\hammerlimit$ and  $\dimnumber$}
%\subsection{Probability ($1-\epsilon$) of achieving saturation $\rratio$ with $\hammerlimit$ throws in dimension $\dimnumber$}
\subsection{Probability of achieved saturation}
\label{sec:parameters}
Our measure of saturation is $\rratio = \rcoverage/\rconflict$, where $\rcoverage$ is the maximum distance from a domain point to its nearest sample, and $\rconflict$ is the minimum guaranteed distance between a sample and its nearest sample, the Poisson-disk radius.
Also, $\rratio^*$ is the desired upper limit on $\rratio.$
Besides spoke length, the main control parameter is $\hammerlimit$, the number of successively-failed spokes  before removing a sample from the front.
The higher the $\hammerlimit$, the more spokes we generate and the longer the run-time, but the more saturated the output.
Note $(1-\bigbetachance)$ quantifies the probability that $\rratio^*$ is achieved.
The structure of our guarantee is that, for a given $\hammerlimit,$ with high probability $(1-\bigbetachance)$ the achieved $\rratio$ 
at a sample is less than $\rratio^*$.  
\Cref{eq:m_in_n_body} quantifies the relationship between $\hammerlimit$, $\bigbetachance$, $\rratio^*$, and $\dimnumber$ 
for \linespokesalg.

\begin{equation}
\hammerlimit  = \left\lceil  {(-\ln{\bigbetachance})} (\rratio^*-1)^{1-\dimnumber} \right\rceil
\Leftrightarrow
\rratio^* = 1 + \left(  \frac{-\ln{\bigbetachance}}{\hammerlimit} \right)^{1/(\dimnumber-1)}
\label{eq:m_in_n_body}
\end{equation}

Our main result is that if $\hammerlimit=12$, then with probability $1 - 10^{-5}$ we will get local $\rratio < 2 = \rratio^*$ in \emph{any} dimension, avoiding the curse of dimensionality.
In general, one can pick any three of $\{\hammerlimit, \bigbetachance, \rratio^*, \dimnumber\}$ and the fourth is determined.
For example, one can pick $\hammerlimit$ and $\rratio^*$ and bound the probability $\bigbetachance$ that $\rratio^*$ was exceeded: 
$\bigbetachance < \exp{(-\hammerlimit (\rratio^* - 1)^{\dimnumber - 1})},$
where $\rratio^* > 1,$ and $-\ln{\bigbetachance} > 0,$ and $\hammerlimit \ge 1.$

\begin{itemize}
	\item \Linespokesalg\ produces $\rratio < 2$ with high probability.
	\item \Twospokesalg\ produces $\rratio < 4$ with high probability.\\(The price of a more uniform spectrum is lower saturation.)
\end{itemize}

We provide some intuition for \Cref{eq:m_in_n_body} here; the derivations are in\nothing{ the supplementary material,} \Cref{sec:boundproofs}.
Let us suppose that the algorithm has terminated and a sample has a far Voronoi vertex.
Consider the empty ball centered at this vertex and tangent to the sample's disk; it lies outside all other samples' disks.
We may expand this ball into a  ``void,'' a larger connected region bounded by sample disks.
We have thrown at least $\hammerlimit$ spokes from each of the void's bounding disks. 
Each of these spokes must have missed this void; otherwise we would have inserted a sample into the void, a contradiction.
%The chance of a \emph{single} spoke \emph{missing} this void is the area the void shares with the spoke's disk, divided by the surface area of the disk, because the spoke's anchor point lies on the disk surface.
%The chance of getting $\hammerlimit$ successive misses is the chance of one miss to the $\hammerlimit$th power.
%Combining this for all disks bounding the void shows that the natural log of the chance that they all missed is (at most) proportional to 
%the area of one disk divided by the total area of the void boundary.
Since the void was not hit, if we add up its surface area across all bounding disks, its surface area is probabilistically-guaranteed to be small compared to the surface area of a single bounding disk.
Thus the surface area of the Voronoi-vertex ball inside the void is also small, which bounds its radius.
The exponential-in-$(\dimnumber-1)$ dependence on $\rratio^*$ in \Cref{eq:m_in_n_body} is precisely the dependence of the surface area of a $\dimnumber$-ball on its radius.
Selecting $\rratio^* = 2$ says we only care about voids with at least the surface area of a single sample disk.
The exponential dependencies on surface areas cancel, and we are left with a Voronoi ball radius at most our sampling radius, meaning $\rratio\le2.$

% samitch: this next paragraph strays into practice, but is a reasonable summary to include here
In practice, we achieve a much better saturation than the guarantee, $\rratio \ll \rratio^*$ for all $\hammerlimit$.
This is expected because the proof is not tight: e.g., the void surface area might be much larger than that of an empty ball inside it, and we ignored chains of misses less than $\hammerlimit$.

While the bound on the probability of achieving $\rratio <2$ is dimension independent, the probability of achieving other $\rratio$ \emph{does} depend on the dimension. 
Re-arranging \Cref{eq:m_in_n_body}, we see 
$(\rratio^* - 1)^{\dimnumber-1} = ({-\ln{\bigbetachance}})/{\hammerlimit}$.
For example, the probability of achieving $\rratio < 1.5$ \emph{decreases} rapidly with dimension, and the probability of achieving $\rratio < 2.5$ \emph{increases} with dimension. Thus, as $\dimnumber$ increases, we should expect the distribution of local betas to narrow and converge towards $2$,
or perhaps to some lesser constant due to the slack in the proof.
This is what we observe in practice; 
see \Cref{fig:betamedian_body} in \Cref{sec:exp}, and also \Cref{fig:comparison_histograms} in \Cref{sec:additional_results}.

\subsection{Bound proofs}
\label{sec:boundproofs}

%===================
\begin{figure} [tb] 
	\centering
	
	\subfloat[shared area of a void and a disk]
	{
		\includegraphics[height=0.9in]{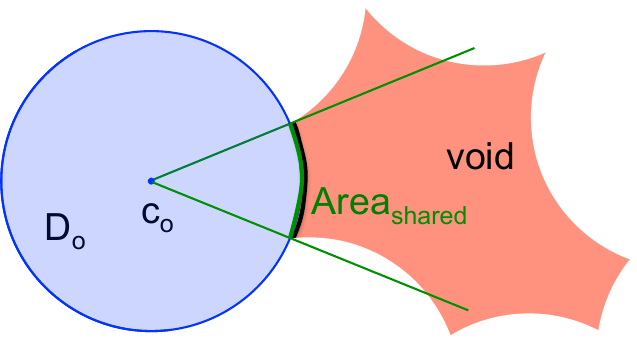}
		\label{fig:shared_area}
	}%\hspace{12pt}%
	\subfloat[a ball with less surface area than its enclosing void] %: $\vol(\void) = \vol(\dvoid)$, $\rvoid < \rdvoid$, and $\area(\void) > \area(\dvoid)$.
	{
		\includegraphics[height=0.9in]{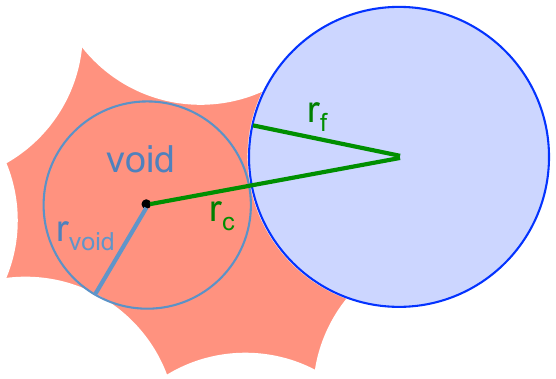}
		\label{fig:ball_void}
	}
	\Caption{Hitting a void from a neighboring disk.}
	
	\label{fig::void_by_neighbor}
\end{figure}
%====================

Here we prove the bounds on $\hammerlimit$, $\rratio$, and $\bigbetachance$ in terms of $\dimnumber$ 
from  \Cref{eq:m_in_n_body}. 
A \emph{void} is an uncovered region. It is bounded by some disks.
The chance of hitting a void will depend on its \emph{surface area} 
$\area(\void)$, the $\dimnumber$-1 dimensional volume of its boundary.
It will also depend on the surface area of any disk on its boundary, $\area(\disk)$.
% We first consider \linespokesalg.
\note{Here we only consider constant-radius balls.}

\subsubsection{Chance of missing the void from one disk}
Let us quantify the chance $\pmissi{1}$ that a \linespoketext{} from disk $\disk_1$ missed a void.
See \Cref{fig:shared_area}. Let $\vdarearatio_1 =  \area(\void \cap \disk_1) / \area( \disk_1 ).$
Since line-spokes are chosen uniformly from the surface area of the disk,
$\phiti{1} = \vdarearatio_1,$ and $\pmissi{1} = 1 - \vdarearatio_1.$
The chance of missing $\hammerlimit$ times consecutively is then
$\pmissmi{1} = \prod_{j=1}^{\hammerlimit} (1 - \vdarearatio_1) = (1 - \vdarearatio_1)^\hammerlimit.$
Using the well-known inequality $e^{-x} = \exp(-x) > 1 -x,$ we have 
$\pmissmi{1} < \exp(-\hammerlimit \vdarearatio_1).$

\subsubsection{Chance of missing the void from all disks}
\label{sec:all_misses}
The chance of missing $\hammerlimit$ times consecutively from all $\neighbors$ bounding disks is then
$\pmissmi{\textrm{all}} = \prod_{i=1}^\neighbors \pmissmi{i} < \exp\left( -\hammerlimit \sum_{i=1}^\neighbors \vdarearatio_i \right) = \exp( -\hammerlimit \vdarearatio),$
where all sample disks have the same radius so we can drop their subscripts and $\vdarearatio = \area(\void) / \area(\disk).$

If we wish this miss chance to be less than $\bigbetachance,$ then it is sufficient to have
$\exp( -\hammerlimit \vdarearatio) < \bigbetachance,$
meaning 
$\hammerlimit \vdarearatio > -\ln( \bigbetachance ) > 0.$

\subsubsection{Bound in terms of $\rratio$}
Now we bound $\vdarearatio$ in terms of $\rratio.$
Suppose there is a domain point $v$ in the void at distance $\rcoverage$ from all samples.
Then a ball at $v$ of radius $\rvoid = \rcoverage - \rconflict$ is strictly inside the void, and $\area(\void) > \area(\disk(\rvoid));$
see \Cref{fig:ball_void}.
Since we are in $\dimnumber$ dimensions and $\rratio = \rcoverage / \rconflict,$ 
\[\vdarearatio = \frac{\area(\void)}{\area(\disk)} > \frac{\rvoid^{\dimnumber-1}}{\rconflict^{\dimnumber-1}} = (\rratio - 1)^{\dimnumber-1}\]

\scott{There is a huge non-tightness to this proof and the subsequent bounds. It assumes that the void shape is a ball, and that all the void volume could be in one big connected void. That single-void thing is a global argument that should be a local one. I thought the second-proof would fix that, but the area subtended by a neighboring ball shrinks as the dimension increases, and the numbers I got out of that weren't any better. If only there were good spatial statistics results or I knew that stuff...}

Hence a sufficient condition is 
$\hammerlimit(\rratio-1)^{\dimnumber-1} > - \ln \bigbetachance,$ or

\[
\hammerlimit  = \left\lceil  {(-\ln{\bigbetachance})} (\rratio-1)^{1-\dimnumber} \right\rceil
\Leftrightarrow
\rratio = 1 + \left(  \frac{-\ln{\bigbetachance}}{\hammerlimit} \right)^{1/(\dimnumber-1)}
%\label{eq:m_in_n}
\]

%\subsubsection{Example $\hammerlimit$ Values}
%\label{sec:m_div_n_table}
%\Cref{tab:m_div_n} gives example $\hammerlimit$ values using \Cref{eq:m_in_n}.
%There is considerable slack in the inequalities and we achieve a much smaller $\rratio$ in practice.
%\input{m_div_n_body_tab}

\subsubsection{\Twospokesalg}
If the first spoke finds an uncovered point, the second spoke always places a sample, so we need only consider the chance of the first spoke missing the void.
%In Appendix~\ref{sec:softbluenoise} we explain that 
The first spoke extends from $2r$ to $4r$.
If we consider the subset of a void that is at least $2r$ from any disk, then propagating these values through the prior analysis shows $\rratio < 4$ within that subset. The uncovered regions between $r$ and $2r$ have local $\rratio<2$ and are subsumed, so the bound holds for the entire void.
Hence $\hammerlimit=12$ gives $\rratio < 4$ with probability $1 - 10^{-5}.$

\subsubsection{Subtleties}
%\pargrph{Boundary caveats}
The reader may have noticed that we made no mention of the domain boundary.
For bounded domains, we assumed that the void was bounded by disks only.
%This may be guaranteed by throwing \spokestext{} on or near the domain boundary first.
For periodic domains, several analysis steps are only guaranteed to hold when the Voronoi-vertex ball spans less than the domain period. 
These issues may be finessed, e.g.\, by initializing with a few well-spaced samples.
%Then we need only considering the subset of a void within one domain period.
% two issues: $\area(\void) > \area(\disk(\rvoid))$ is only guaranteed when the void diameter is small compared to the period of the domain. And if the void touches a disk and its periodic copy, only one copy gets to count the 12 successive misses

%\pargrph{Order independence}
%There is a statistical subtly in \Cref{sec:all_misses}.
There is another statistical subtly concerning the order of spokes.
The consecutive misses from one bounding disk are not guaranteed to be consecutive with the misses from another disk.
But this does not matter, because the misses for each disk is independent of whether the void was hit and reduced by some \spokestext{} from a later front disk. 
The important thing is that no \spoketext{} ever hit the boundary of the void that remains after the algorithm terminated.

% this a note to myself. leave it out of the paper
\scott{
	\pargrph{Slack in $1-\vdarearatio_1 < \exp(-\vdarearatio_1)$}
	While there is slack in this inequality, it becomes tight in the limit as a void is bounded by an infinite number of balls.
	It may be possible to use the kissing number to get a slightly better bound, but the difference will be very small. 
}

\section{Experimental Results}
\label{sec:result}

%%%%%%%%%%%%%%%%%%%%%%%%%%%%%%%%%%%%%%%%%%%%%%%%%%%%%%%%

\subsection{Distribution comparisons}
\label{sec:exp}
\label{sec:bridsonexperiments}

%%%%%%%%%%%%%%%%%%%%%%%%%%%%%%%%%%%%%%%%%%%%%%%%%%%%%%%%

% implementation(s)
We compare the distributions of our methods and \bridsonmethod\ experimentally.
We provide open source software on the github repository SpokeDartsPublic~\cite{spokedartspubliccode}, which may be used to verify the results. 
%In this section we focus on comparisons to \bridsonmethod~\cite{Bridson:2007:FPD}.
The original version of \bridsonmethod~\cite{Bridson:2007:FPD} does not support periodic domains,
so we re-implemented it as \bridsonmethodours\ for periodic domains in our framework.

\Cref{fig:rdf_spectra_comparison} shows the spectra, radial distance function (RDF), and radial power (RP) for \bridsonmethodours, line-spokes, and two-spokes, over 2--10 dimensional periodic domains.
We conducted experiments in dimensions up to $30$, but the figures for higher dimensions reveal no new structure or trends.
 Anisotropy is negligible because the algorithms do not depend on the choice of axes, e.g.\ all spoke directions are random, and sample neighborhoods are spheres. The only possible contribution to anisotropy is the fact that the domain's periodicity is axis aligned.

Many blue noise methods produce an RDF spike at $\rnumber$.
However, for \bridsonmethod, we were surprised to discover a discontinuity in the RDF at the outer annulus radius, $2\rnumber$, regardless of periodicity or  implementation.
For our methods, we notice a slight rise in RDF at $2\rnumber$ for \linespokesalg\ and some non-constantness even for \twospokesalg.
These artifacts tend to decrease with dimension.

\begin{figure}[tbh]
  \centering
%    \subfloat[Bridson]
    {
       \includegraphics[width=0.47\linewidth]{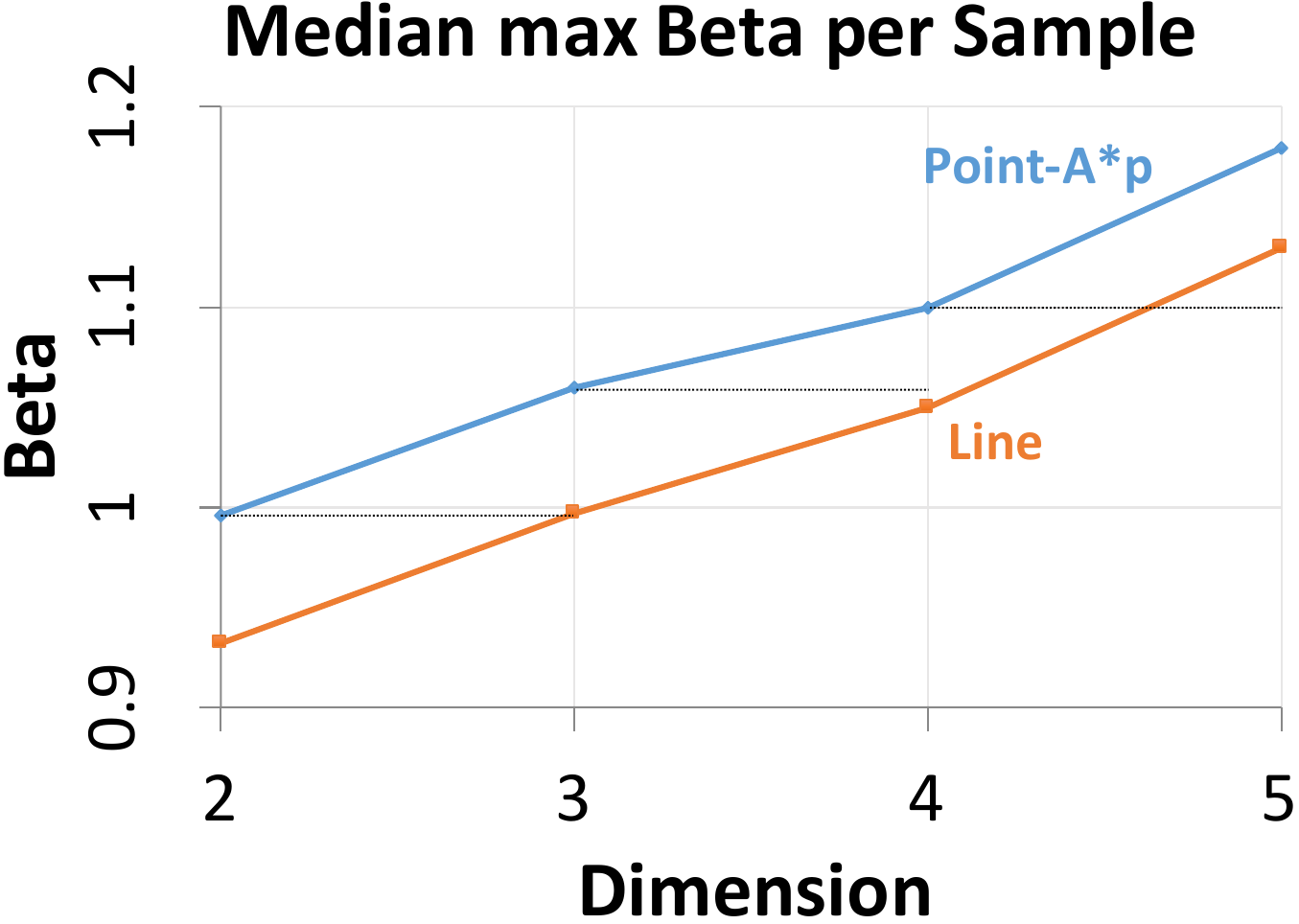}
       \quad
       \includegraphics[width=0.47\linewidth]{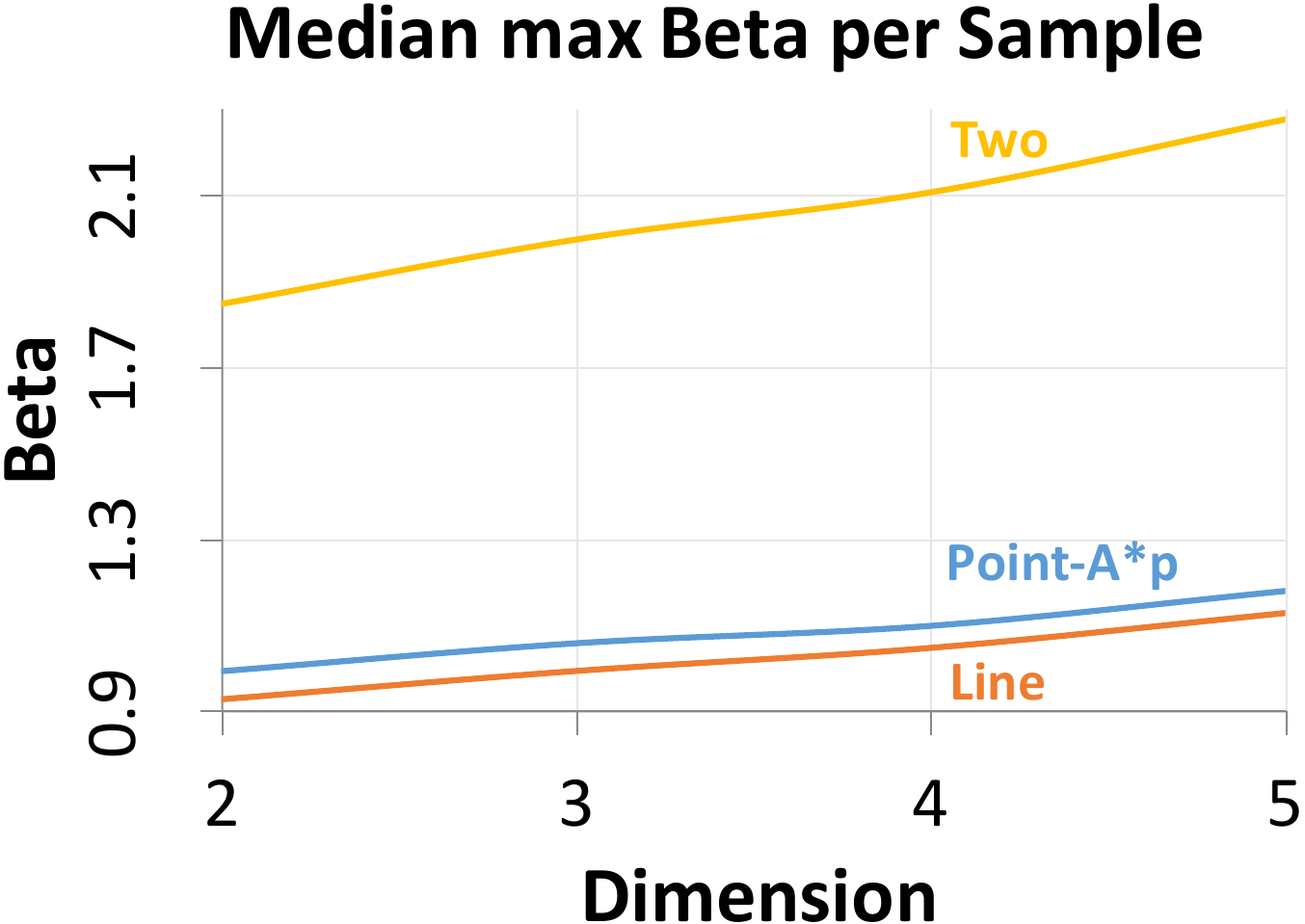}
       \\
       \vspace{2mm}
       \begin{tabular}{cccc}
       	 \includegraphics[width=0.212\linewidth]{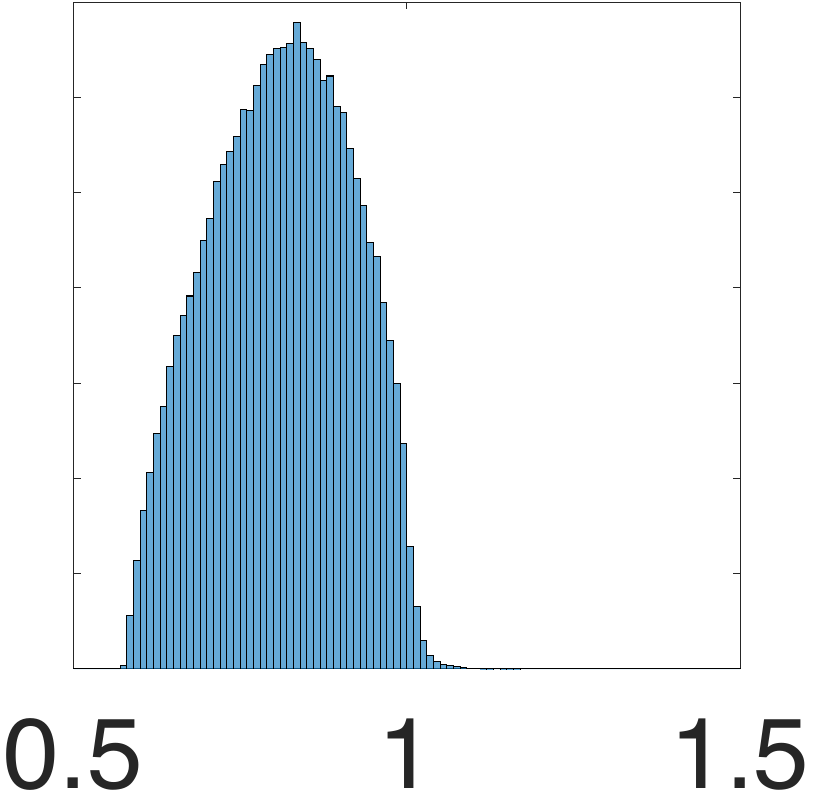}
       	&\includegraphics[width=0.212\linewidth]{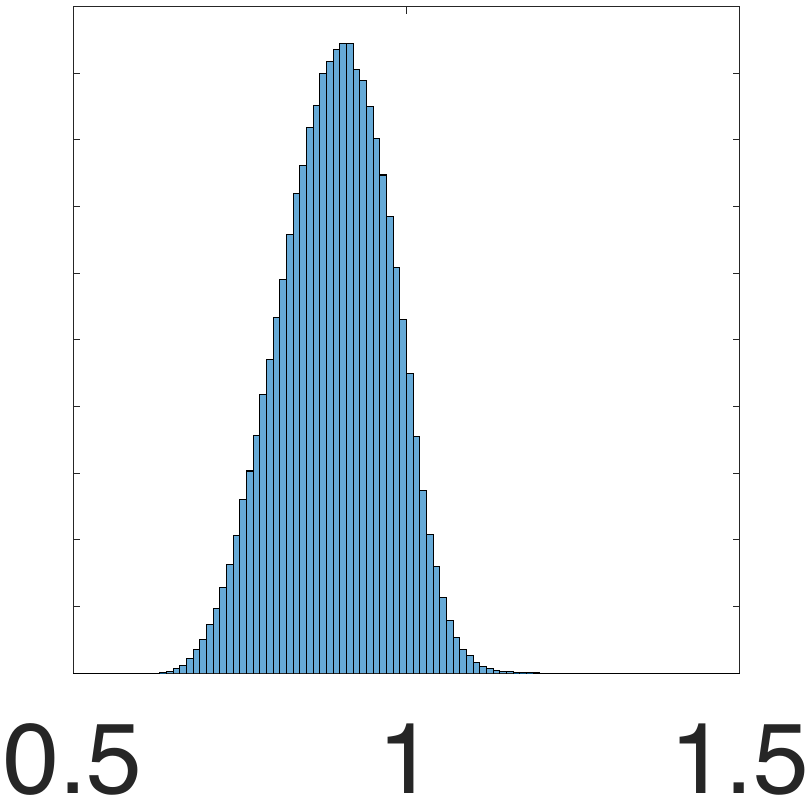}
       	&\includegraphics[width=0.212\linewidth]{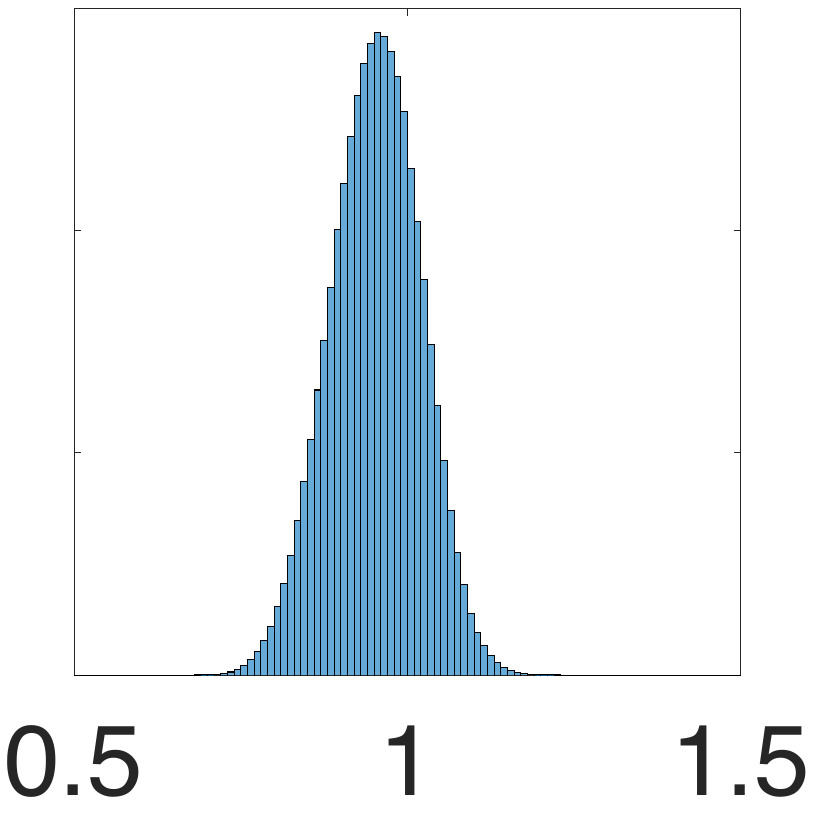}
       	&\includegraphics[width=0.212\linewidth]{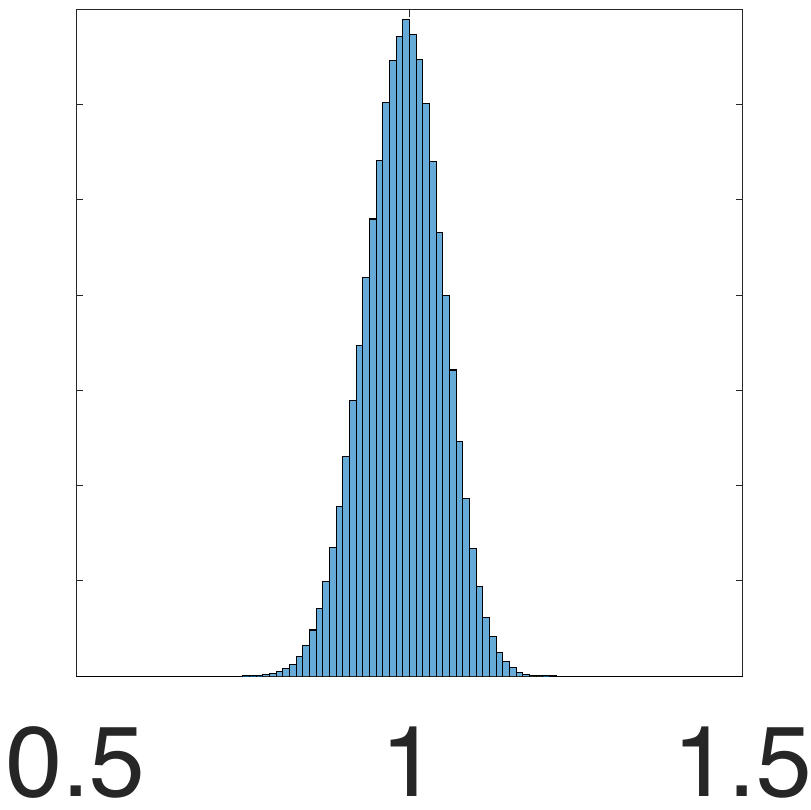}
       	\\
       	$\dimnumber=2$ &$\dimnumber=3$ &$\dimnumber=4$ &$\dimnumber=5$
       \end{tabular}       
    }
    \Caption{Trends in $\rratio$ in practice.}{Top, \linespokesalg\ gives about the same saturation as \bridsonmethodours\ in one dimension lower. \Twospokesalg\ has about twice the $\rratio$ of \Linespokesalg\ by design. 
    %Despite the observed rise in small dimensions, $\rratio$ is expected to converge to a fixed value as $\dimnumber \rightarrow \infty$. 
    Bottom, the distribution of $\rratio$ (Voronoi-vertex to nearest-sample distances) narrows by dimension, and converges around a fixed value.   
    We only show dimensions 2--5 because the available tools for computing Voronoi vertices, e.g.\ Qhull, run out of time and memory in higher dimensions.
\nothing{
\liyi{(October 31, 2017) Better use vector graphics (e.g. svg or eps and convert to pdf) instead of raster images for the bottom 4 distributions.}%liyi
\scott{(November 2, 2017) done}%scott
}%nothing
 }
    \label{fig:betamedian_body}
\end{figure}

%Their spectra do not vary significantly as the dimension increases.
%See \Cref{fig:rdf_spectra_ld,fig:rdf_spectra_fd,fig:rdftwo} in \Cref{sec:additional_results} for spectra examples in $\dimnumber=3, 4, 5$.
%\Cref{fig:rdf} shows the Radial Distance Function (RDF) for $\dimnumber=6$.
By design, the RDF and spectra of line- and two-spokes differ significantly. However, they have similar $\rratio$ distributions, after scaling by $\rratio_{two} \approx 2 \rratio_{line}$, as described in \Cref{sec:analysis}.
\Cref{fig:betamedian_body} shows the median $\rratio$ by method and dimension.
\begin{itemize}
	\item \Linespokesalg\ has median $\rratio \approx $ 0.9--1.2 as $d=$ 2--5. %\\but stays below $2$ as $\dimnumber \rightarrow \infty$.
	\item \Twospokesalg\ has median $\rratio \approx $ 1.8--2.4 as $d=$ 2--5. %\\but stays below $4$ as $\dimnumber \rightarrow \infty$.
	%\item \Favoredspokesalg\ has median $\rratio \approx 1.5$ regardless of $d$.
\end{itemize}
The median $\rratio$ rises with dimension in part because of the increase in the number of Voronoi vertices around each sample, so the probability of at least one being far increases.
However, recall from \Cref{sec:parameters} that the distribution of achieved $\rratio$ narrows as the dimension increases, and should stay below a fixed value ($<2$) as $\dimnumber \rightarrow \infty$. 
Additional data presented in \Cref{fig:comparison_histograms} in \Cref{sec:additional_results} bear this out. 
%See \Cref{sec:additional_results} for additional results.
%
% \input{rdf_spectra_comparison_highd_fig} % not interesting, no rdf and rp because radius is too big. spectra are white noise.
%

{
%\noindent %noindent causes extra vspace when the figure is input
\begin{figure*}
	\centering
	\setlength{\tabcolsep}{2pt}  %2_P_198_0spoke
	\resizebox{1.0\linewidth}{!}{
		\begin{tabular}{cccccccc}
					& dimension  & 2 & 3 & 4 & 6 & 8 & 10 \\
					\hline \\
%			\multirow{ 3 }{*}{\rotatebox[origin=l]{90}{\textbf{{\color{blue} $\Longleftarrow$ Point-Annulus $\Longrightarrow$}}}}
			\multirow{ 3 }{*}{\rotatebox[origin=l]{90}{\textbf{{\color{blue}  \Large Point-Annulus$^*$p }}}}
      & \rotatebox[origin=l]{90}{{\hspace{16pt}spectra}} 
    \ifthenelse{\equal{\isarxiv}{1}}
    {
			& \includegraphics[height=1in]{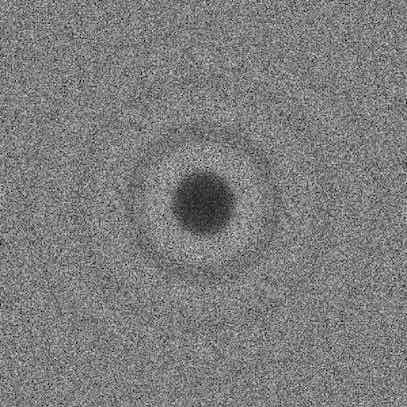}
			& \includegraphics[height=1in]{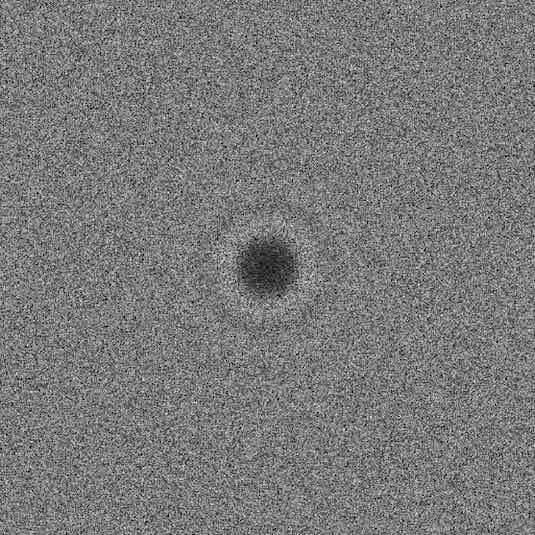}
			& \includegraphics[height=1in]{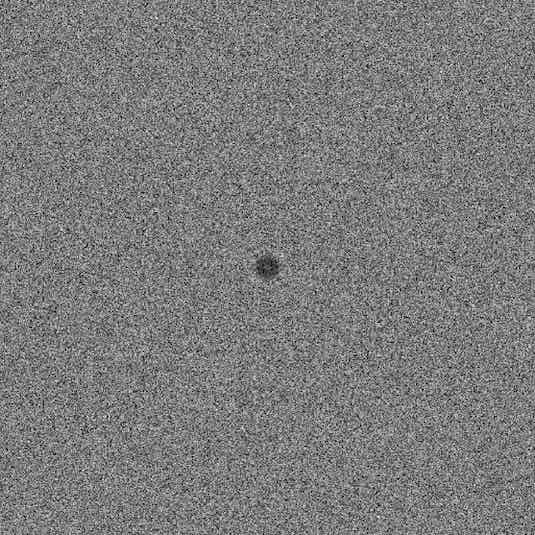}
      			& \includegraphics[height=1in]{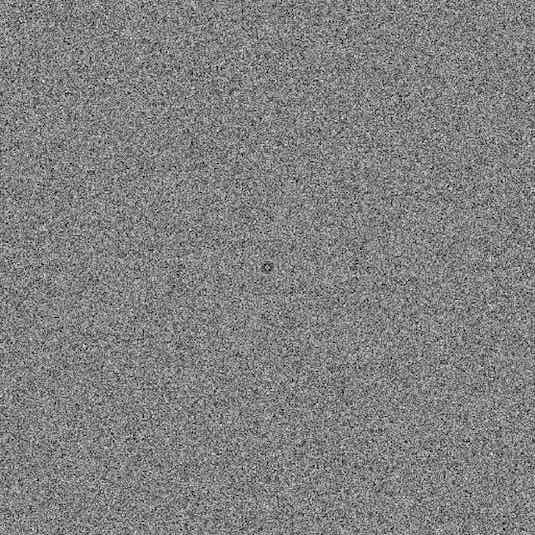}
      			& \includegraphics[height=1in]{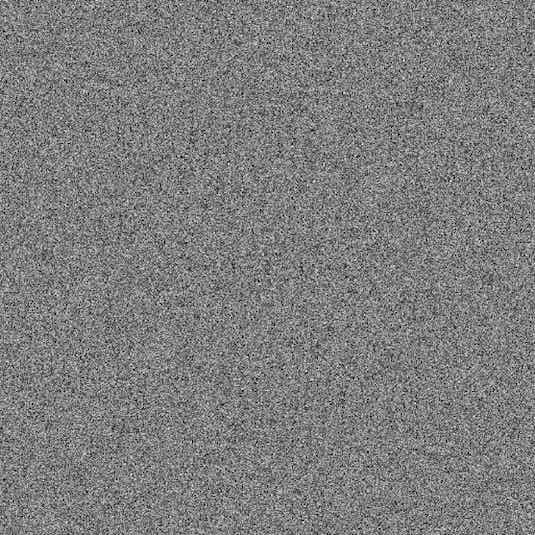}
		     	&\includegraphics[height=1in]{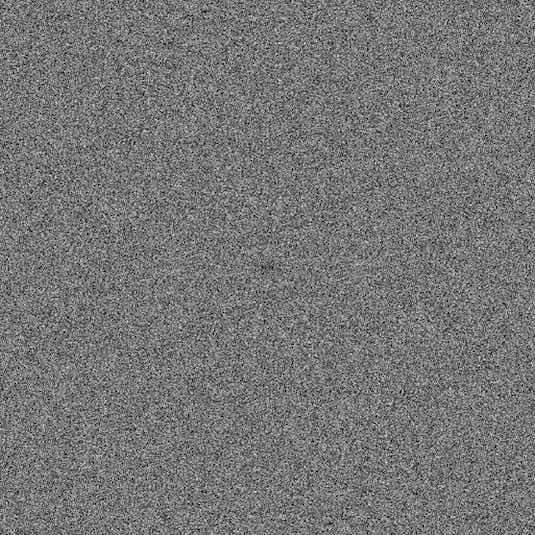}
    }
    {
			& \includegraphics[height=1in]{figs/bridsonexperimentsour/2d/fftslice.png}
			& \includegraphics[height=1in]{figs/bridsonexperimentsour/3d/fftslice.png}
			& \includegraphics[height=1in]{figs/bridsonexperimentsour/4d/fftslice.png}
      			& \includegraphics[height=1in]{figs/bridsonexperimentsour/6d/fftslice.png}
      			& \includegraphics[height=1in]{figs/bridsonexperimentsour/8d/fftslice.png}
		     	&\includegraphics[height=1in]{figs/bridsonexperimentsour/10d/fftslice.png}
    }
			\\
      & \rotatebox[origin=l]{90}{{\hspace{28pt}RDF}}
			&\includegraphics[height=1in]{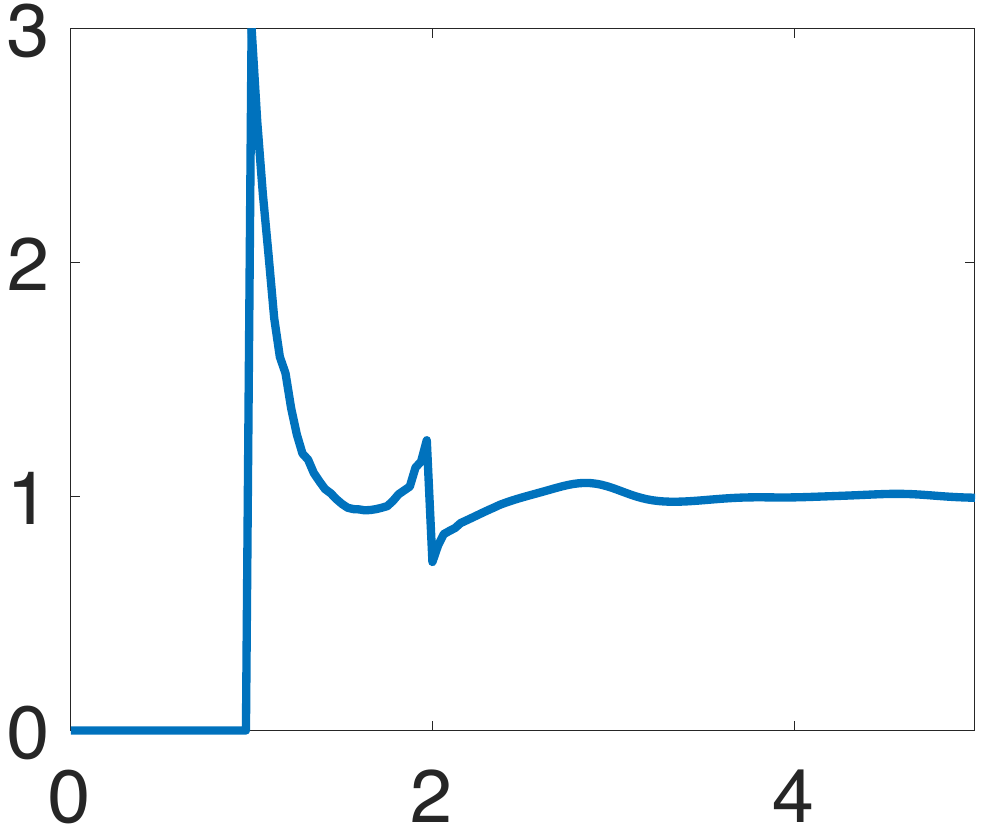}
			&\includegraphics[height=1in]{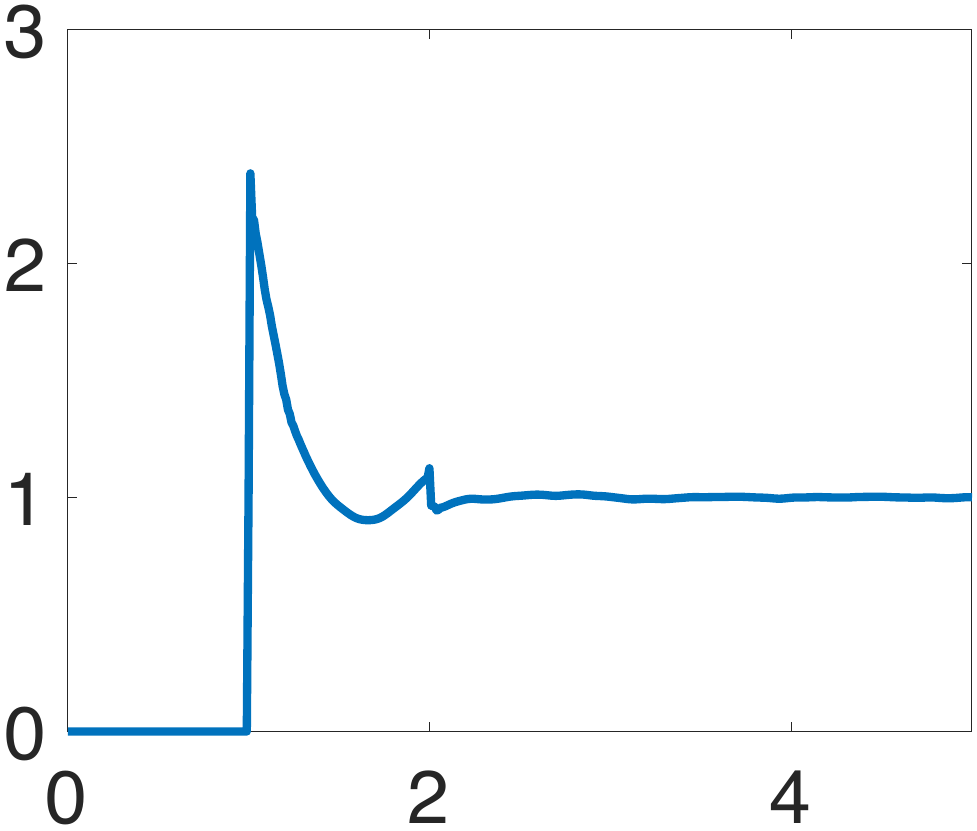}
			&\includegraphics[height=1in]{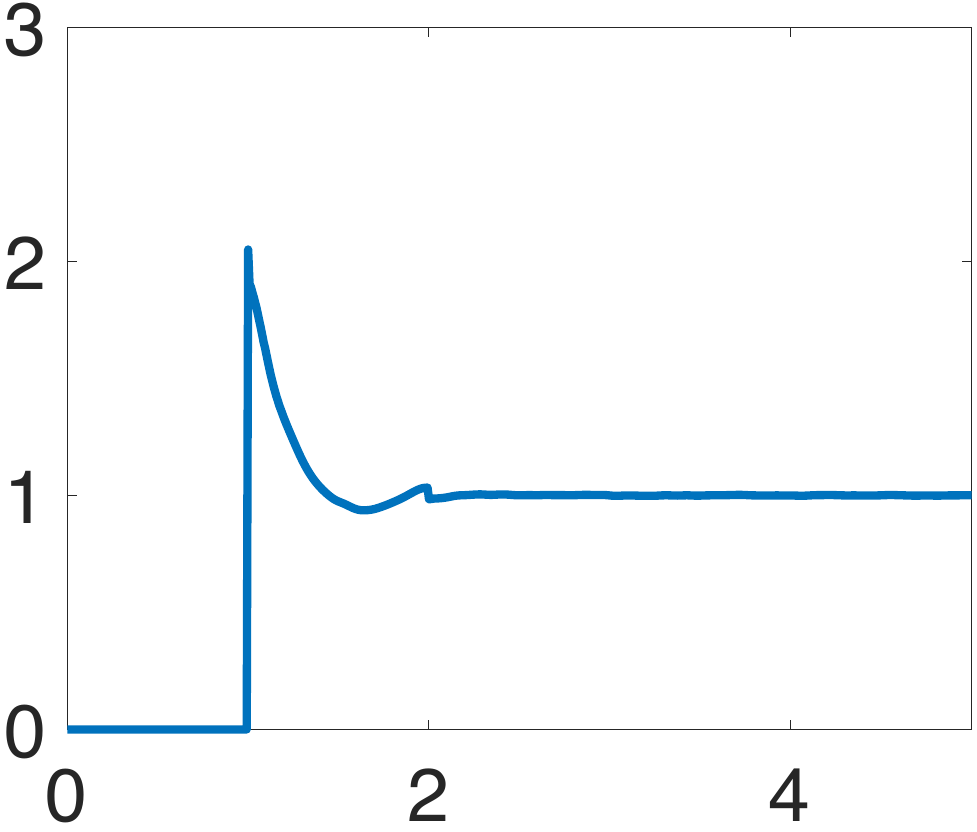}	
      			&\includegraphics[height=1in]{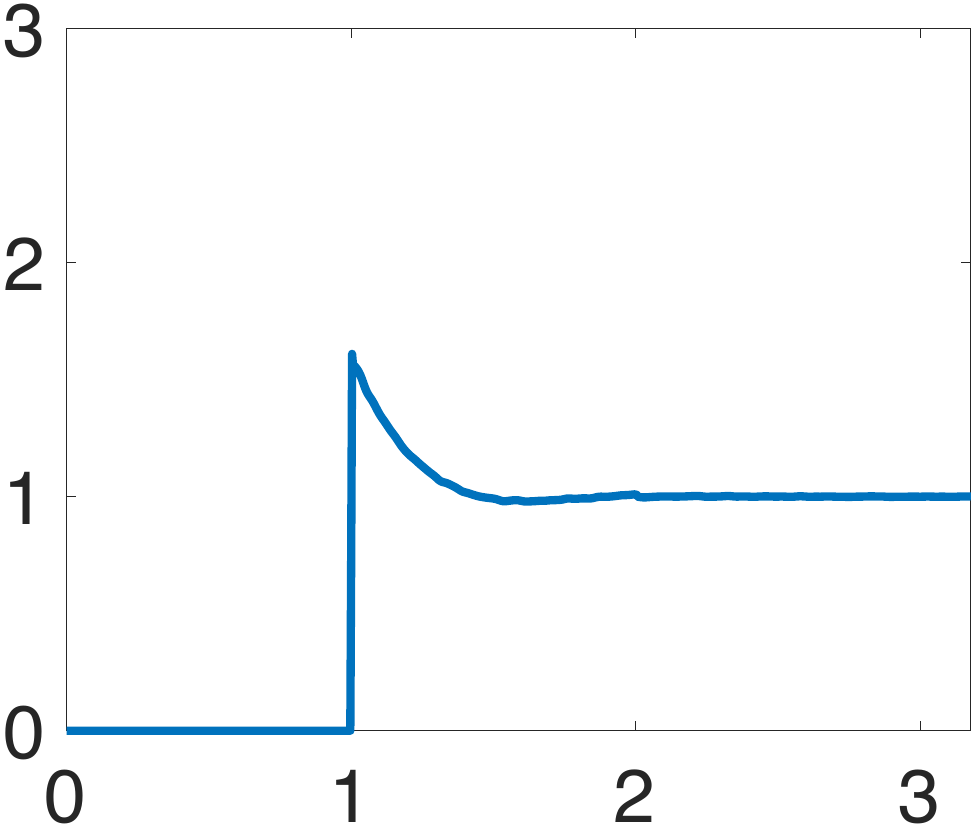}
      			&\includegraphics[height=1in]{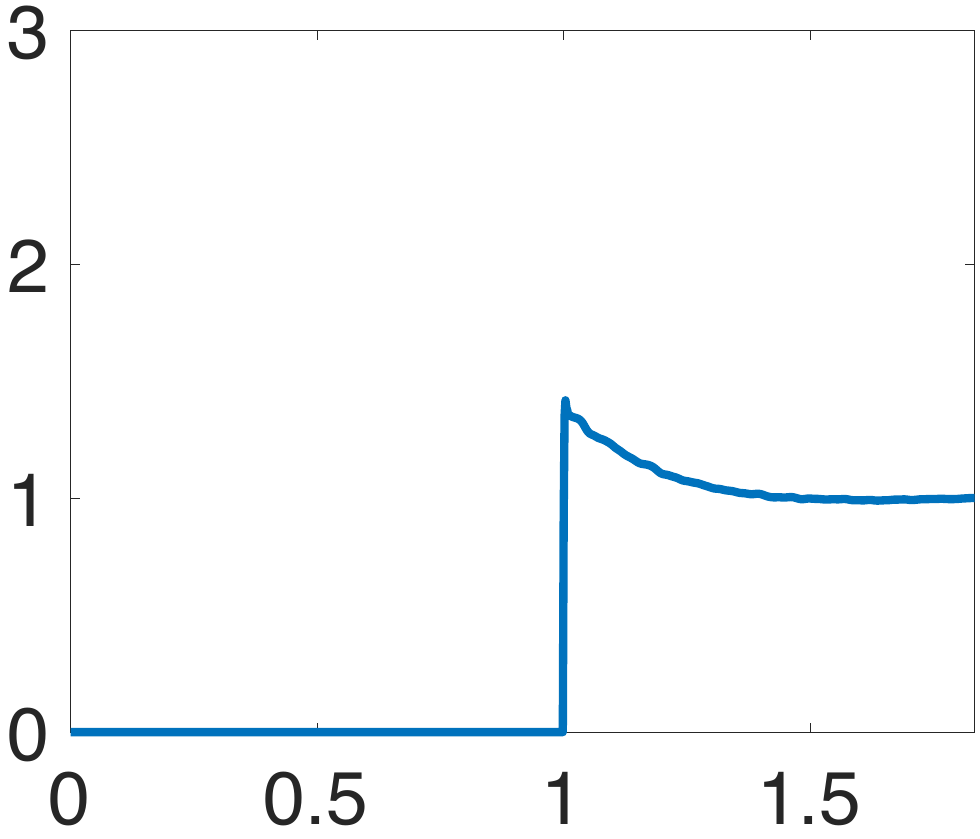}
 		        &\includegraphics[height=1in]{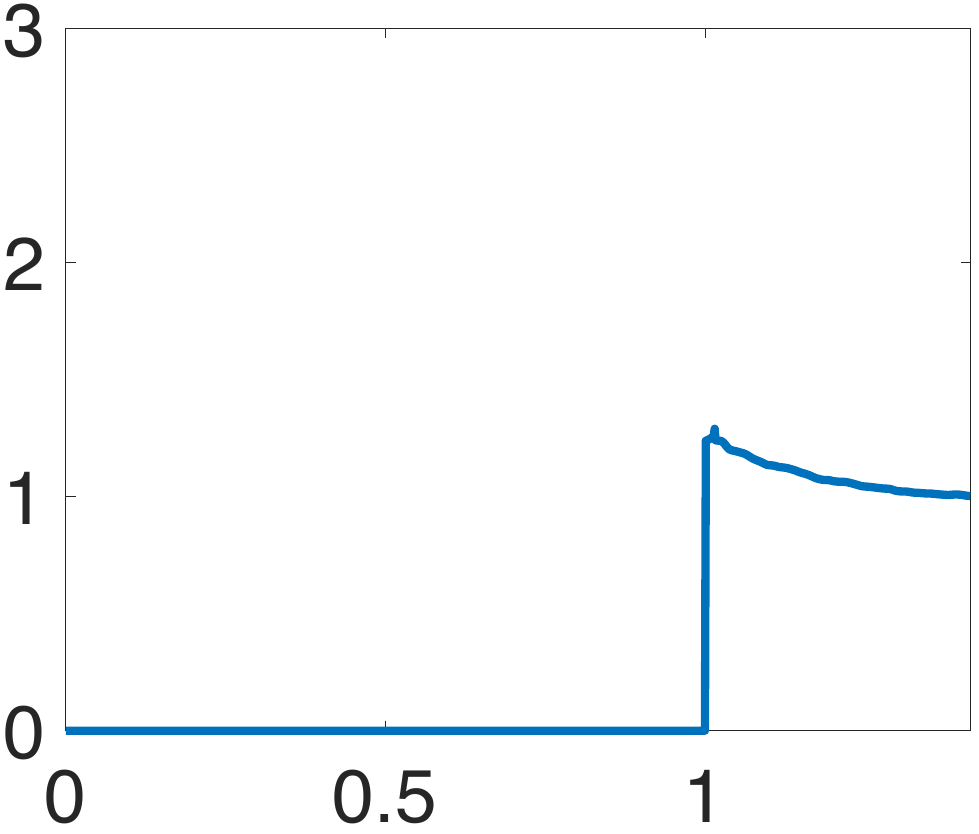}
			\\
      &\rotatebox[origin=l]{90}{{\hspace{28pt}RP}}
      			&\includegraphics[height=1in]{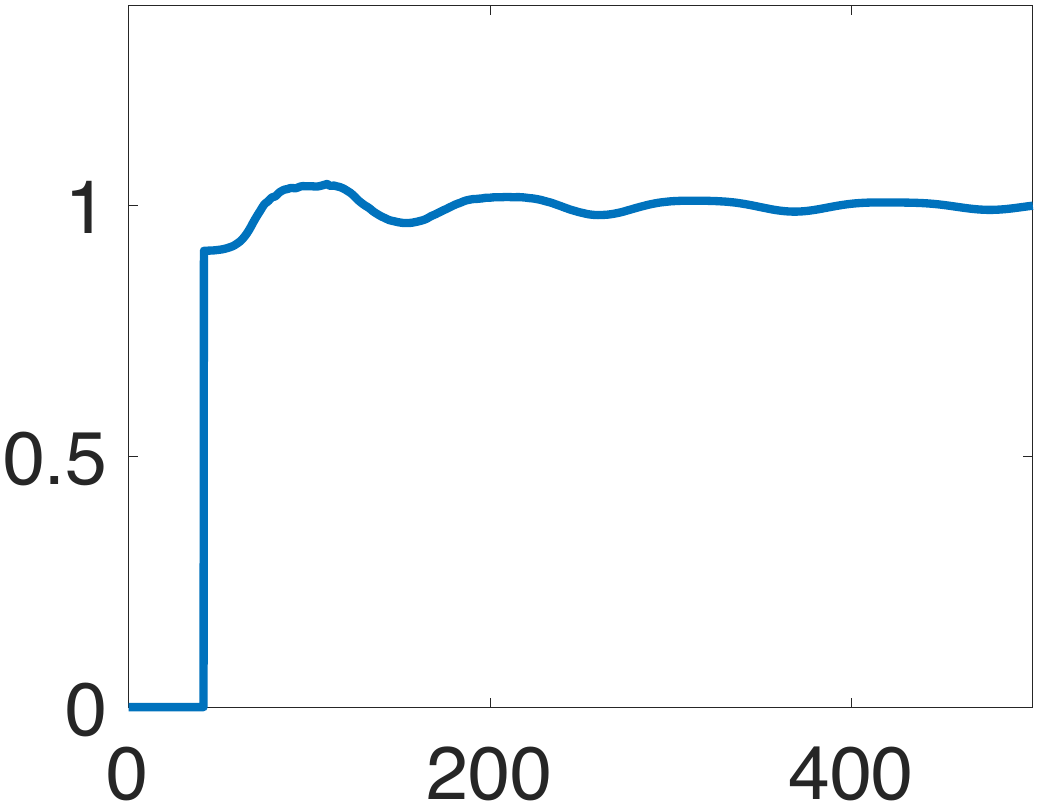}
      			&\includegraphics[height=1in]{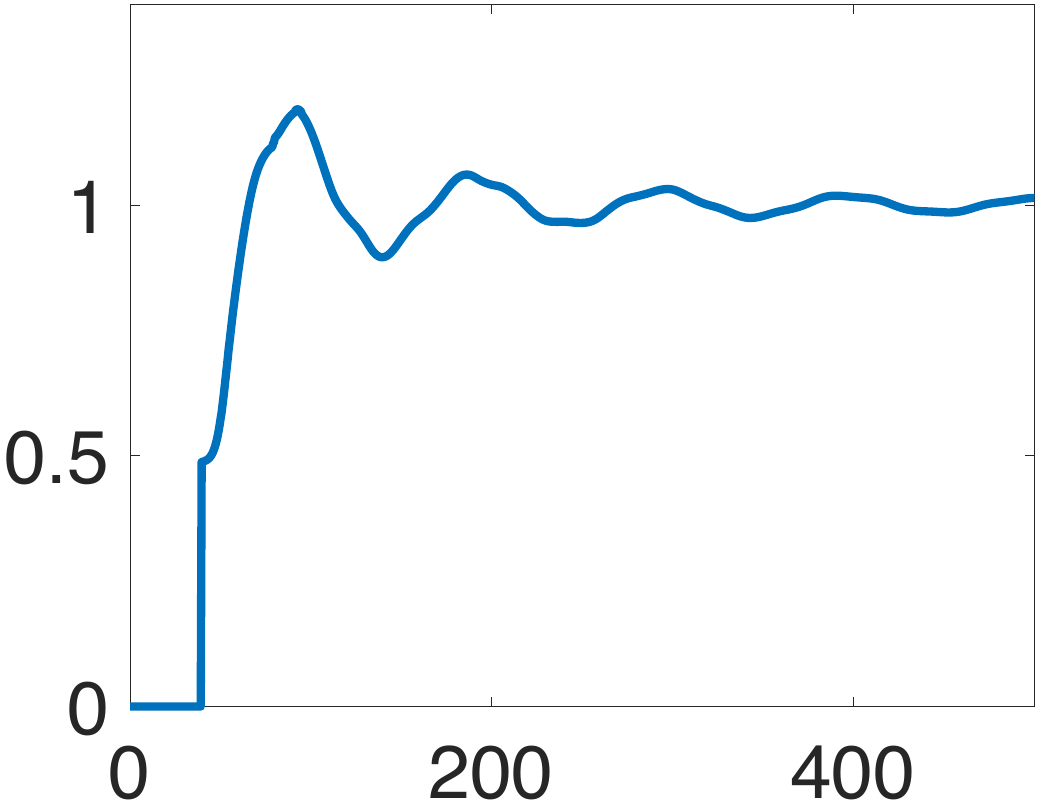}
			&\includegraphics[height=1in]{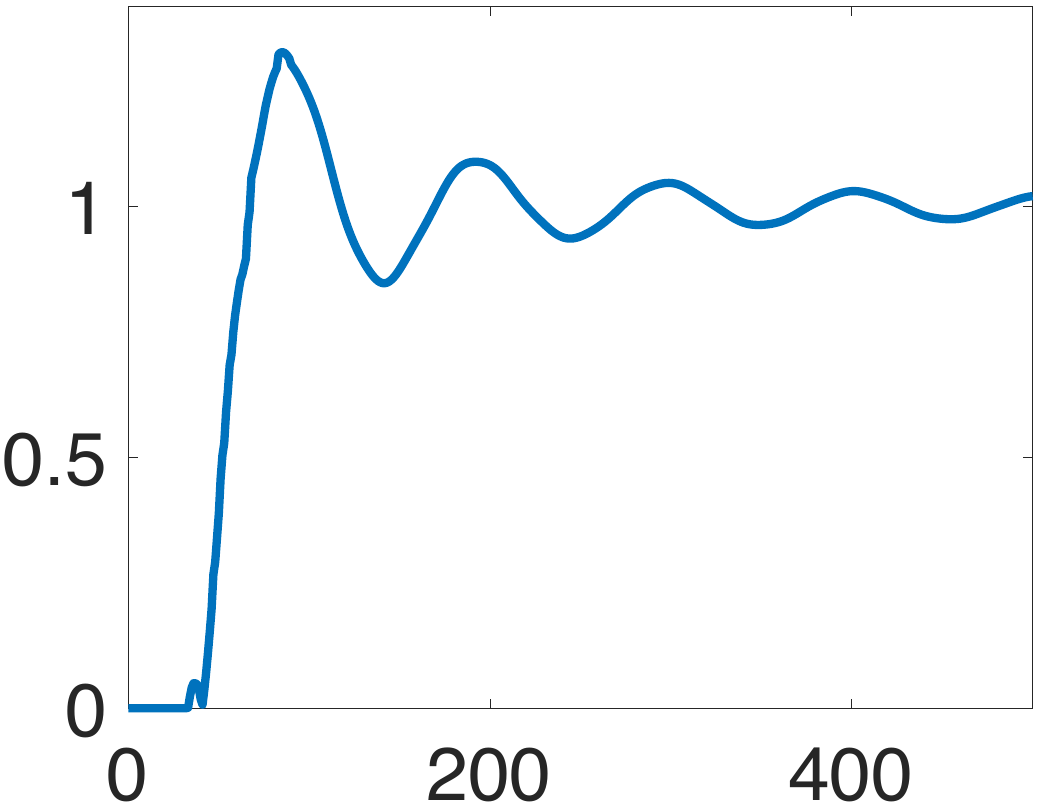}
      			&\includegraphics[height=1in]{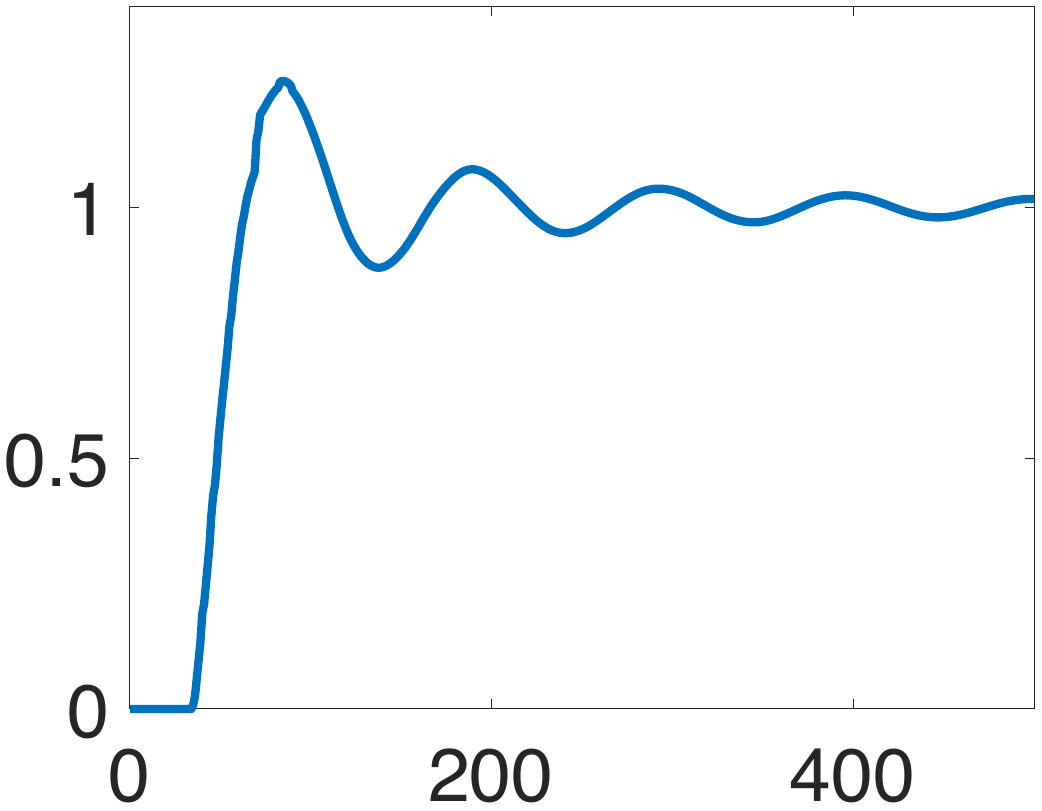}
      			&\includegraphics[height=1in]{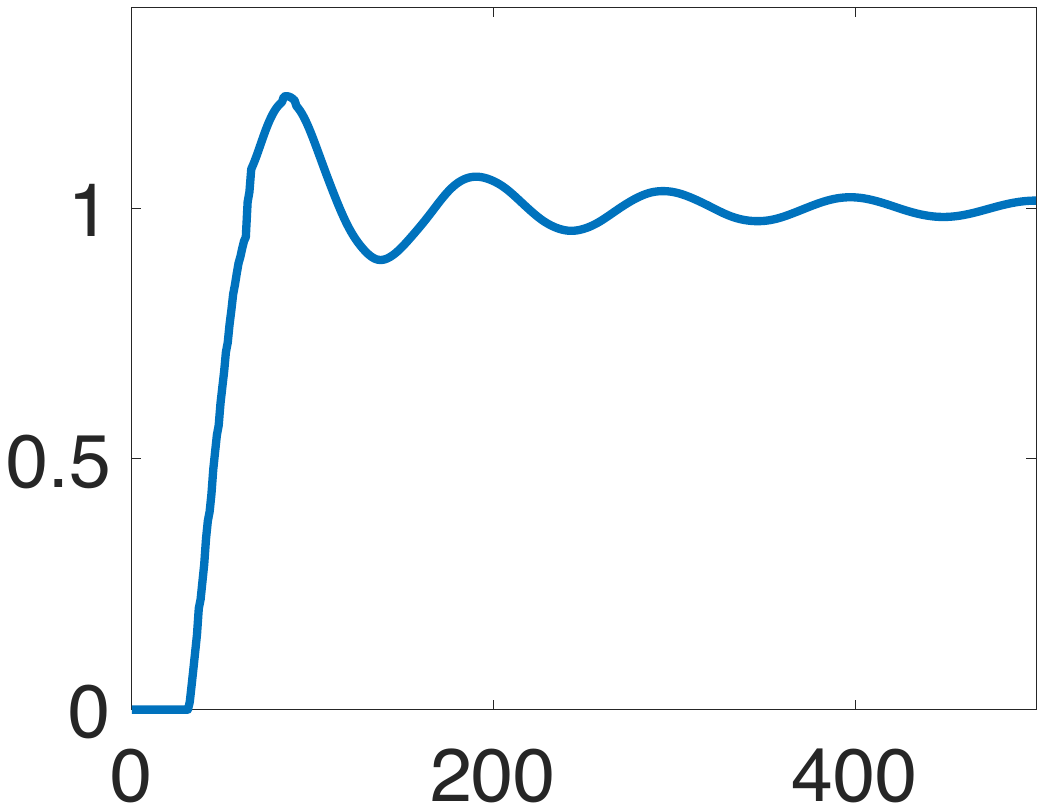}
			&\includegraphics[height=1in]{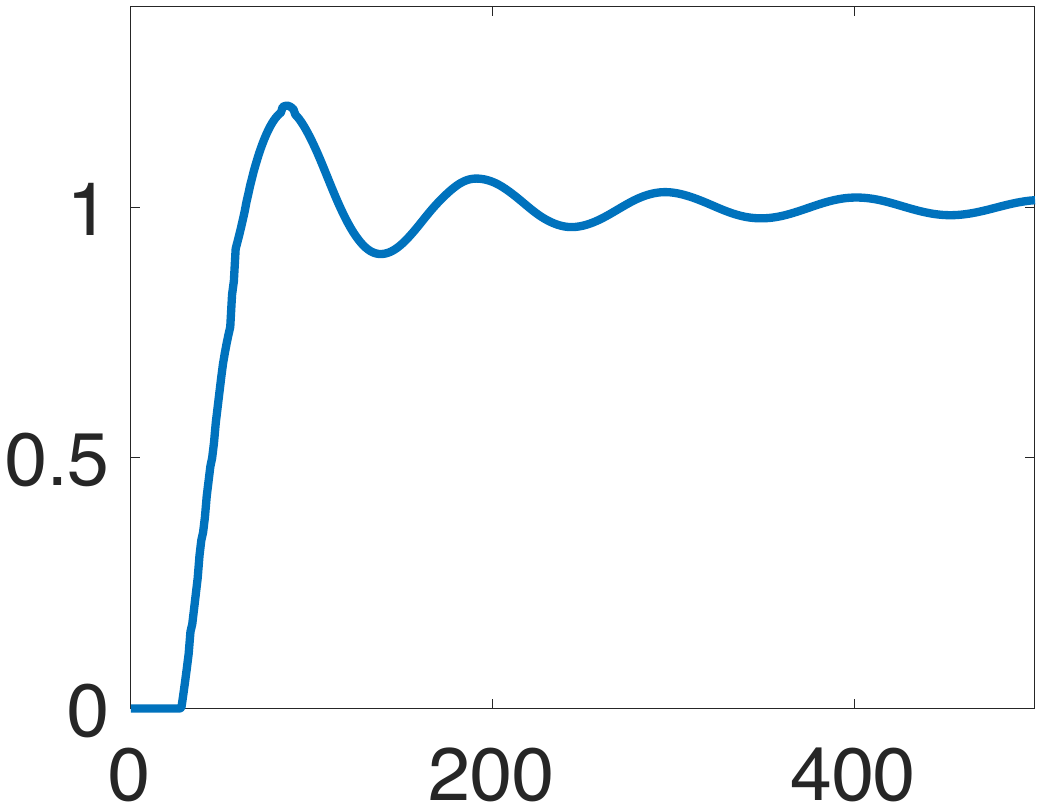}
			\\
			\hline
			\\
			\multirow{ 3 }{*}{\rotatebox[origin=l]{90}{\textbf{{\color{red} \Large Line-Spokes }}}}
      &\rotatebox[origin=l]{90}{{\hspace{16pt}spectra}}
%			& \includegraphics[height=1in]{figs/distributions/spoke_spec.png}
    \ifthenelse{\equal{\isarxiv}{1}}
    {
			&\includegraphics[height=1in]{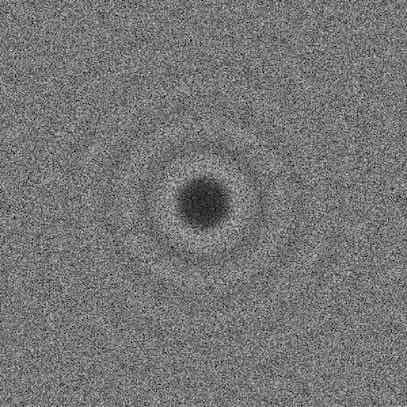}
			&\includegraphics[height=1in]{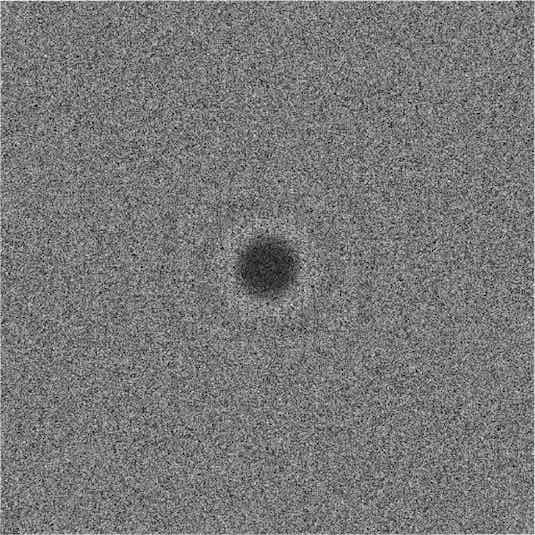}
			&\includegraphics[height=1in]{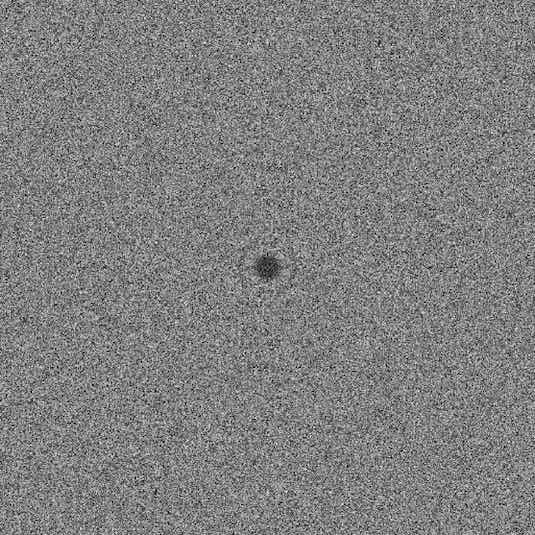}
      			&\includegraphics[height=1in]{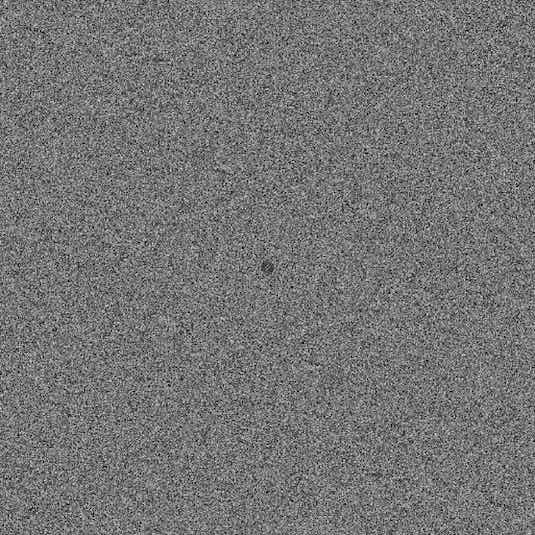}
      			&\includegraphics[height=1in]{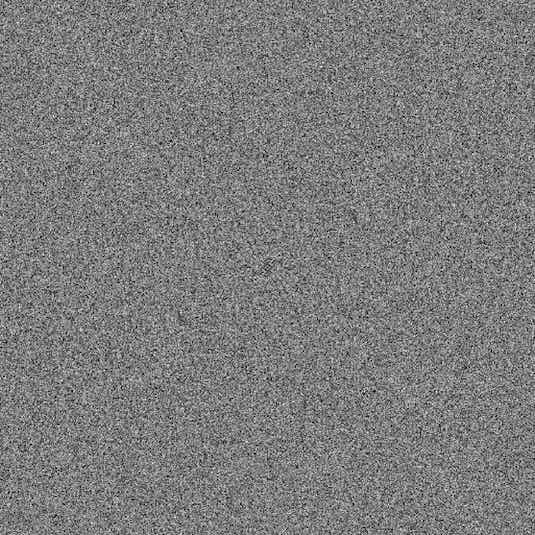}
		      	&\includegraphics[height=1in]{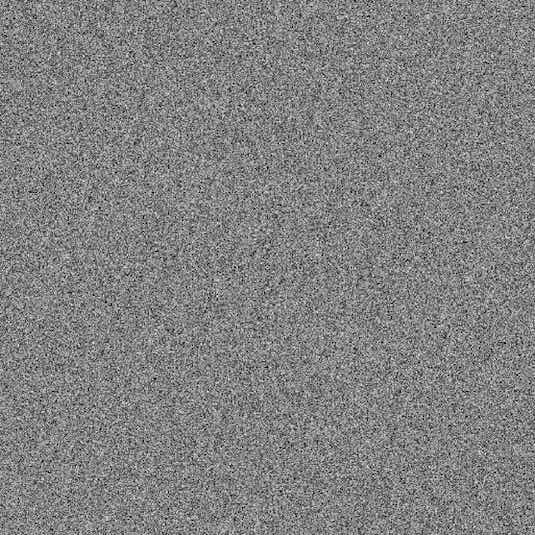}
    }
    {
			&\includegraphics[height=1in]{figs/distributions-p-midd/02d/slice513_1_1655.png}
			&\includegraphics[height=1in]{figs/distributions-p-midd/03d/fftslice.png}
			&\includegraphics[height=1in]{figs/distributions-p-midd/04d/fftslice.png}
      			&\includegraphics[height=1in]{figs/distributions-p-midd/06d/fftslice.png}
      			&\includegraphics[height=1in]{figs/distributions-p-midd/08d/fftslice.png}
		      	&\includegraphics[height=1in]{figs/distributions-p-midd/010d/fftslice.png}
    }
      \\
      &\rotatebox[origin=l]{90}{{\hspace{28pt}RDF}}
%			&\input{figs/distributions/2_P_198_0spoke_411222_beta_points_rdf}
			&\includegraphics[height=1in]{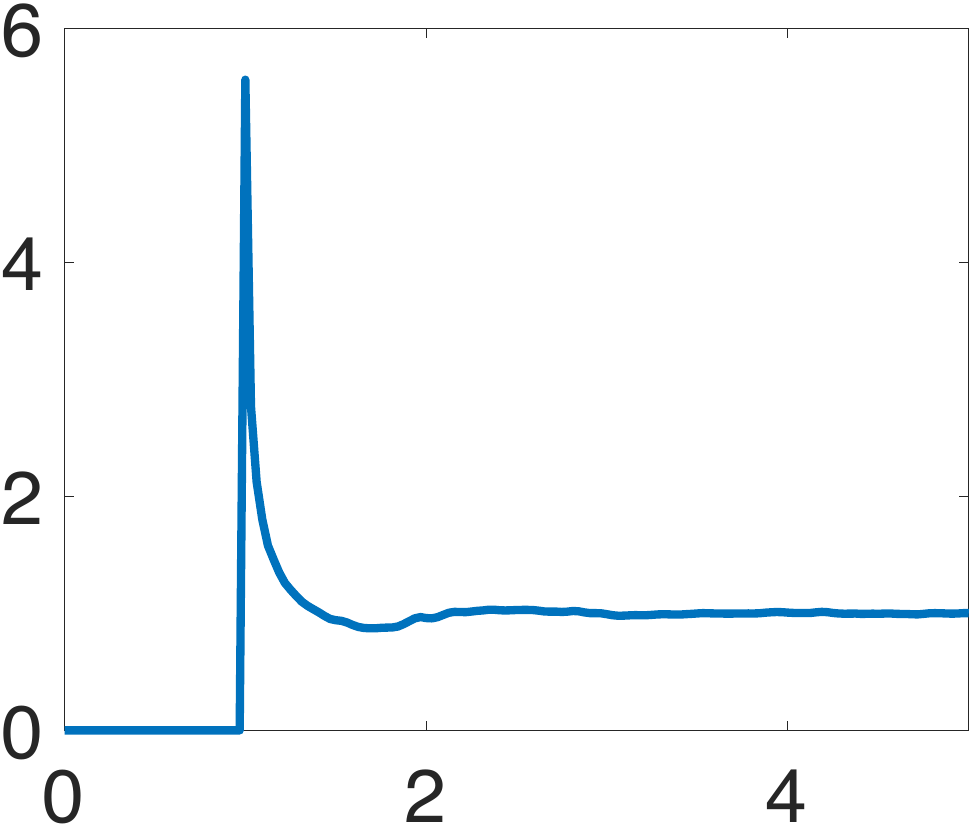}
			&\includegraphics[height=1in]{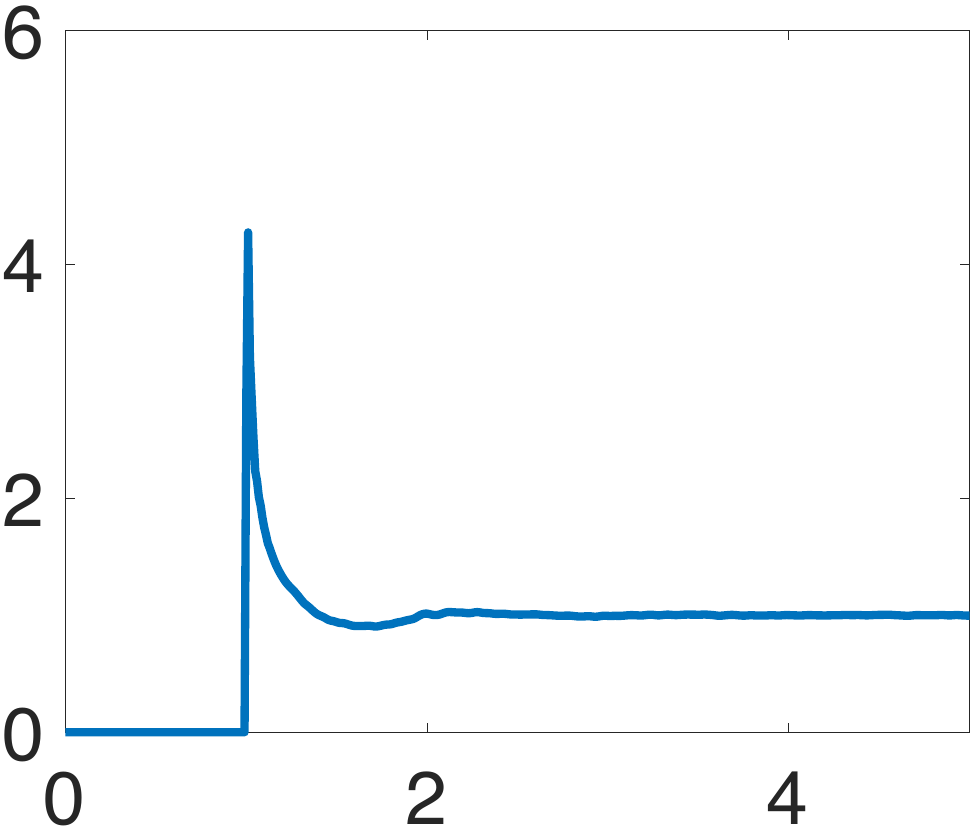}
			&\includegraphics[height=1in]{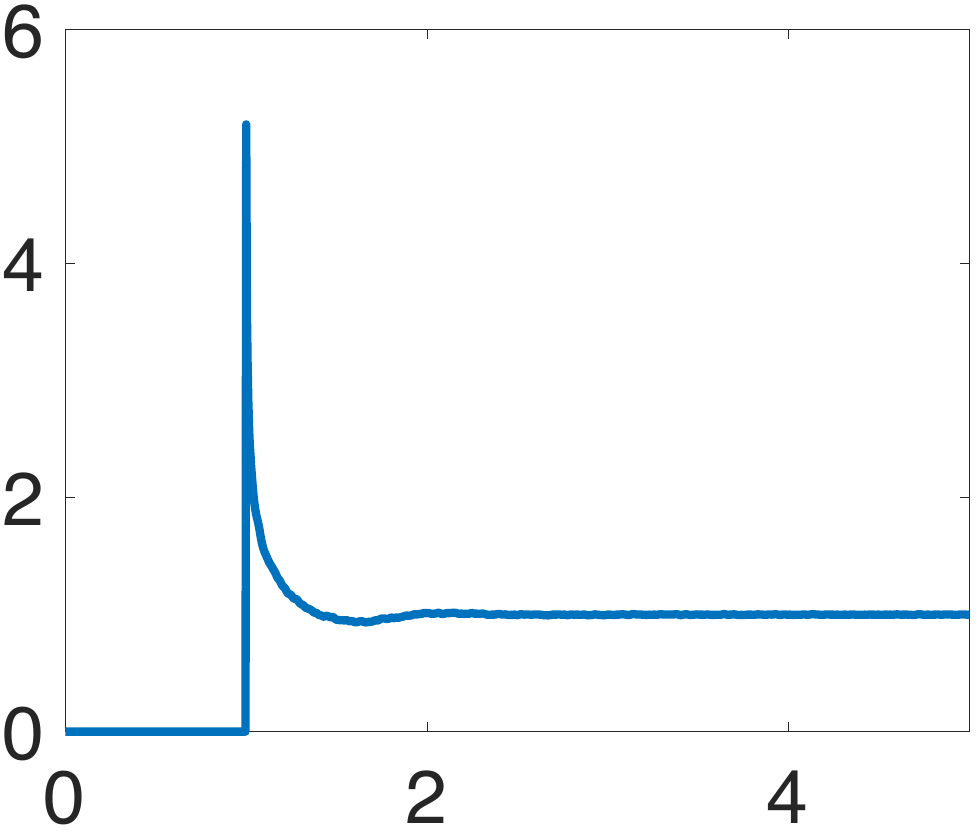}			
			&\includegraphics[height=1in]{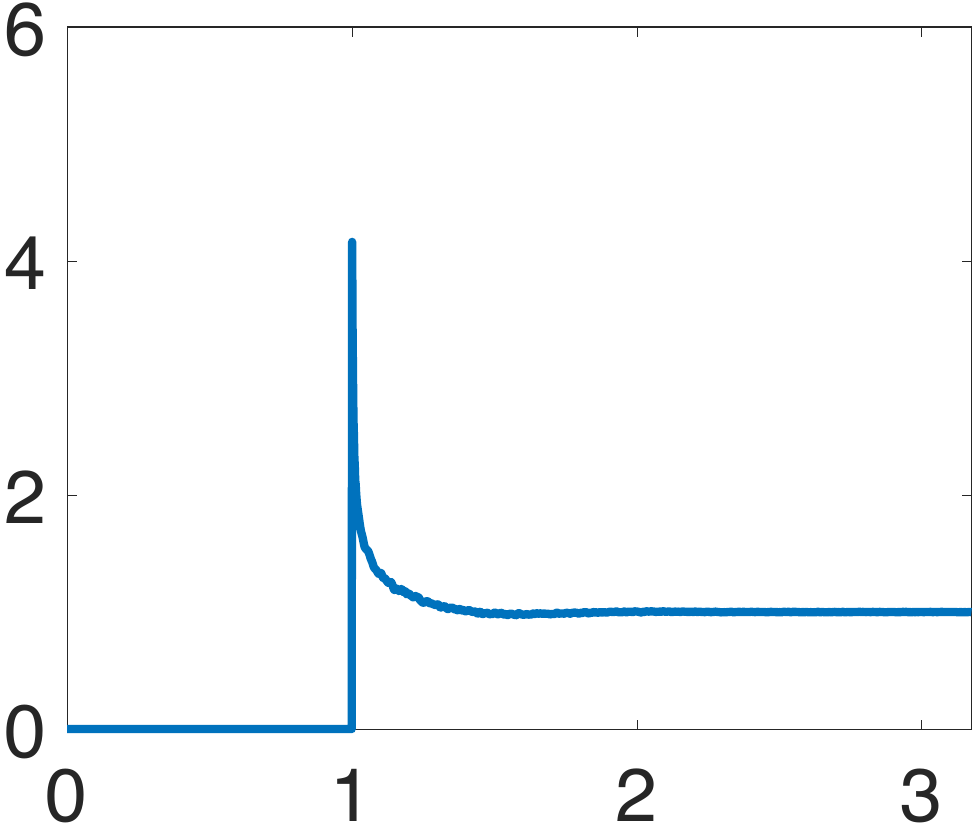}
			&\includegraphics[height=1in]{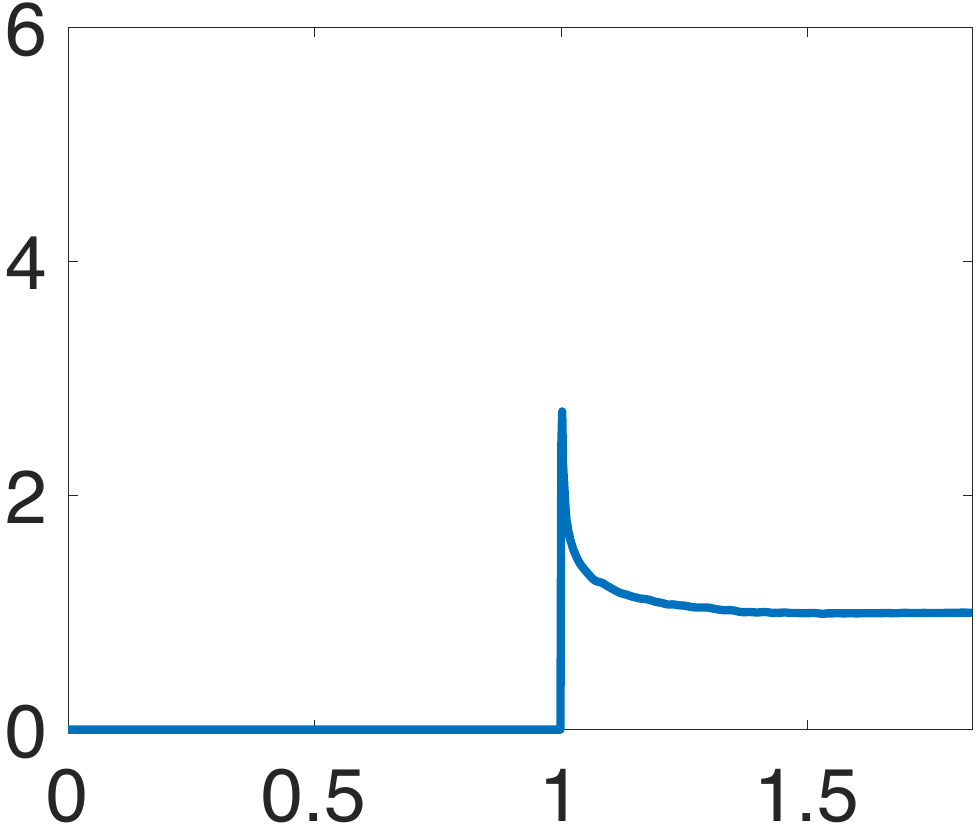}
			&\includegraphics[height=1in]{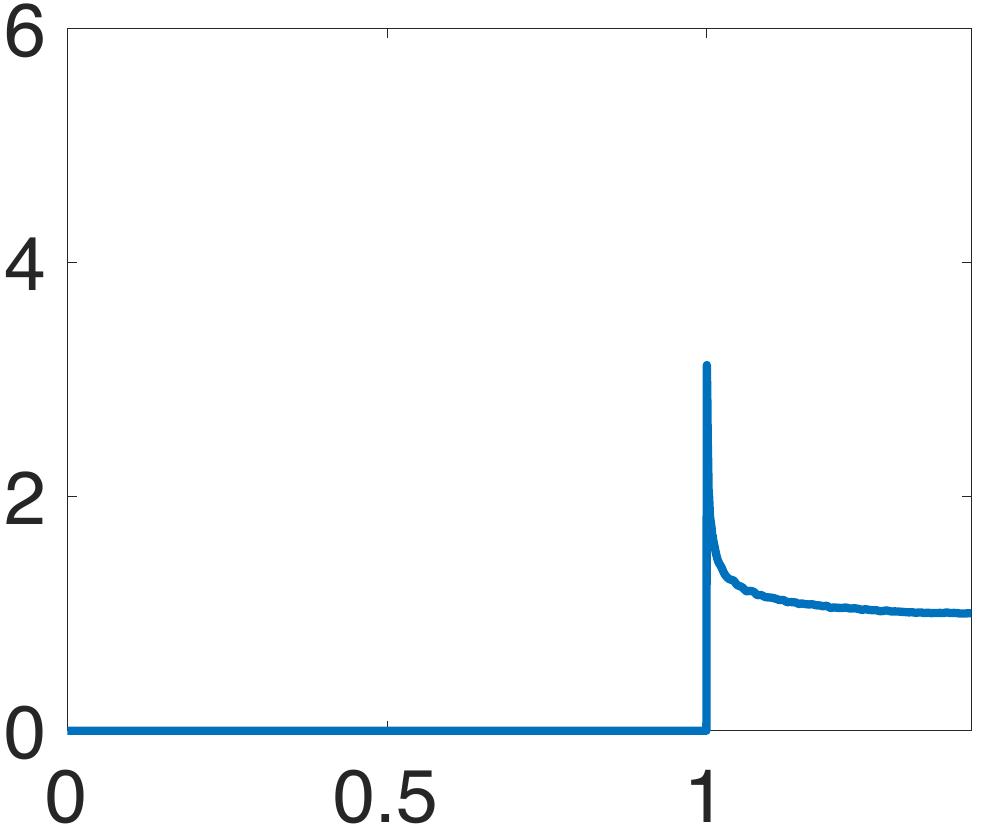}
			\\
      &\rotatebox[origin=l]{90}{{\hspace{28pt}RP}}
%			&\input{figs/distributions/spoke_rp}
			&\includegraphics[height=1in]{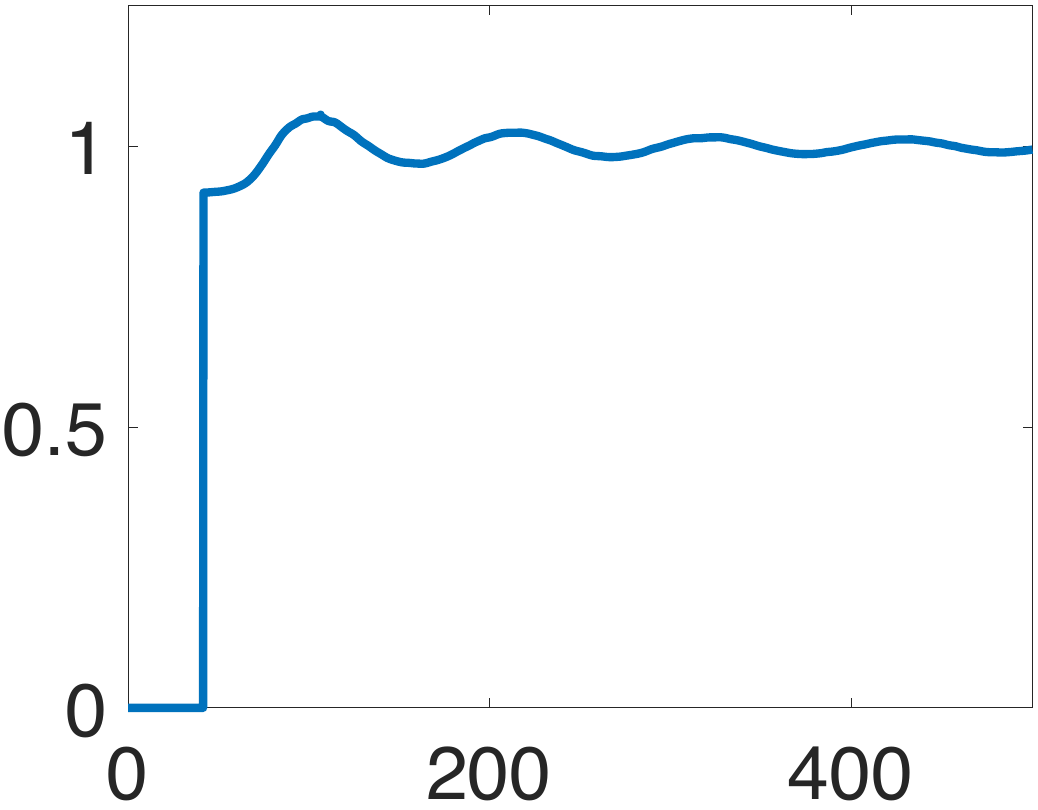}
			&\includegraphics[height=1in]{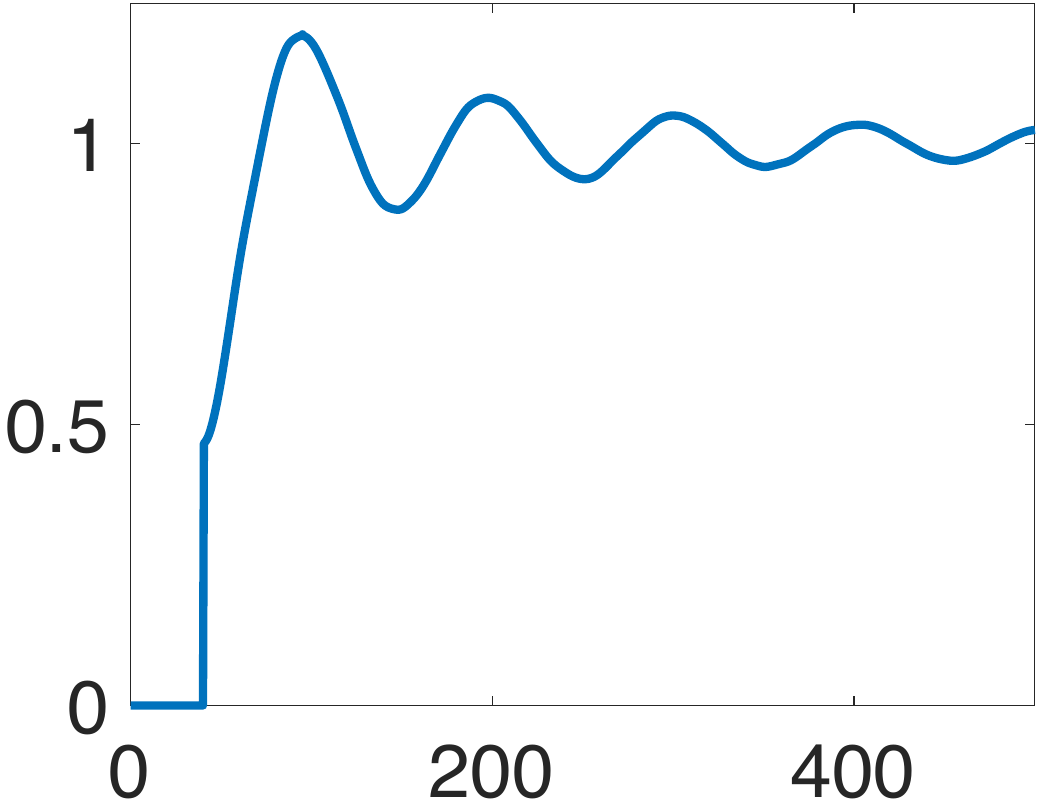}
		 	&\includegraphics[height=1in]{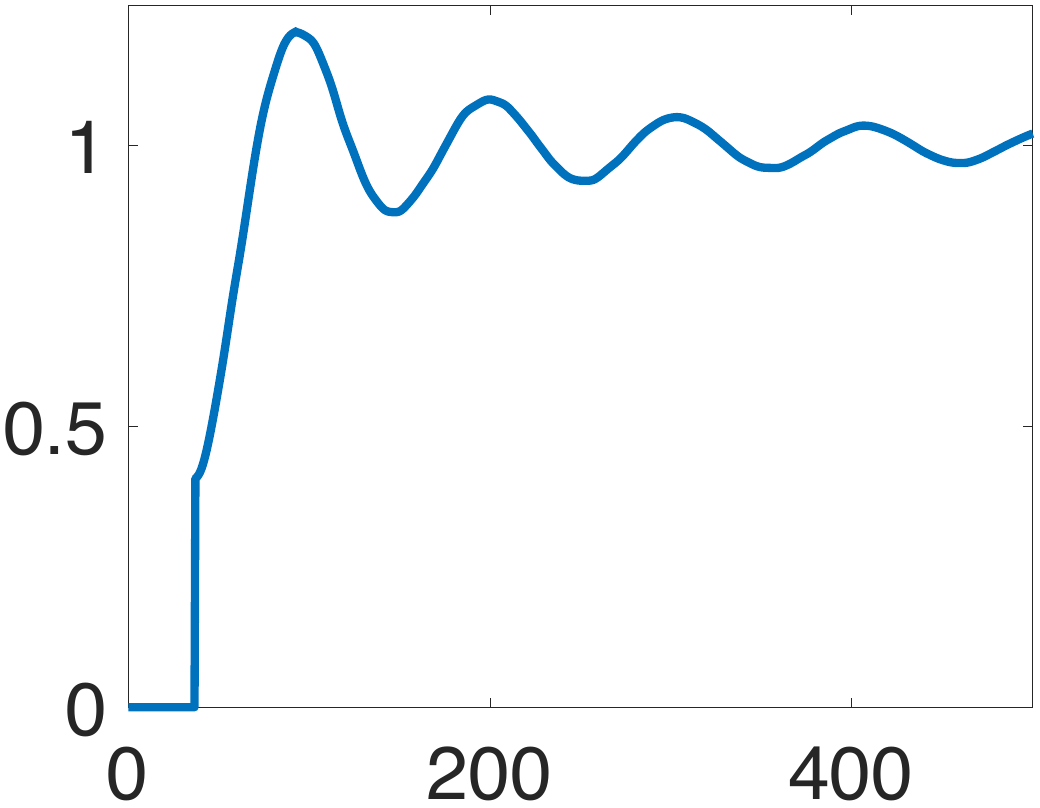}
      			&\includegraphics[height=1in]{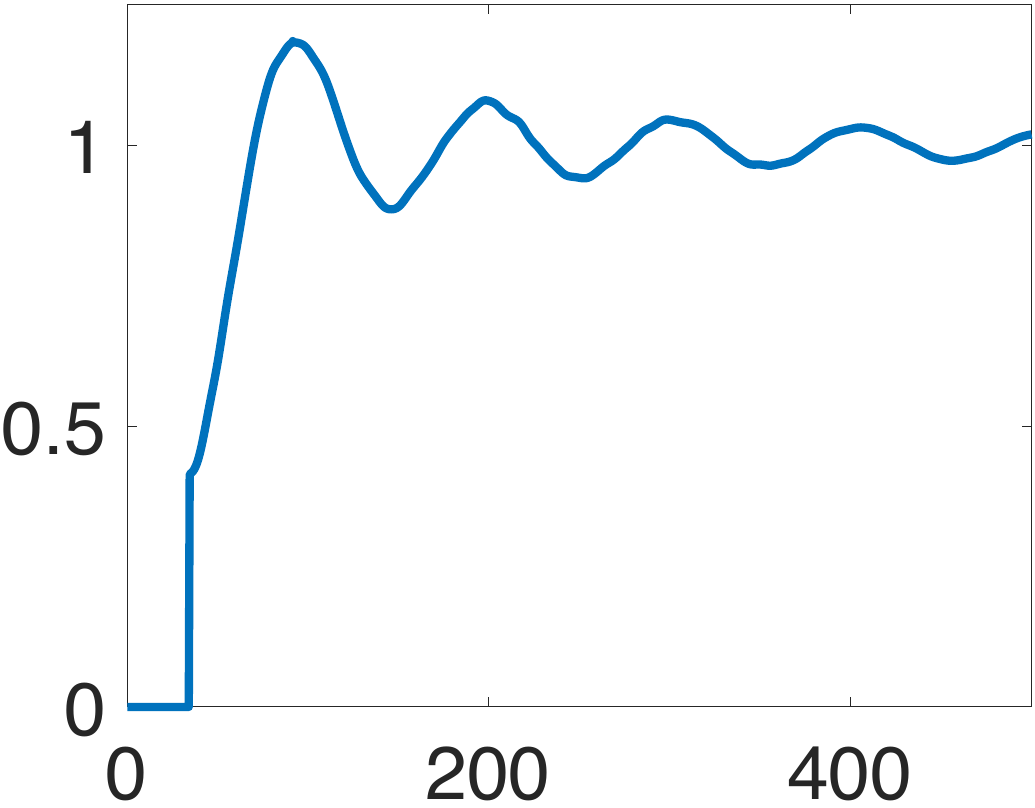}
     			&\includegraphics[height=1in]{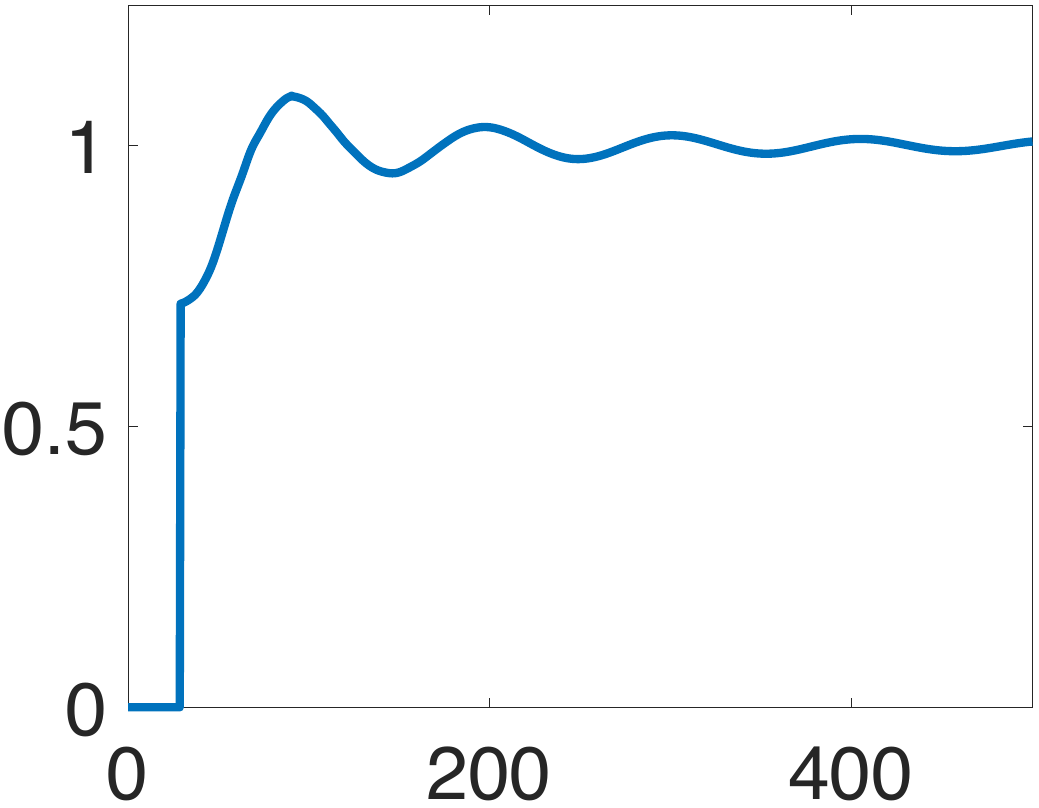}
			&\includegraphics[height=1in]{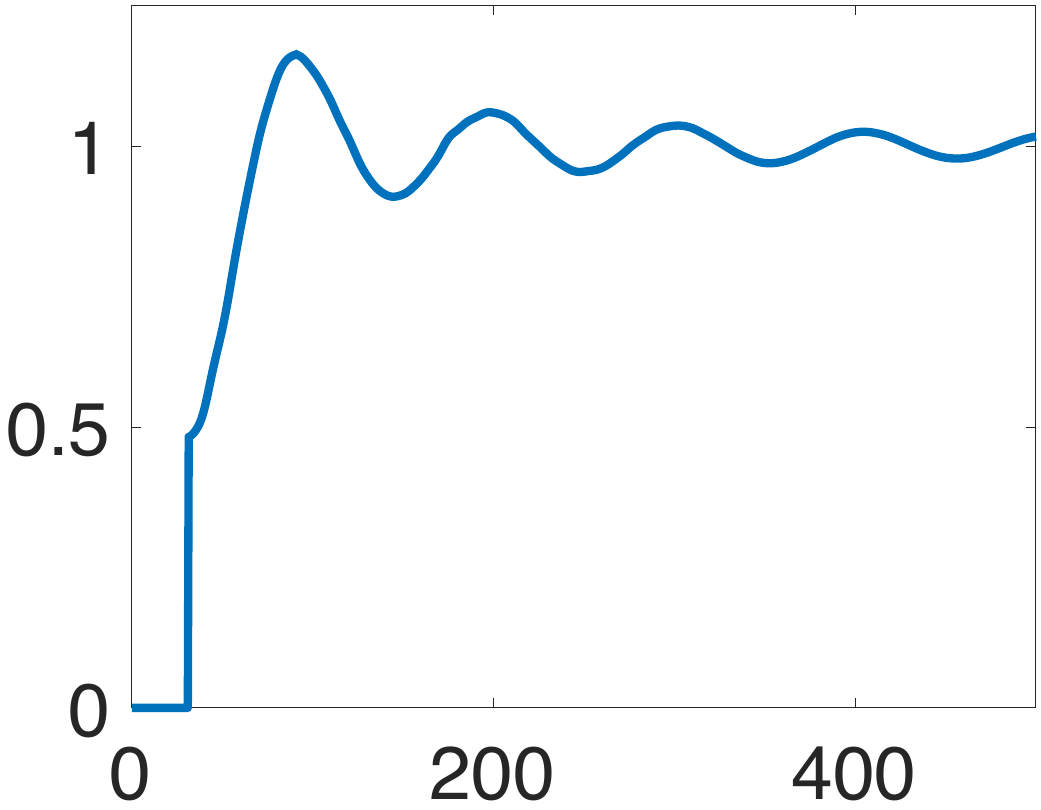}
			\\
			\hline
			\\
			\multirow{ 3 }{*}{\rotatebox[origin=l]{90}{\textbf{{\color{magenta} \Large Two-Spokes }}}}
      &\rotatebox[origin=l]{90}{{\hspace{16pt}spectra}} 
%			& \includegraphics[height=1in]{figs/distributions/two_spec.png}
    \ifthenelse{\equal{\isarxiv}{1}}
    {
			& \includegraphics[height=1in]{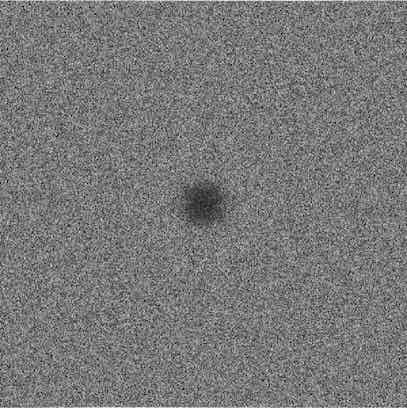}
			& \includegraphics[height=1in]{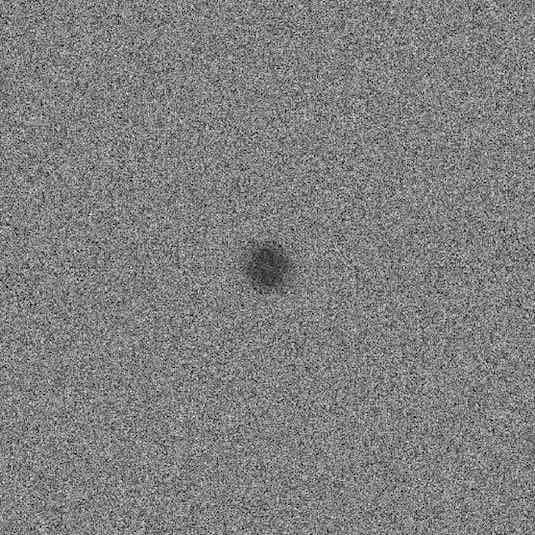}
			& \includegraphics[height=1in]{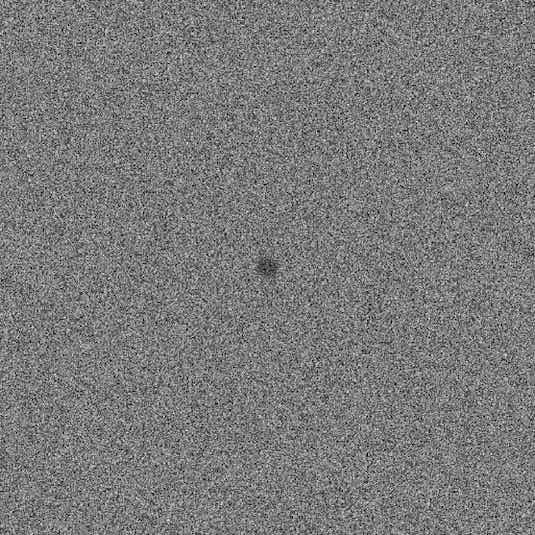}
	  		& \includegraphics[height=1in]{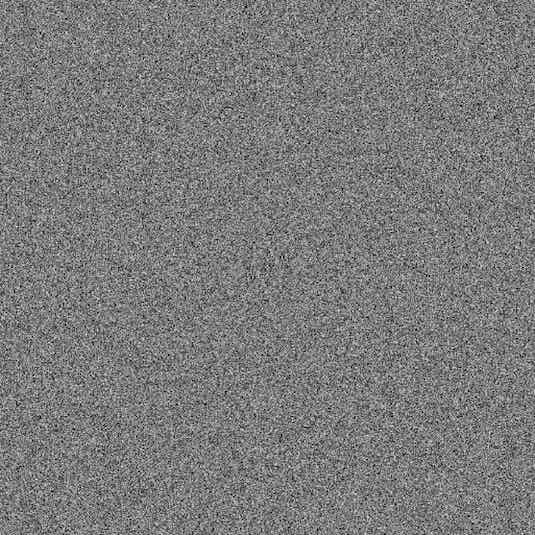}
		        & \includegraphics[height=1in]{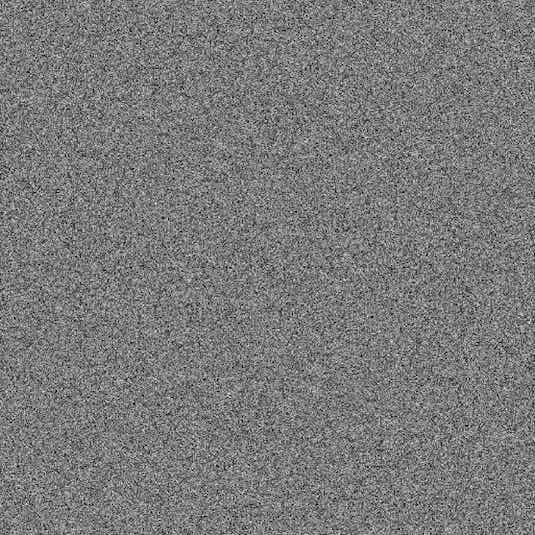}
		        & \includegraphics[height=1in]{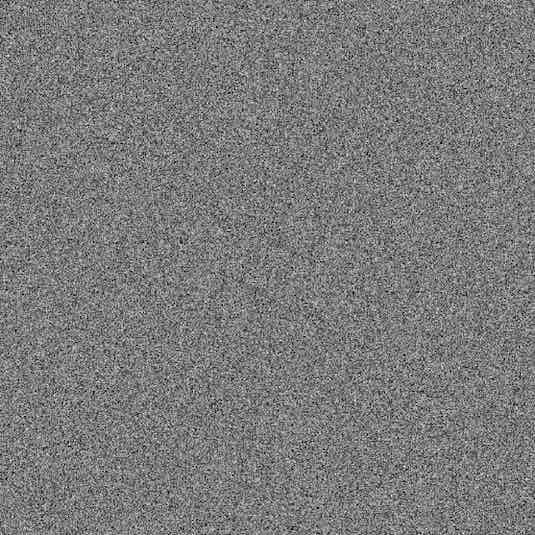}
   }
   {
			& \includegraphics[height=1in]{figs/distributions-p-midd/22d/slice513_1_1913.png}
			& \includegraphics[height=1in]{figs/distributions-p-midd/23d/fftslice.png}
			& \includegraphics[height=1in]{figs/distributions-p-midd/24d/fftslice.png}
	  		& \includegraphics[height=1in]{figs/distributions-p-midd/26d/fftslice.png}
		        & \includegraphics[height=1in]{figs/distributions-p-midd/28d/fftslice.png}
		        & \includegraphics[height=1in]{figs/distributions-p-midd/210d/fftslice.png}
  }
			\\
      &\rotatebox[origin=l]{90}{{\hspace{28pt}RDF}}
%			&\input{figs/distributions/2_P_397_2two_459569_beta_points_rdf}
			&\includegraphics[height=1in]{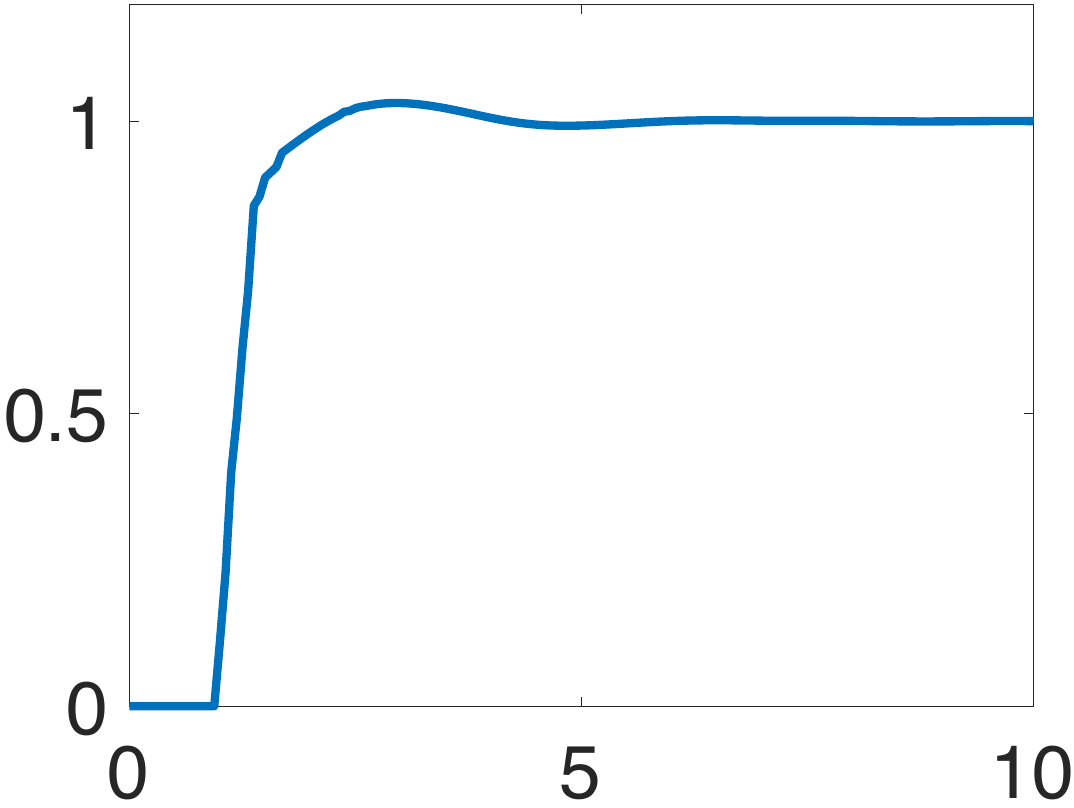}
			&\includegraphics[height=1in]{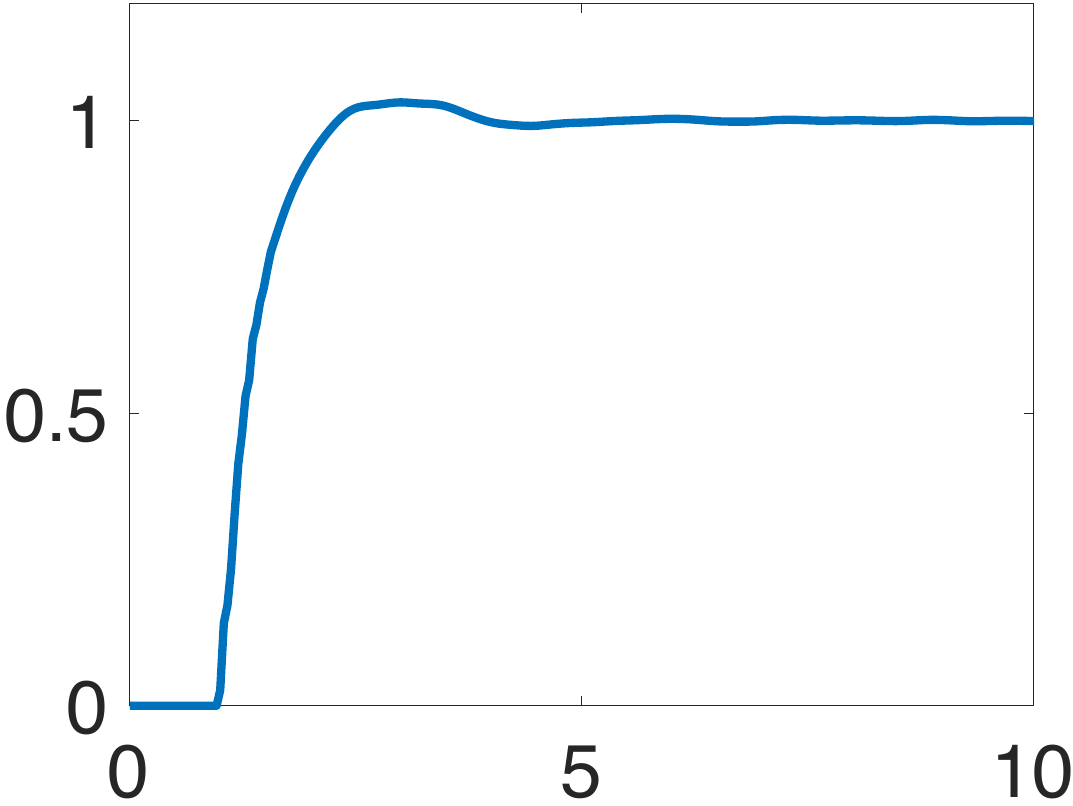}
			&\includegraphics[height=1in]{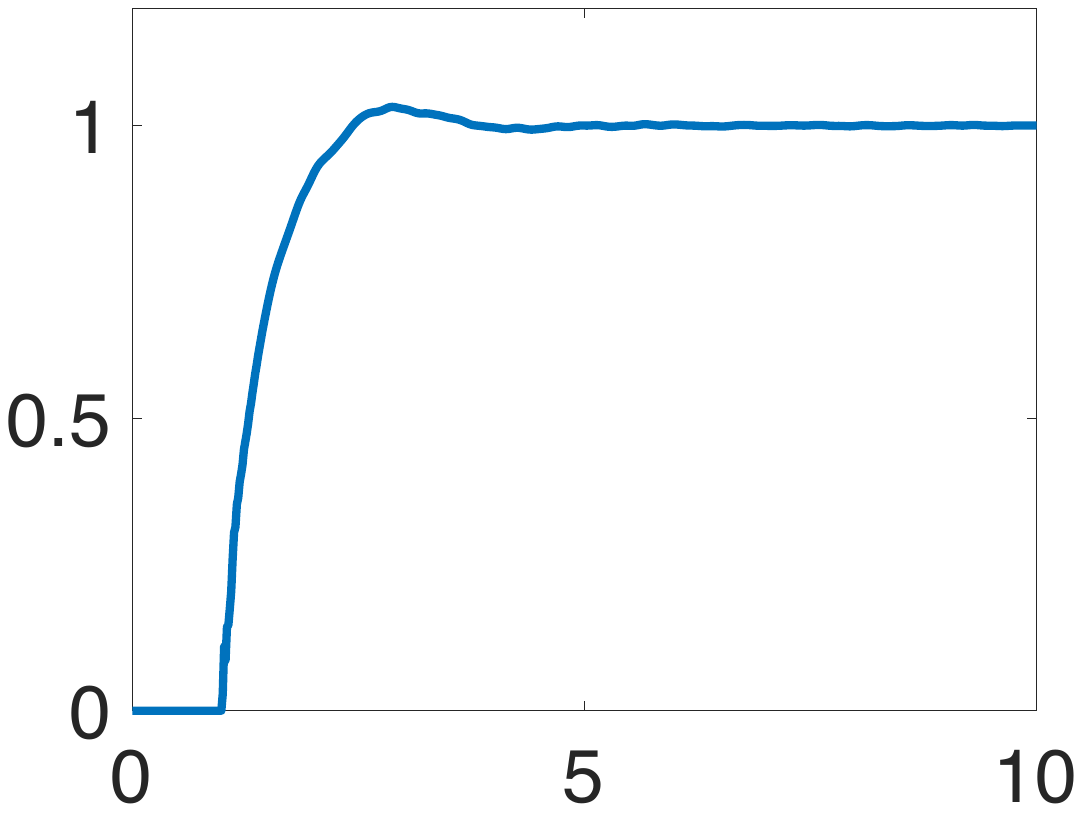}
			&\includegraphics[height=1in]{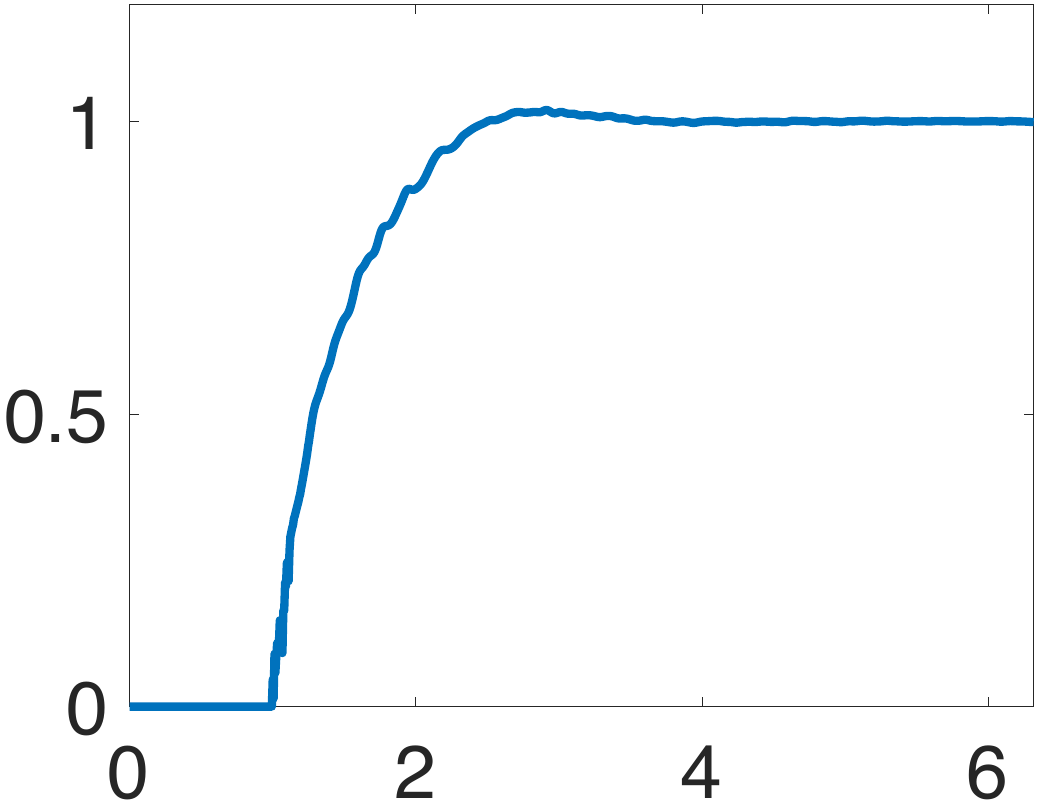}
			&\includegraphics[height=1in]{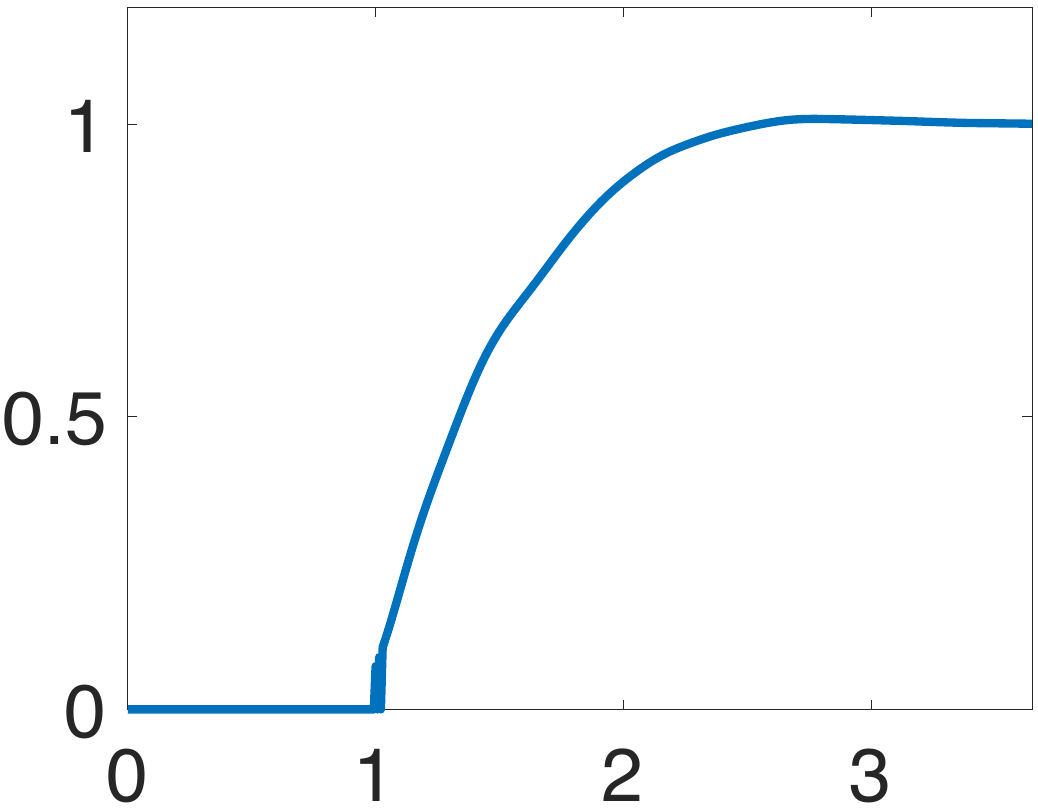}
			&\includegraphics[height=1in]{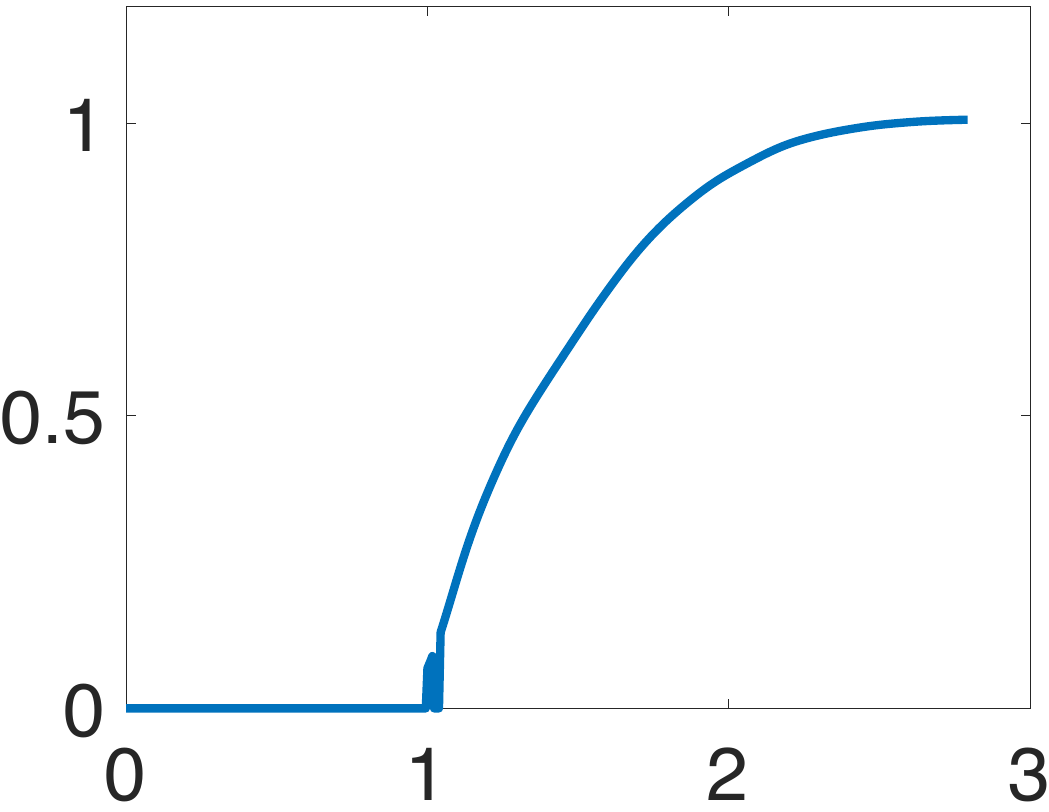}
			\\
      &\rotatebox[origin=l]{90}{{\hspace{28pt}RP}}
%			&\input{figs/distributions/two_rp}
			&\includegraphics[height=1in]{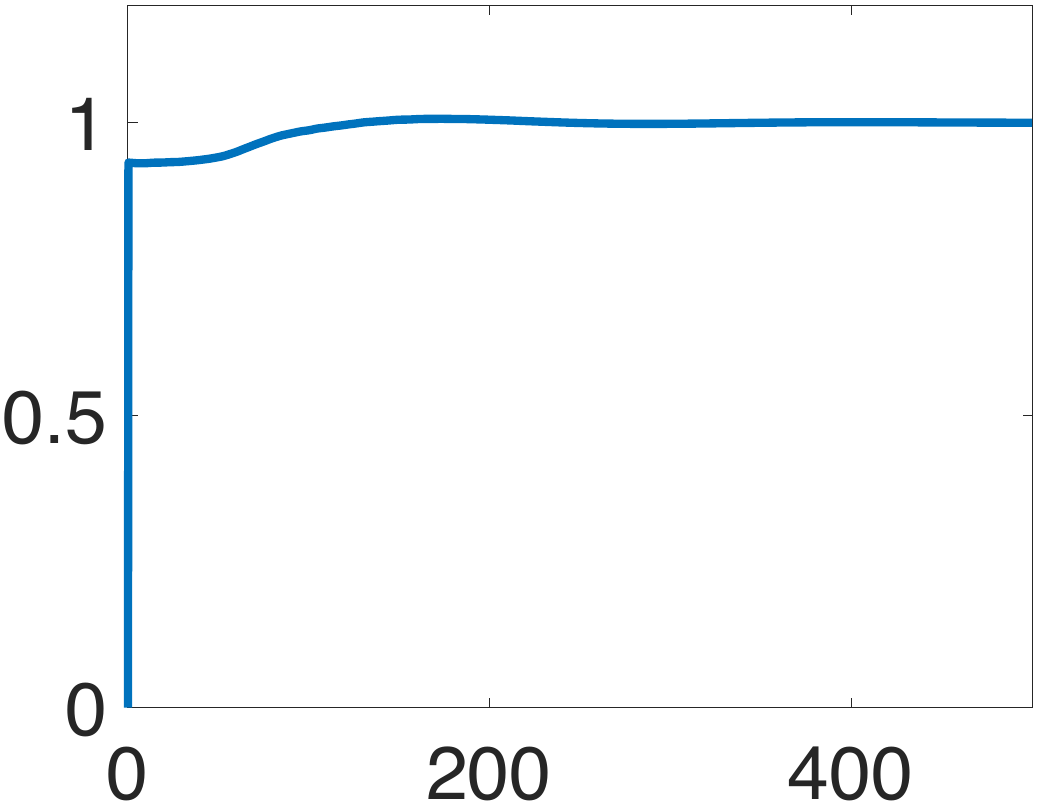}
			&\includegraphics[height=1in]{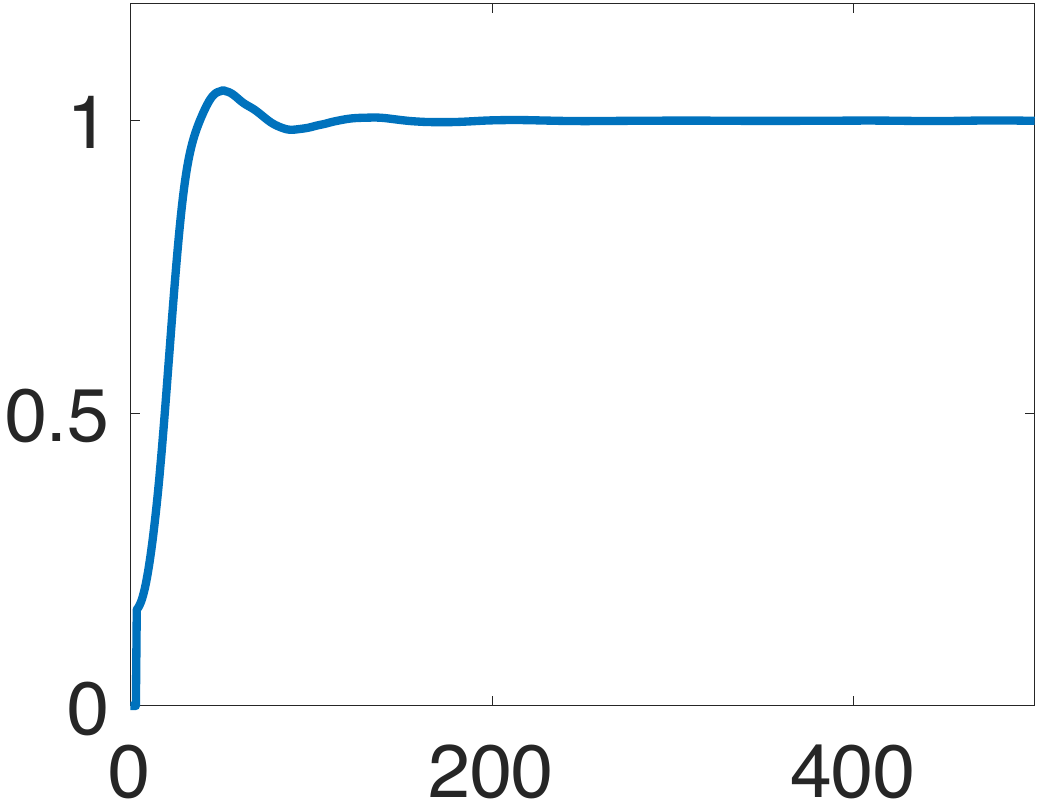}
			&\includegraphics[height=1in]{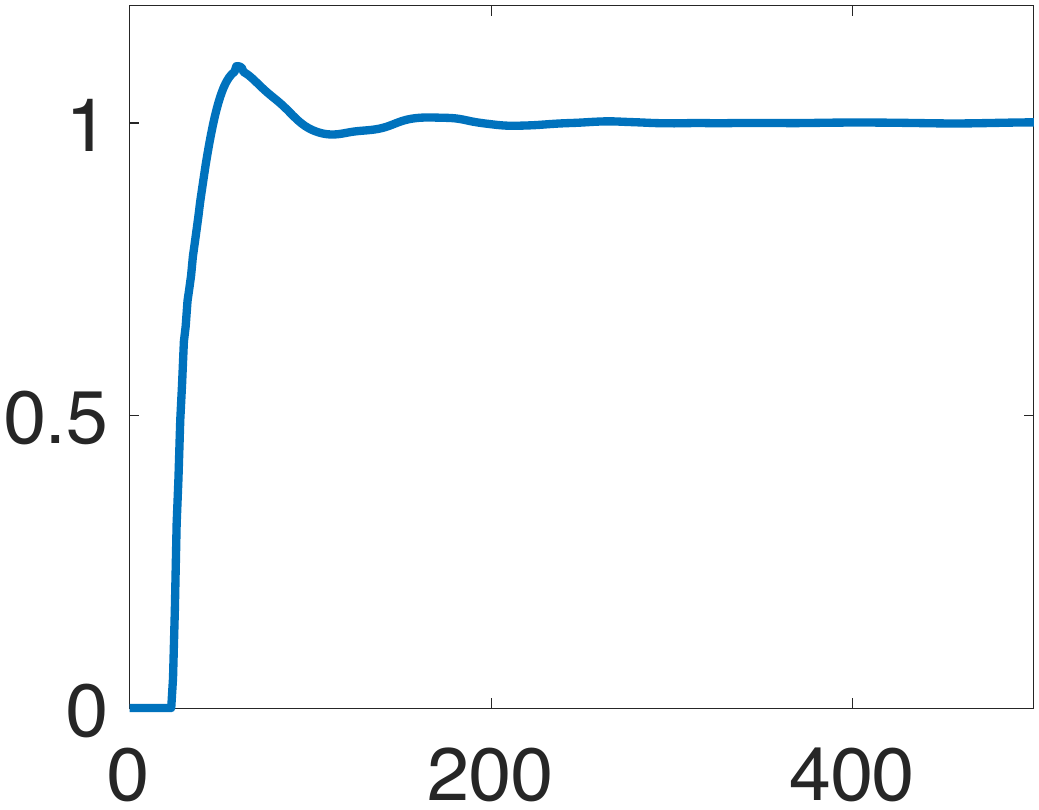}
		   	&\includegraphics[height=1in]{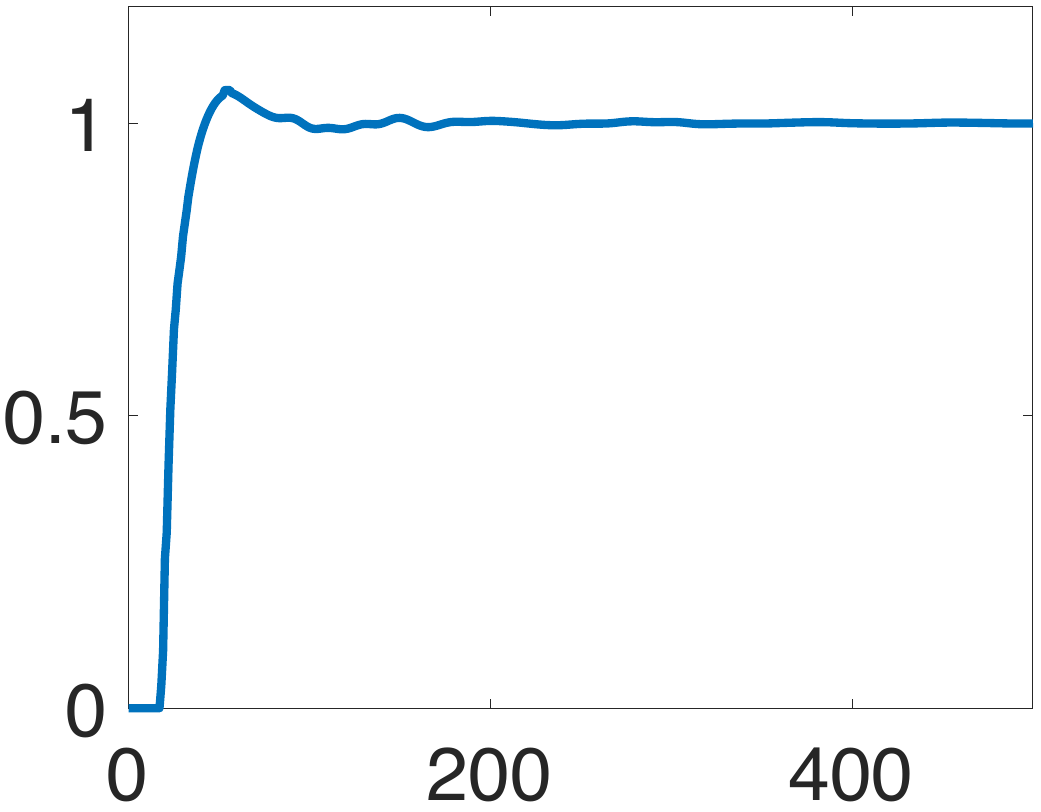}
            		&\includegraphics[height=1in]{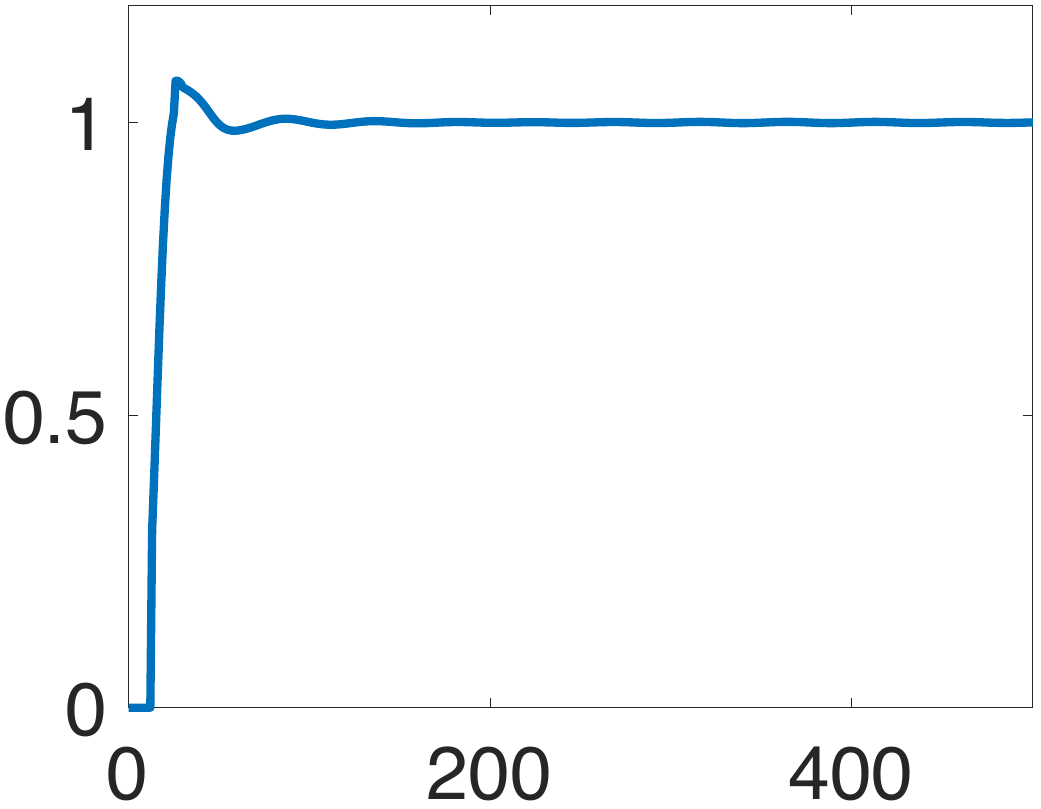}
            		&\includegraphics[height=1in]{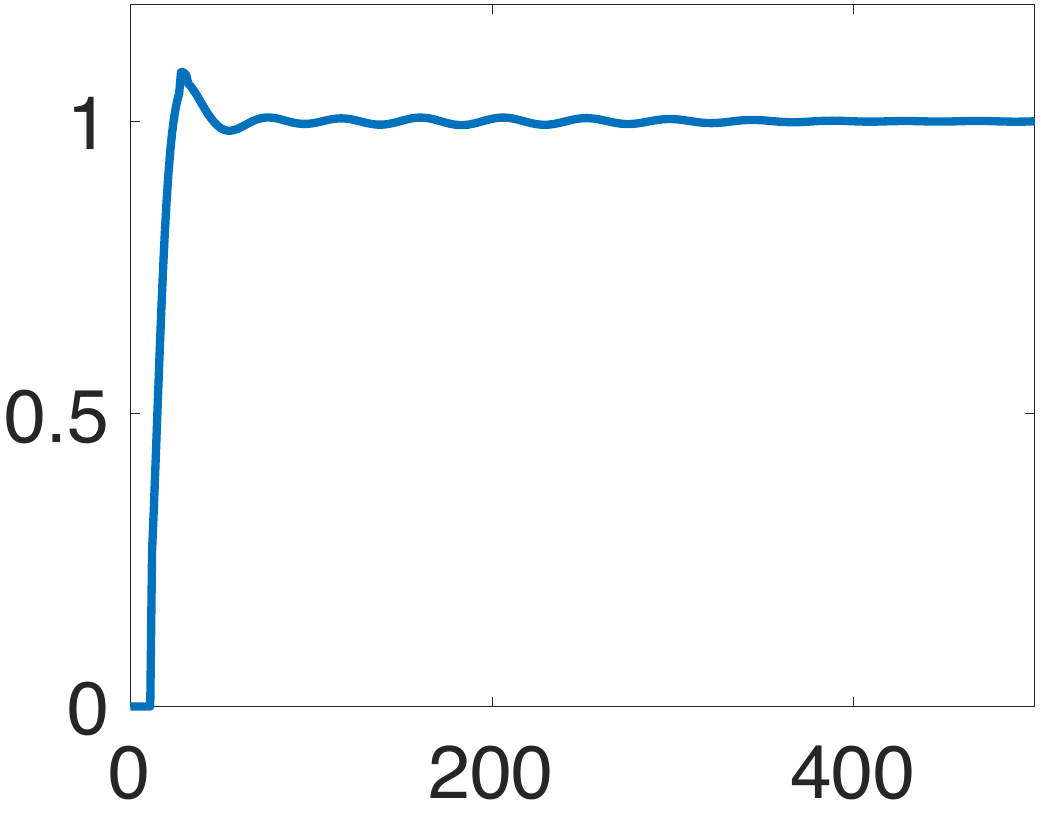}
		\end{tabular}
	}
	\Caption{Spectra, RDF, and RP for dimensions 2--10.}
        {%
 Spectra for $\dimnumber=$ 11--30 are similar to $\dimnumber=10.$ To keep the memory requirements tractable, spectral slices are computed directly in 2D using the Project-Slice Theorem (see, e.g.\ \cite{1451555}). All plots use $\samplenumber \approx$ 32,000, except $\dimnumber=2$ uses more for smoother figures. RDF and RP were produced using the TargetRDF software~\cite{Heck:2013:BNS}. For RP the DC component was filtered. Both RP and RDF were selectively smoothed and scaled. In RDF, ``1'' is scaled to $\rnumber$, the minimum distance between samples, and plots are truncated at absolute distance 0.5, to avoid the complication of the domain periodicity. For  \bridsonmethod\ and especially \linespokesalg, the RDF spike at $\rnumber$ is sharp, and the plotted heights depend on the width and alignment of histogram bins, so exercise caution in drawing conclusions.
\nothing{
\liyi{(October 22, 2017) IMPORTANT: use vector graphics, not raster images, for all RDF and RP plots, for better clarify-to-file-size ratio.}%liyi 
\scott{(Nov 1, 2017)done.}%scott
}%nothing
        }
 \label{fig:rdf_spectra_comparison}
\end{figure*}
}

\subsection{Output size}
\label{sec:outputsize}

%%%%%%%%%%%%%%%%%%%%%%%%%%%%%%%%%%%%%%%%%%%%%%%%%%%%%%%%

We describe the relationship between the output number of samples $\samplenumber$ and the sampling radius $\rnumber$ to allow the user to select the necessary $\rnumber$ to achieve the desired $\samplenumber,$ for example.
A sample point inhibits the introduction of nearby samples in a neighborhood related to distance $\rnumber$, so the volume of this neighborhood is roughly proportional to $\rnumber^\dimnumber.$ 
%Thus we should expect $\samplenumber \appropto 1/\rnumber^\dimnumber$ for fixed $\dimnumber.$ 
\[ \samplenumber \approx \knr / {\rnumber^d} \]
As $\dimnumber$ varies, the constant of proportionality $k$ will vary, depending on the inherent packing density of the dimension~\cite{Weisstein:1999:HP}, and also because of our achieved $\rratio.$ Experimentally, for \linespokesalg, 
\[
% \knrlinep % no other k are mentioned in the main body
\knr = (0.46 \dimnumber + 1.8 ) 1.04^\dimnumber \volunit,
\]
where $\volunit$ is the volume of a unit $\dimnumber$-sphere.

For \twospokesalg, the neighborhood around a point is roughly twice as large as \linespokesalg\ for the same $\rnumber$, so we expect $\samplenumber$ to be a factor of about $1/2^d$ smaller. In practice, 
$\knrtwo = (0.45\dimnumber + 2.5)(1.04/2)^d\volunit$.
% $\samplenumber_{\textrm{two}} \approx (1.1 + 0.004\dimnumber)(1/2)^\dimnumber \samplenumber_{\textrm{line}}.$
%
See \Cref{sec:outputsize_appendix} for additional details.

\subsection{Runtime scaling}
\label{sec:runtime}
We have three main observations:
\begin{itemize}
\item Runtime is linear in $d$ for high dimensions, using exhaustive neighbor search. Albeit runtime is quadratic in $n$: $\mathcal{O}(dn^2)$.
\item Runtime is $\approx \mathcal{O}(n\log n)$ for fixed $\dimnumber$, using $k$-d trees.
\item The crossover is about $d=7$  for $n=200,000$, meaning $k$-d trees are faster than exhaustive search for $d < 7$. The crossover dimension increases as $n$ increases.
\end{itemize}

\subsubsection{Complexity analysis}
The runtime is $T=\mathcal{O}( \samplenumber F + \dimnumber \hammerlimit \neighbors \samplenumber),$  where $\samplenumber$ is the number of samples generated, and $F$ is the time to find the $\neighbors$ neighbors of a single sample.
The  $\dimnumber \hammerlimit \neighbors \samplenumber$  term represents the time to throw and trim spokes, including those that miss the void.
Short sequences ($<m$) of miss spokes are charged to the next spoke that hits, just as in Bridson~\shortcite{Bridson:2007:FPD}.
% all
%Recall $m=12$ is constant.
% high
For large $d$, we have $N\le n$ and exhaustive search has $F = \mathcal{O}(dn)$;  
thus $T=\mathcal{O}(dn^2)$.
% low
For small fixed $d$, with $n\gg 2^d$ and $n\gg N$,
using $k$-d trees $F\approx \mathcal{O}(\log n + \neighbors)$ and $\neighbors = \mathcal{O}(1)$;
thus $T\approx\mathcal{O}(n \log n)$.

\subsubsection{Experiments}
\begin{figure}[!tbh]
  \centering
   \subfloat[exhaustive array search]
   {
      \label{fig:scaling_na_body}
       \includegraphics[width=0.49\linewidth]{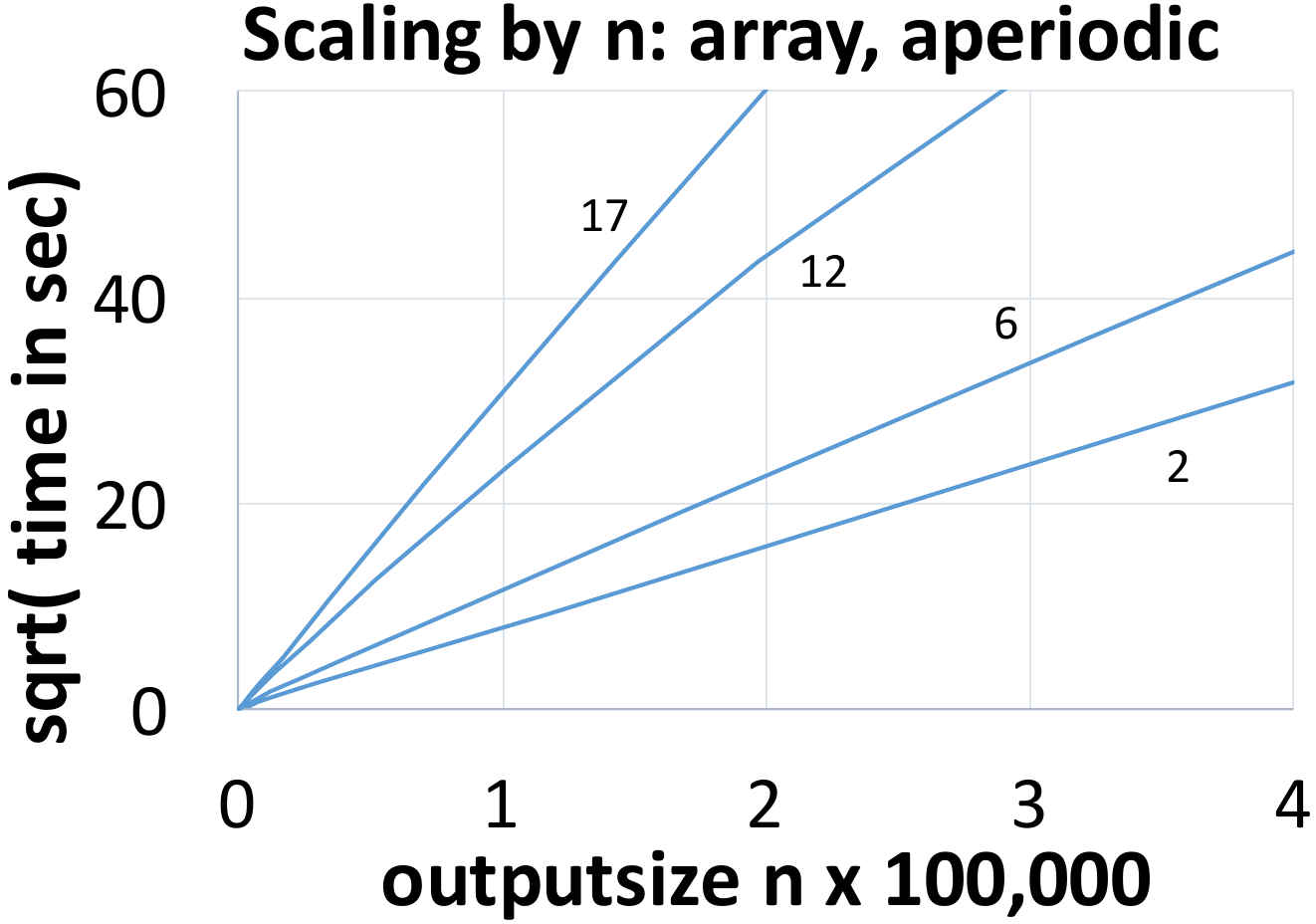}
    }
   \subfloat[$k$-d tree neighbor search]
   {
      \label{fig:scaling_pt_body}
       \includegraphics[width=0.49\linewidth]{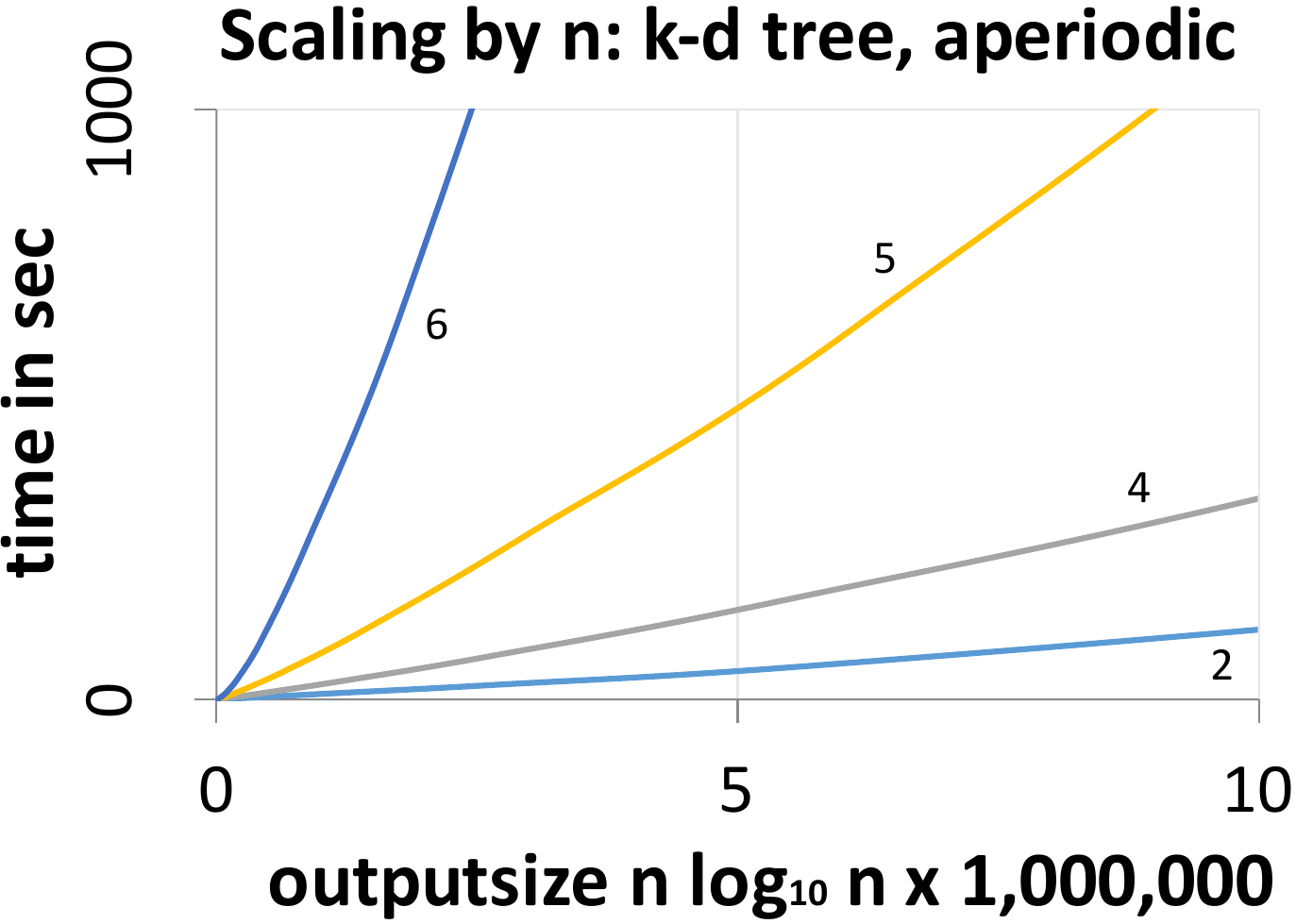}
    }
   \Caption{\Linespokesalg\ scaling by $\samplenumber$ for an aperiodic domain. }
   {Each trendline is labeled by the fixed dimension of the domain in that study. Left, straight trendlines illustrate $\mathcal{O}(n^2)$ runtime for fixed $d$ using exhaustive ``array'' search. Right, straight trendlines would illustrate perfect $\mathcal{O}(n\log n)$ scaling for $k$-d trees.}
  \label{fig:scaling_t_body}
\end{figure}

\begin{figure}[!tbh]
  \centering
  \subfloat[Runtime cross-over]
   {
    \label{fig:crossover_body}
       \includegraphics[width=0.49\linewidth]{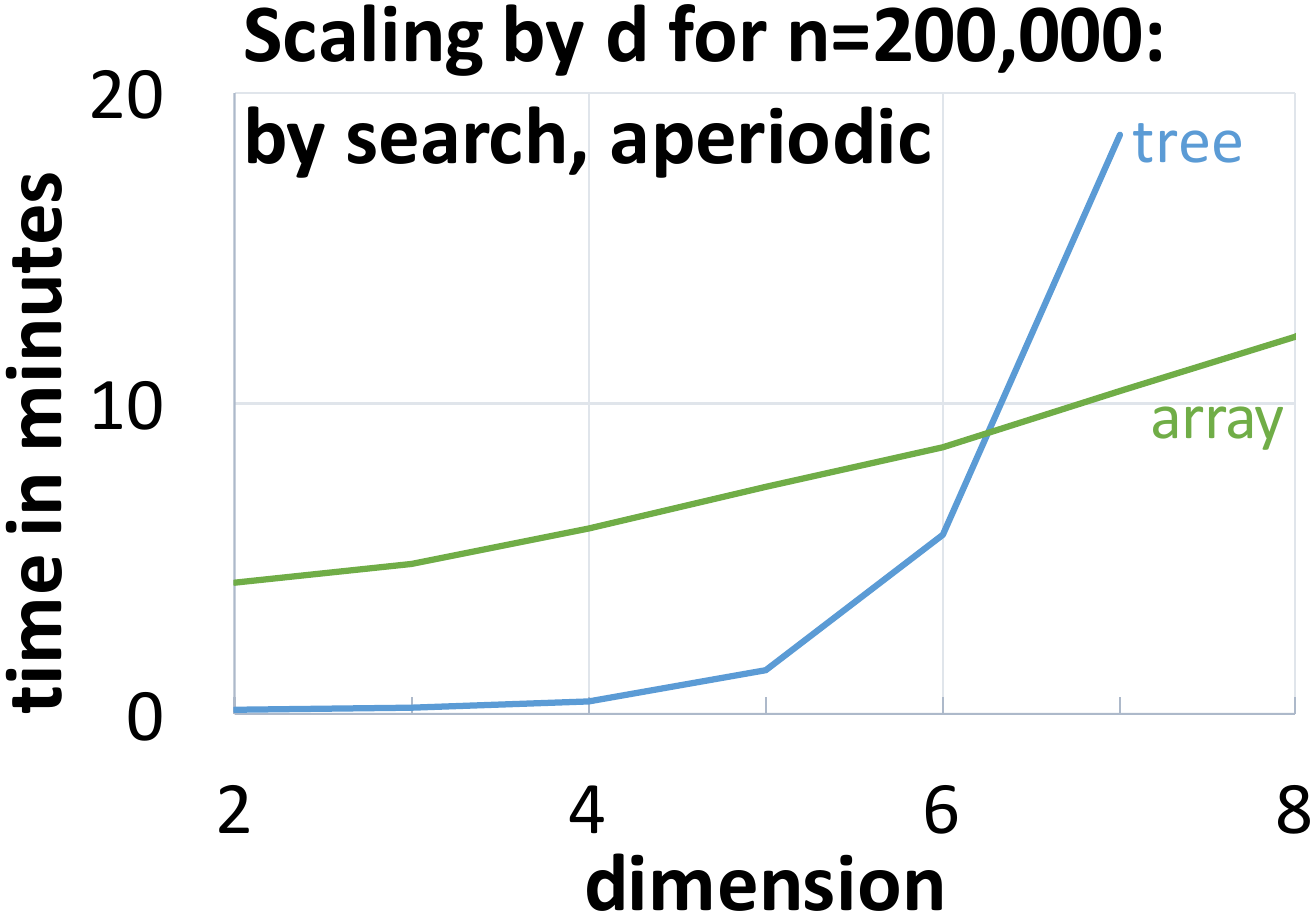}
    }
  \subfloat[Linear runtime in $\dimnumber$]
  {
     \label{fig:scaling_by_d_fixed_N_body}
       \includegraphics[width=0.49\linewidth]{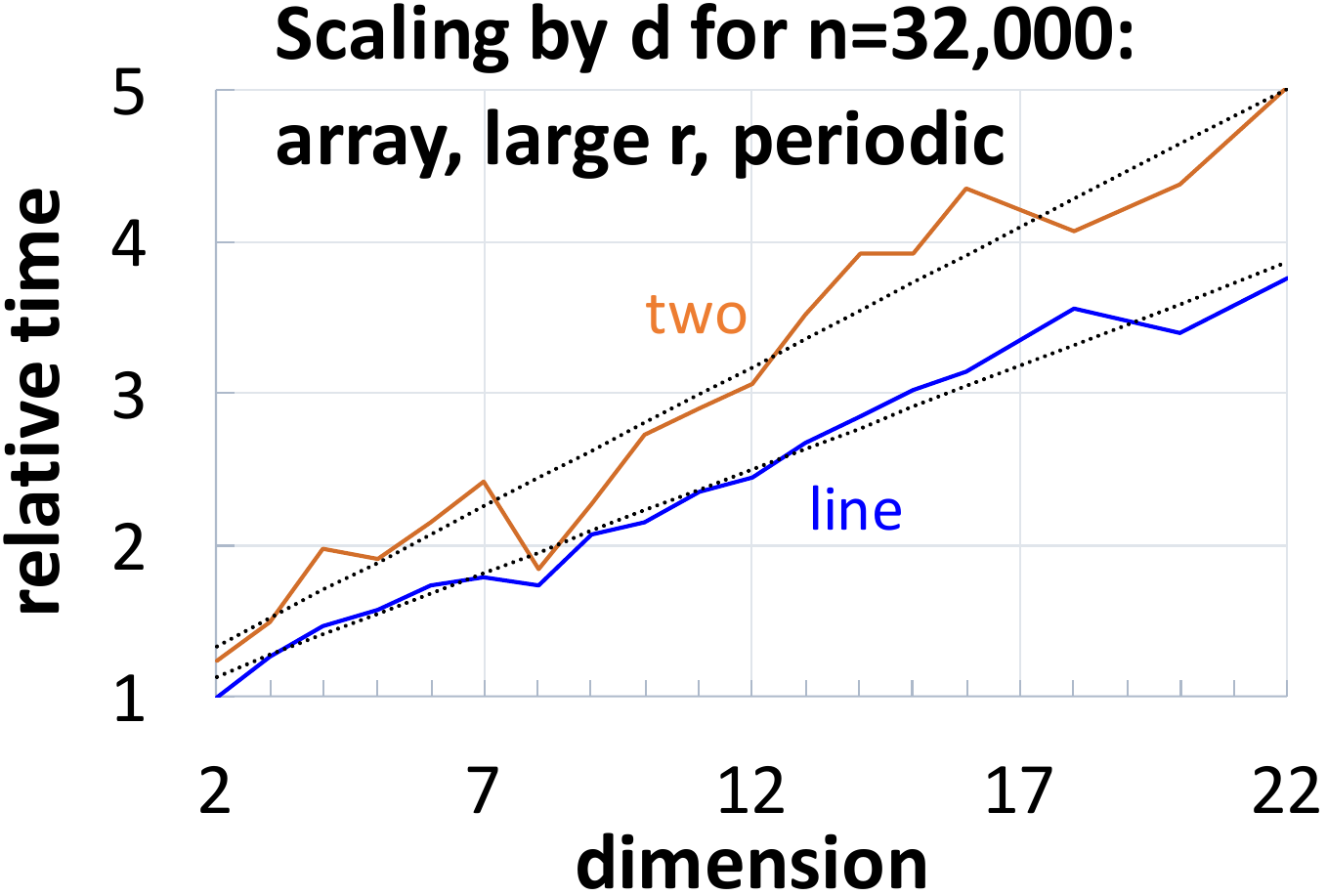}
   }
   \Caption{Fixed-$\samplenumber$ scaling by $\dimnumber$. Left shows that $k$-d trees save time in moderate dimensions. Right illustrates that runtime is linear in $d$ for exhaustive array search. The right graphs are not smooth because we used only a few trials, and perhaps because of dimensional-dependent memory layout and machine issues, e.g.\ $\dimnumber=8$ appears particularly efficient.}
   {}
  \label{fig:scaling_fixed_n_body}
\end{figure}

We verified these complexities experimentally.
\Cref{fig:scaling_t_body} shows the predicted $\mathcal{O}(n^2)$ and $\mathcal{O}(n\log n)$ runtimes.
\Cref{fig:scaling_fixed_n_body} demonstrates linear runtime in $d$ using exhaustive search.
Experimentally, the \linespokesalg\ runtime $T$ using exhaustive search (array) over aperiodic domains is  about
$\num{2.0E-09}	(	1	+	0.81	d	) n^2 +
\num{5.5E-08}	(	1	+	0.05	d	) Nn +
\num{2.4E-04}	(	1	+	0.05	d	) n$.
Experimentally, the runtime for $k$-d tree search over periodic domains is  about
$\num{7.8E-07}  dn (	0.12	\log_{10} n +	N).$
See \Cref{sec:runtime_details} for additional analysis and experiments, including higher dimensions, 
 periodic vs.\ aperiodic domains, and the number of neighbors by $d$.

%FIGURES for RUNTIME SCALING

\section{Applications}
\label{sec:application}
% ========== new =======
% Messages
% method enables two new algs, 
% approx Delaunay graph - radial searches provide significant neighbors found, without constructing geometry of cells
% global opt is new algorithm, improvement over classic, most significant result
% uses approx Delaunay graph, advancing front enables Vor cells, better subdivision of search space than rectangles
% high-dimensional spoke-darts blue noise used directly in robotic motion planning
% rendering: proof-of-concept for high-dimensional rendering. Used blue noise, but no apparent advantage over low discrepancy sequences
% high-dimensional used
%
% structure: motivation paragraph on why, 
% paragraph subsubsection on each application on *how*
%
We demonstrate the versatility of the \hidmethod\ approach, and the utility of its blue noise output.
We briefly summarize each application below; additional details are in the Appendices.
Delaunay graphs span $\dimnumber =$ 6--14, optimization $\dimnumber=$ 6--100, rendering $\dimnumber=$ 4--8, and motion planning $\dimnumber =$ 6--23.

The advancing-front \hidmethod\ process provides new algorithms for approximate Delaunay graphs and global optimization.  
Our algorithm for approximate Delaunay graphs is significant because it avoids the curse of dimensionality
and is dynamic. (By dynamic, we mean it can be updated quickly when inserting points, in contrast to some other known fast algorithms~\cite{Dwyer:1989:HVD:73833.73869}.)
We propose Opt-darts, a modification of the DIRECT global optimization algorithm~\cite{shubert1972sequential,jones1993lipschitzian}.
Opt-darts uses the dynamic approximate Delaunay graph, and produces a well-spaced random output distribution of samples.
For two standard test functions, we show that Opt-darts needs fewer function evaluations, and this speedup increases as the dimension increases.

Rendering and motion planning use our high-dimensional blue noise output directly as input.
We show that high-dimensional rendering is possible, but using blue noise provides no apparent improvement over standard inputs.
Being able to produce high-dimensional blue noise makes it feasible to run motion planning in high dimensions.

\subsection{Approximate Delaunay graph}
A Delaunay graph is just the edges in the Delaunay complex of a set of vertices (samples). These edges are dual to the $(d-1)$-dimensional facets of the Voronoi diagram. We find some of these facets, along with a point inside the facet. We shoot a spoke from a vertex, and trim it by each hyperplane separating the vertex from another vertex, retaining the hyperplane that trimmed it the most. The final spoke endpoint is a point inside a Voronoi facet, a witness to the fact that the facet exists in the Voronoi diagram. We tend to find the facets that subtend a large solid angle at the vertex, but miss some small facets.
See \Cref{sec:meshing} for details.

\subsection{Global optimization}
\label{sec:gopt}

\reviewer[AE]{
Please significantly tone down your achievements in the optimization section / application.
}%reviewer

The global-optimization algorithm DIRECT~\cite{shubert1972sequential,jones1993lipschitzian} is a classical and still-used  method for optimizing expensive black-box functions, such as finite element simulation runs.
It evaluates the objective function at each sample, and partitions the domain into hyperrectangles around each sample. A rectangle is recursively chosen for refinement if it is possible for the global minimum to lie inside it, assuming a fixed but unknown Lipschitz constant.
%This is determined by the function value at its sampled center, the distance from the center to a farthest point of the rectangle, and an unknown Lipschitz constant. 
Our variant, Opt-darts, partitions by Voronoi cells around each sample, instead of rectangles; these cells are implicit and only approximations are constructed. Opt-darts refines by adding new samples and updating nearby cell approximations. The new samples are chosen from among the spoke endpoints produced during the approximate Delaunay graph construction. Thus opt-darts uses both an approximate Delaunay graph, and generates an adaptive advancing-front random sampling.
%
% \Cref{fig:gopt_demo_fspace_body} contrasts the progression of samples.
%\input{gopt_demo_fspace_fig_body}

In our tests, Opt-darts more accurately represents sample neighborhoods, and new samples are more well-spaced, so fewer of them are needed.
This advantage becomes more pronounced as the dimension increases.
The disadvantage of Opt-darts is the higher computational cost in managing cells, but in the applications of interest the cost of the function evaluation at each sample dominates.
In \Cref{tab:gopt_performance}, we evaluate our method using community-standard high-dimensional test functions~\cite{jamil2013}.
These were designed to be challenging for global optimization by having many local minima or a small gradient  over most of the domain.
Most difficult global optimization problems have some combination of these two features.
For the Easom test function, in 6--10D speedups are 4--25$\times$. 
For the Bohachevsky test function, in 20--100D speedups are 5--27$\times$.
%Speedups increase with dimension.
See \Cref{sec:optimization} for details.

% ------------------------------------------------
\begin{table}[!htb]
	\centering
	\tabtextsize
	\resizebox{\columnwidth}{!}{%
		\begin{tabular}{| c || c | c | c | c |}
			\hline
			Benchmark $f$      & dimension      & DIRECT         & Opt-darts  & Speedup\\
			\hline
			\multirow{ 4}{*}{ \rotatebox[origin=l]{90}{\tiny{\hspace{10pt} Easom}} 			    \ifthenelse{\equal{\isarxiv}{1}}
{\includegraphics[width=0.35\linewidth,height=0.53in]{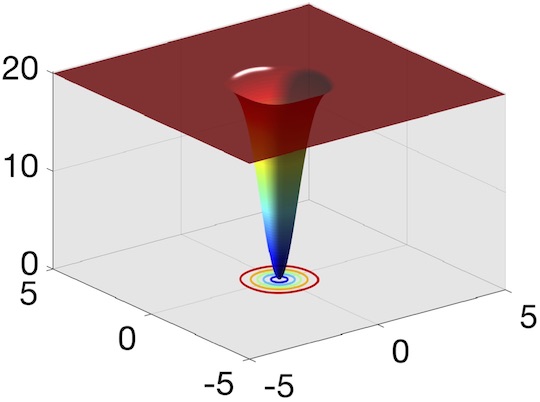}}
{\includegraphics[width=0.35\linewidth,height=0.53in]{easom.pdf}} } 		  
			& 6     &     5657 &    1320   &   4.3 $\times$ \\	
			& 7     &   20987 &    3276    &   6.4 $\times$ \\	  
			& 8     &   71677 &    4814    & 14.9 $\times$ \\			
			& 9     & 257539 &     14258           &  18.1       $\times$ \\					
			& 10	  & 837203 &  33852    & 24.7 $\times$ \\
			\hline	
			\multirow{ 4}{*}{\rotatebox[origin=l]{90}{\tiny{ Bohachevsky}}
			    \ifthenelse{\equal{\isarxiv}{1}}
			{\includegraphics[width=0.35\linewidth,height=0.53in]{bohachevsky-small.jpg}}     
			{\includegraphics[width=0.35\linewidth,height=0.53in]{bohachevsky.pdf}}}     
			& 20   &     5689 & 1269 &   4.5 $\times$ \\					
			& 40	  &   25807 & 2633 &   9.8 $\times$ \\
			& 60   &   63765 & 4345 & 14.7 $\times$ \\
			& 80	  & 122503 & 6246 & 19.6 $\times$ \\
			& 100 & 208185 & 7802 & 26.7 $\times$ \\
			\hline			
			
		\end{tabular}%
	}
	\Caption{Speedup of Opt-darts over DIRECT, 
	measured by the number of function evaluations needed to find an approximation 
	$f^\ast$ close to the true global minimum $\hat{f}$.  
	That is: $|f^\ast - \hat{f}| < 10^{-4}$
	where $f \in {[}0,20{]}$.
	}
	{Since Opt-darts is random, results are averages over 20 runs.}
	\label{tab:gopt_performance}
\end{table}
% ------------------------------------------------

\subsection{Rendering}
We integrated \hidmethod{} into the Mitsuba renderer~\cite{Mitsuba}. The torus-in-glass image (\figref{fig:teaser_rendering}) demonstrates a bidirectional path tracer using 8D samples corresponding to 2D for the sky emitter, 2D for the camera screen, and 2D for each bounce along each camera and light path. \Hidmethod, stratified sampling, and low-discrepancy sequences all produced images of similar quality.
See \Cref{sec:rendering} for details.

% just how, not why or its significance
\subsection{Motion planning}
Motion planning explores the high-dimensional configuration space of robots to find collision-free paths between the given starting and desired ending configurations. In the ``parallel RRT'' algorithm~\cite{park2016ieeetro}, this space is pre-sampled by blue noise, and multiple threads explore connect-the-dots paths. Sometimes, because the configuration space has narrow regions and fine features, the pre-sampling is too coarse to determine if the path between two nearby points is collision-free. In that case, fine blue-noise samples are adaptively added. We did motion planning for a challenging 23D problem (\figref{fig:teaser_motion_planning}), and a well-known suite of  6D benchmark scenarios~\cite{sucan2012the-open-motion-planning-library}. 
See \Cref{sec:motion_planning} for details.

% ========== old =======
%
%We apply spoke-darts to Delaunay graph generation (Appendix~\ref{sec:meshing}), global optimization (Appendix~\ref{sec:optimization}), rendering (Appendix~\ref{sec:rendering}), and motion planning (Appendix~\ref{sec:motion_planning}).
%In particular, we use our spoke-dart \emph{techniques} for constructing approximate Delaunay graphs of prescribed sites.
%We use our spoke-dart \emph{blue noise outputs} for the paths in rendering and the allowable configurations in motion planning.
%Global optimization uses the well-spaced outputs of a modified spoke-dart technique.
%% All these applications rely on the samples saturating the domains. (but Delaunay graph spokes are not well spaced, so the samples are not saturated. Need blue noise on sphere surface.
%Delaunay graphs span $d =$ 6--14, optimization $d=$ 6--15, rendering $d=$ 4--8, and motion planning $d > 20$.
%

%%  (\Cref{sec:high_d_meshing}), ($d=$ 6--15, \Cref{sec:optimization})  ($d > 20$, \Cref{sec:motion_planning})
%
%\nothing{
%\liyi{(December 3, 2013)
%
%Need to emphasize/highlight why blue noise is needed for these applications.
%
%(December 25, 2013)
%
%I phrased above.
%Let me know if accurate.
%}%liyi
%
%\mohamed{Yes, that's correct.}
%\scott{I modifed the statements.}
%}%nothing

\section{Conclusions and Future Work}
\label{sec:conclusion}
\label{sec:limitation}
\label{sec:future_work}

We present \hidmethod{} 
as a new framework for high dimensional sampling.
The method combines the advantages of state-of-the-art methods: the locality of advancing-front and the dimension-mitigation of \kddarts, specifically line-samples.
We provide \hidmethod\ as open-source software\nothing{, SpokeDartsPublic}~\cite{spokedartspubliccode}.
%We demonstrate its usefulness by generating well-spaced blue noise distributions.
To our knowledge, we provide the first algorithm for high dimensional blue noise with provable guarantees of local saturation.
\Linespokesalg\ uses the same advancing front approach as \bridsonmethod, but, by using line samples, it produces a median saturation about the same as \bridsonmethod{} in one dimension lower.
We also produce blue noise with less significant artifacts.
We have the option to avoid the traditional distribution spike at the disk radius and corresponding oscillations in the spectra.
We demonstrate \hidmethod's generality by adapting it for a variety of applications, including generating high-dimensional adaptive blue noise for global optimization.
%We provide three distribution variations: \linespokesalg, \favoredspokesalg\ and \twospokesalg. Each has their advantages.
%\Linespokesalg\ is the fastest and simplest way to produce traditional MPS-like blue noise in high dimensions. 
%\Favoredspokesalg\ provides step blue noise, and its distributions are the most invariant across dimensions.
%\Twospokesalg\ produces a softer step in the RDF, directly producing ideal spectra for anti-aliasing~\cite{Heck:2013:BNS,Subr:2013:FAS:2461912.2462013}.
%Local saturation is probabilistically guaranteed, and the user may increase parameter $\hammerlimit$ to achieve higher saturation.
Our algorithm uses linear memory, and is computationally efficient in high dimensions, up to the efficiency of finding nearby neighbors.
We speculate that approximate nearest neighbors may improve scalability in moderate dimensions, but not high dimensions.

%The algorithms for \favoredspokesalg{} and \twospokesalg{} sampling are manually crafted to achieve the target step and smooth noise characteristics.

A potential future work is a universal algorithm that can automatically tune for a continuum of properties, analogous to Jiang et al.~\shortcite{Jiang:2015:BNS}.
%nothing
% Scott: skip this 
%
%Our analysis can provide only a probabilistic, not a deterministic, guarantee that the distribution has reached saturation.
%Even though this guarantee is still a vast improvement over prior dimensionality-agnostic methods such as brute-force dart throwing \cite{Cook:1986:SSCG}, we would like to investigate further on the theory side for deterministic guarantees.
It may be possible to produce a closer approximation to the true Delaunay graph 
by searching in a blue noise set of spoke directions.
This could be generated by point-sampling the surface of a unit sphere, using spokes that are great-circle arcs. 
We speculate that approximate Delaunay graphs may be better than $k$-nearest neighbors for some computational topology and manifold learning problems, especially when data are non-uniformly spaced.
%, because they explicitly consider all directions.
%The benefit is that our Delaunay graph considers all directions;
%in contrast, for a point near a dense cluster, $k$-nearest neighbors can miss significant neighbors in directions opposite to the cluster. 
% Scott: skip this, too hard and too general of a problem
%We seek to find proximity search algorithms that are effective in our setting.
%
%\nothing{\liyi{(January 13, 2014) I am not sure what is the point of this paragraph.}
%\Spokedarts{} may also benefit numerical integration.
\nothing{
\laura{I feel that it is important to make the point that \spokedarts{} used in the optimization could have a significant influence on global optimization in high dimensions which has always been a challenging problem
}%laura
\liyi{This is an excellent fit for future work.}
}%nothing
% scott, agreed, but AE asks for language to be "toned down", so just say we improve DIRECT.
High-dimensional global optimization is challenging,  
and Opt-darts demonstrates an improvement over DIRECT for two well-known test problems. 
Future research directions include cell selection criteria and parallelization.
%
%Further analysis and applications of this novel approach will be a topic on future research.
%
%In addition to sampling applications in geometry and optimization that we have explored, a potential future direction is to apply
\nothing{\liyi{(June 23, 2016) We have rendering in \Cref{sec:rendering}.}
A potential application for
high dimensional blue noise sampling is rendering.
}%nothing
We briefly touched on using high-dimensional blue noise for rendering; there is the potential for future work in Monte Carlo integration \cite{pilleboue2015variance} and low discrepancy sequences \cite{Keller:2012:AMC}. 
In our current implementation for motion planning we precompute all samples.
We are investigating the possibility of adaptive sampling by exploiting the similarity between our method and tree growth.

\nothing{
\liyi{(December 26, 2013)
I know numerical integration is not part of our applications, but I wonder if it can be worthwhile to talk about it, either here or as a future work in \Cref{sec:future_work} (already done).
}%liyi

\mohamed{
Spokes are awesome for integration on a local neighborhood. See my Early career proposal for a comparison against $k$-d darts for estimating volumes of overlapping sphere. We have a separate paper for this application. We may mention it with some figures briefly here if that's Ok with Scott and you think it would help this submission. We will publish another paper about global optimization yet we have it here as an application that introduce Opt-darts so mentioning integration results should be OK as well, I guess! I just don't want to go to argument like .. Oh run experiments for integrating some problems using spokes with various dimensions. We got that request for $k$-d darts and delayed publication for about a year. In addition I don't think higher dimensional spoke are efficient for integration .. just line spokes. I think recursive $k$-d darts is the ultimate solution for the integration problem.
}

\liyi{(December 28, 2013) OK for simplicity let us just mention this as a future direction.}
}%nothing

\section*{Acknowledgements}
%We are grateful to Gamito and Maddock for freely providing their software.
%
We thank the authors of \bridsonmethod\ \cite{Bridson:2007:FPD}, TargetRDF \cite{Heck:2013:BNS}, and PSA \cite{Schlomer:2011:ASA} for making their software available, and the reviewers for their helpful feedback and suggestions.

{
\fontsize{6}{6}\selectfont
% \small
% include this if ASCR funding WAS used
This material is based upon work supported by the U.S. Department of Energy, Office of Science, Office of Advanced Scientific Computing Research (ASCR), Applied Mathematics Program. 
% always include this
Sandia National Laboratories is a multi-mission laboratory managed and operated by National Technology and Engineering Solutions of Sandia, LLC., a wholly owned subsidiary of Honeywell International, Inc., for the U.S. Department of Energy's National Nuclear Security Administration under contract DE-NA0003525.

This paper describes objective technical results and analysis. Any subjective views or opinions that might be expressed in the paper do not necessarily represent the views of the U.S. Department of Energy or the United States Government.
}
\normalsize

\bibliographystyle{acmtog}
{
\bibliography{common/paper}
}
\normalsize

% appendix needs to start on a new page, because otherwise some figures in appendix b results_appendix don't get displayed and aren't grouped on a pages right.
%\cleardoublepage

% ----------------------------------

\ifdefined\supp
\ifthenelse{\equal{\supp}{1}}
{
\clearpage
\appendix

\section{\nothing{Favored-Spokes and Two-Spokes for \liyi{(October 22, 2017) Better fit into one line}}Soft Blue Noise}
\label{sec:alg_variants}
%motivation/intro
We seek a better blue noise spectrum than \linespokesalg\ or \bridsonmethod\ produces, where ``better'' is defined in the following sense.
The main drawbacks of those distributions is a large spike in the inter-sample distances near $r$, and corresponding oscillations in the radial power.
To remove this, we first consider changing the distribution for choosing a random sample from a spoke to be more uniform by volume.
This, by itself, proved insufficient to remove the spike.
We experimented with additional rules such as skipping short spokes; see
\favoredspokesalg\ \Cref{sec:stepbluenoise}. 
Although this algorithm has a unique advantage of a median saturation that is invariant by dimension, it should mostly be considered a stepping-stone towards \twospokesalg{} in \Cref{sec:softbluenoise}.
We found that taking a second spoke was simpler and more intuitive than the skip rules, 
and generally produced a distribution with a flatter spectrum.
% ; see \twospokesalg\ \Cref{sec:softbluenoise}.

\subsection{Favored-Spokes}
\label{sec:stepbluenoise}

To generate step blue noise, we use the same top level algorithm: \Cref{alg:spoke}.
However, we seek to \emph{avoid the spike} in the RDF distribution at $\rnumber$.
We use different rules to accept and sample from spokes:
\begin{itemize}
	\item place samples on short trimmed spokes less often,
	\item place samples nearer the center of uncovered intervals.
\end{itemize}
This reduces the number of disks nearly-$\rnumber$ apart, but does not eliminate them because a spoke might be close to a disk without intersecting it.
%The intuition is that if a spoke is trimmed at both ends by disks, then placing a sample near either end creates a nearly-$\rnumber$ distance between disk centers. 
%If such a spoke is short enough, then even the midpoint is near both disks.
%One might think that this would avoid distance-$\rnumber$ disks altogether, but this is not the case.
%Because we do not track how far a spoke is from a \emph{side disk,} an extant disk that comes close to crossing the spoke, we place a sample near one of these side disks often enough to create a reasonable number of almost-$\rnumber$ distances.
%
We must avoid any sharp cut-off values in the rules, because these would create new discontinuities in the RDF. 
%In our experiments, using fixed distance values reduced the discontinuity at $\rnumber$ but some discontinuities appeared at the fixed distances.
We arrived at the following ranges experimentally.
%We next describe the range combinations that worked well.

%\input{favored_greedy_fig}
%\input{favored_skip_short_fig}

\subsubsection{Skip short spokes}
We use spoke interval $\interval=[1,3.8]\rnumber$.
We never place a sample point farther than $3.4\rnumber$, but the spoke extends to $3.8\rnumber$ so we can detect if a sample point would be near an extant disk.

We have two skip rules.
The first rule is if a spoke is trimmed by any disk, then we discard it and treat it as a miss.
We do this until we have 6 successive misses.
After this we reset the miss count to zero and apply the second rule until we again get 6 successive misses.
See the \emph{open} and \emph{closed} sectors in  Figure~\ref{fig:favored_ranges}.

The second rule is spokes are discarded if they are \emph{short}. 
Spokes are discarded if their extent is less than $ 3.2\rnumber$ (length $< 2.2\rnumber$), and randomly discarded with decreasing probability if their extent is in $[3.2,3.5]\rnumber;$
i.e.\ always discarded if within the dark blue ring, and sometimes discarded if in the light ring,
in  Figure~\ref{fig:favored_ranges}.
The discard probability is zero at 3.5 and grows cubically to 1 at 3.2.
Experimentally, a cubic rate produced better output than a linear or dimensional-dependent rate.
\Cref{alg:trim_step_blue_noise} describes TrimSpoke with these rules in place.

\begin{algorithm}
  \Caption{TrimFavoredSpoke, for step blue noise.}{}
  \label{alg:trim_step_blue_noise}
  \begin{algorithmic} [1]
    \REQUIRE line spoke $\linespoke$ anchored at $\anchor$ for sample $\sample$
    \ENSURE uncovered segment of trimmed spoke $\linespokeprime$    
    \IF{TrimAnchor($\anchor$) = empty}
      \RETURN\ empty
    \ENDIF
    \STATE $\linespokeprime \leftarrow$ TrimInterval($\linespoke$) \pcomment not empty
    \IF{SkipTrimmedSpoke and WasTrimmed($\linespokeprime$) }
			\IF{$reject=6$}
			  \STATE SkipTrimmedSpoke $\leftarrow$ false
			  \STATE $reject \leftarrow 0$
			\ENDIF
			\RETURN\ empty
  \ENDIF
    \IF{IsShort($\linespokeprime$)}
	\RETURN\ empty
  \ENDIF
  \RETURN\ $\linespokeprime$
  \end{algorithmic}
\end{algorithm}

%\begin{figure}[tbh]
%  \centering
%
%  \subfloat[Untrimmed spokes]
%  {
%    \label{fig:favored_spokes_spokes}
%    \includegraphics[height=1.2in]{figs/handdrawn/favored-spokes.pdf}
%  }    
%  \subfloat[Short spokes]
%  {
%    \label{fig:favored_short}
%     \includegraphics[height=1.2in]{figs/handdrawn/favored-spokes-skipshort.pdf}
%  }
%  \Caption{\Favoredspokesalg\ geometry and sampling.}
%   {Untrimmed spokes lie in the light red area of \subref{fig:favored_spokes_spokes}. In  \subref{fig:favored_short}, spokes in the medium-light areas have some chance of being considered not-short and generating a sample.
%}
% \label{fig:favored_spokes} 
%\end{figure}

\begin{figure}[tbh]
  \centering
    \label{fig:favored_spokes_spokes}
    \includegraphics[width=0.7\columnwidth]{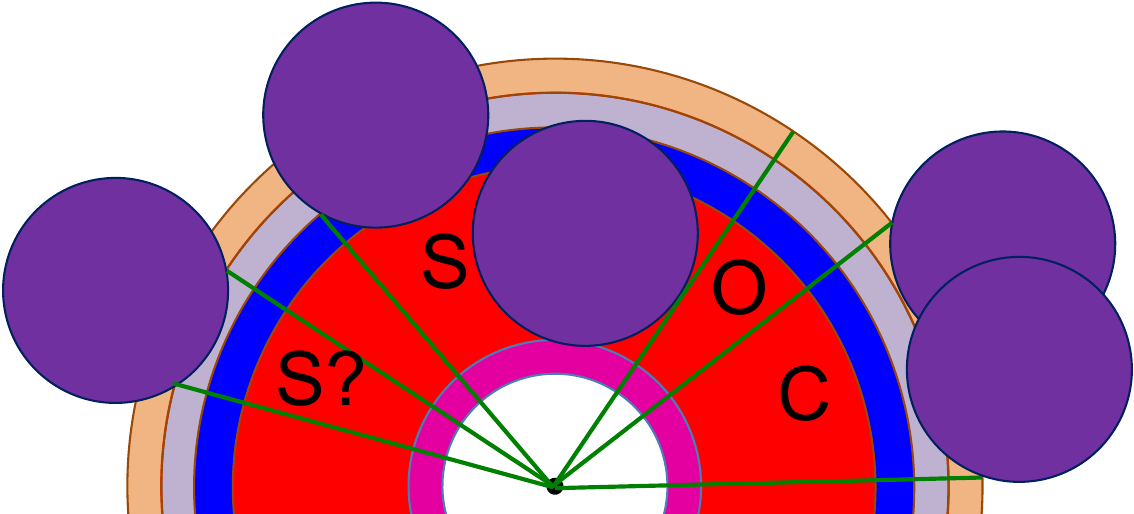}
  \Caption{\Favoredspokesalg\ ranges.}
   {Spokes in the ``O'' sector are open and ``C'' are closed; ``S'' are short, and ``S?'' are considered short with some probability.
   The following are the outer ring radii as a factor of $r$: 
   white = 1,  
   magenta = 1.3,
   red = 2.8,
   blue = 3.2,
    light blue = 3.5,
    brown = 3.8.
   }  \label{fig:favored_spokes} 
   \label{fig:favored_ranges}
\end{figure}

\subsubsection{Randomize spoke endpoints}
We shorten the ends of a spoke by a random amount in $[0.3,0.8]\rnumber$ to avoid sharp cutoff values.

\subsubsection{RandomSample for \favoredspokesalg\ and \twospokesalg}
A spoke actually trimmed by a neighbor is \emph{closed}; one with no disk intersections is \emph{open}.
For closed intervals, we place the sample point approximately uniformly by volume by the distance to the \emph{nearer spoke end}. 
This underweights the volume near the front sphere.
For open segments, we place the sample approximately uniform by volume by distance to the end near the anchor, then flatten off and ramp down; see \Cref{fig:ramp_distribution}.

%In 2D, uniform by volume is the standard ``triangular'' distribution, with a peak about half-way between the ends. 
%n higher dimensions, the sides of the triangle are not linear, but are polynomials in $d-1.$

%Recall that standard MPS places disks uniformly by volume and generates a large spike in the distribution at $r$.
%
%\Cref{fig:ramp_distribution} describes our distribution.

%, and \Cref{alg:ramp_distribution} gives pseudocode.

%\input{ramp_distribution_fig}
%\input{ramp_distribution_alg}

% samitch: for some reason, using width=0.5 breaks each figure into its own line on at least some latex implementations.
\begin{figure}[tbh]
  \centering
  \subfloat[Open spokes]
  {
    \label{fig:open_spokes}
     \includegraphics[height=0.62in]{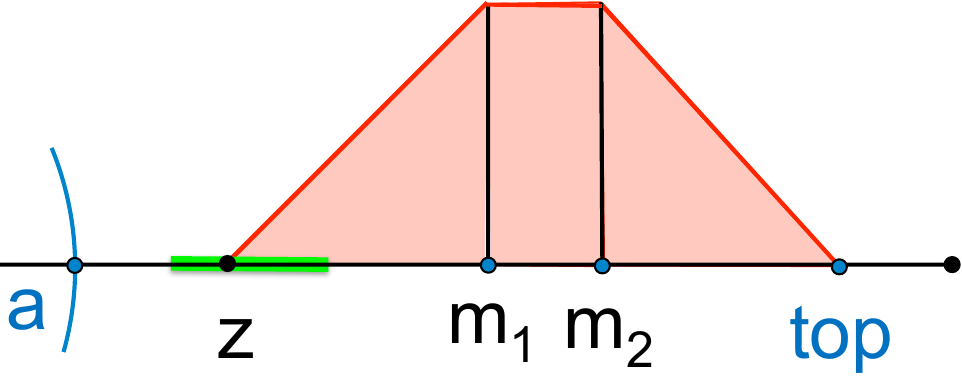}
  }
  \subfloat[Closed spokes]
  {
    \label{fig:closed_spokes}
    \includegraphics[height=0.62in]{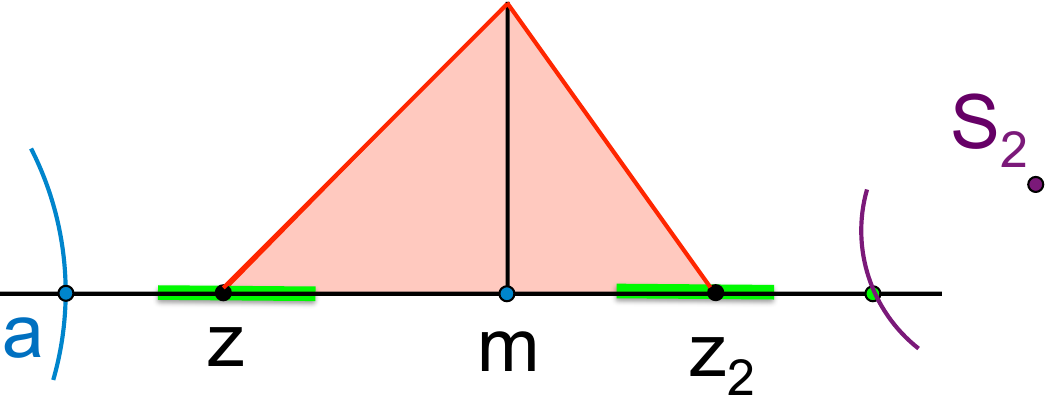}
  }    
  \Caption{Non-uniform sampling for \favoredspokesalg\ and \twospokesalg.}
   {The sample is chosen uniformly by volume under the red curves.  In 2d, from $z$ the curve is linear with slope 1 until $m$. In general, $y=(x-z)^{d-1}$.
 \subref{fig:closed_spokes}   For closed spokes, the midpoint $m$ is 0.6 of the way from $z$ to $z_2.$ 
 \subref{fig:open_spokes}   For open spokes, $m_1=0.54$ and $m_2=0.7$ . 
  For \twospokesalg\ the $z$ and top are the spoke endpoints. For \favoredspokesalg\  the $z$ are chosen uniformly to be distance $[0.3,0.8]r$ from the end of the spoke, on the green segments, and the top at distance $3.4r$ rather than $3.8r$.
}
 \label{fig:ramp_distribution} 
\end{figure}

\subsection{Two-Spokes for soft blue noise}
\label{sec:softbluenoise}
\Twospokesalg\ also uses the same top level algorithm: \Cref{alg:spoke}.
We generate an uncovered point as with \linespokesalg, without the skip rules of \favoredspokesalg.
We pick the final sample by taking a second spoke through this uncovered point; see \Cref{fig:two_spokes}.
This mimics our prior Two Radii~\cite{Mitchell:2012:VRPD} sampling:
the first spoke mimics finding a point uncovered by prior $2r$ disks, and the second spoke mimics the larger admissible region for centers of disks that can cover it.

\subsubsection{First spoke}
The first spoke extends from $2r$ to $4r$, with anchor at $2r$, and is trimmed by $2r$ disks.
We select the uncovered point $\sampleprime$ from it using the RandomSample of \favoredspokesalg.

\subsubsection{Second spoke}
The second spoke is centered on, and anchored by, $\sampleprime$ from the first spoke, with $I=[-2,2]r.$ 
After it is trimmed by $1r$ disks, we split it by $\sampleprime$ into two sides.
We pick one side uniformly by length, and select the final sample from that side's segment using RandomSample from \favoredspokesalg.
See \Cref{alg:secondspoke}.

\begin{algorithm}
  \Caption{SecondSpoke, generating a random uncovered point near the input uncovered point.}{}
  \label{alg:secondspoke}
  \begin{algorithmic} [1]
    \REQUIRE uncovered point $\sampleprime$
    \ENSURE output nearby uncovered point $\sampleprimeprime$
        \STATE $\spoke \leftarrow \funct{RandomSpoke}(\sampleprime, \interval=[-2,2]r)$
	\STATE $\spoke \leftarrow \funct{TrimSpoke}(\spoke, \anchor=\sampleprime, \neighbors(\sampleprime))$  \pcomment{never empty}
    \RETURN\ $\sampleprimeprime \leftarrow \funct{RandomSample}(\spoke)$
  \end{algorithmic}
\end{algorithm}

\subsubsection{Generalization}
\Twospokesalg\ may be generalized to give the option to soften the step in the RDF distribution by different amounts,
by parameterizing  by $\alpha \in[0,\inf)$ and $\gamma \in [0,1]$. 
\Linespokesalg\ corresponds to $\alpha=0$ and $\gamma=0$, and \twospokesalg\ as previously described corresponds to $\alpha=1$ and $\gamma=1.$
The first spoke has $\interval=(1+\alpha)r[1,2]$ and $\anchor=(1+\alpha)r$ and is trimmed by radius $(1+\alpha)r$ sample disks. The second spoke has $\interval=\gamma(1+\alpha)r[-1,1]$ and $\anchor=0$ and is trimmed by radius $r$ sample disks. 
This will ensure that the first spoke finds an uncovered point, at least $(1+\alpha)r$ away from all prior samples, and the chosen sample will cover it by its $(1+\alpha)r$ disk. The chosen sample will be at least $\rconflict \ge r$ away from any other sample. For small $\gamma$, the intersample distance will be even larger, by the triangle inequality: $\rconflict \ge (1-\gamma)(1+\alpha)$.  Thus $\rconflict \ge \max(r, (1-\gamma)(1+\alpha))$.
The form of the $\rratio^*$ guarantee is that $\rcoverage$ is likely at most twice the anchor distance of the first spoke:
 $\rcoverage \le 2(1+\alpha)r$.
 Thus for $\hammerlimit=12$, with probability $1 - 10^{-5}$ we have 
 \begin{equation}
 \rratio < \rratio^*=2 \min\left( (1 + \alpha), \frac{1}{1-\gamma} \right)
 \end{equation}

\section{Additional experimental results}
\label{sec:additional_results}
In this appendix, we show further comparisons between Point-Annulus, \linespokesalg, \favoredspokesalg, and \twospokesalg{}.
%\section{Bound Proofs for Section~\protect\ref{sec:parameters}}
%\newpage
\subsection{Output size data}
\label{sec:outputsize_appendix}

\Cref{sec:outputsize} gave approximate formulas for the number of output samples $\samplenumber$ by radius $\rnumber$ and dimension $\dimnumber$ for \linespokesalg\ and \twospokesalg\ for periodic domains.
Recall
$ \samplenumber \approx \knr / {\rnumber^d} $ with a different $\knr$ for each algorithm.
For \linespokesalg\ over periodic domains of dimensions 2--22, we have
$ 
\knrlinep = (0.46 \dimnumber + 1.8 ) 1.04^\dimnumber \volunit,
$
where $\volunit$ is the volume of a unit $\dimnumber$-sphere.
For \twospokesalg\, we have $\knrtwo = (0.45\dimnumber + 2.5)(1.04/2)^d\volunit$.

For bounded domains, part of a sample's neighborhood falls outside the domain, so we expect the same $\rnumber$ to produce a larger $\samplenumber$.
In practice, for \linespokesalg\ over dimensions 2--30, we observe 
$ \knrlinea \approx (0.0004 \dimnumber^4 - 0.027 \dimnumber^3 + 0.52 \dimnumber^2 -2.5 \dimnumber + 6.2 ) \volunit.$

The output size trends for \favoredspokesalg\ are similar to \linespokesalg, but with the constant of proportionality $\knr$ having less dimensional dependence. Experimentally, we observe 
\[
\knrfavored = (0.035\dimnumber + 1.15) 1.04^\dimnumber \volunit,
\]
See \figref{fig:kvsd} for a summary.
In addition, \Cref{fig:scaling_r} shows how the radius varies across dimensions for fixed $\samplenumber$ for our algorithm variants.

\begin{figure}[tbh]
  \centering
       \includegraphics[width=0.6\linewidth]{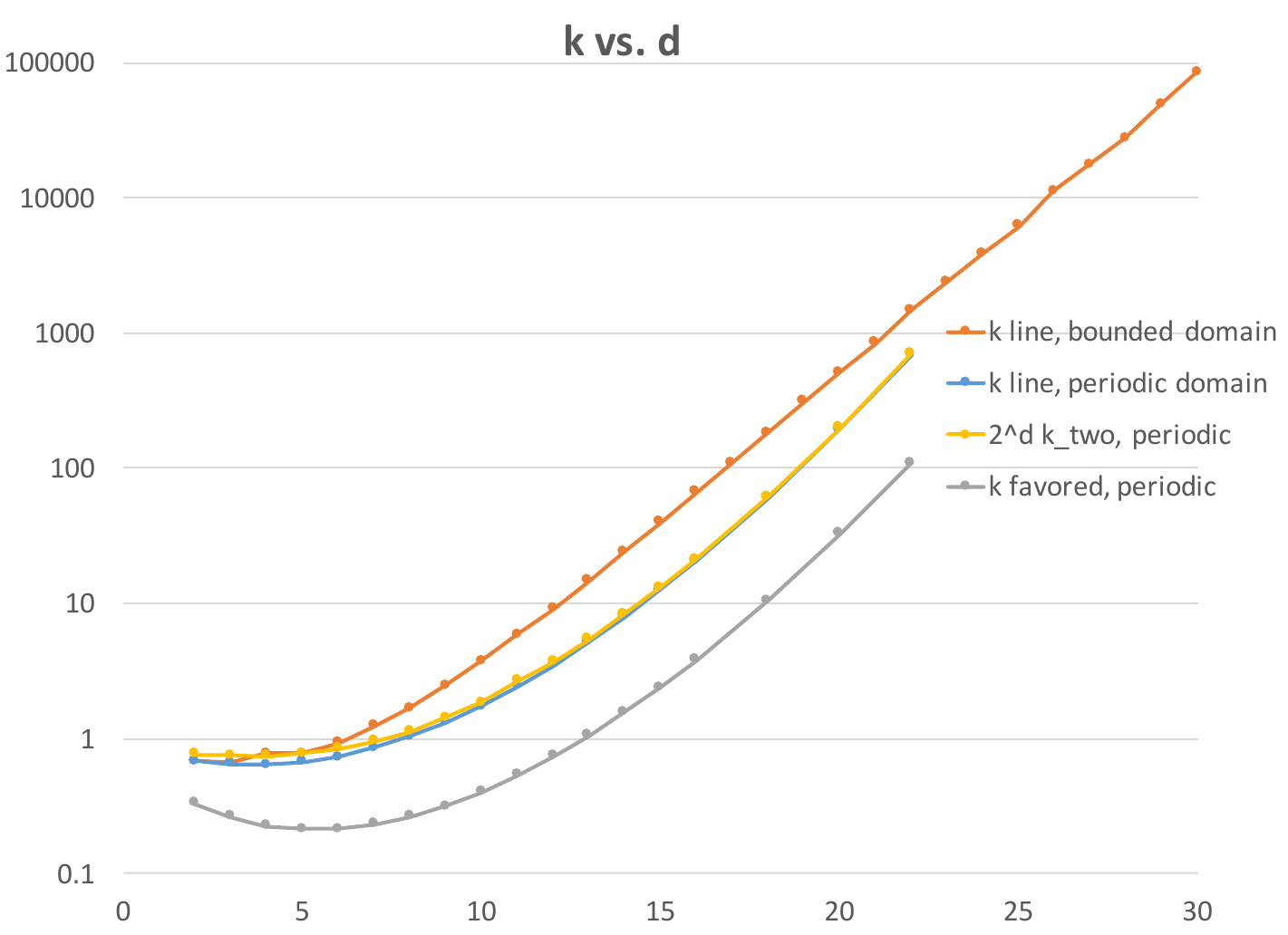}
   \Caption{Output size constants of proportionality. Number of samples $\samplenumber \approx \knr / \rnumber^\dimnumber$ where $\knr$ is plotted vs.\ $\dimnumber$.}{}
  \label{fig:kvsd}
\end{figure}

\subsection{Distribution}
\Cref{fig:coverage_ext} shows how increasing $\hammerlimit$ affects the output saturation.
As the dimension increases the distributions tend to resemble a Gaussian and become more sharply peaked.
While the theory guarantees are invariant in dimension, in practice \linespokesalg\ and \twospokesalg\ produce slightly larger $\rratio$ in higher dimensions.

\Cref{fig:betamedian} shows how the median beta varies by dimension, for $\hammerlimit=12$ and all methods. \Cref{fig:comparison_histograms} provides distribution details beyond the median. Specifically, it shows the distribution of the distance from each sample to its farthest Voronoi vertex and the distribution of distances from each Voronoi vertex to its nearest samples, using our algorithm's variations as well as Bridson's for $d=2,3,4,5$.

%We show spectra output for our reimplementation \bridsonmethodours\ of Bridson's \bridsonmethod\ in \Cref{fig:ourbridsonextra} over periodic domains.
For aperiodic domains the boundary significantly affects the distribution characteristics, especially for coarse samplings, so these results are mostly omitted. The exception is \Cref{fig:point_annulus_aperiodic}, which shows the spectra for Bridson's implementation run over aperiodic domains. 

Besides supporting periodic domains and $k$-d tree search, the implementations differ in the order in which the front is advanced. 
In our implementation samples become the active front disk in the order in which they were created. We continue to sample around the active disk until 30 consecutive misses, then remove the disk from the front. In contrast, \bridsonmethod\ visits front disks in random order: remove a random point from the queue as the active front disk, throw exactly 30 darts, then reinsert the front disk into the queue if any new sample was accepted. In our comparisons, this algorithmic difference does not affect the output distribution significantly.

Although we can create blue noise distributions in higher dimensions, our ability to analyze their $\rratio$ is limited by the challenge of computing Voronoi vertices. We use QHull, and for large $\dimnumber$ and reasonable $\samplenumber,$ QHull runs out of memory (and time).

%%%%%%%%%%%%%%%%%%%%%%%%%%%%%%%%%%%%%%%%

%% New figures --

\begin{figure}[tbh]
\centering
\subfloat[Total inserted points by $\hammerlimit.$]
  { \label{fig:histogram2_1}
     \includegraphics[width=0.48\linewidth]{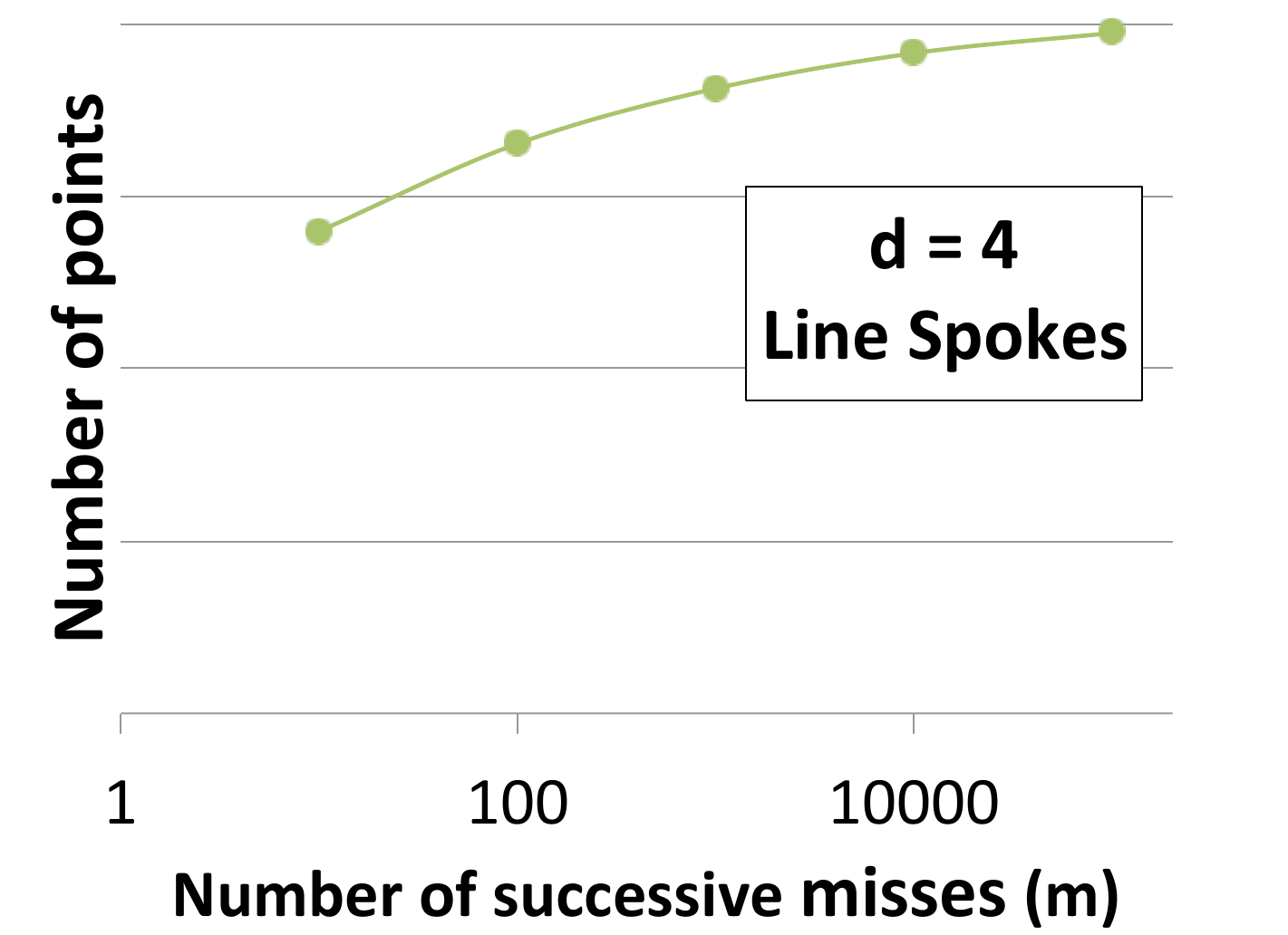} 
  }
\subfloat[$\rratio$ for different $\hammerlimit$]
{
    \label{fig:coverage}
     \includegraphics[width=0.48\linewidth]{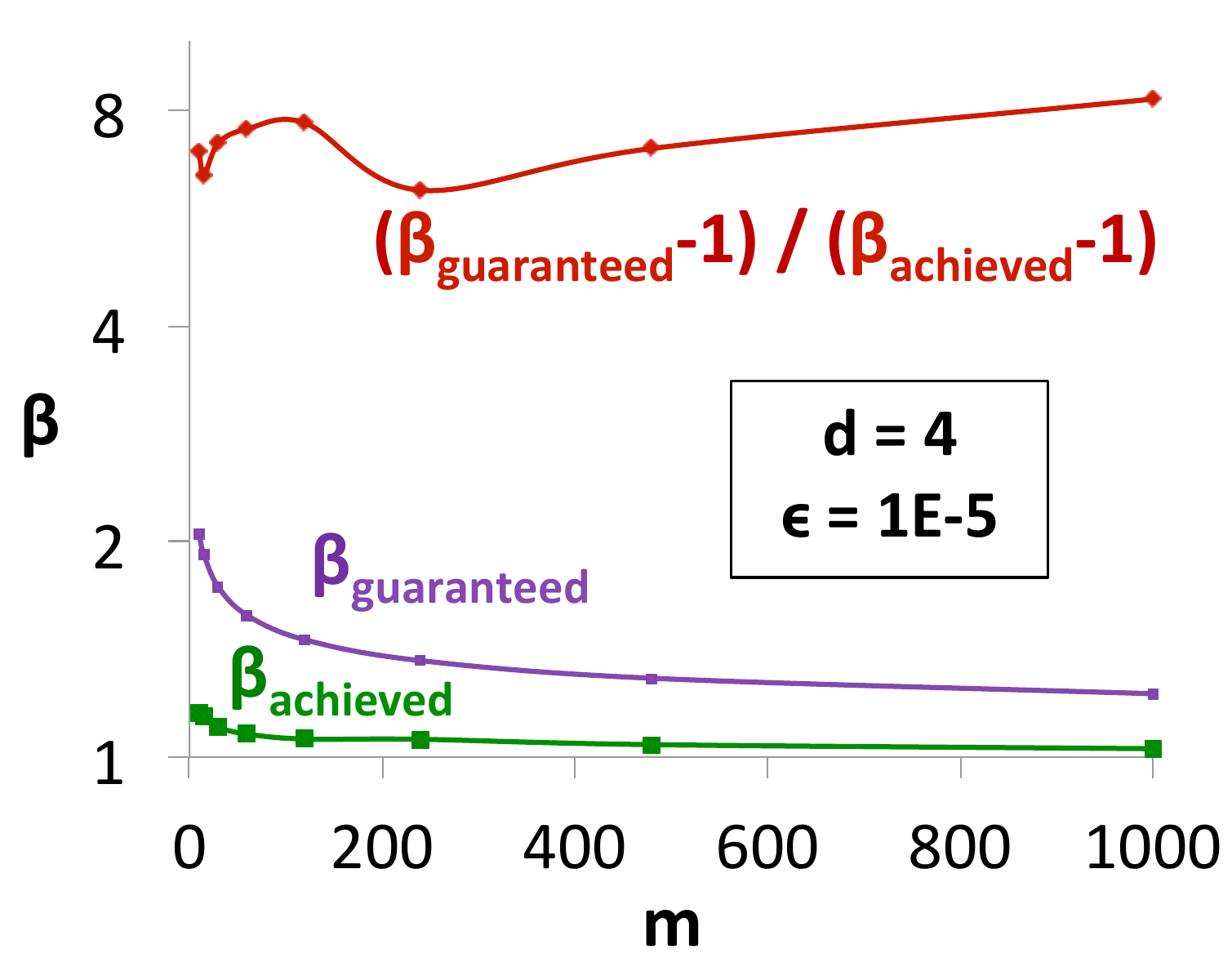}
}
 \Caption{Local saturation for \linespokestext{} in theory and practice by $\hammerlimit$.}
{Here $\rratio_{guaranteed} = \rratio^*,$ the probabilistically-guaranteed saturation upper-bound in theory. And $\rratio_{achieved}$  is the $\rratio$ observed in experiments.
In practice $\rratio$ is about $7 \times$ closer to 1 than the theory guarantee: e.g.\ $\rratio \approx 1.15$ for $m=12$.
%and are the coverage distances and inter-sample distances achieved in practice, and $\rratio_{Achieved}$ their ratio. 
%Recall $\rnumber$ is the disk radius.
%
%For example, for $\hammerlimit=60$ we have $\rratiopractice \approx 1.08$, almost maximality, whereas $\rratiotheory \approx 1.6.$
%Since $\rconflict \approx r$ in practice, it is $\rcoverage$ that is determining 
%$\rratio = {\rcoverage}/{\rconflict}$. 
%
%
%We see that many fewer samples are needed in practice than in theory. 
}
 \label{fig:coverage_ext}
 %BND_AdvFront\Results\CenterSearch\4d\CentreSearch_coverage_properties.xlsx
\end{figure}

%nothing

\begin{figure}[tbh]
  \centering
%    \subfloat[Bridson]
    {
       \includegraphics[width=0.65\linewidth]{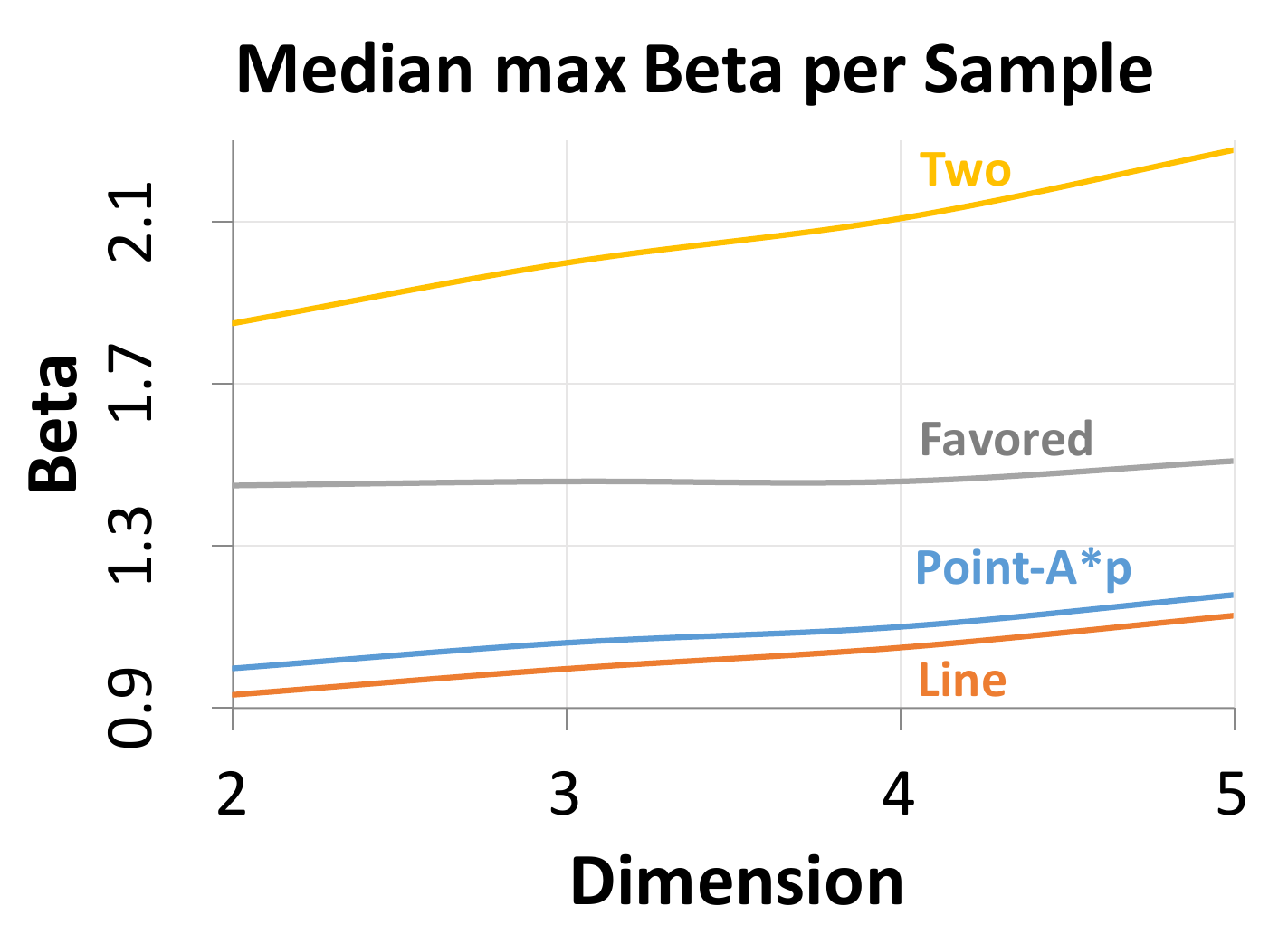}
    }
    \Caption{Trends in median $\rratio$ for $\hammerlimit=12$ by $\dimnumber$.}{}
    \label{fig:betamedian}
\end{figure}

{
	%\noindent %noindent causes extra vspace when the figure is input
	\begin{figure*}
		\centering
		\setlength{\tabcolsep}{3pt}
		\resizebox{\linewidth}{!}{
			\begin{tabular}{ccccccc}
				& &$\dimnumber=2$ &$\dimnumber=3$ &$\dimnumber=4$ &$\dimnumber=5$ &Percentiles \\
								\hline
								\\
				\multirow{ 4 }{*}{\rotatebox[origin=l]{90}{\textbf{\Large Sample to farthest Voronoi vertex distance}}}
				&\rotatebox[origin=l]{90}{{\hspace{16pt}\color{blue}Point-Annulus}}
				&\includegraphics[width=0.23\linewidth,height=1.2in]{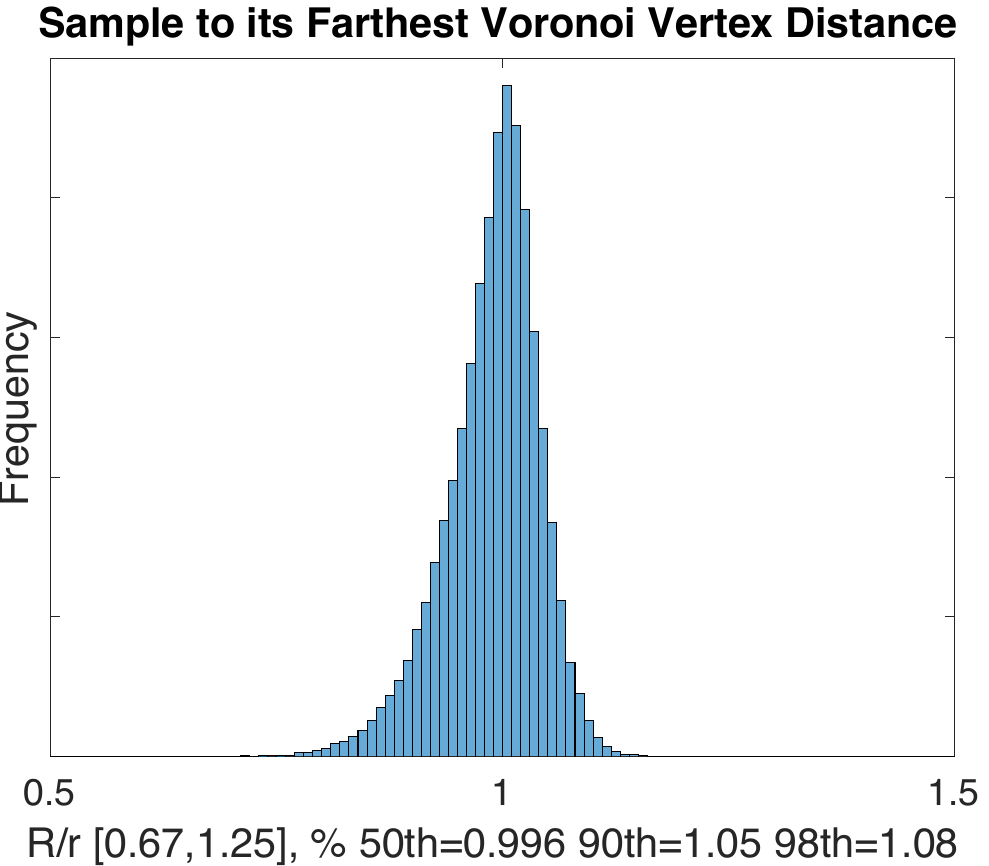}
				&\includegraphics[width=0.23\linewidth,height=1.2in]{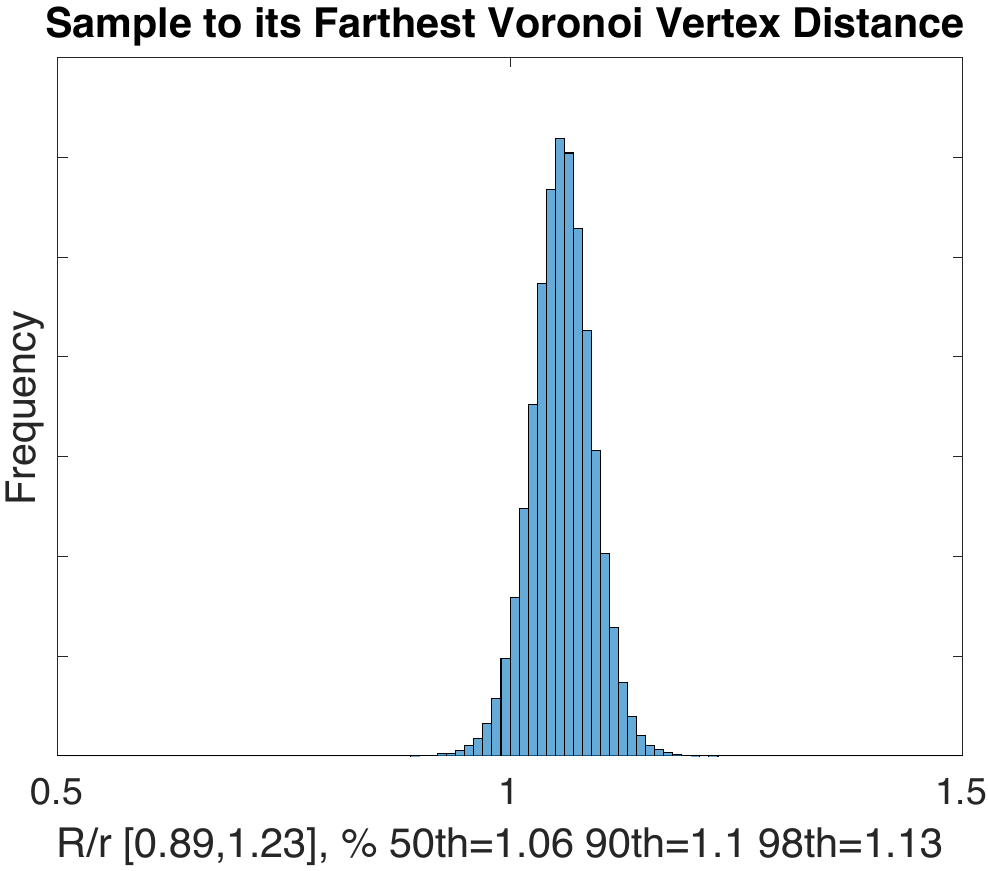}
				&\includegraphics[width=0.23\linewidth,height=1.2in]{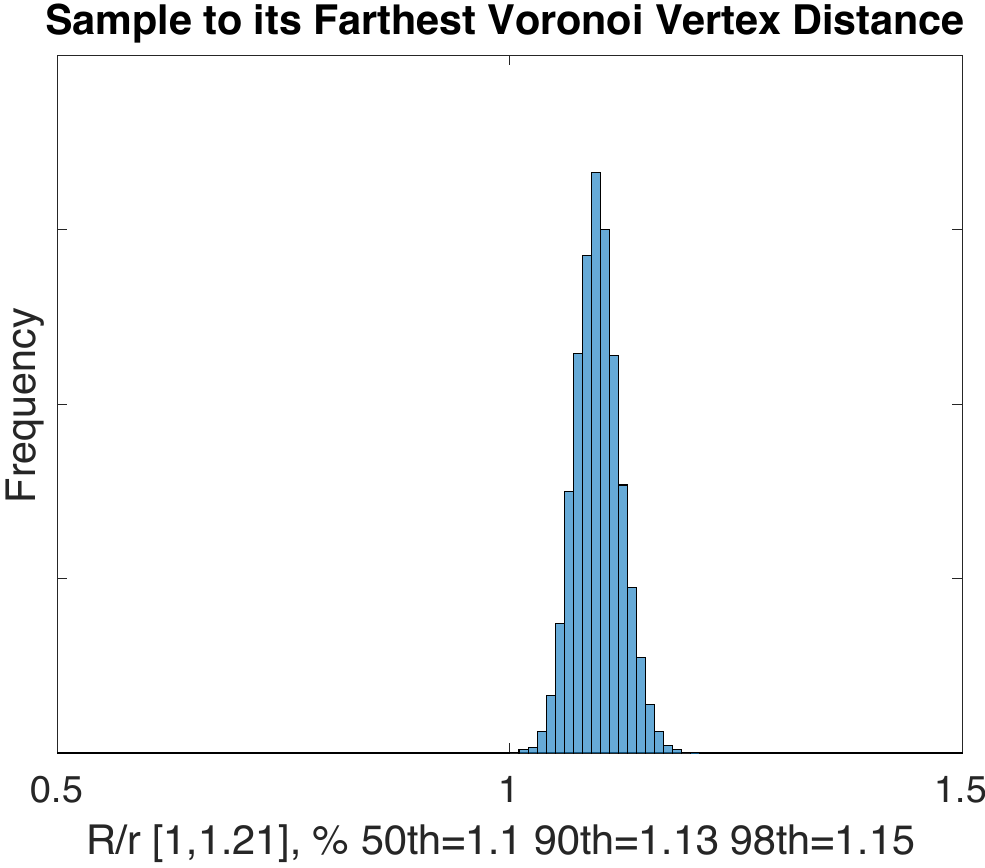}
				&\includegraphics[width=0.23\linewidth,height=1.2in]{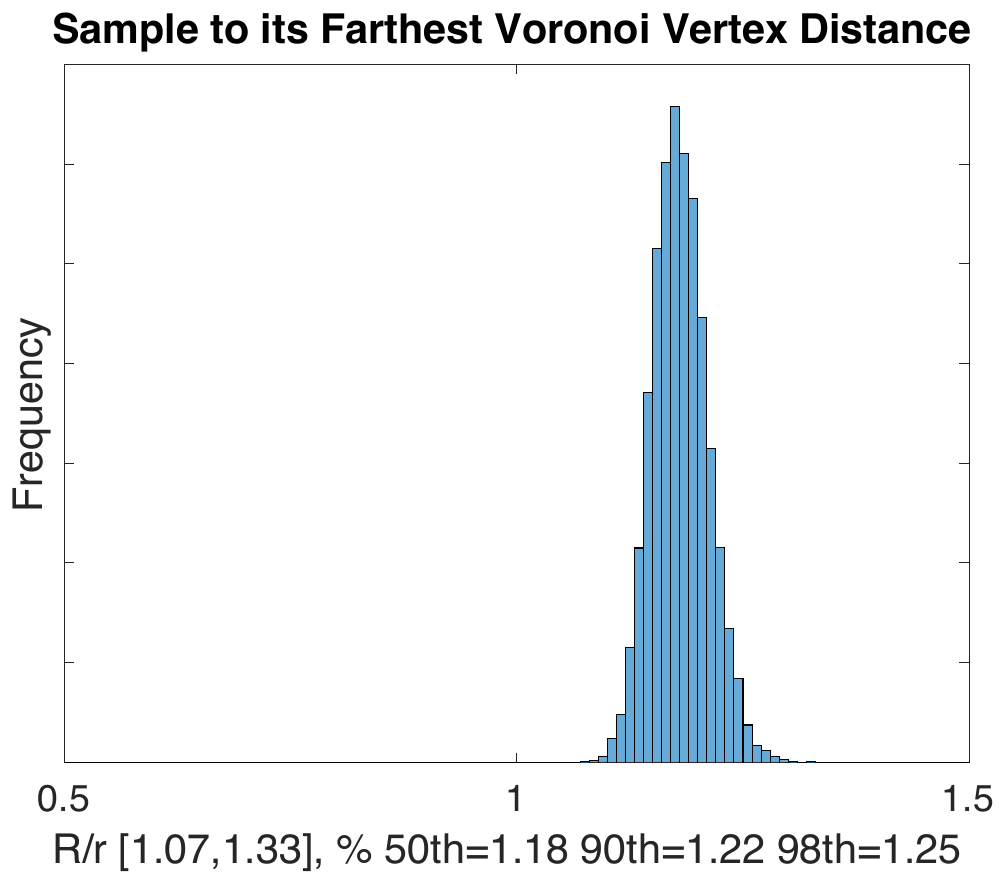}
				&\includegraphics[width=0.23\linewidth,height=1.2in]{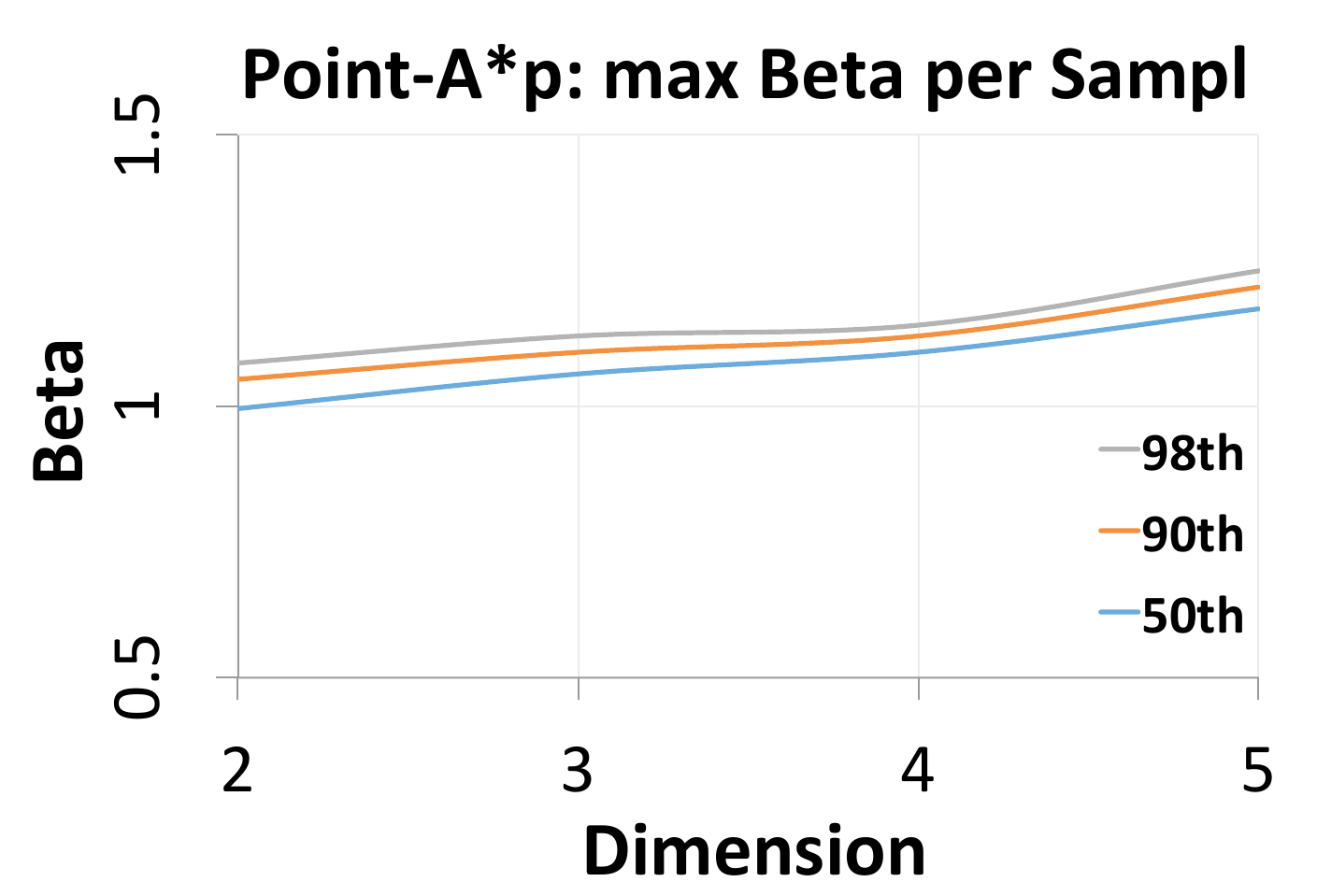}
				\\
				&\rotatebox[origin=l]{90}{{\hspace{24pt}\color{red}Line-Spokes}}
				&\includegraphics[width=0.23\linewidth,height=1.2in]{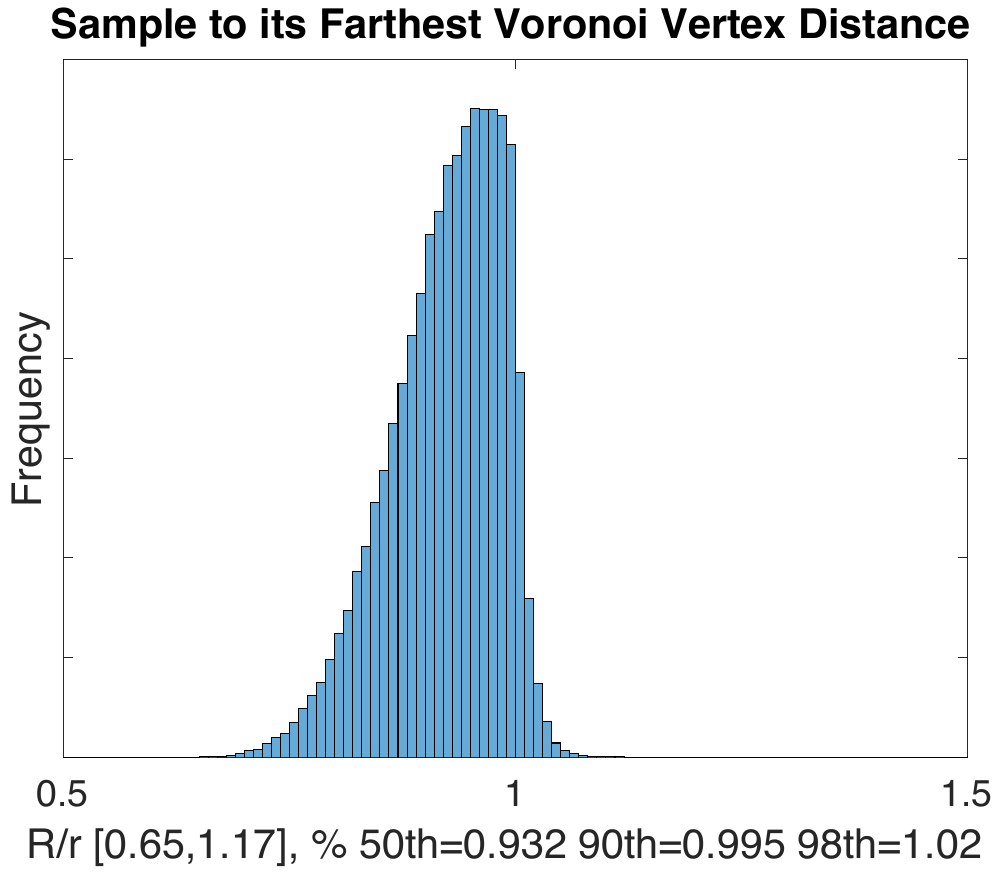}
				&\includegraphics[width=0.23\linewidth,height=1.2in]{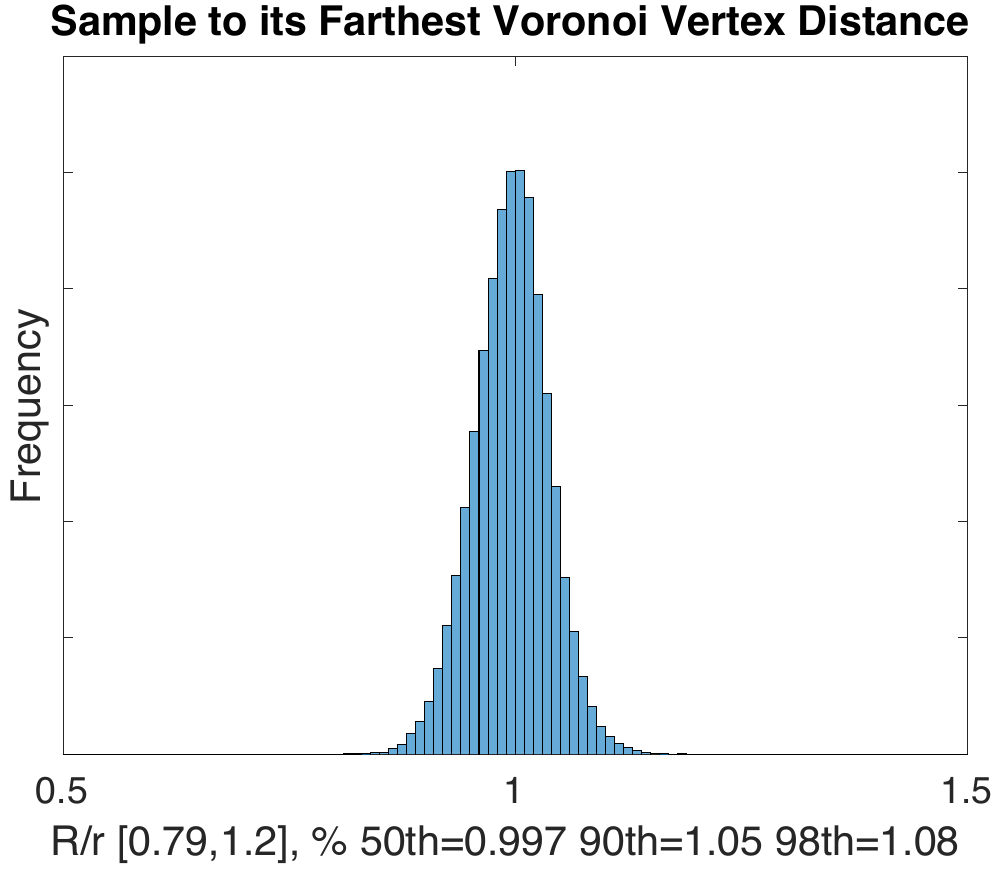}
				&\includegraphics[width=0.23\linewidth,height=1.2in]{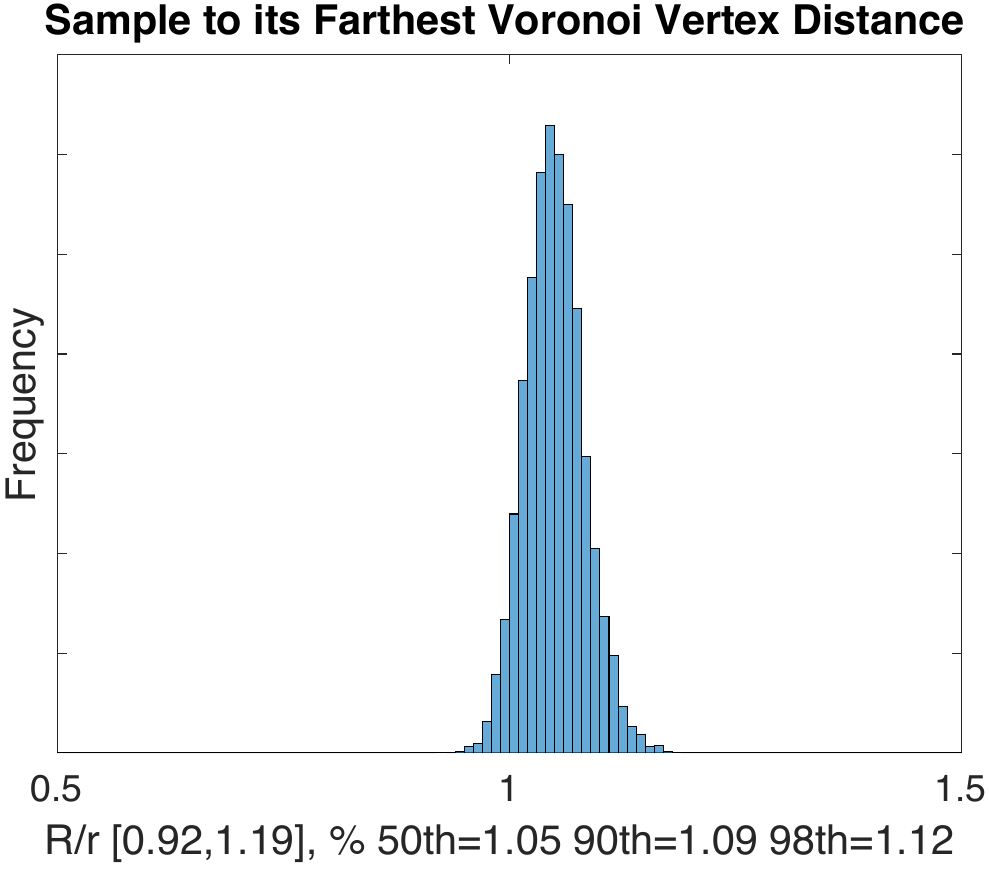}
				&\includegraphics[width=0.23\linewidth,height=1.2in]{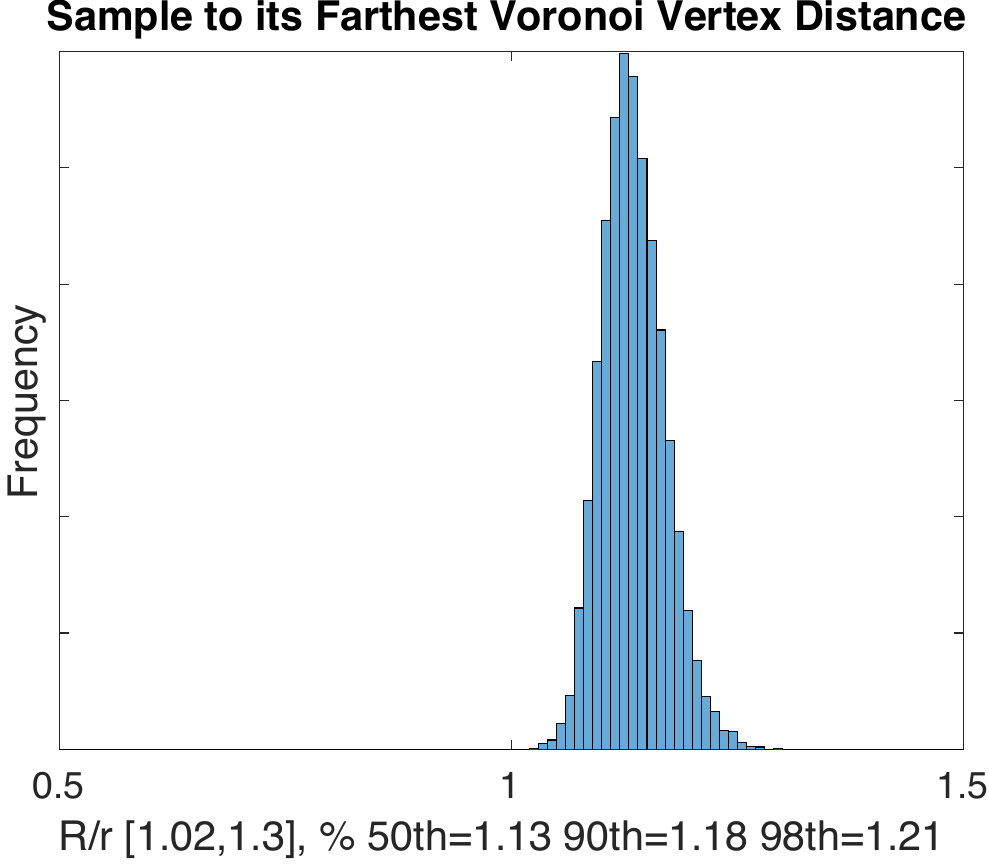}
				&\includegraphics[width=0.23\linewidth,height=1.2in]{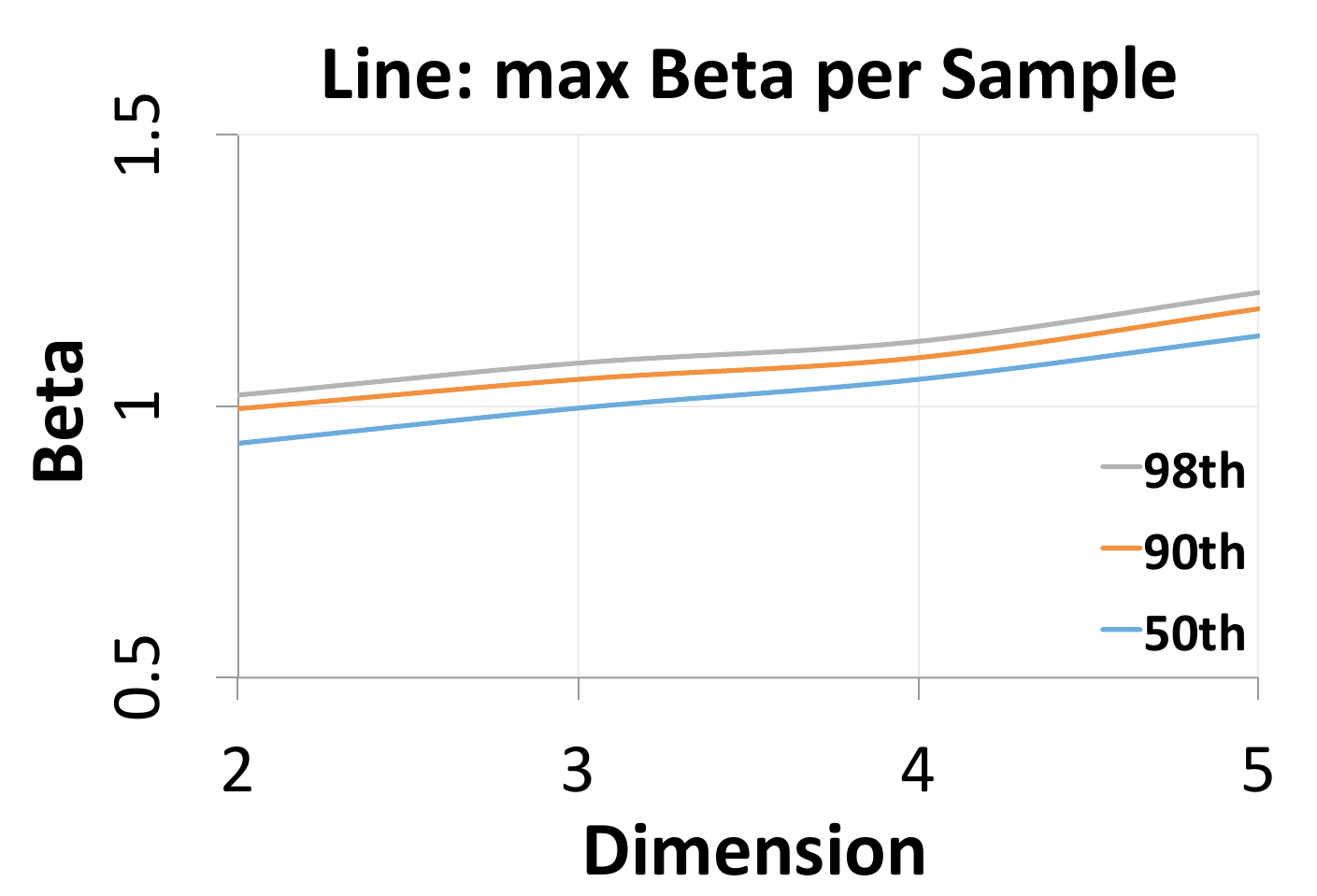}
				\\
				&\rotatebox[origin=l]{90}{{\hspace{16pt}\color{brown}Favored-Spokes}}
				&\includegraphics[width=0.23\linewidth,height=1.2in]{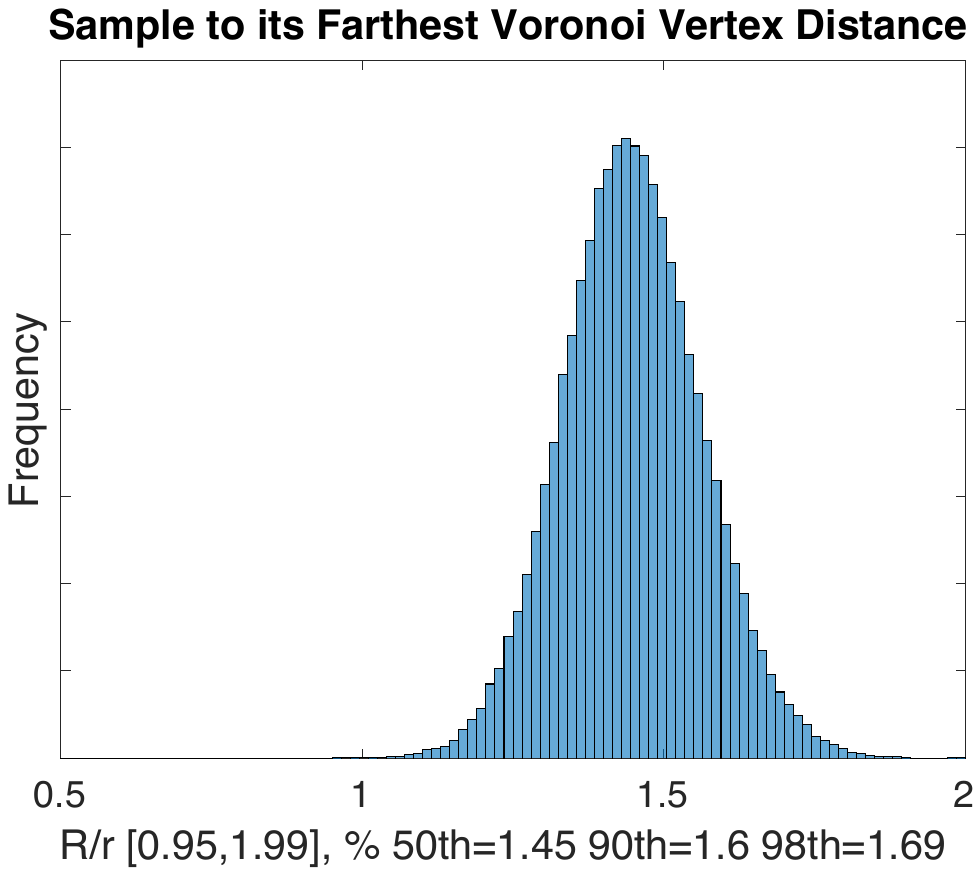}
				&\includegraphics[width=0.23\linewidth,height=1.2in]{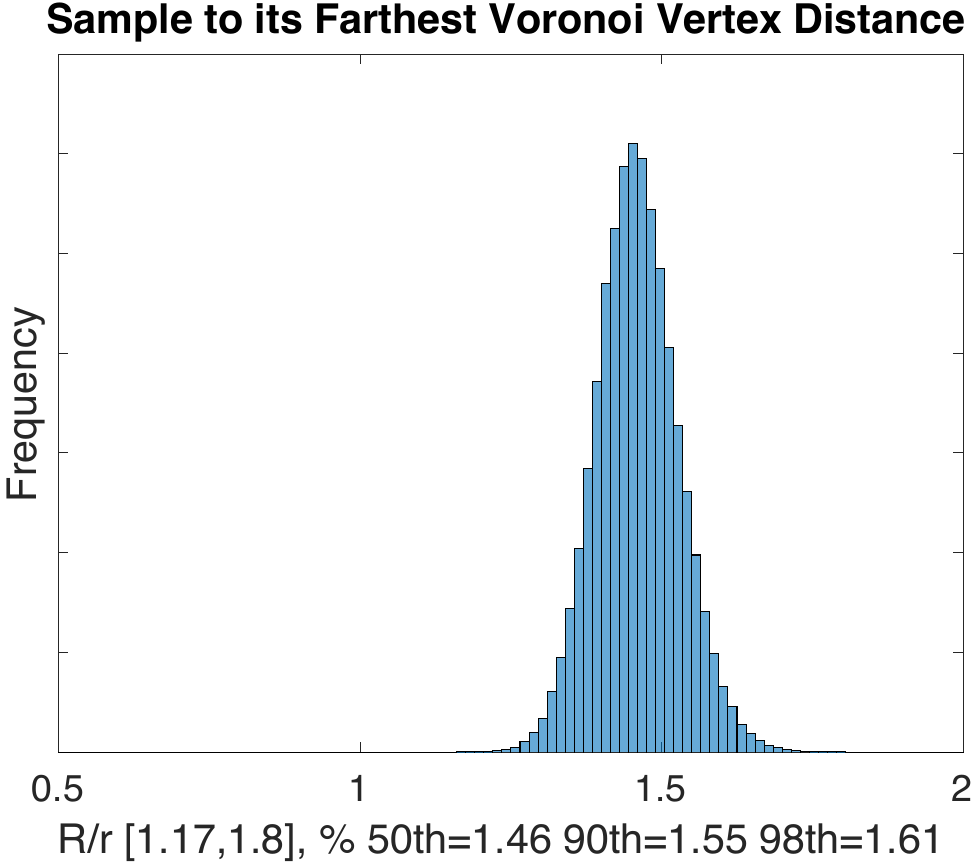}
				&\includegraphics[width=0.23\linewidth,height=1.2in]{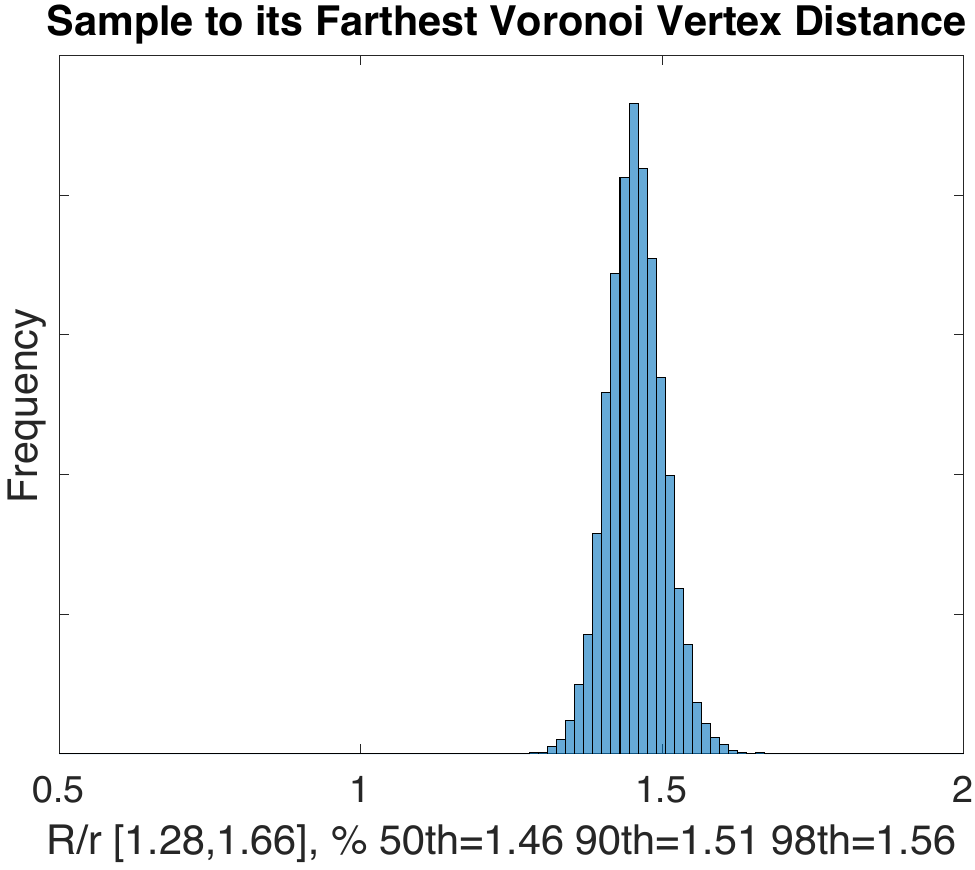}
				&\includegraphics[width=0.23\linewidth,height=1.2in]{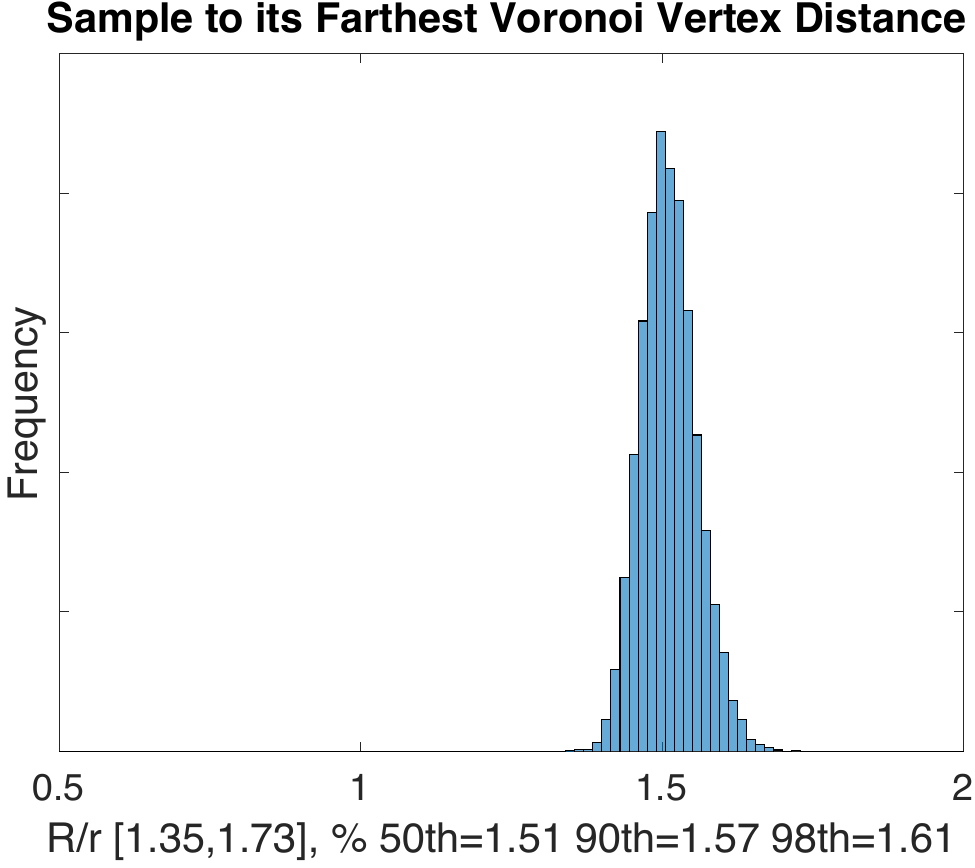}
				&\includegraphics[width=0.23\linewidth,height=1.2in]{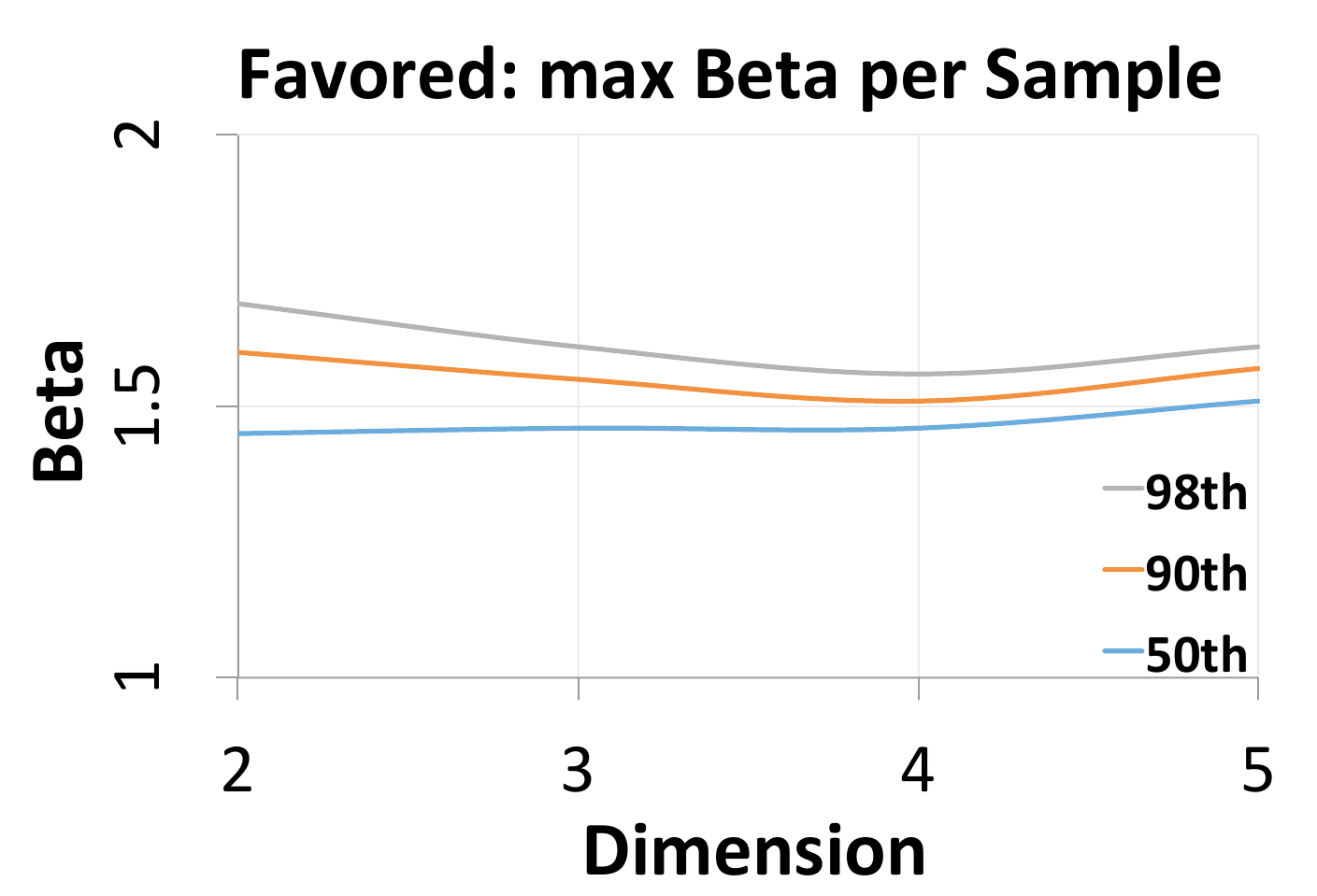}					
				\\
				&\rotatebox[origin=l]{90}{{\hspace{24pt}\color{magenta}Two-Spokes}}
				&\includegraphics[width=0.23\linewidth,height=1.2in]{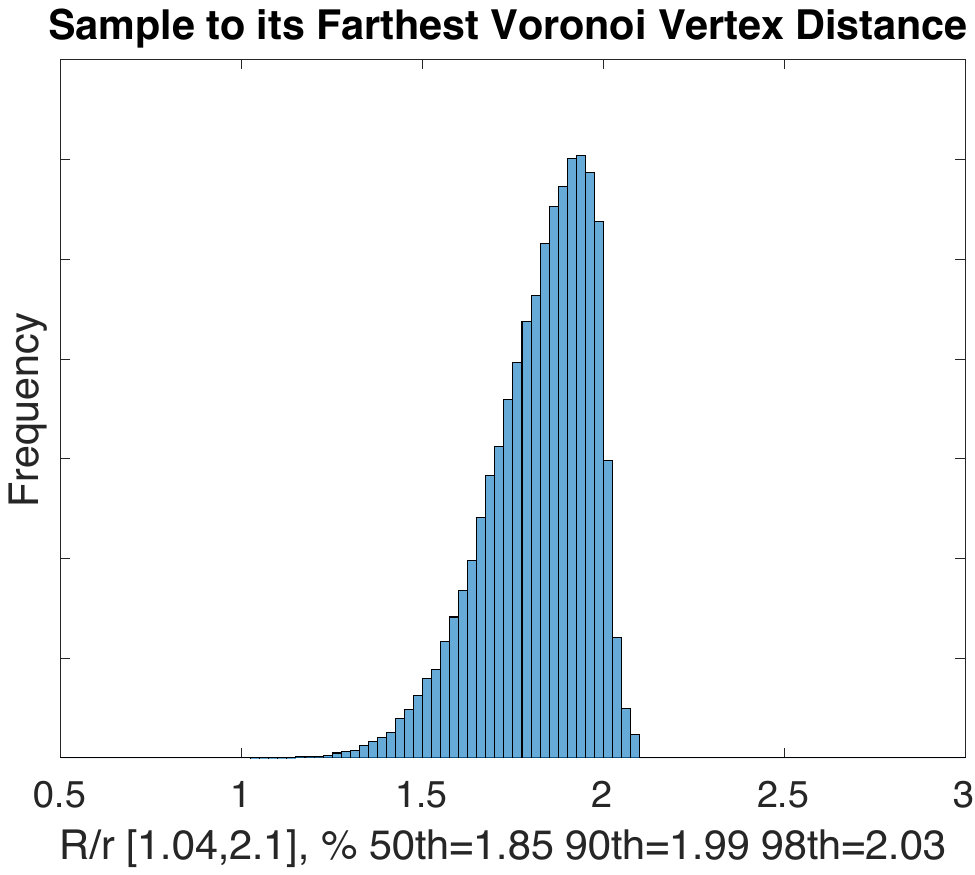}
				&\includegraphics[width=0.23\linewidth,height=1.2in]{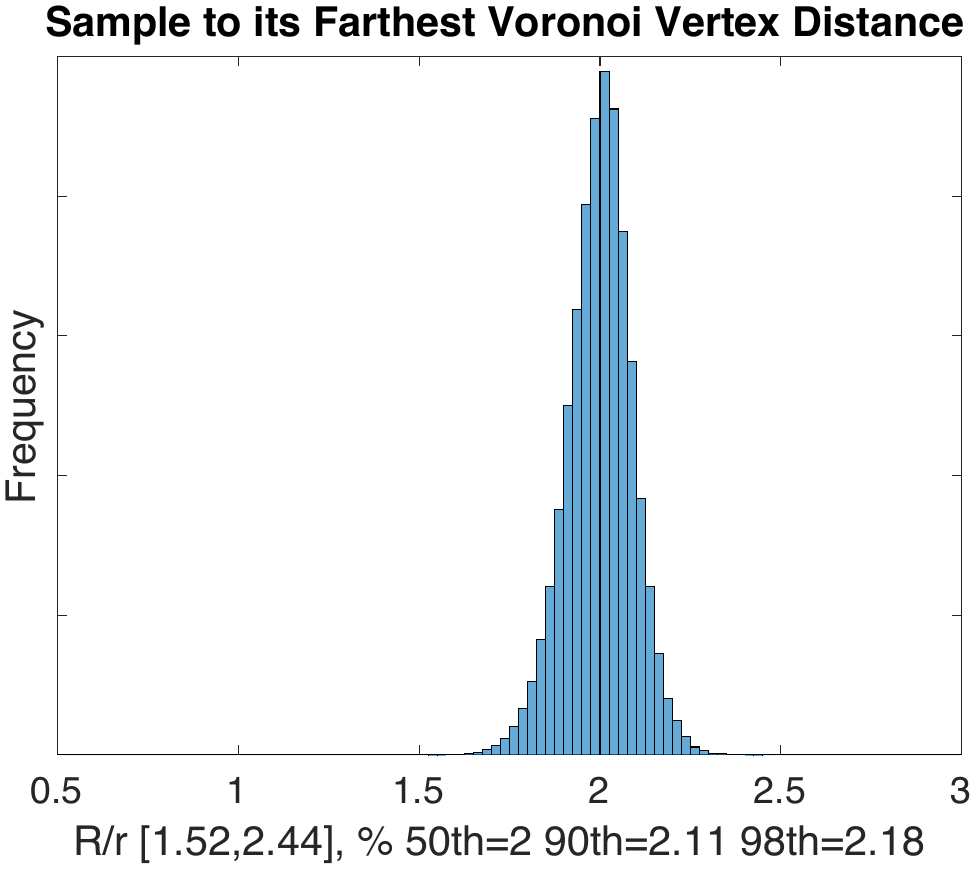}
				&\includegraphics[width=0.23\linewidth,height=1.2in]{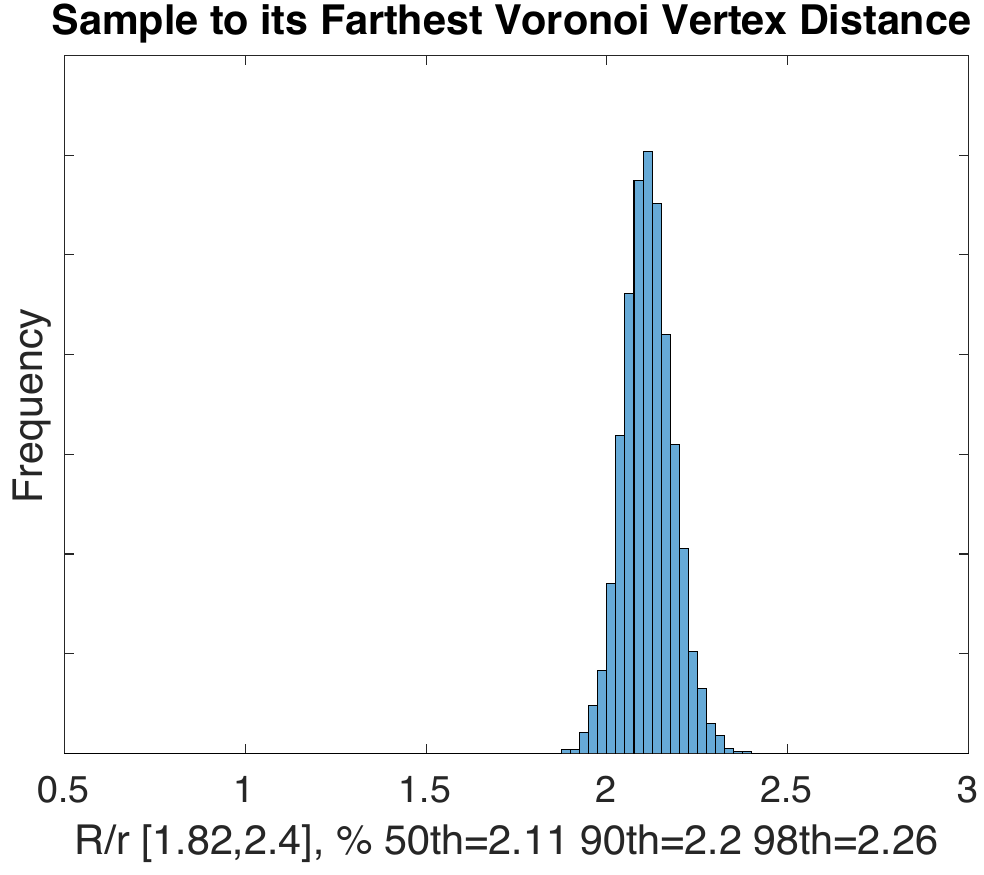}
				&\includegraphics[width=0.23\linewidth,height=1.2in]{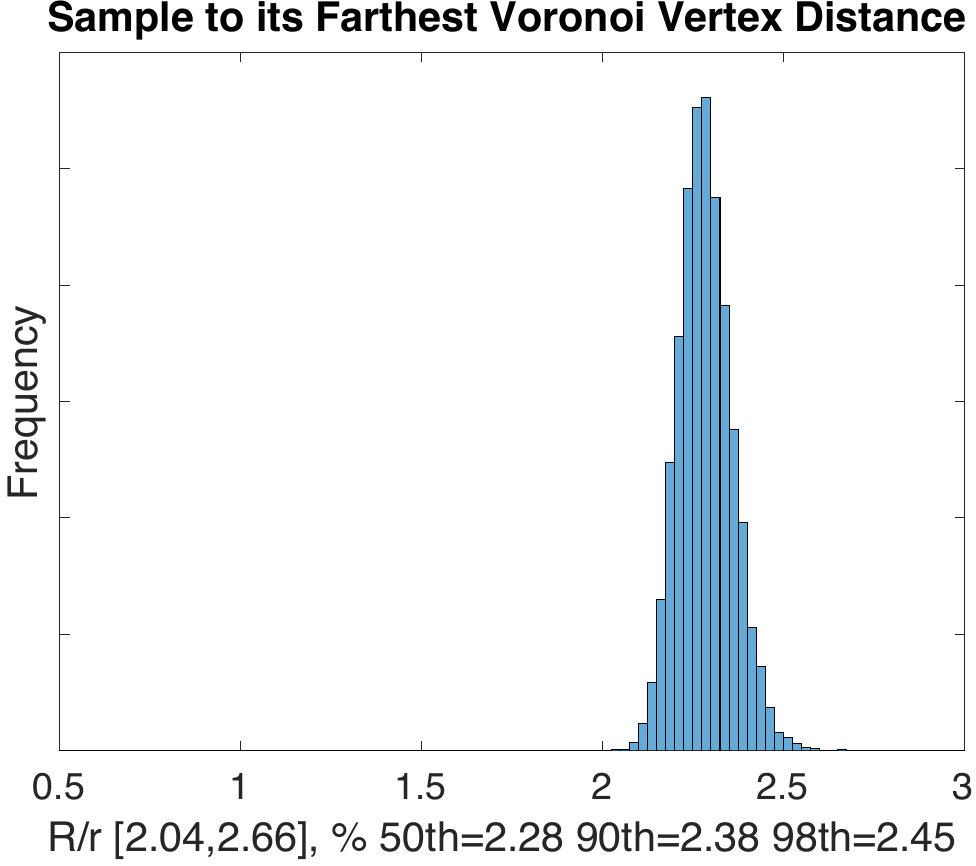}		
				&\includegraphics[width=0.23\linewidth,height=1.2in]{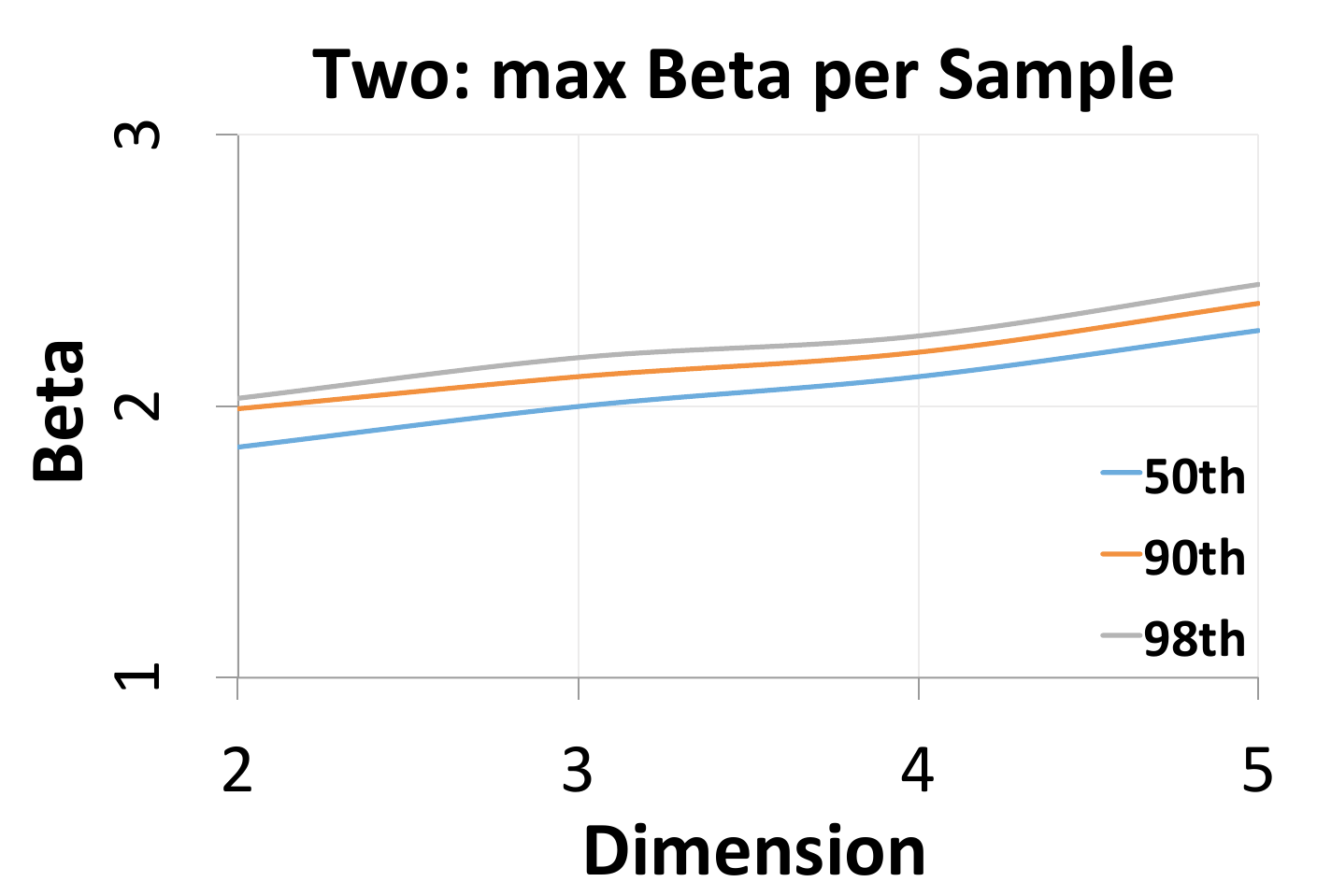}
				\\
				\hline
				\\
				\multirow{ 4 }{*}{\rotatebox[origin=l]{90}{\textbf{\Large Voronoi vertex to nearest sample distance}}}
				&\rotatebox[origin=l]{90}{{\hspace{16pt}\color{blue}Point-Annulus}}
				&\includegraphics[width=0.23\linewidth,height=1.2in]{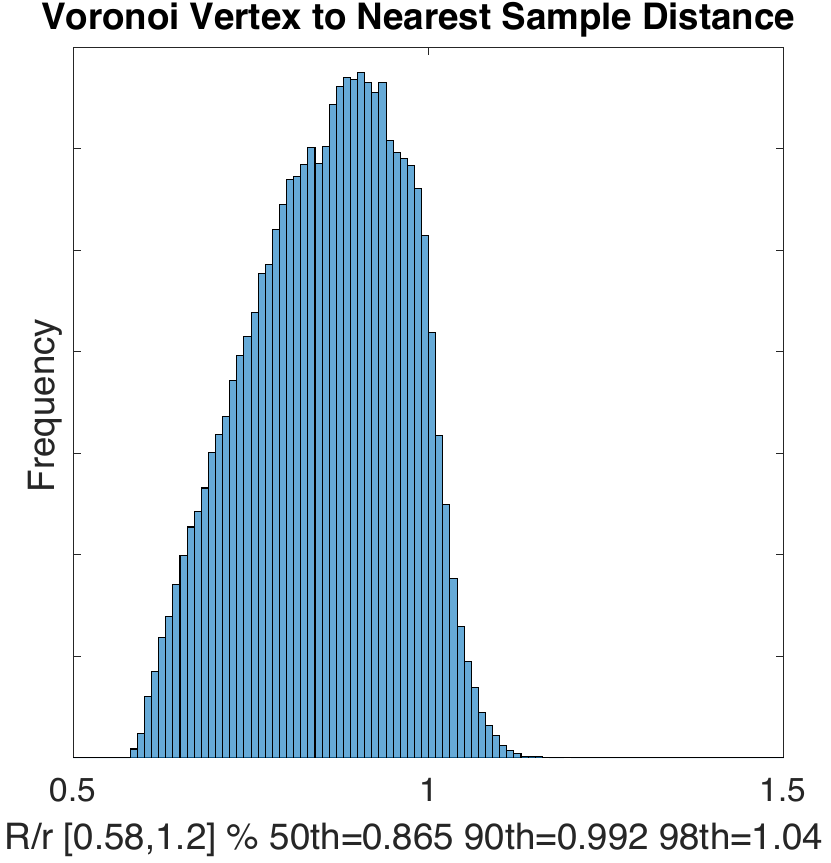}
				&\includegraphics[width=0.23\linewidth,height=1.2in]{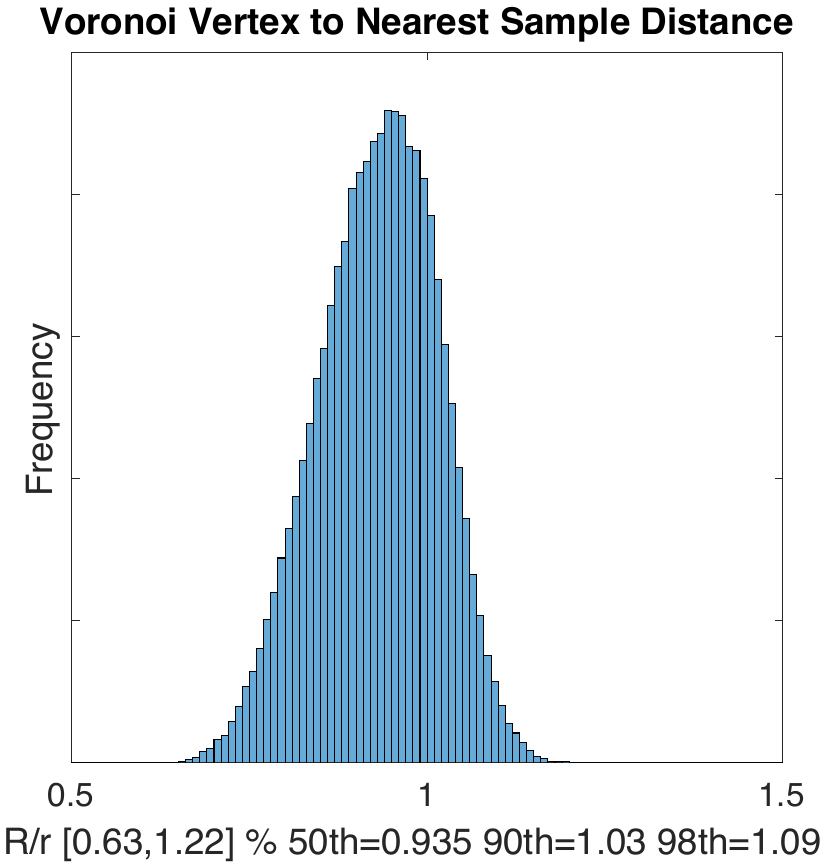}
				&\includegraphics[width=0.23\linewidth,height=1.2in]{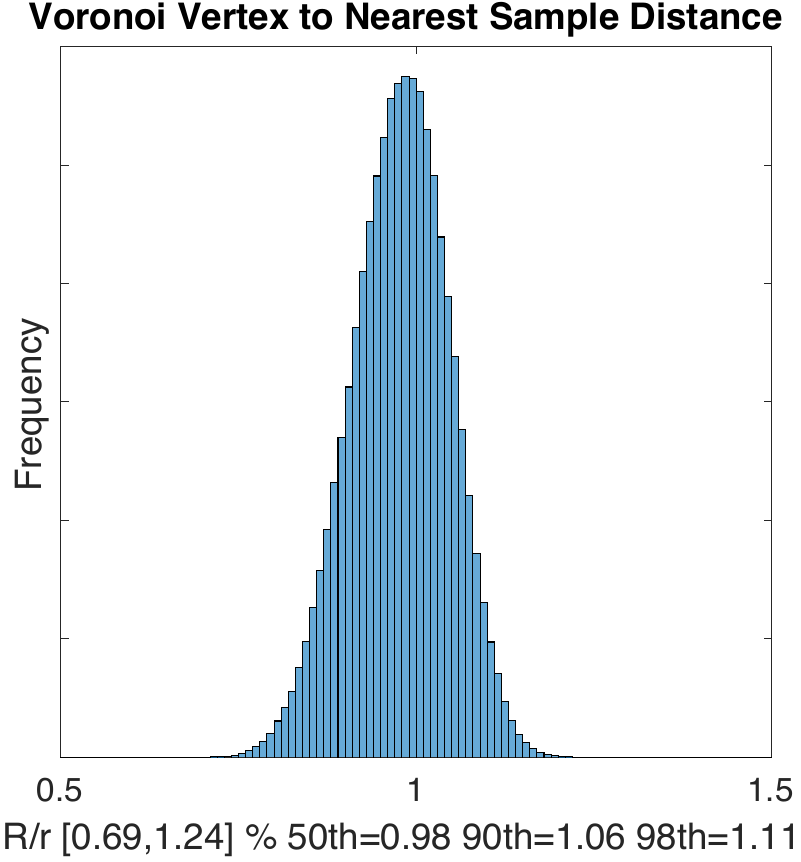}
				&\includegraphics[width=0.23\linewidth,height=1.2in]{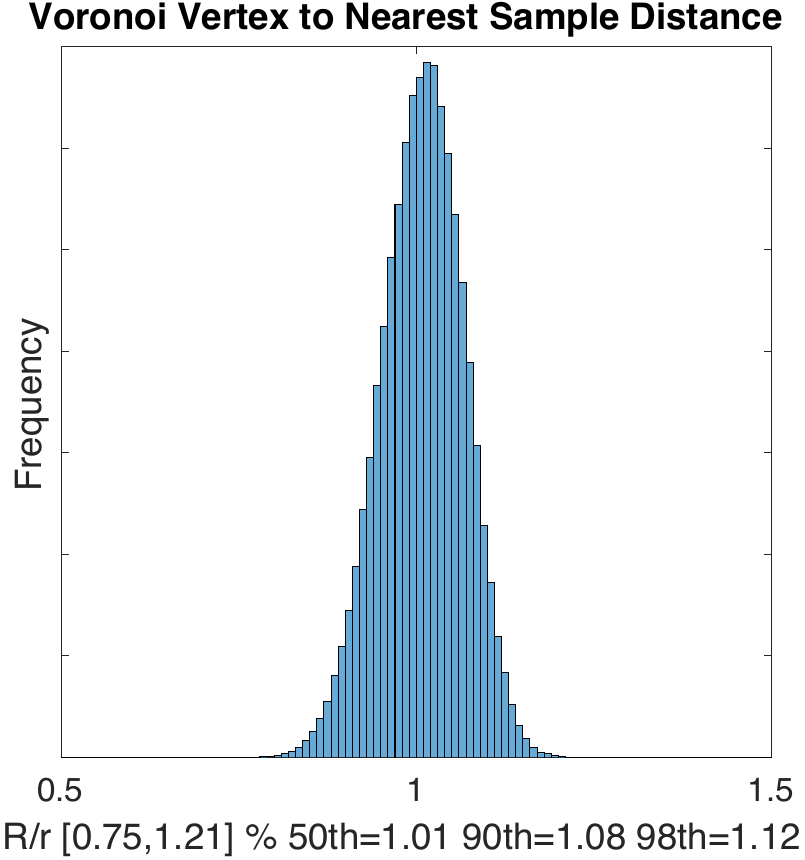}
				&\includegraphics[width=0.23\linewidth,height=1.2in]{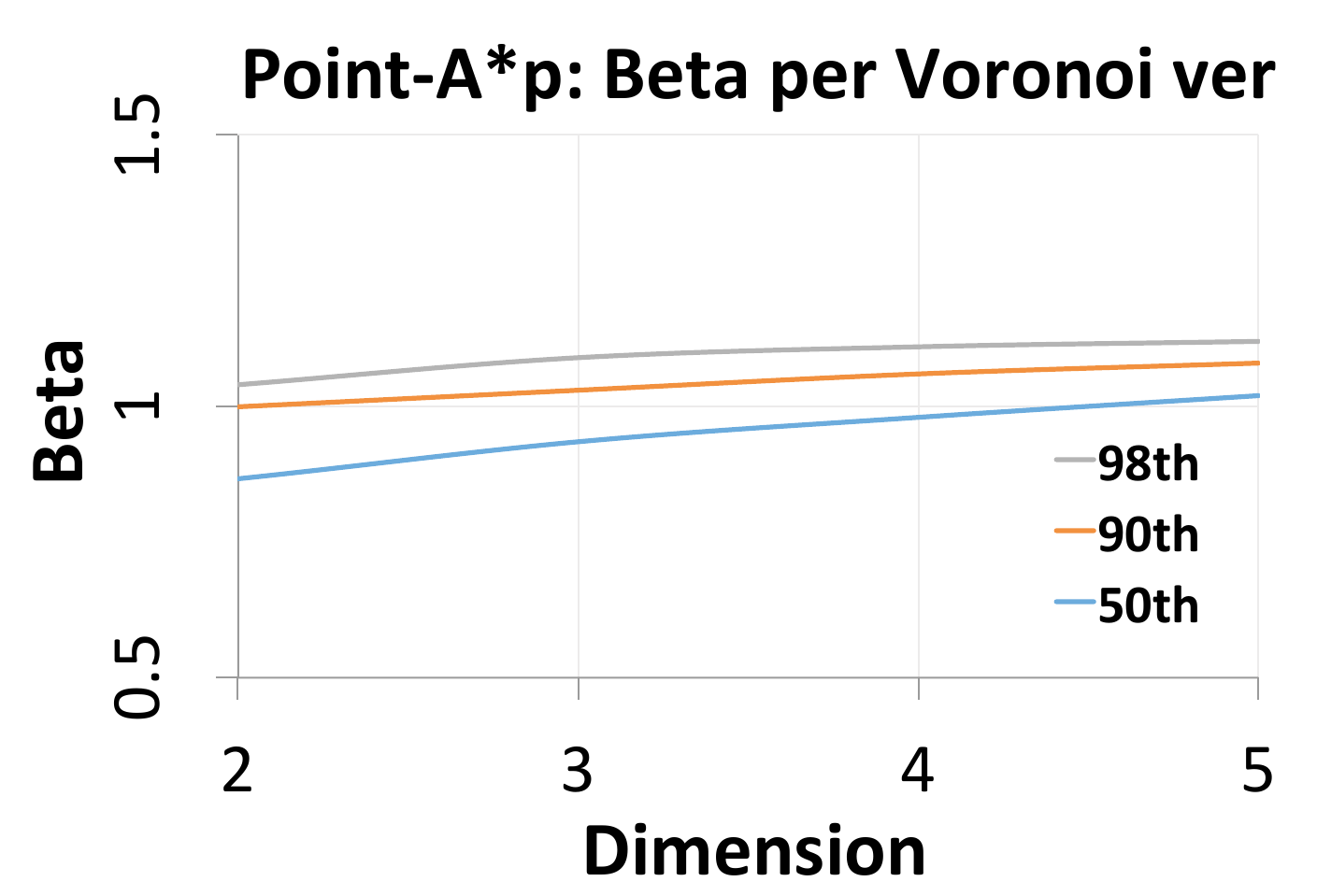}
				\\
				&\rotatebox[origin=l]{90}{{\hspace{24pt}\color{red}Line-Spokes}}
				&\includegraphics[width=0.23\linewidth,height=1.2in]{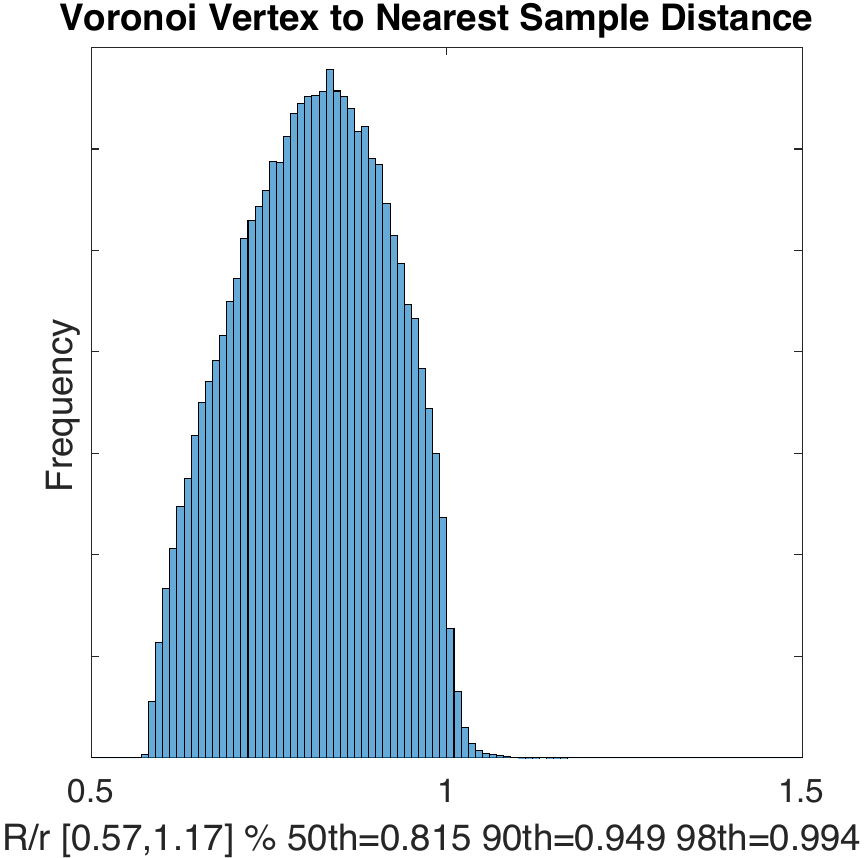}
				&\includegraphics[width=0.23\linewidth,height=1.2in]{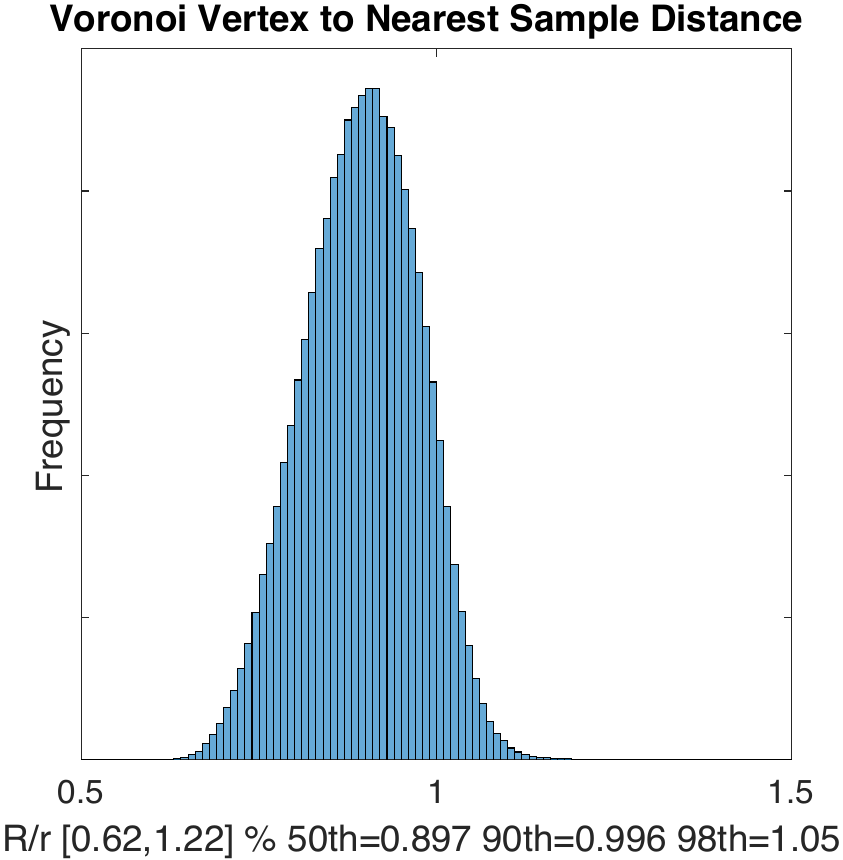}
				&\includegraphics[width=0.23\linewidth,height=1.2in]{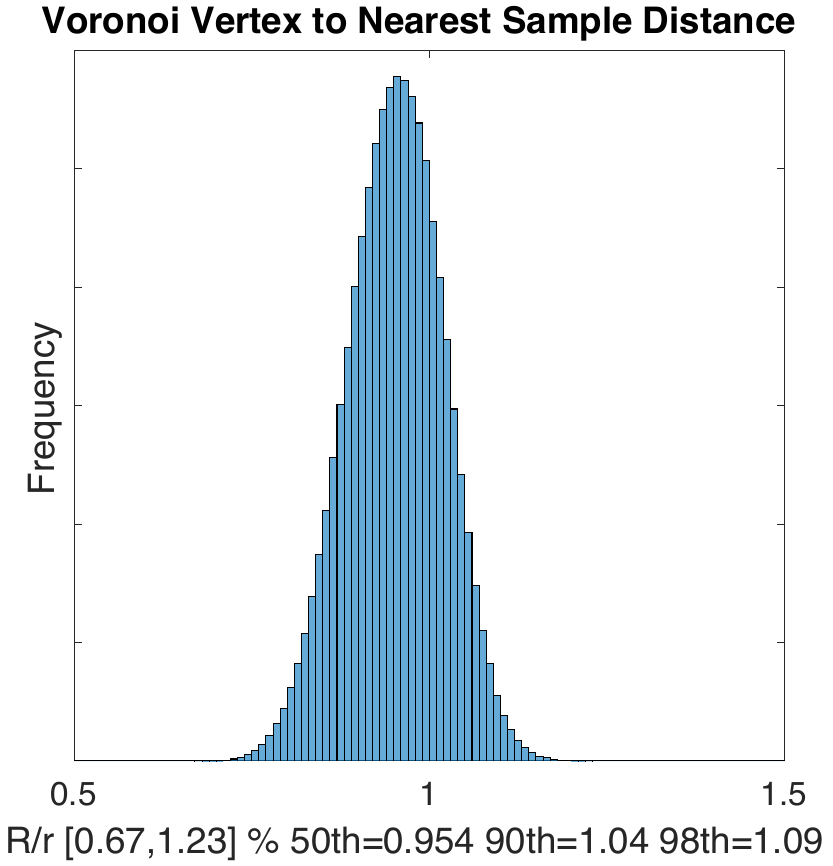}
				&\includegraphics[width=0.23\linewidth,height=1.2in]{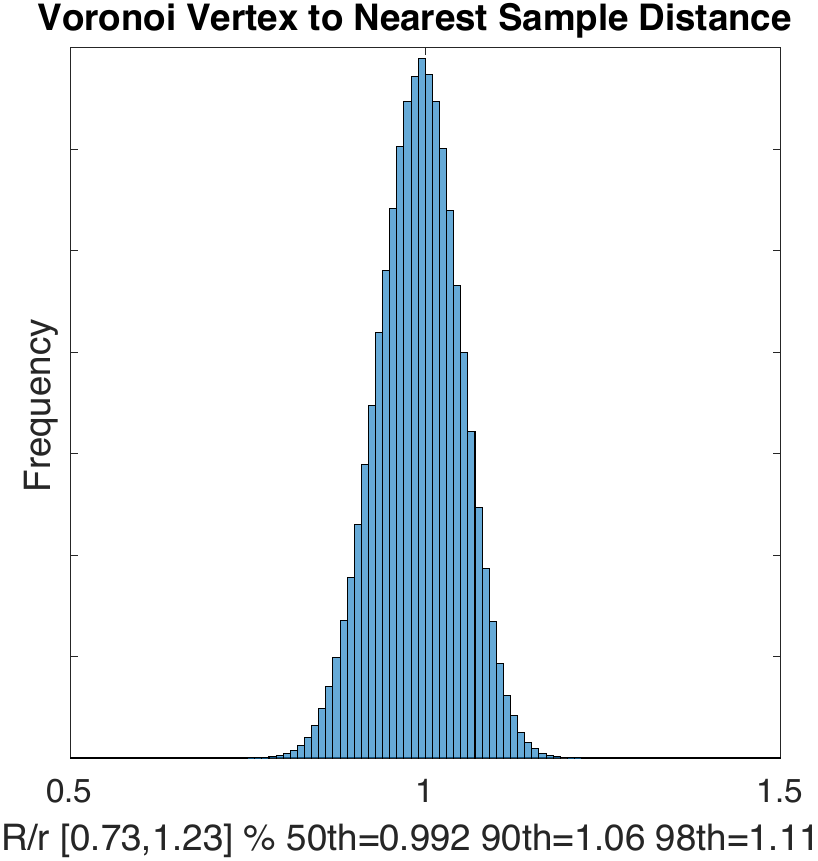}
				&\includegraphics[width=0.23\linewidth,height=1.2in]{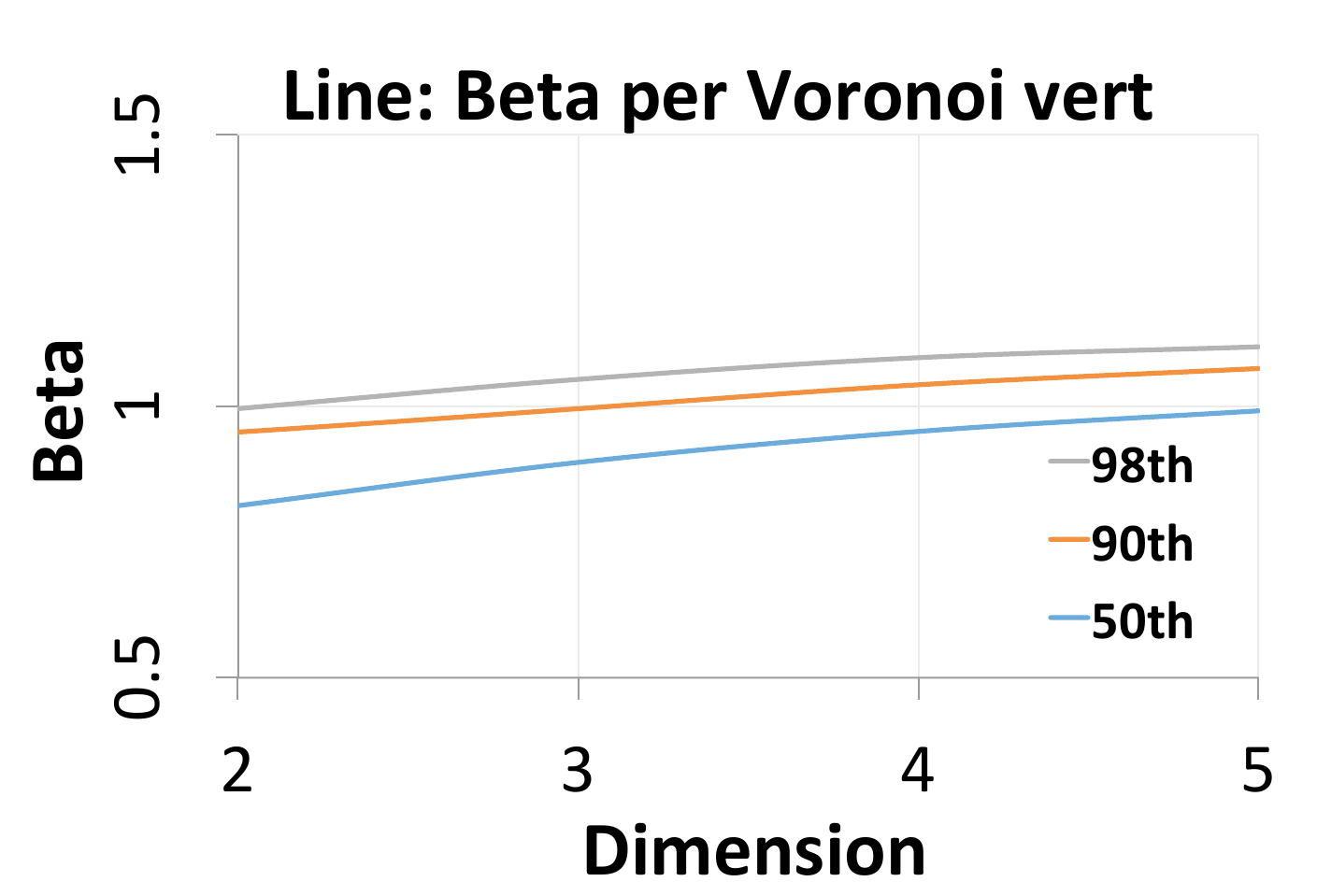}
				\\
				&\rotatebox[origin=l]{90}{{\hspace{16pt}\color{brown}Favored-Spokes}}
				&\includegraphics[width=0.23\linewidth,height=1.2in]{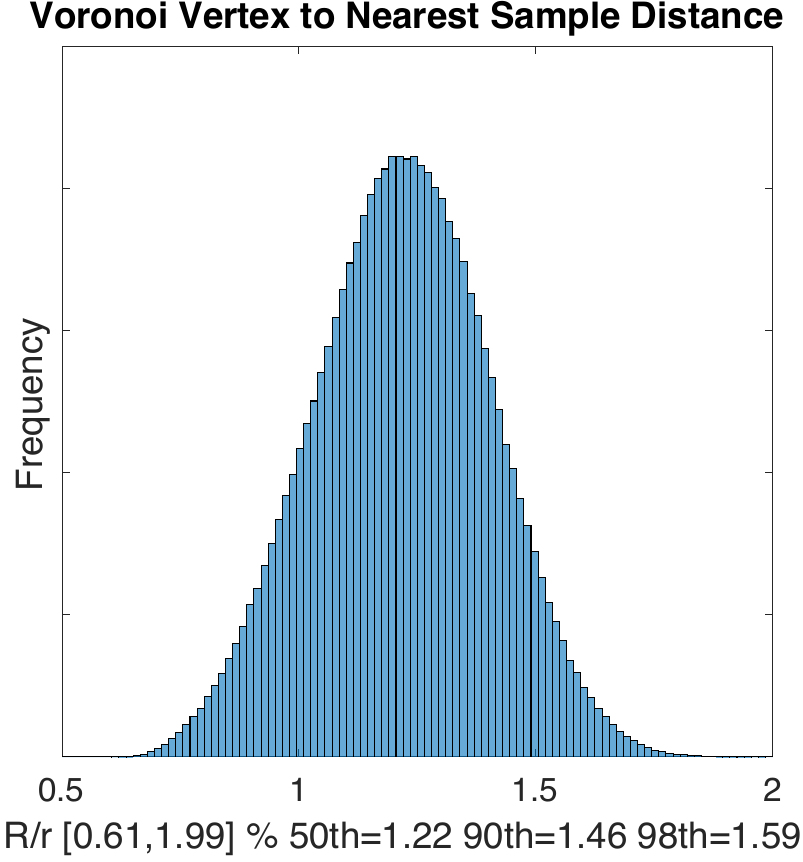}
				&\includegraphics[width=0.23\linewidth,height=1.2in]{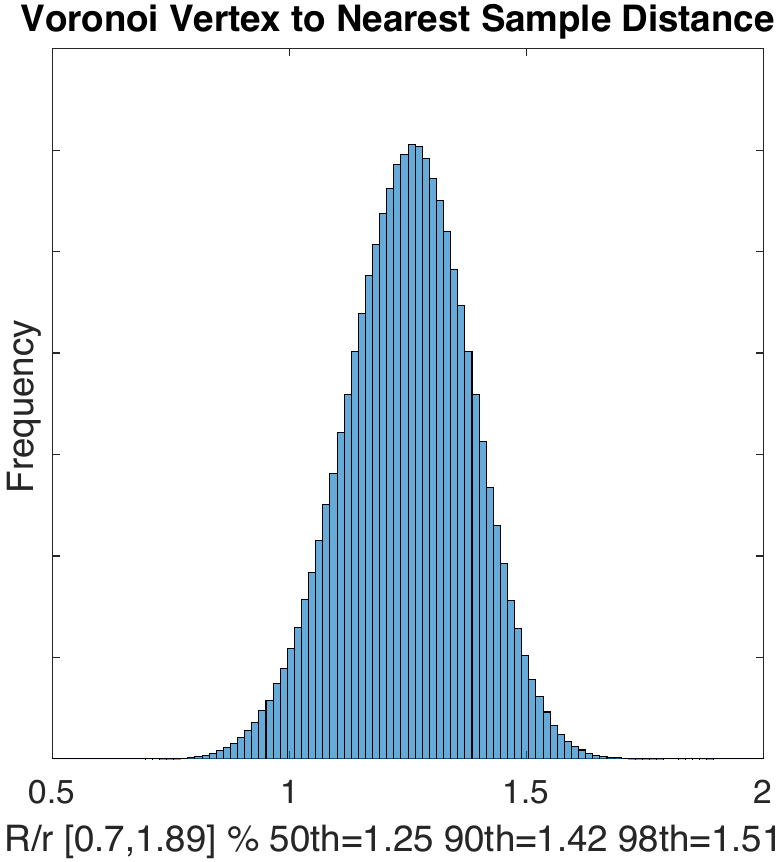}
				&\includegraphics[width=0.23\linewidth,height=1.2in]{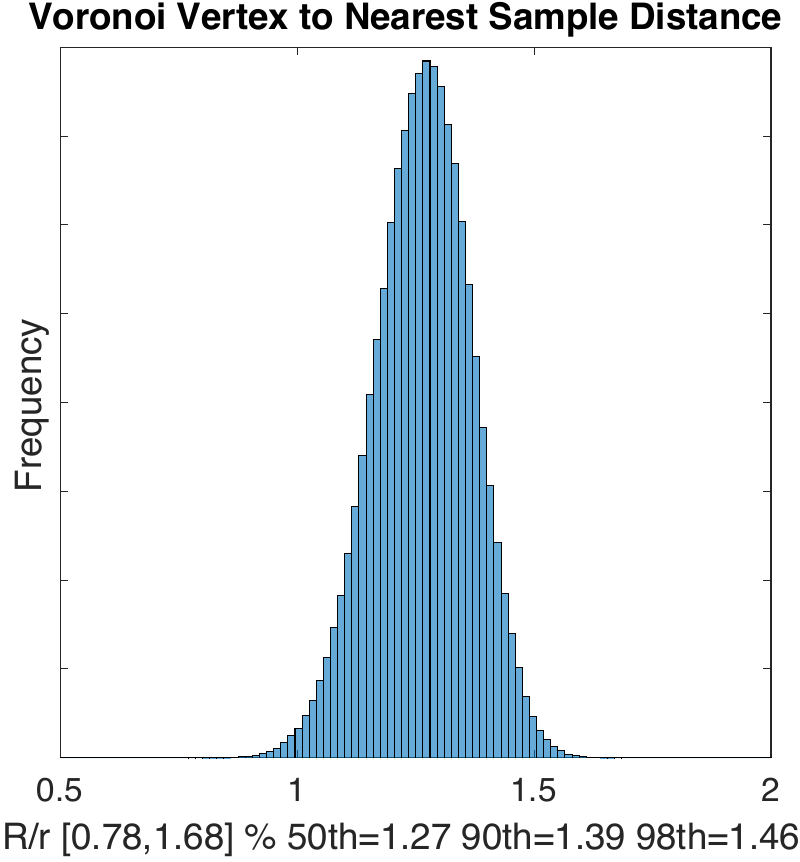}
				&\includegraphics[width=0.23\linewidth,height=1.2in]{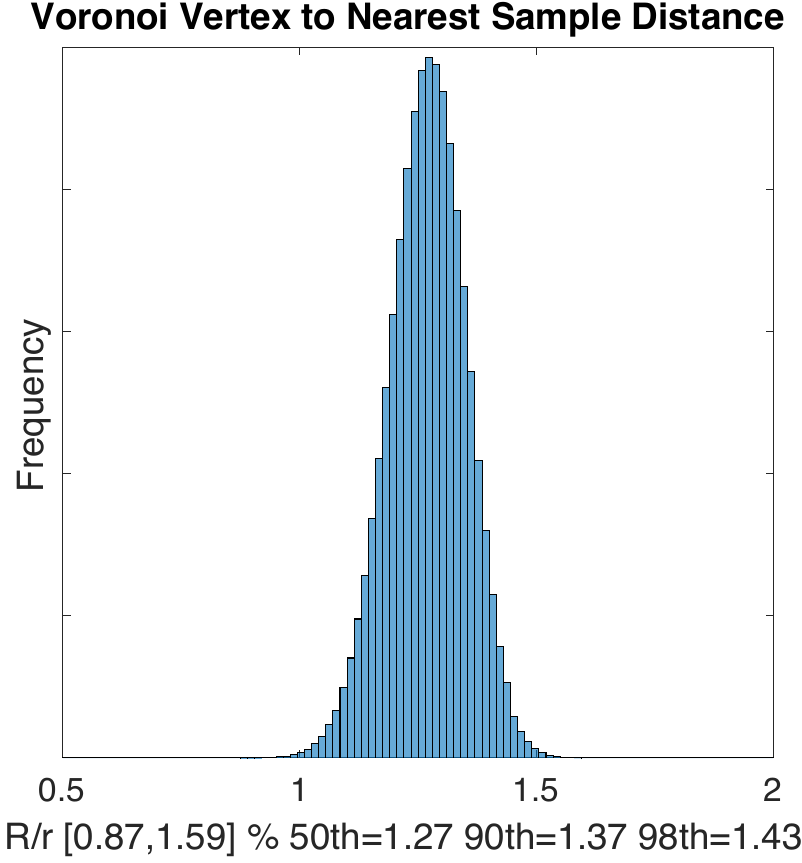}
				&\includegraphics[width=0.23\linewidth,height=1.2in]{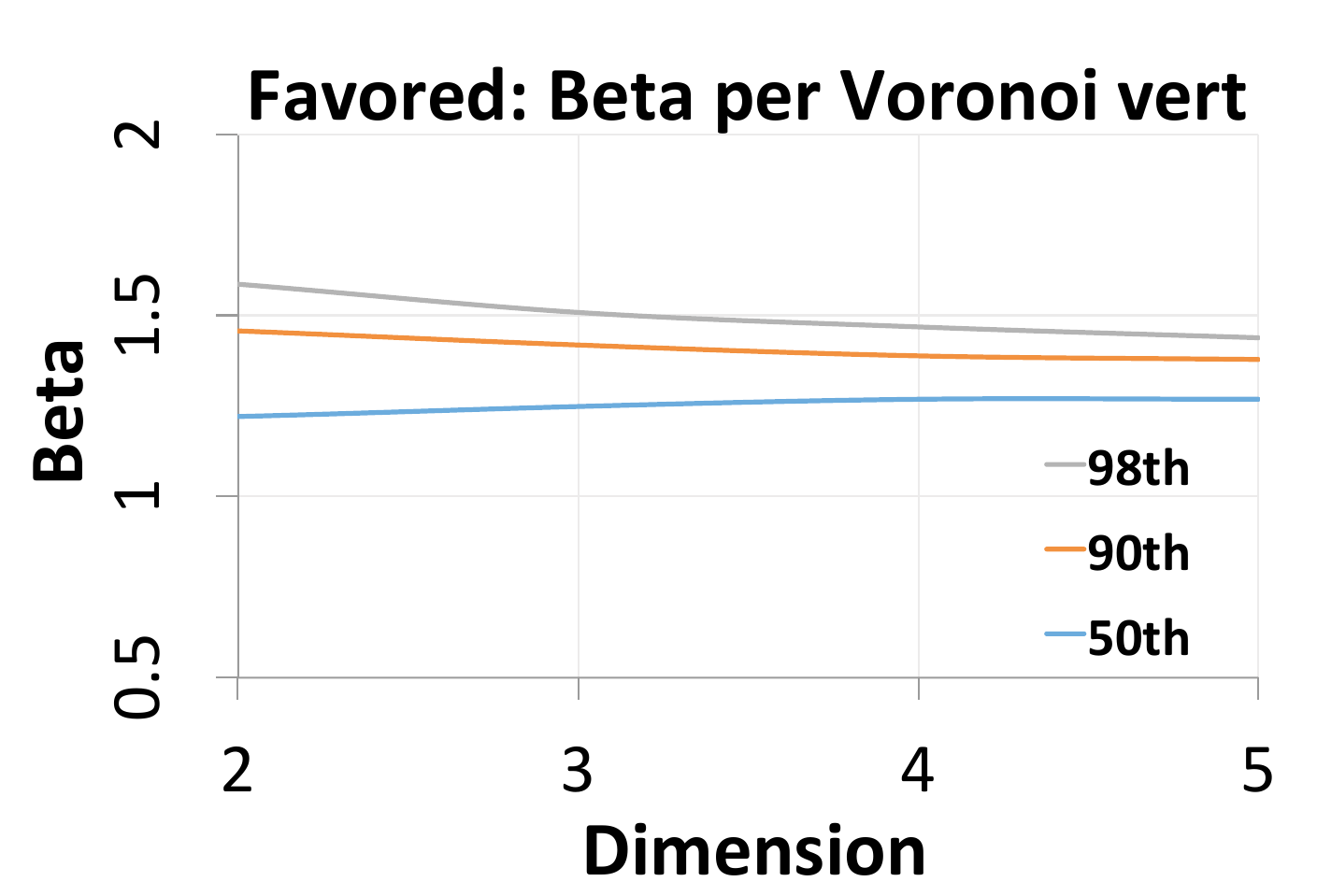}					
				\\
				&\rotatebox[origin=l]{90}{{\hspace{24pt}\color{magenta}Two-Spokes}}
				&\includegraphics[width=0.23\linewidth,height=1.2in]{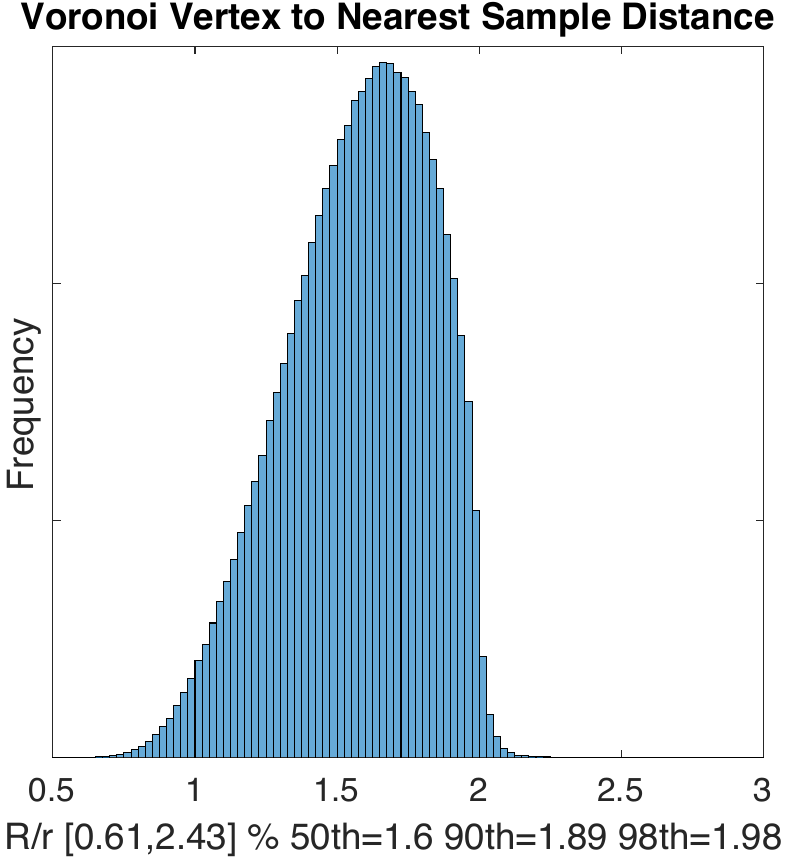}
				&\includegraphics[width=0.23\linewidth,height=1.2in]{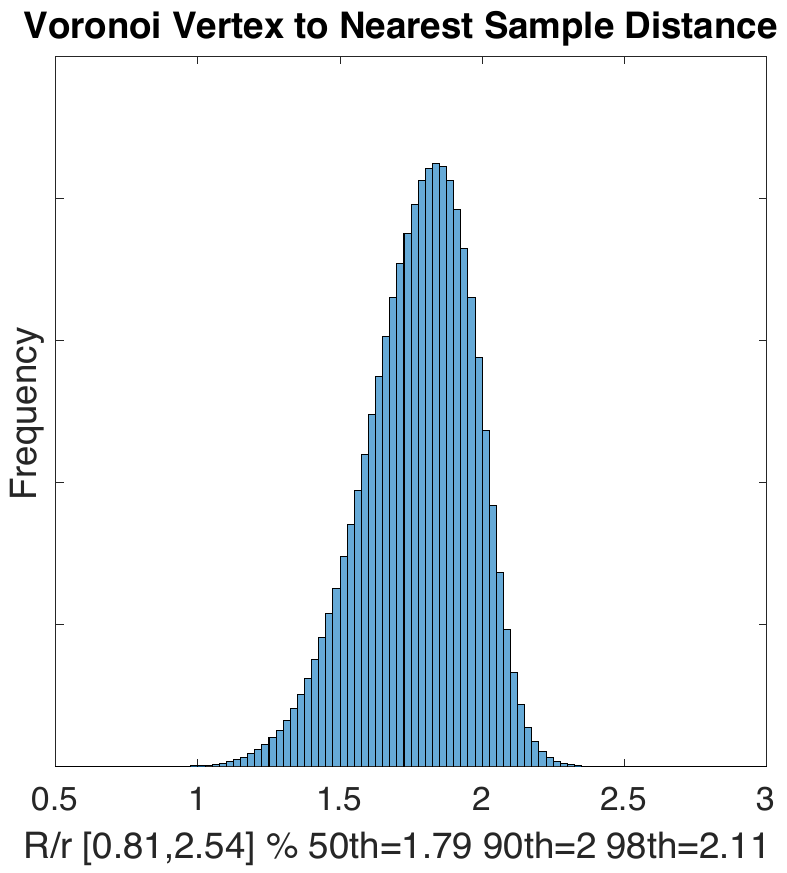}
				&\includegraphics[width=0.23\linewidth,height=1.2in]{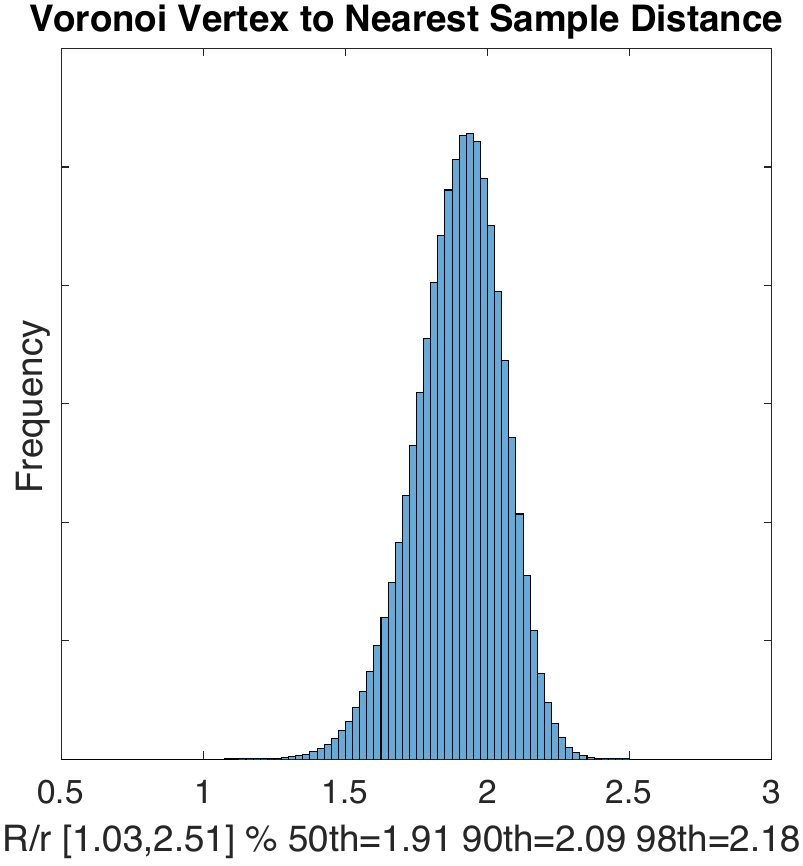}
				&\includegraphics[width=0.23\linewidth,height=1.2in]{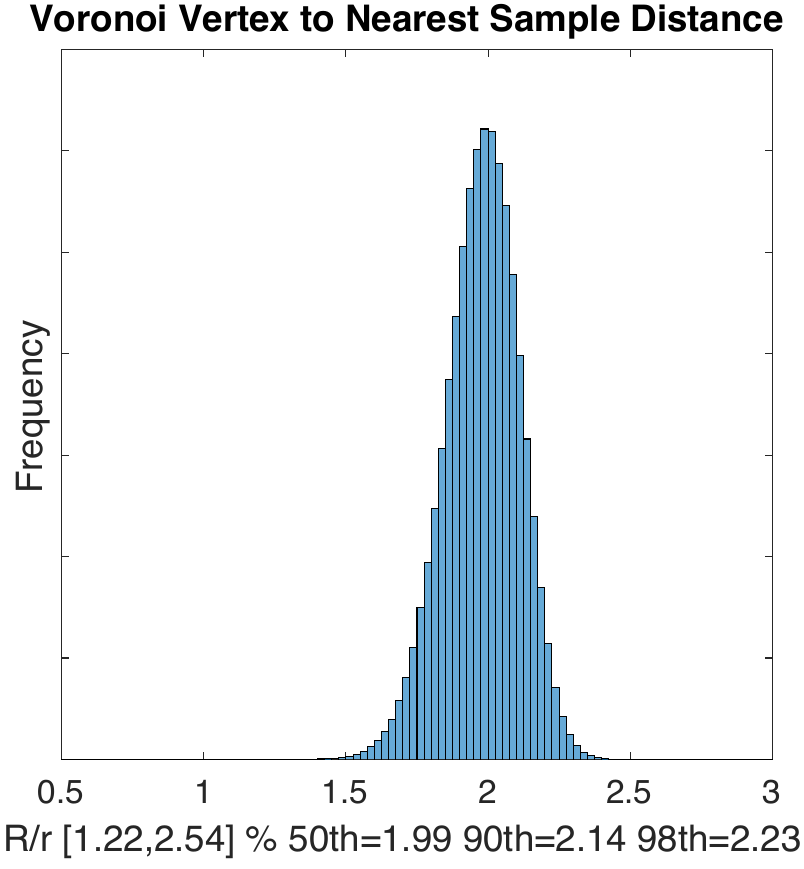}
				&\includegraphics[width=0.23\linewidth,height=1.2in]{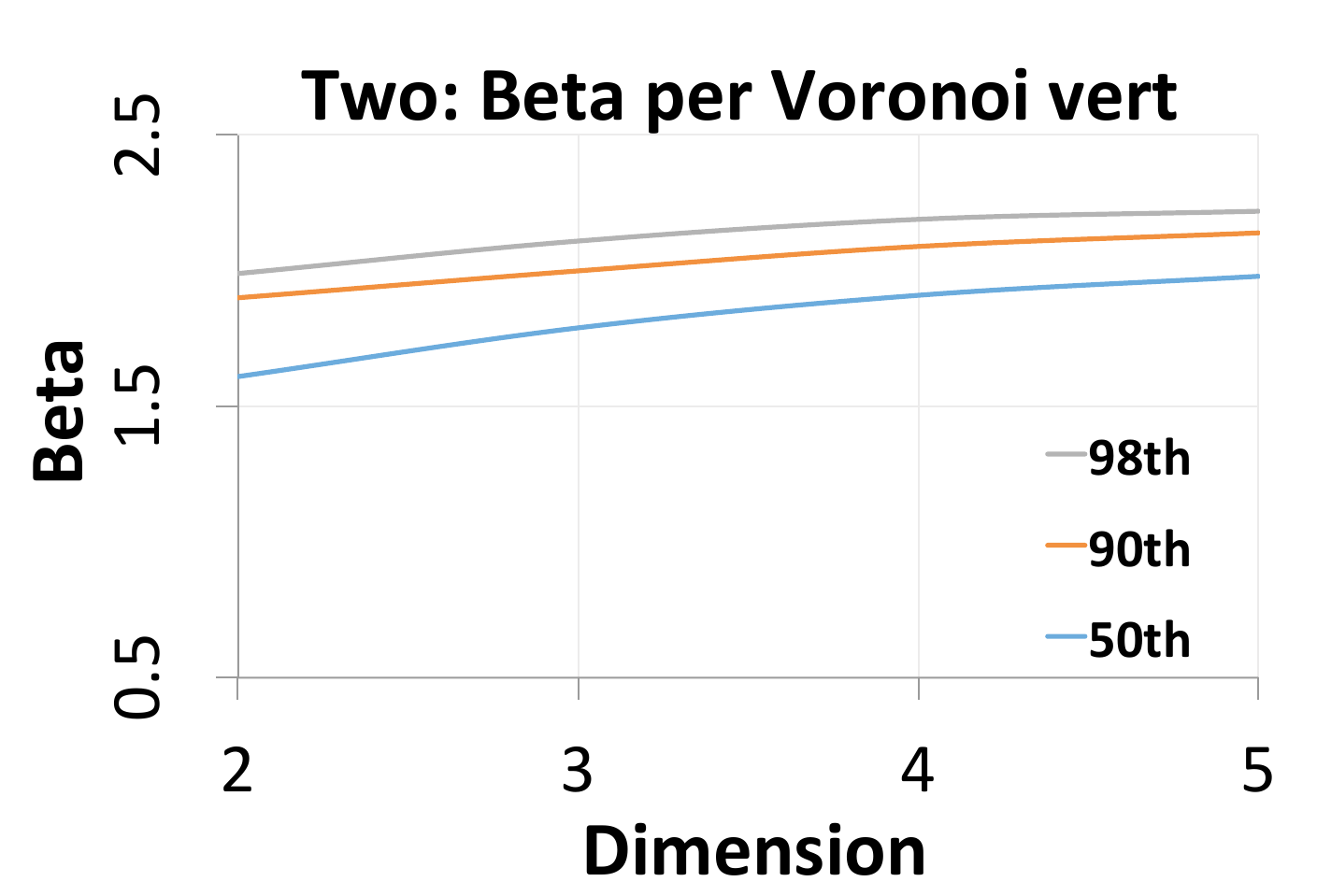}
%				\\
%				\hline
%				\\
			\end{tabular}
		}
	\caption{$\rratio$ distribution histograms: (top) sample to farthest Voronoi vertex distance, and (bottom) Voronoi vertex to nearest sample distance, for $\dimnumber=$2--5. The bottom-half rightmost-column illustrates that, in practices, $\rratio$ narrows and converges to a fixed constant as $\dimnumber$ increases, as theory predicts.}\label{fig:comparison_histograms}
	   %\caption{Sample to farthest Voronoi vertex distance histograms. For each algorithm, we ran many trials: 100,000 in $d={2,3}$, and 10,000 in $d={4,5}$. For each trial, we sampled a spherical neighborhood around a single point, and determined that point's farthest Voronoi vertex distance. In $d\ge5,$ this distance was approximated using a deterministic dense uniform set of rays around each sample, as the exact calculation of Voronoi vertices using Qhull became intractable.} \label{fig:local_beta_farthestvoronoi}
	   %\caption{Voronoi vertex to nearest sample distance histograms. We created one large sampling and found the distance from each Voronoi vertex to its nearest sample. For $\dimnumber=\{2,3\}$,  we have $\samplenumber \approx$  110,000. For $\dimnumber=4$,  $\samplenumber \approx$  70,000. For $\dimnumber=5$,  $\samplenumber \approx$  6,000; beyond this $\samplenumber$ and $\dimnumber$, the exact calculation of all Voronoi vertices using Qhull becomes intractable.}\label{fig:local_beta_extra}
	\end{figure*}
}

{
%\noindent %noindent causes extra vspace when the figure is input
% height = 0.93in
% width=0.18\linewidth
\begin{figure*}
	\centering
	\setlength{\tabcolsep}{3pt}
	\resizebox{0.9\linewidth}{!}{
		\begin{tabular}{cccccc}
			& $\dimnumber=2$ &$\dimnumber=3$ &$\dimnumber=4$ &$\dimnumber=5$ &$\dimnumber=6$ \\
		        \rotatebox[origin=l]{90}{{\hspace{28pt}spectra}} 
		            \ifthenelse{\equal{\isarxiv}{1}}
    {
			&  \includegraphics[width=0.18\linewidth]{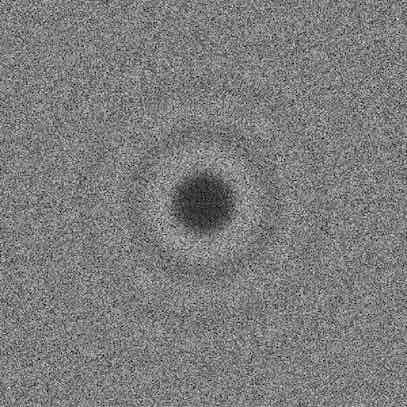}
			&\includegraphics[width=0.18\linewidth]{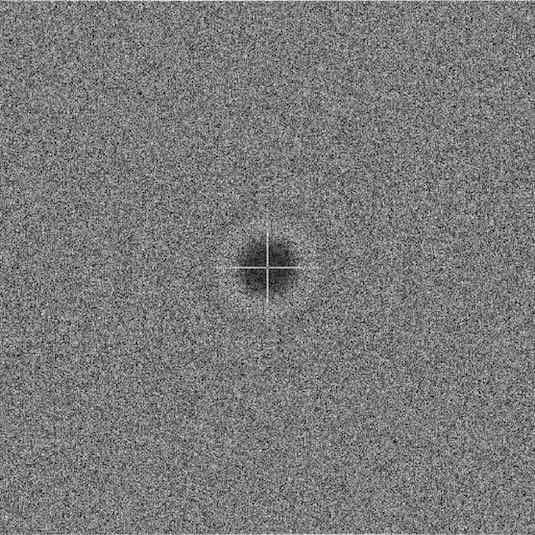}
			&\includegraphics[width=0.18\linewidth]{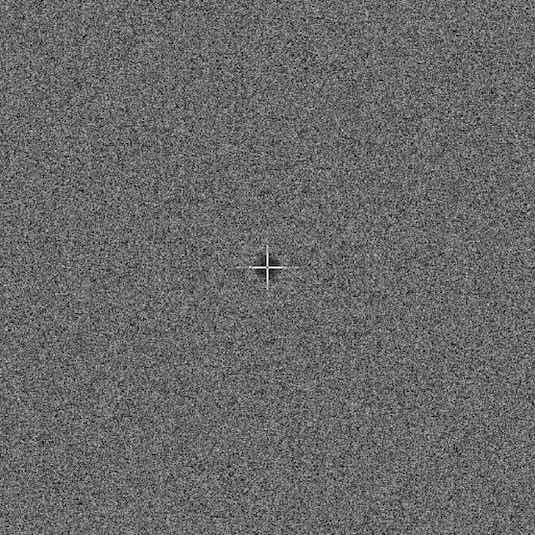}
			&\includegraphics[width=0.18\linewidth]{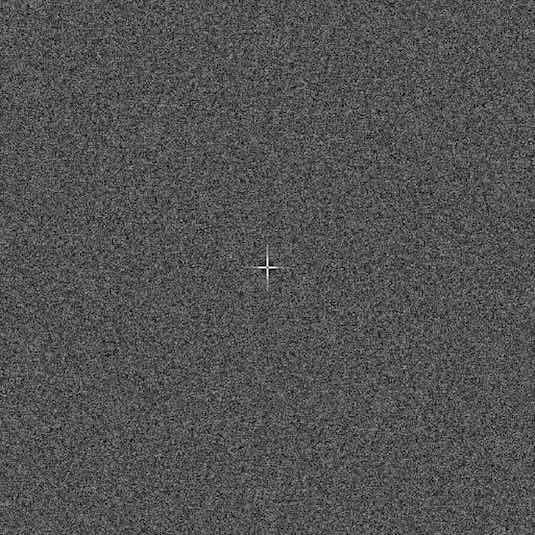}
			&\includegraphics[width=0.18\linewidth]{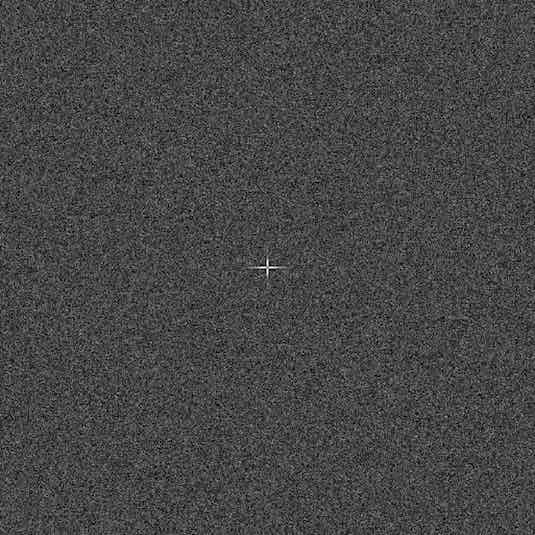}
    }
    {
			&  \includegraphics[width=0.18\linewidth]{figs/bridsonexperiments/2d/fftslice.png}
			&\includegraphics[width=0.18\linewidth]{figs/bridsonexperiments/3d/fftslice.png}
			&\includegraphics[width=0.18\linewidth]{figs/bridsonexperiments/4d/fftslice.png}
			&\includegraphics[width=0.18\linewidth]{figs/bridsonexperiments/5d/fftslice.png}
			&\includegraphics[width=0.18\linewidth]{figs/bridsonexperiments/6d/fftslice.png}
    }
			\\
			  \rotatebox[origin=l]{90}{{\hspace{28pt}RDF}}
			&\includegraphics[width=0.18\linewidth]{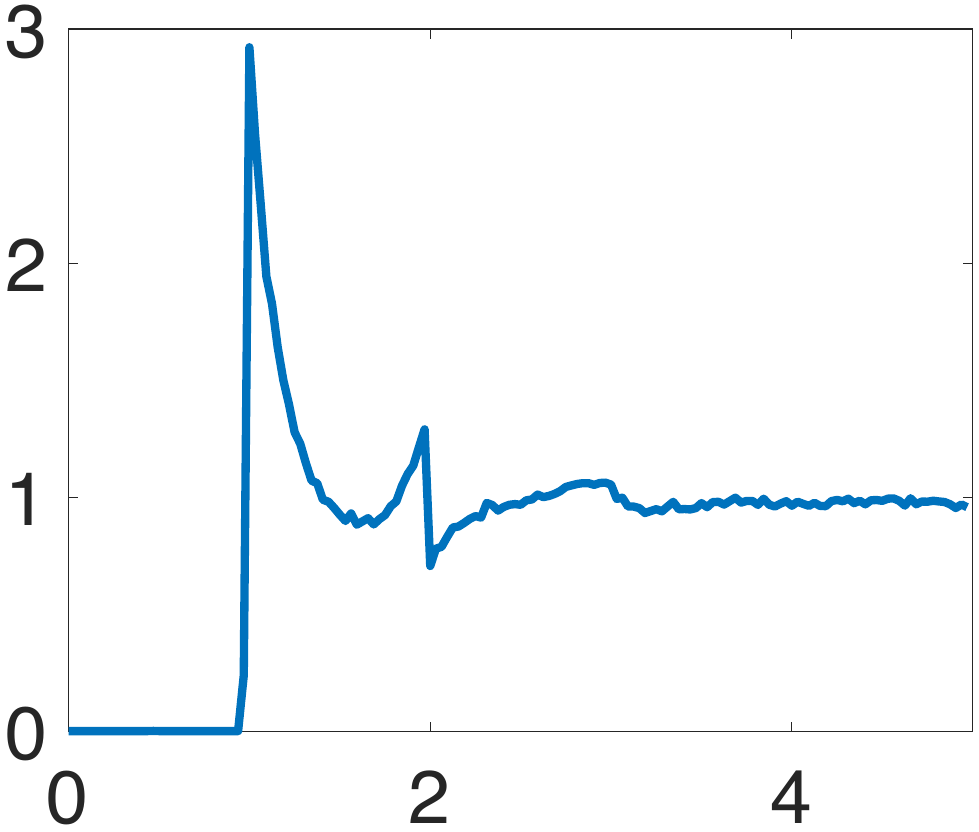}
			&\includegraphics[width=0.18\linewidth]{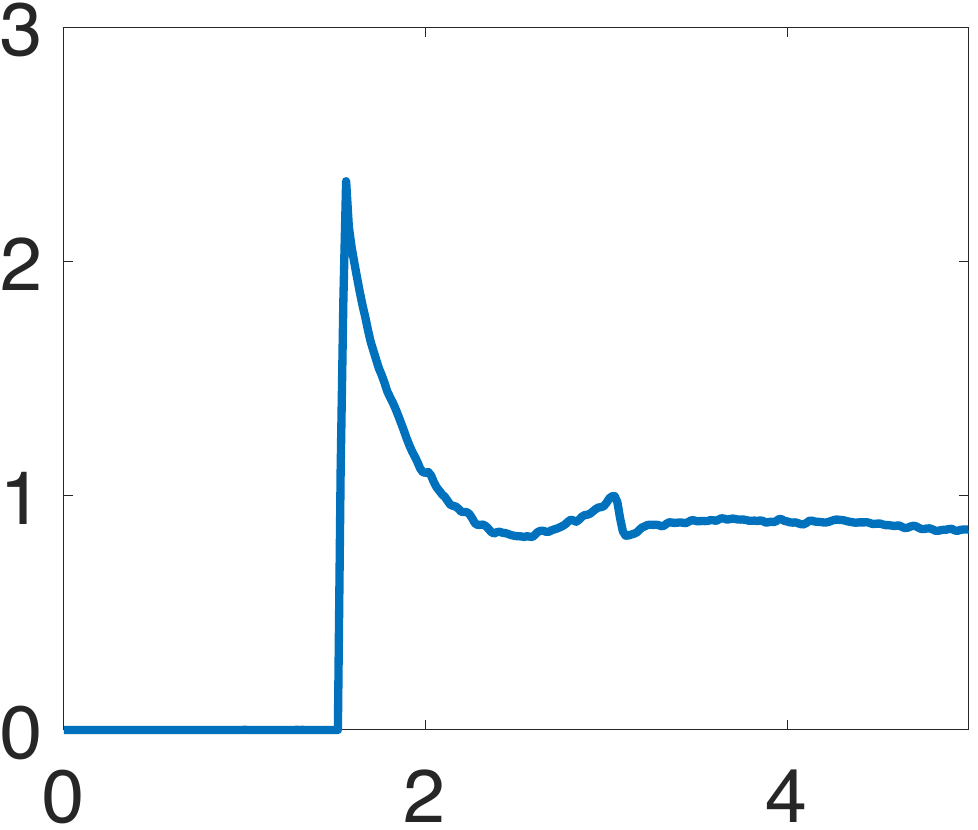}
			&\includegraphics[width=0.18\linewidth]{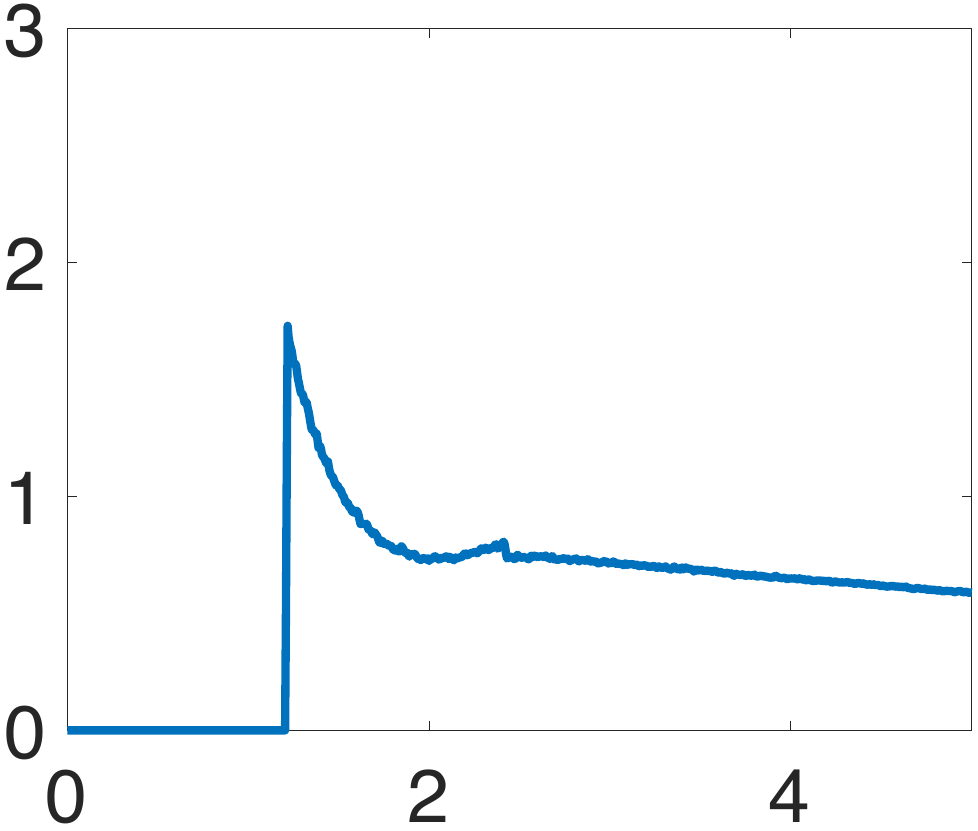}		
			&\includegraphics[width=0.18\linewidth]{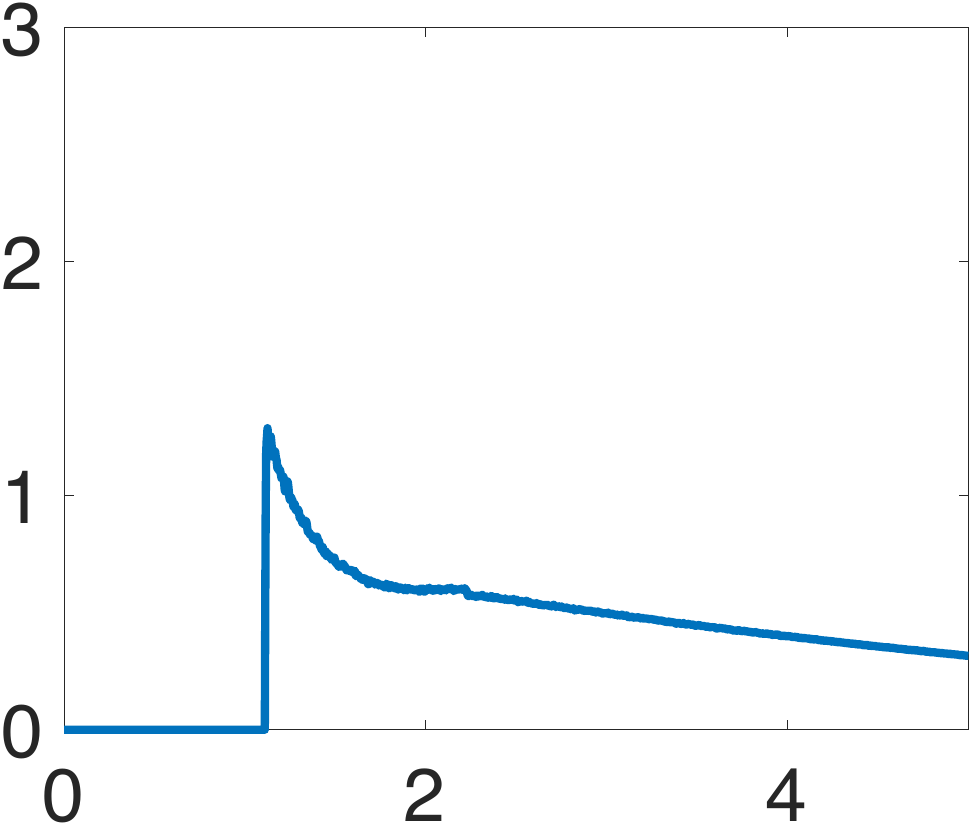}
			&\includegraphics[width=0.18\linewidth]{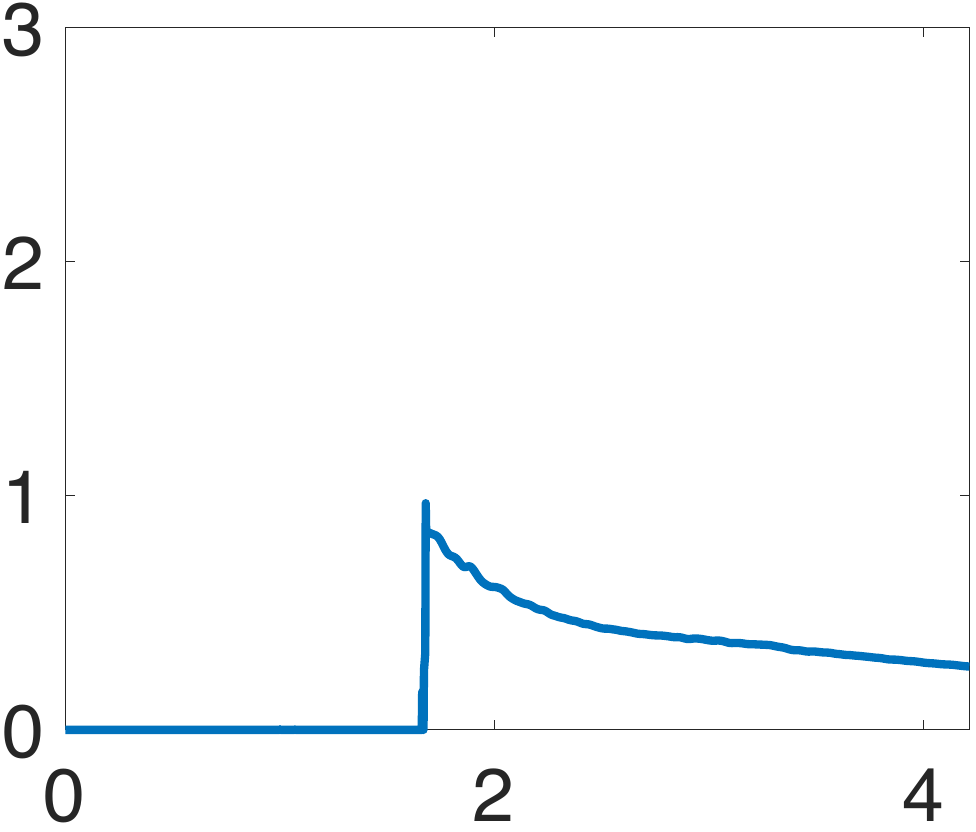}			
			\\
			  \rotatebox[origin=l]{90}{{\hspace{28pt}RP}}
			&\includegraphics[width=0.18\linewidth]{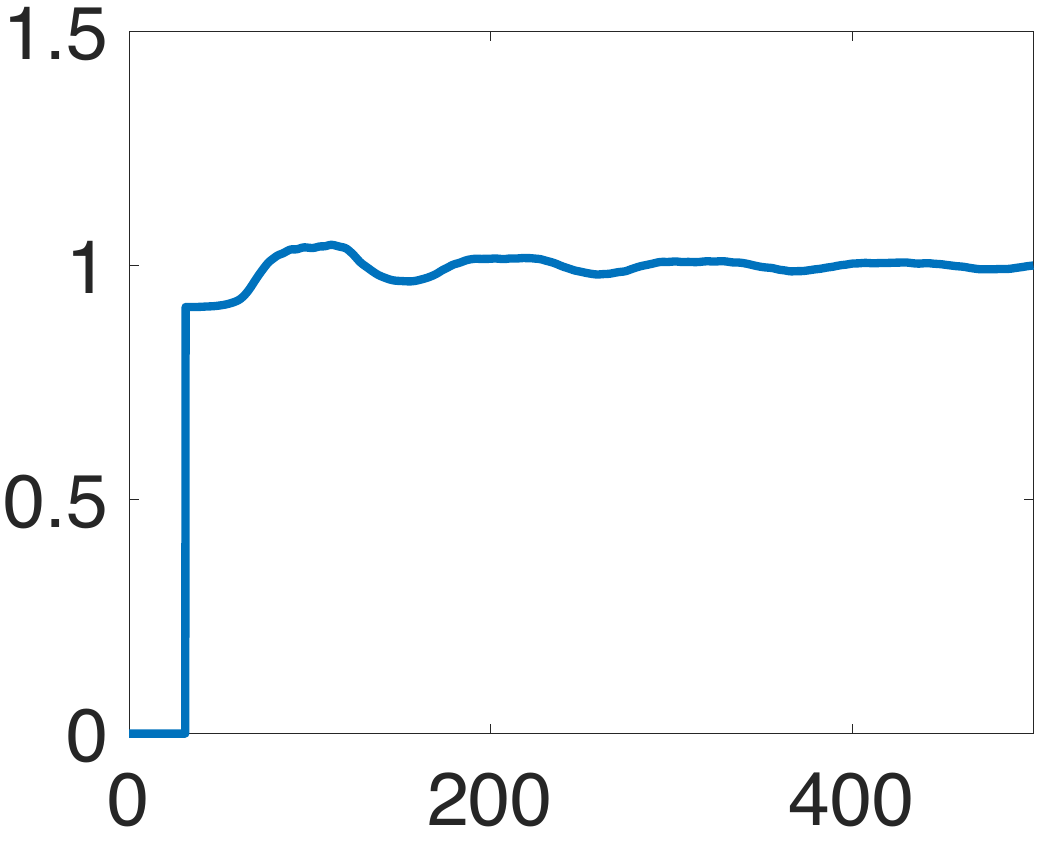}
			&\includegraphics[width=0.18\linewidth]{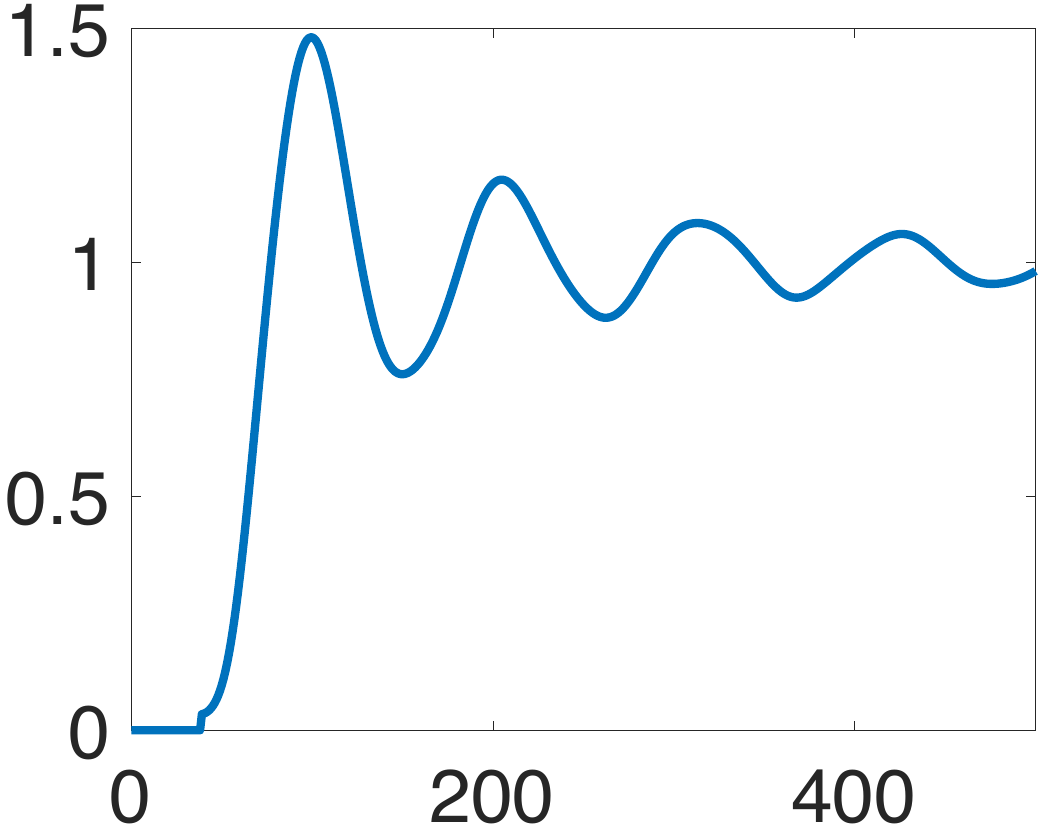}
			&\includegraphics[width=0.18\linewidth]{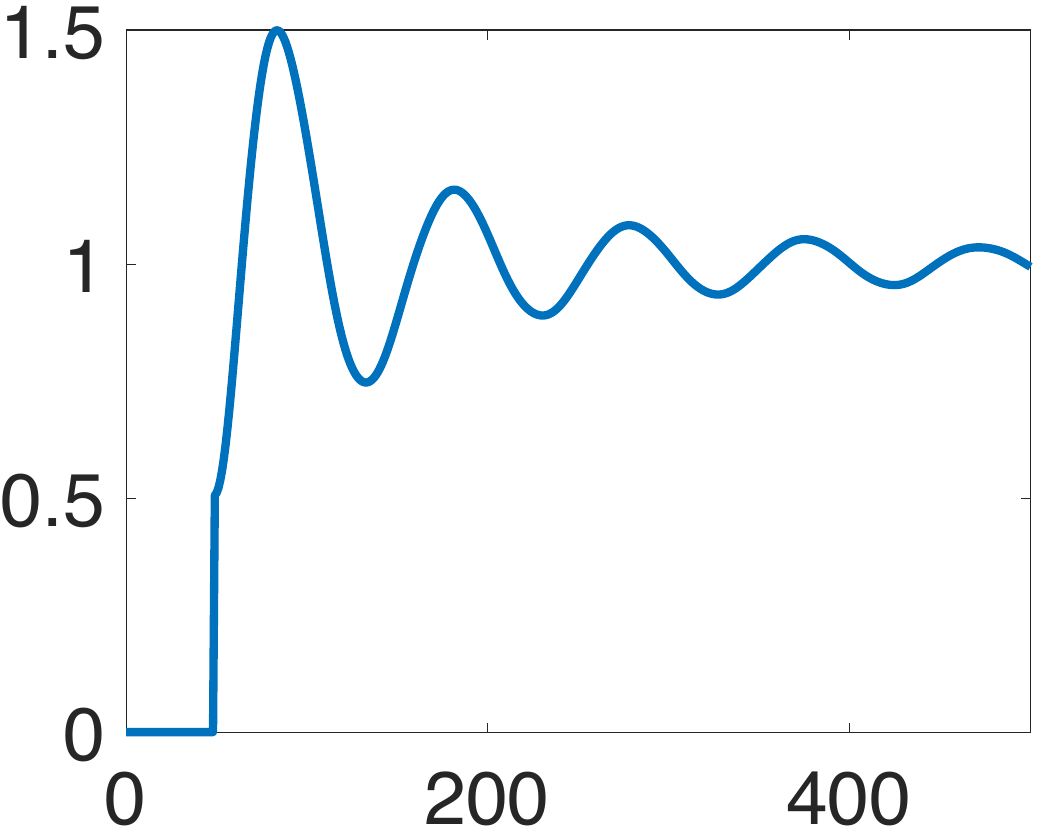}
			&\includegraphics[width=0.18\linewidth]{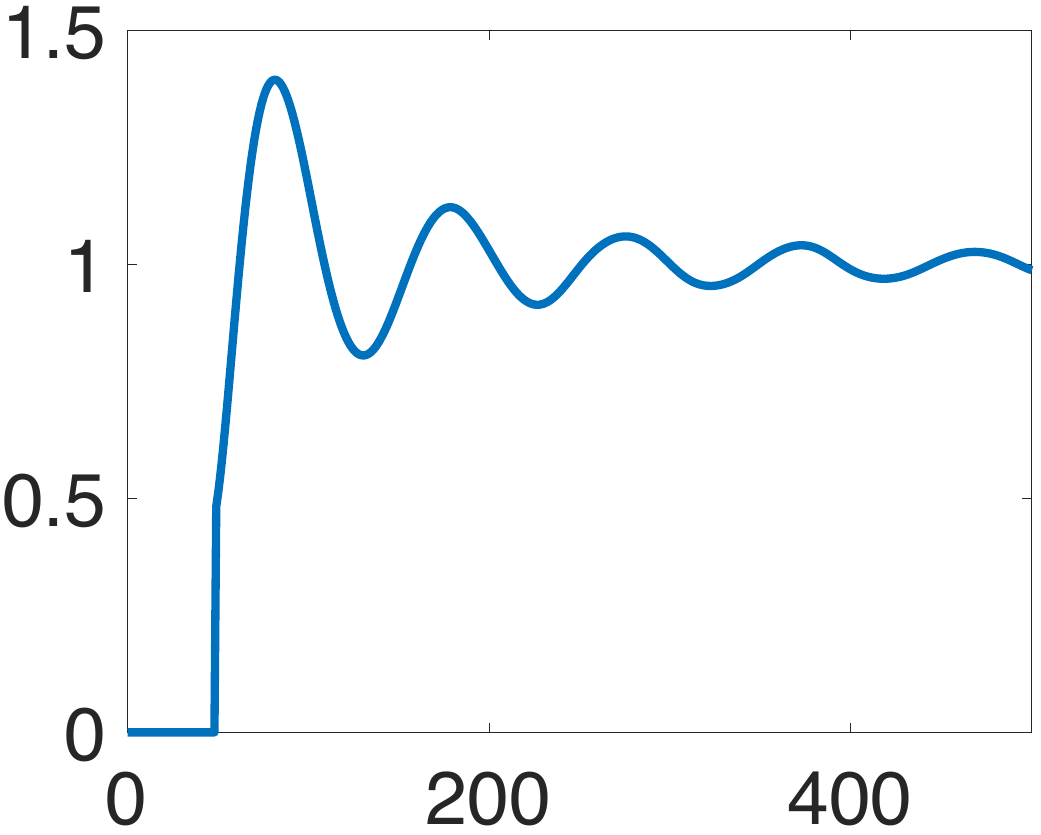}
			&\includegraphics[width=0.18\linewidth]{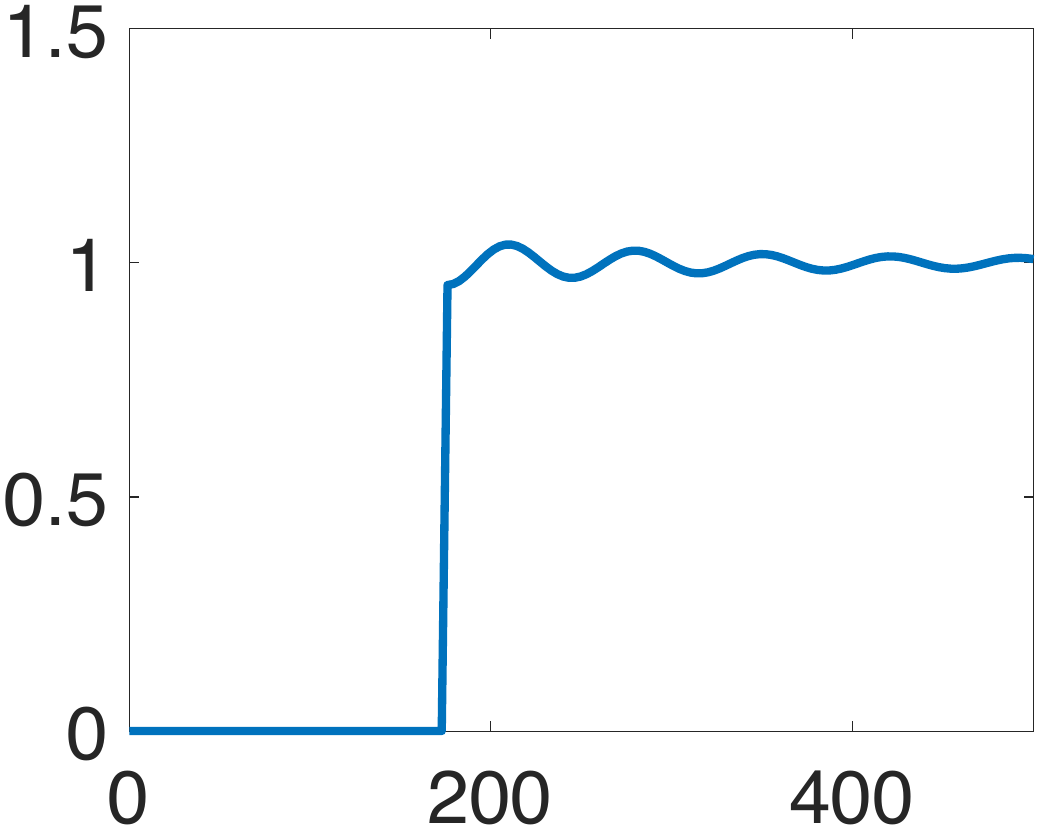}
			\\
 			  \rotatebox[origin=l]{90}{{\hspace{16pt}Vor vert to sample}}
			&\includegraphics[width=0.18\linewidth]{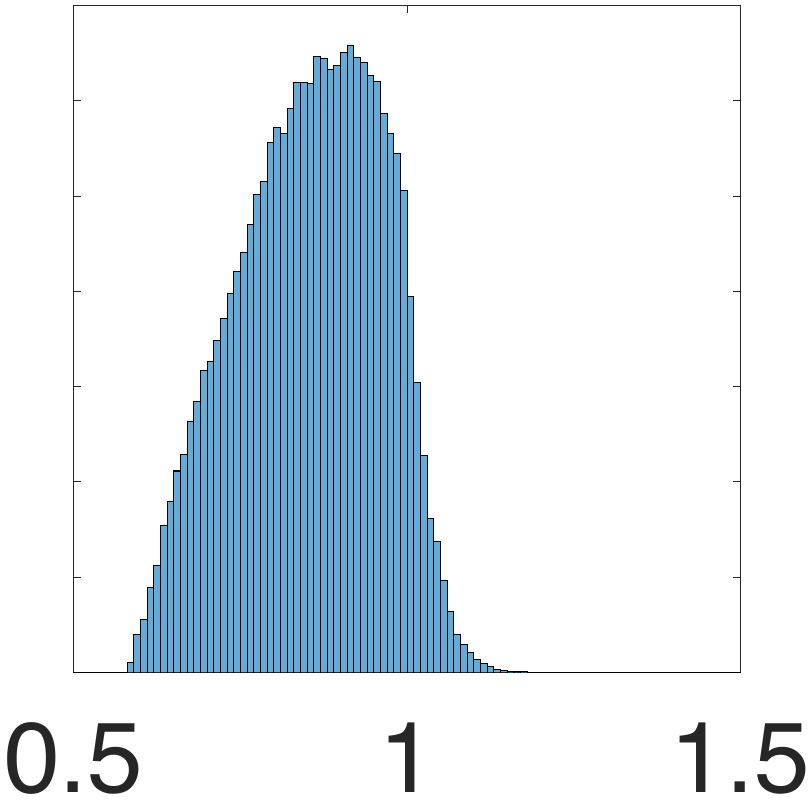}
			&\includegraphics[width=0.18\linewidth]{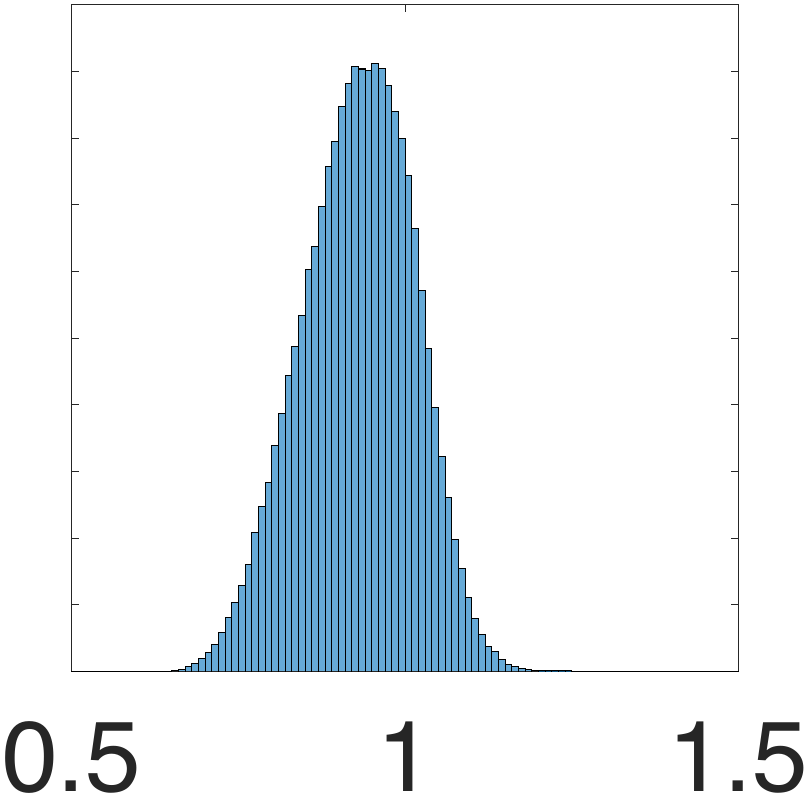}
			&\includegraphics[width=0.18\linewidth]{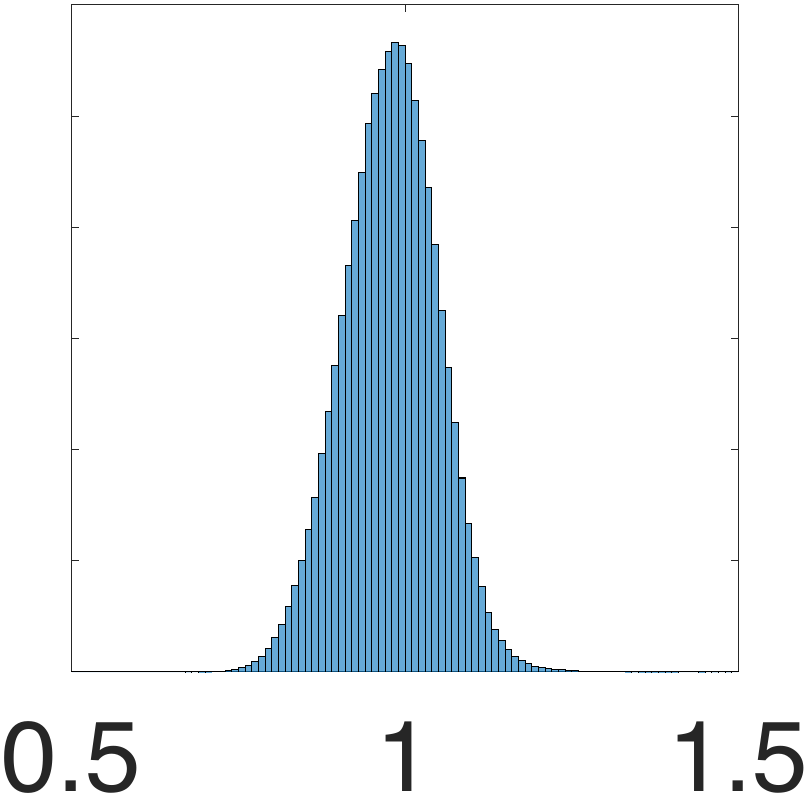}
			&\includegraphics[width=0.18\linewidth]{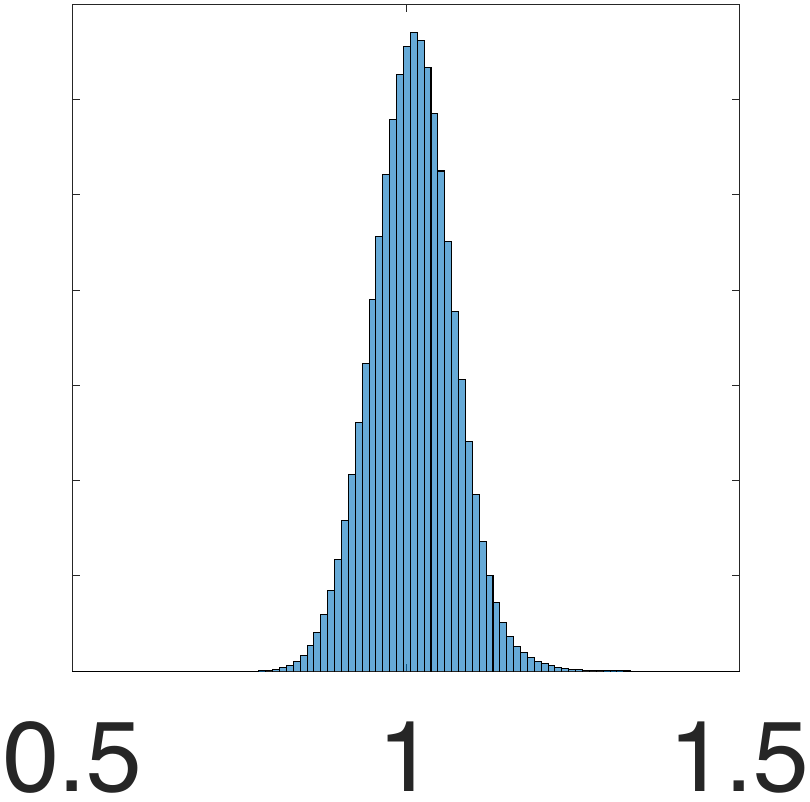}
			&\includegraphics[width=0.18\linewidth]{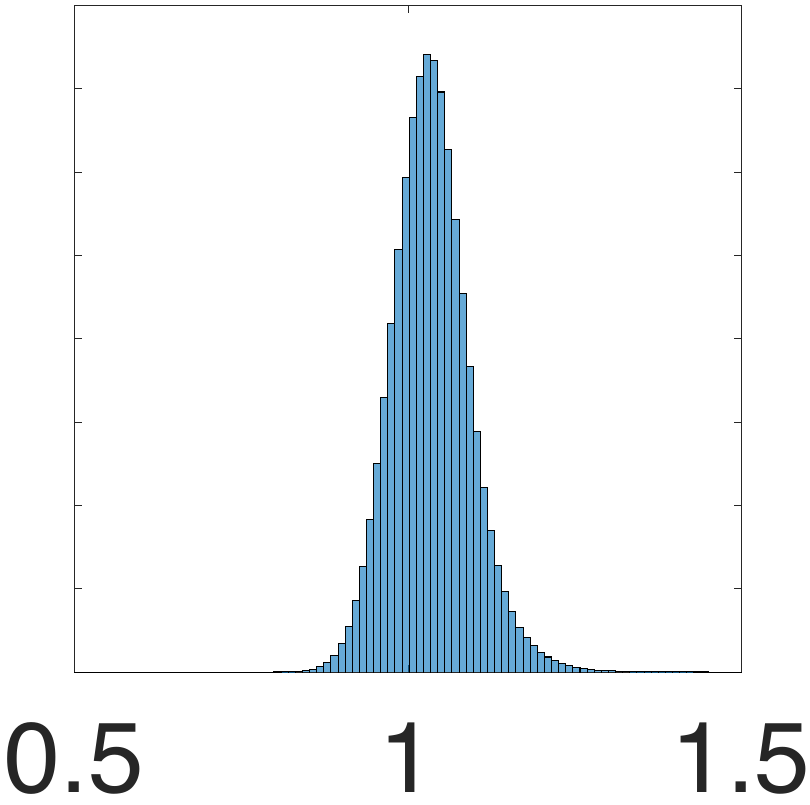}
%			\\
%			\hline
%			\\
		\end{tabular}
	}
    \Caption{\bridsonmethod\ output across dimensions 2--6 for aperiodic domains.}{}
    % ", using Bridson's implementation."
    % originally this used Bridson's implementation, but in Oct 2017 Scott regenerated the files using our implementation in order to generate vector graphics.
	%\Caption{Point-Annulus spectra across dimensions for aperiodic domains. Here we are using Bridson's implementation [Bridson 2007]. Some artifacts, such as power falling with radius, and the white `+', arise because the samplings are aperiodic. The spectra and reported dimensions are limited because of the computational complexity of the associated analysis tools, and because the available implementation of Point-Annulus uses a background grid for neighbor searching.}{}
\label{fig:point_annulus_aperiodic}
\end{figure*}
}

\subsection{Runtime scaling details}
\label{sec:runtime_details}

\pargrph{Complexity analysis details} The runtime complexity is $\mathcal{O}( \samplenumber F + \dimnumber \hammerlimit \neighbors \samplenumber),$  where $\samplenumber$ is the number of samples generated, and $F$ is the time to find $\neighbors$ neighbors.
All primitives such as computing distances and trimming spokes are linear in $\dimnumber$.
Note $\neighbors$ has dimensional dependence, but this is bounded by observing $\neighbors < \samplenumber.$
Also $F$ has additional dimensional dependence if one uses $k$-d trees or other techniques, but this is bounded by brute force searching $F = \mathcal{O}( \dimnumber \samplenumber )$.
Hence runtime is $\mathcal{O}( \dimnumber \hammerlimit \samplenumber^2 ).$
The $\mathcal{O}( \dimnumber \hammerlimit \neighbors \samplenumber)$ term arises from generating spokes and trimming. 
For each sample we have one chain of $\hammerlimit$ consecutive ``miss'' spokes that do not generate a new sample, 
plus some shorter miss chains that end with a spoke that generates a sample.
To bound the overall run-time, we must account for these small chains.
We assign the cost of a small chain to the sample disk insertion following it, not the front disk generating the chain. 
Thus each sample accounts for the ($< \hammerlimit-1$) misses preceding it, the spoke that created it, plus its own $\hammerlimit$ final successive misses, for a total of at most $2\hammerlimit$ spokes. 
Thus in the entire algorithm we throw at most $2\hammerlimit \samplenumber$ spokes, each of which takes $\mathcal{O}(\dimnumber \neighbors)$ to trim.

\pargrph{Experiments set up} We verified this complexity experimentally. 
Calculations were performed on a mid-2010 Mac, with 3.33 GHz 6-Core Intel Xeon processor, and 16 GB RAM.

\pargrph{Experiments by output size} \Cref{fig:scaling} shows the scaling of \linespokesalg\ by $\samplenumber$.
Experimentally, the \linespokesalg\ runtime using exhaustive search (array) over aperiodic domains is  about
$\num{2.0E-09}	(	1	+	0.81	d	) n^2 +
\num{5.5E-08}	(	1	+	0.05	d	) Nn +
\num{2.4E-04}	(	1	+	0.05	d	) n$.
Experimentally, the runtime for $k$-d tree search over periodic domains is  about
$\num{7.8E-07}  dn (	0.12	\log_{10} n +	N).$

\pargrph{Experiments across search type}
\Cref{fig:crossover} compares the runtime of all the line-spoke variations in small dimensions. 
Dimension 3 is close to as fast as dimension 2, because in all cases neighbor searching is a small fraction of the total time.
The advantages of a tree vs.\ exhaustive search disappear at dimension 6 or 7 for $\samplenumber=200,000$.
For smaller $\samplenumber$, the advantages disappear sooner.

\pargrph{Neighbors and periodicity}
\Cref{fig:neighbors_by_d} shows the exponential increase in the number of neighbors by dimension for periodic domains. 
It also shows the effect this has on the runtime of the favored- and two-spoke $k$-d tree variants, mainly due to their longer spokes.
% of the variants that use longer spokes when using $k$-d trees.
%, a fundamental fact independent of algorithm. 
For aperiodic domains, the boundary strongly effects $N$ as $\samplenumber$ varies, and is not illuminating.

Neighbor searching with exhaustive search is not much more expensive for periodic domains than aperiodic domains, but for $k$-d tree search periodic domains are increasingly expensive as the dimension increases. 
For $k$-d tree search, in the worst case one must do a tree search for each periodic translation of the query point, multiplying the query time by $2^d$. 
This happens more frequently for small $\samplenumber$ (large $\rnumber$). However, with larger $\samplenumber$, there is additional expense in rebalancing the tree.

%large r algorithm
For exhaustive search and small radius, the only extra step is to find the periodic translation of each coordinate of the test point that is closest to the query point, only adding a small $O(d)$ term. For large radius, a spoke may cross more than one periodic copy of a sphere. To trim a spoke, we march numerically along the spoke from the anchor, ensuring that we are trimming by the closest periodic copy of each sphere at each step. This increases the runtime by a factor of about 50; this is large, but still a constant across $d$ and $n$.
%; see \Cref{fig:scaling_by_d_fixed_N}.

\pargrph{Experiments by dimension} 
\Cref{fig:scaling_by_d_fixed_N} shows linear scaling in $d$ for fixed $\samplenumber$ using exhaustive search.
%
%
%
% QUANTIFY THE RATIO

%FIGURES for RUNTIME SCALING

% samitch: for some reason, using width=0.5 breaks each figure into its own line on at least some latex implementations.
\begin{figure}[tbh]
  \centering
   \subfloat[aperiodic domain]
   {
      \label{fig:scaling_na}
       \includegraphics[width=0.46\linewidth]{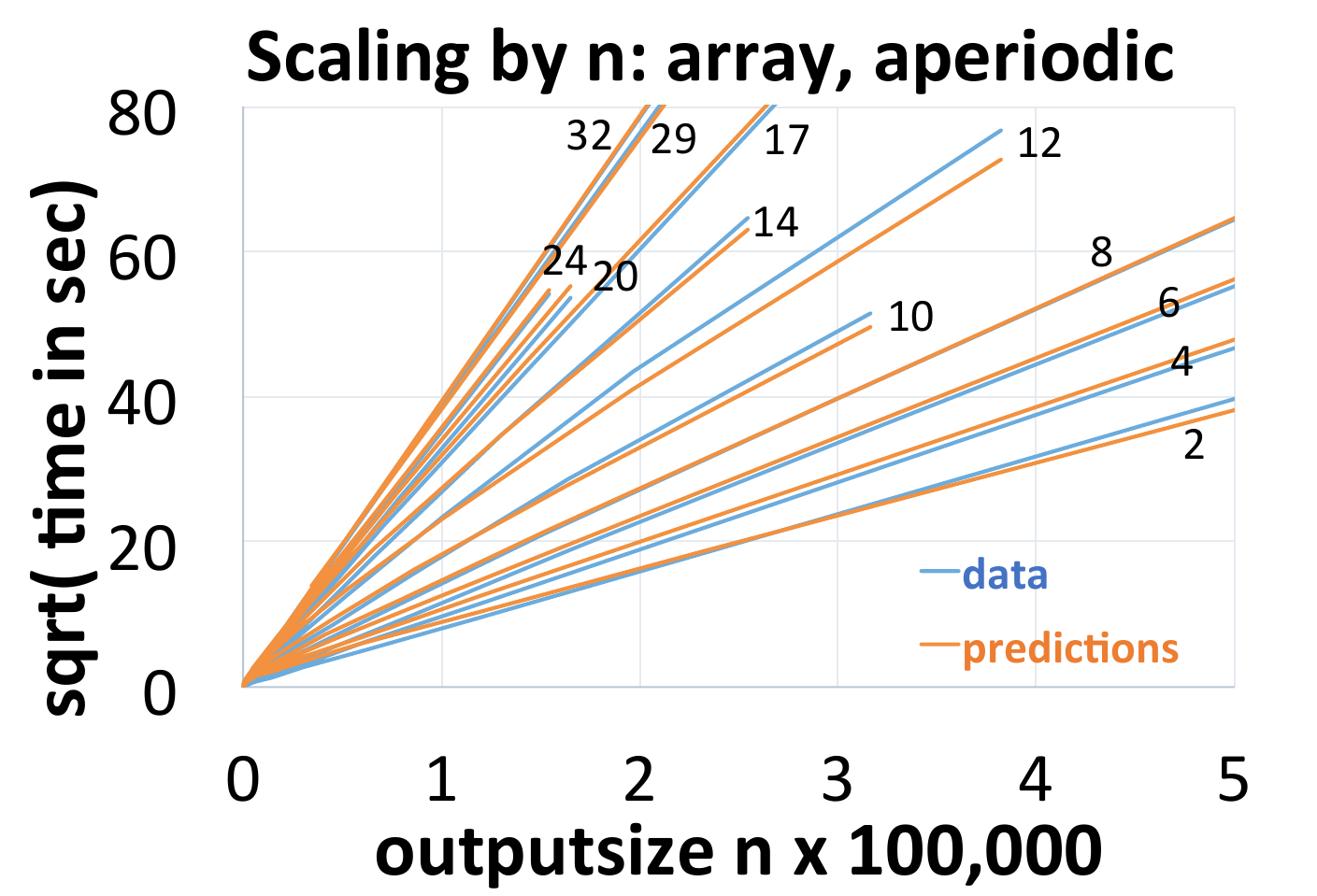}
    }
   \subfloat[periodic domain]
   {
  \label{fig:scaling_pa}
       \includegraphics[width=0.46\linewidth]{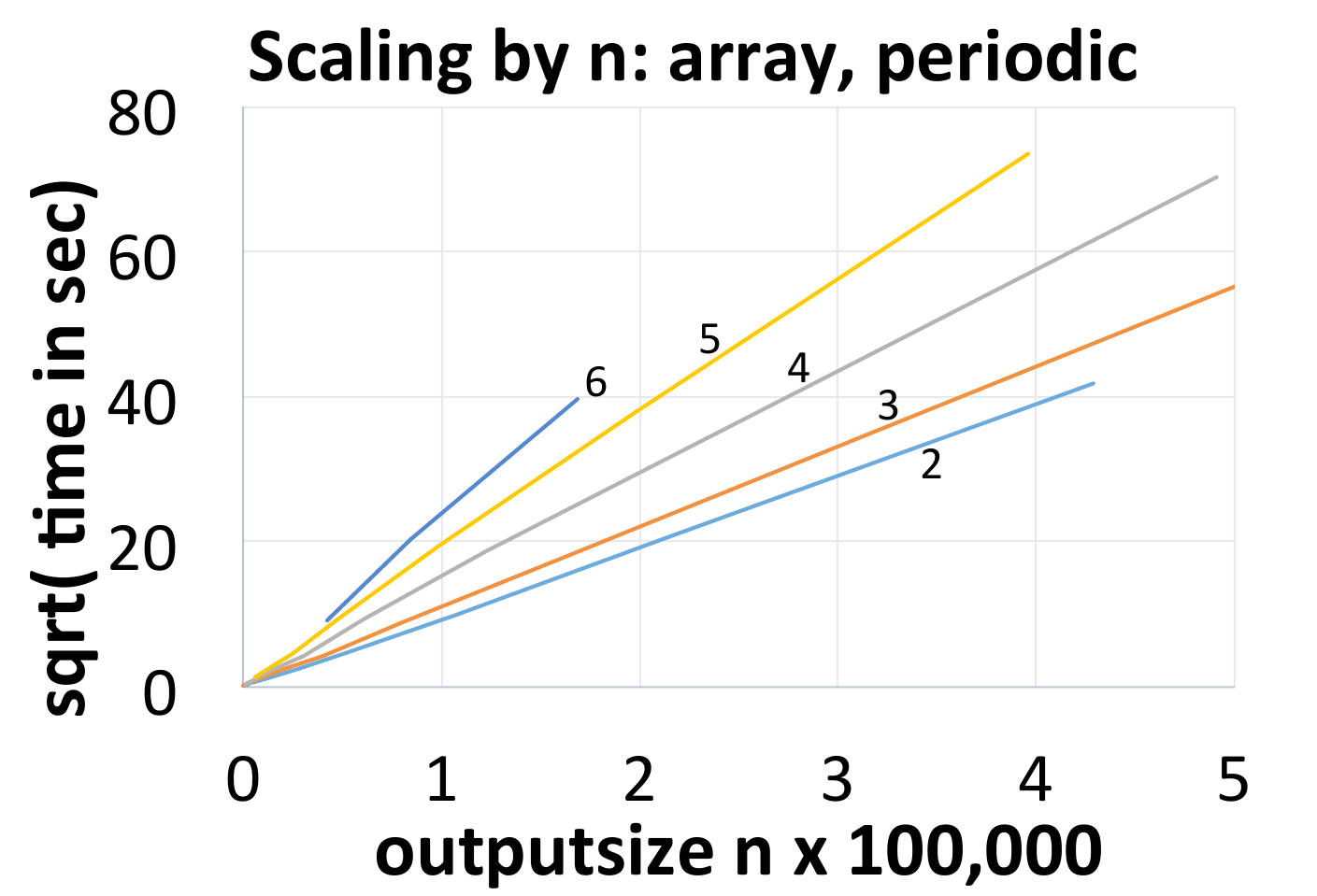}
    }
%   \caption{%
%\Linespokesalg\ with exhaustive neighbor search by $\samplenumber$.}
  \label{fig:scaling_a}
%\end{figure}
%
%
%\begin{figure}[tbh]
%  \centering

   \subfloat[aperiodic domain]
   {
      \label{fig:scaling_nt}
       \includegraphics[width=0.46\linewidth]{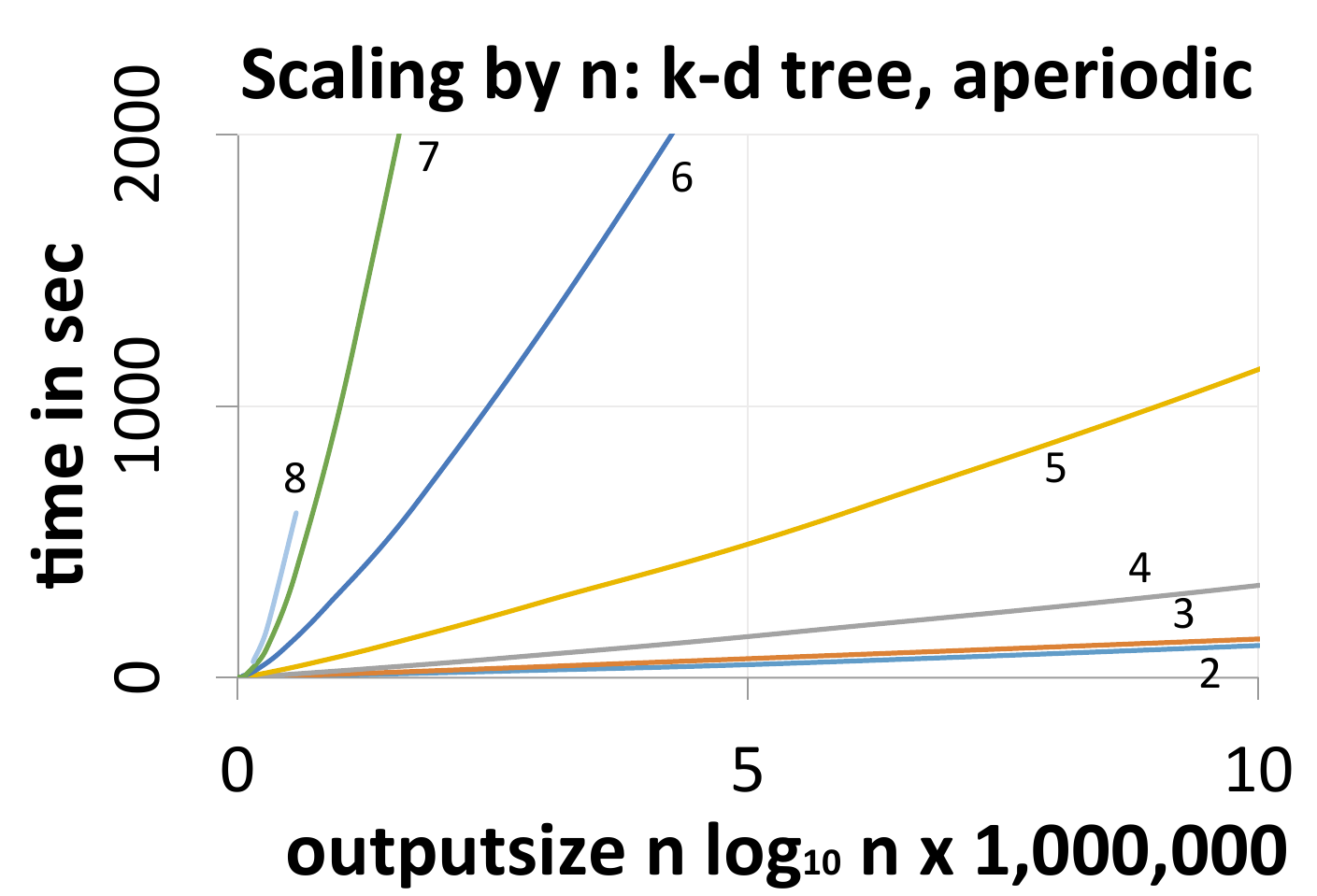}
    }
   \subfloat[periodic domain]
   {
  \label{fig:scaling_pt}
       \includegraphics[width=0.46\linewidth]{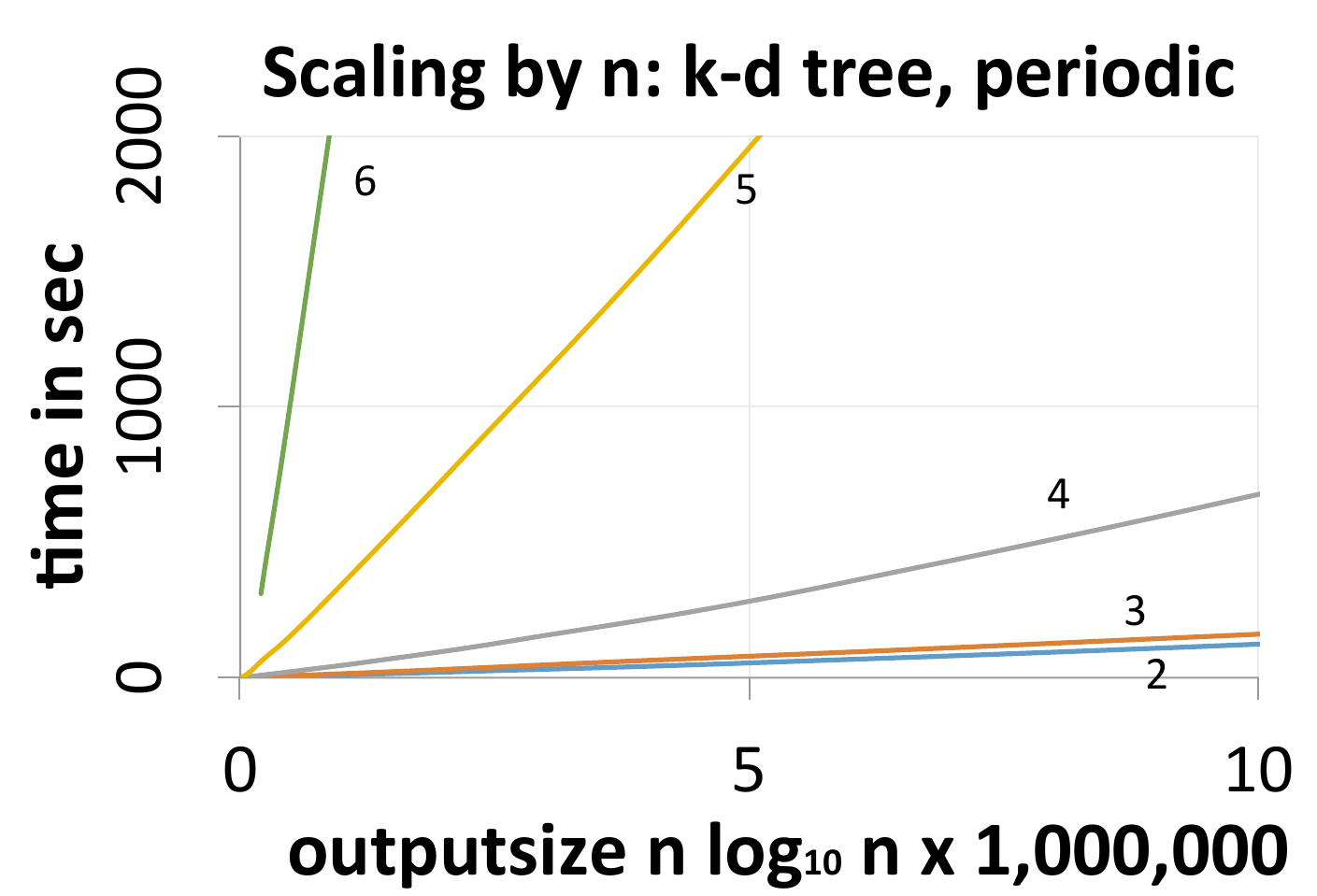}
    }
   \Caption{\Linespokesalg\ scaling by output sample number $\samplenumber$. }
   {Top: exhaustive array neighbor search. Bottom: $k$-d tree search. Each trendline is labeled by the dimension of the domain in that study. In the top, the trendlines being straight illustrate that using exhaustive search runtime is $\mathcal{O}(n^2)$ for large but fixed $d$. In the bottom, straight trendlines would illustrate perfect $\mathcal{O}(n\log n)$ scaling.}
  \label{fig:scaling_t}
  \label{fig:scaling}
\end{figure}

% samitch: for some reason, using width=0.46 breaks each figure into its own line on at least some latex implementations.
\begin{figure}[tbh]
  \centering
  \subfloat[Runtime cross-over]
   {
    \label{fig:crossover}
       \includegraphics[width=0.46\linewidth]{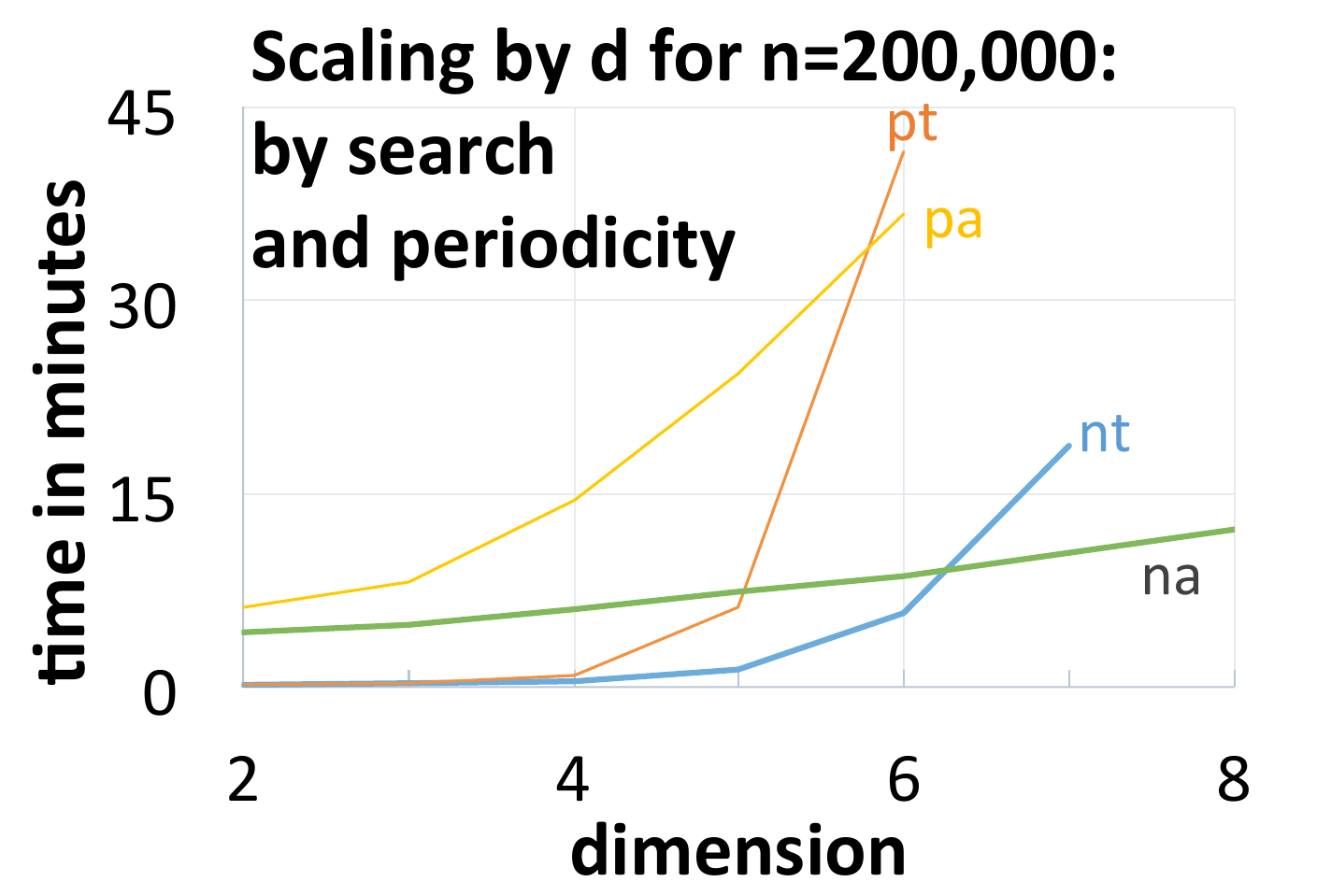}
    }
   \subfloat[Neighbor scaling]
   {
      \label{fig:neighbors_by_d}
       \includegraphics[width=0.46\linewidth]{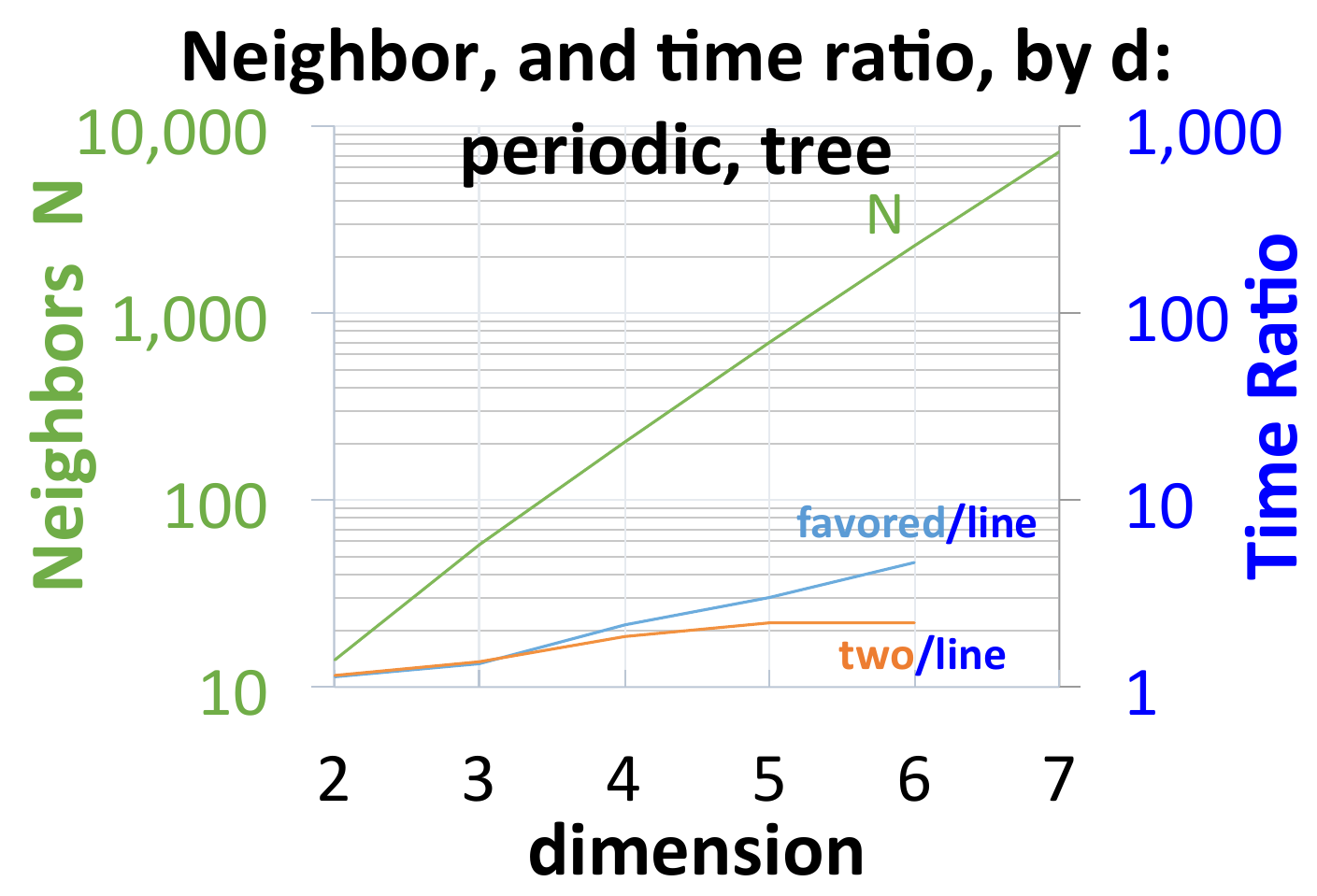}
    }   
    
  \subfloat[Linear runtime in $\dimnumber$]
  {
     \label{fig:scaling_by_d_fixed_N}
       \includegraphics[width=0.46\linewidth]{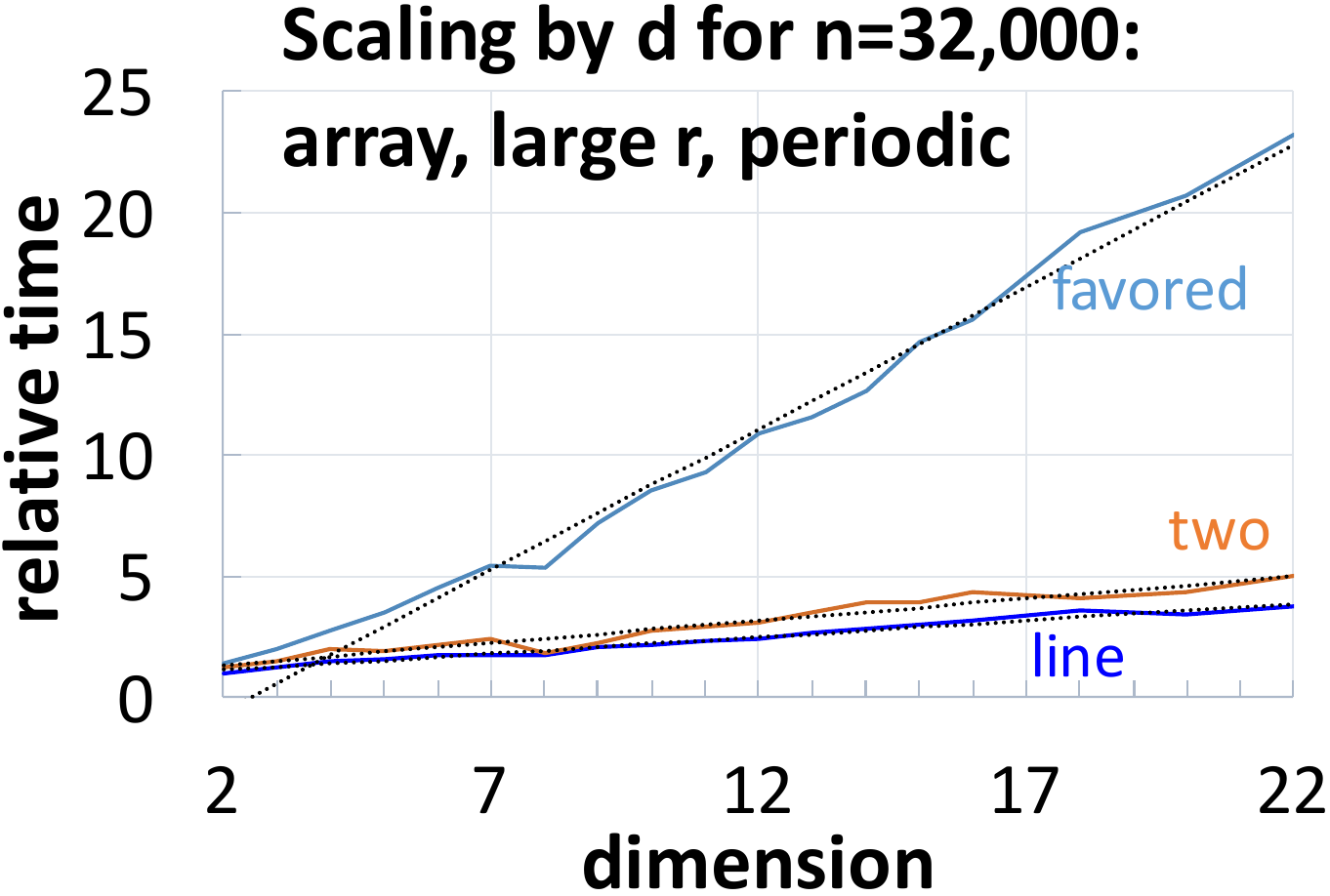}
   }
  \subfloat[Radius for fixed $\samplenumber$ by $\dimnumber$]
  {
     \label{fig:scaling_r}
       \includegraphics[width=0.46\linewidth]{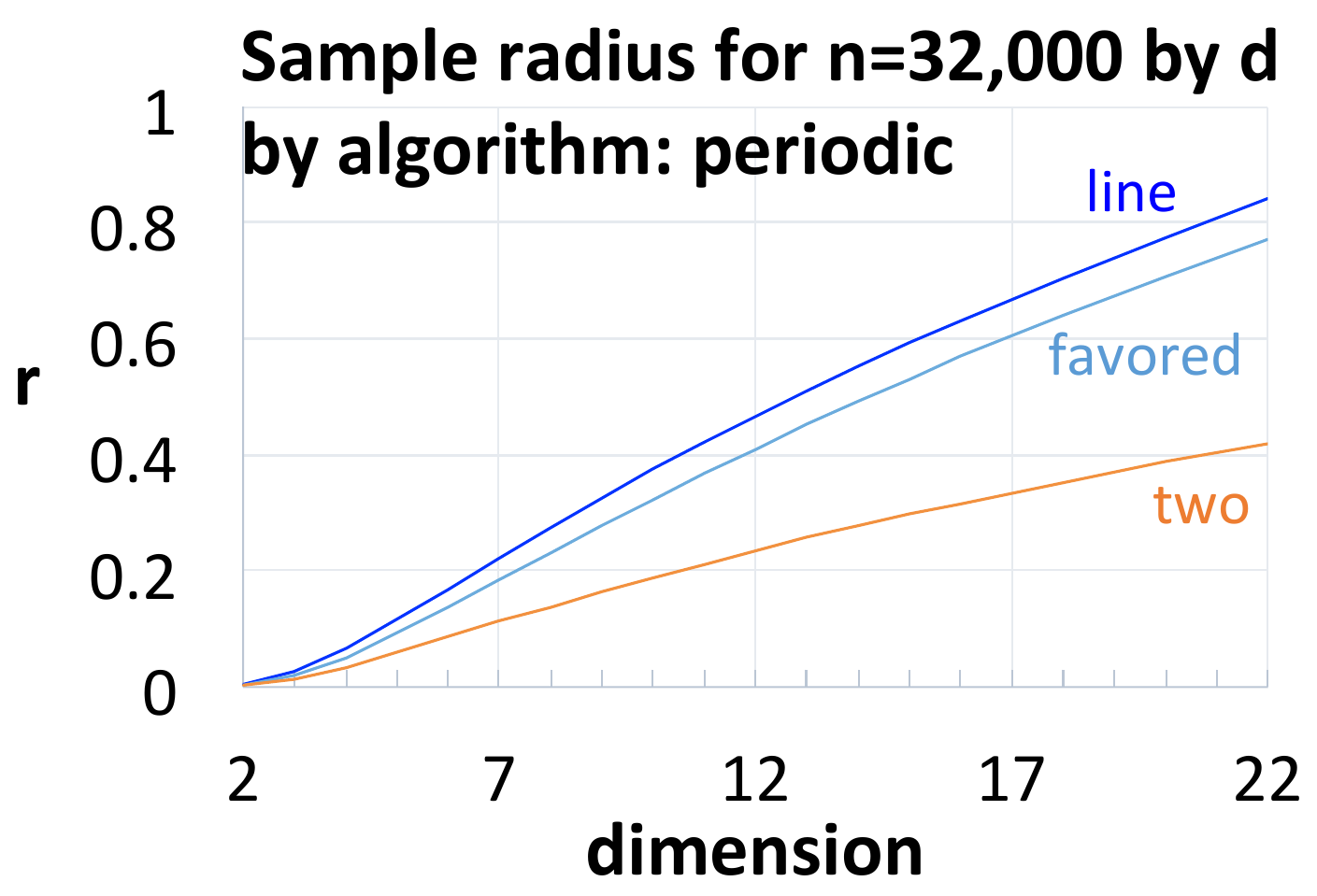}
   }
   \Caption{Fixed-$\samplenumber$ scaling by dimension.}
   {In \subref{fig:crossover}, ``pt'' is periodic domain with tree search, and ``na'' is non-periodic with exhaustive array search, etc. The straight lines in \ref{fig:scaling_by_d_fixed_N} demonstrate linear runtime in $d$ for array search.}
  \label{fig:scaling_fixed_n}
\end{figure}

%\input{scaling_method_fig}
 % figures are appearing many pages late. Need to fix

\section{Application: Approximate Delaunay graphs} 
\subsection{Motivation}
%\scott{I would title this "High-dimensional Delaunay Graphs." This is not meshing, because meshing includes inserting new points and attempting to control the shape of the elements. We don't even have any elements beyond edges.}
\label{sec:high_d_meshing}
\label{sec:meshing}

\nothing{
\liyi{(December 24, 2013) 
This is an application of \spokedarts\ but not blue noise sampling.
}%liyi
}%nothing

Many applications rely on knowing the nearby neighbors of points.
Often it is not enough to know just the nearest point or those within some threshold.
A Delaunay graph describes both nearness and directionality; intuitively it provides all points that are nearest in some general direction.
%If directionality is also important, then a Delaunay graph can help.
The size of the full Delaunay tessellation (including faces of all dimension) is inherently exponential in dimension, and computing it becomes intractable~\cite{barber1996quickhull}. However, the number of edges is at most $O(\samplenumber^2)$. For some types of random input, the number of edges is expected to be linear, $\expected(\samplenumber)$, and these can be found in $\expected(\samplenumber)$ time for static input~\cite{Dwyer:1989:HVD:73833.73869}.
For meshing and other algorithms, the input is not static and as points are adaptively added the Delaunay graph must be dynamically updated.
Some recent theoretical papers \cite{Miller:2013:NAO,Miller:2013:FAW} have considered dynamic approximate graphs and the challenges of high dimensions from the standpoint of complexity analysis, although no implementations or experimental results for these algorithms are available, and they require an over-approximation rather than an under-approximation as we provide.
%nothing

% punchline 
To our knowledge, we present the first practical implementation for dynamic approximate Delaunay graphs in high dimensions.
\nothing{
\scott{Li-Yi, does the above punchline work?}
\liyi{Awesome. See if happy with my update.}
}%nothing
% how
Graph $\dgraphast$ contains with high probability those edges whose dual Voronoi faces subtend a large solid angle with respect to the site vertex.
We call these edges {\em significant Delaunay edges}, and the corresponding $\dgraphast$ a {\em significant Delaunay graph}.
%
%  witness rewrite
As a further benefit, our method produces a {\em witness} for each edge, a domain point on its true Voronoi face, which can be used to estimate the radial extent \nothing{$\cellradius$ }of the Voronoi cell.
This is used in our global optimization application, \Cref{sec:optimization}.

The significant edges are a subset of the true Delaunay edges, and the Voronoi cell defined by the significant neighbors geometrically contains the true Voronoi cell.
Many high-dimensional applications can accept approximate Delaunay graphs; the 
effect of the missing edges is application dependent.
\nothing{
\scott{``Hence our method over-estimates the cell volumes, which gives an upper bound on the achieved $\rratio$ of the distribution.'' I'm leaving this out because we can't calculate the volume, and we can't calculate the Voronoi vertices, so we can't calculate a bound on $\rratio$.}
}%nothing
For global optimization, \Cref{sec:optimization}, the approximation affects efficiency and not correctness.
The neighborhood sizes around sample points determine the order in which new samples are generated.
Prior approaches use rectangles which usually grossly overestimate the neighborhood sizes.
The significant Delaunay graph provides a more accurate estimate of neighborhood sizes.
The true Voronoi vertices of a true Delaunay graph would give the most accurate sizes, but there are an exponential number of them.
Such compromises are commonly necessary for high-dimensional problems. For example, the state-of-the-art high-dimensional nearest neighbor (and $k$-nearest) query~\cite{muja2009fast} often returns the wrong nearest point, but its distance is probably not much greater than the distance to the true nearest neighbor. 
%This algorithm is widely used.
As algorithms for high-dimensional graphs have improved, their use has increased in fields such as uncertainty quantification~\cite{witteveen2012simplex} and computational topology~\cite{gerber2010visual}.
\nothing{including the Morse-Smale complex and its application to visualization~}
\nothing{, and segmentation~\cite{paris2007topological}\liyi{\cite{paris2007topological} is not high dimensional as far as I can see.}}
\nothing{
\liyi{(December 3, 2013)
Need a connection/transition from high dimensional meshing above and approximate Delaunay graph below.
(January 14, 2014)
I did a try; see if correct.
}%liyi
}%nothing
%\subsubsection{Approximate Delaunay Edges}
%

%
%
% whose two closest sample points are the cell center and the neighbor, which is very valuable and yet difficult to compute in high dimensions. 
%
%
% Many high dimensional applications, such as classic approximate nearest neighbor problem, accept approximate Delaunay graphs.
%
%An edge in the Delaunay graph connects two points that shares a Voronoi face.  
%An approximate Delaunay graph $\mathcal D^\ast$ may have some missing edges, but these are dual to facets with relatively small solid angles, and small areas for well-spaced point sets because of the aspect ratio bound.
\nothing{\liyi{simplified into above}
The classic approximate nearest neighbor problem in high-dimensions might also make use of our under-approximation, because either the true nearest neighbor is significant, or it is obscured by other almost-as-near neighbors.
}%nothing
\nothing{\liyi{not important; the witness point can simply be computed as the midpoint from each Delaunay edge, no?}
\scott{The witness is important. They are hard to compute. If you could just check the midpoint then computing the exact Delaunay graph would be an $n^3$ operation in every dimension, and it isn't. There is no guarantee that the midpoint of a Delaunay edge is closer to the two endpoints than any other vertex. For example, in two-dimensions this is only true about the midpoint when all angles of the triangulation are non-obtuse, which is rare. The witness is extremely valuable, because it avoids the need to go searching for one later when, for example, we want to add a sample point in our version of Direct. Saves about half the time of the algorithm. I mentioned this to Mohamed before but I'm not sure it is in the current write-up. Also, having a witness of a claimed fact is extremely useful for showing the correctness of an algorithm or implementation. The witness is a classic concept in computer science.  I don't understand why you think it is not important or uninteresting.} 
}%nothing 
%nothing
\nothing{
\liyi{(January 14, 2014)
I cannot understand how \spokedart{} got applied to high dimensional meshing from the following paragraph.

(January 17, 2014)
Mohamed, I cannot understand this part at all, neither the what part nor the why part.
It does not really say anything. No reviewer will understand it.
Almost every single sentence is vague hand-waving.
To begin with, it is not even clear what is the output.
Are we just trying to find the edges connecting the vertices, or also the dual Voronoi regions, for which we need to find all simplexes (not just edges but also Voronoi vertices, faces, etc.)?
For people with difficulty expressing themselves, I usually find it very effective to force them write down the pseudo-code.
(Assuming they know the algorithm, they should be able to do so.)
I can then translate that into English.
Please draft \Cref{alg:meshing}, and let me know when you are done.
I have done pseudo-codes for other parts of the paper and you can look at them as examples.
Another thing that I highly recommend is to add a 2D illustration analogous to what you have done in \Cref{fig:spokes} and \Cref{fig:gopt_main}.
}%liyi
}%nothing

\subsection{Algorithm}
\begin{algorithm}
  \begin{algorithmic}[1]
   
    \REQUIRE $\sample$, graph $\dgraphast$, NeighborCandidates $\neighborcandidates$, RecursionFlag $R$
    \ENSURE $\dgraphast$ with $\sample$ added
    \STATE \pcomment{$R=\textrm{true}$ for a new vertex $\sample$}
    \STATE $\neighborset = \emptyset$ \pcomment{approx.\ Delaunay neighbors of $\sample$}
    \STATE $\cellradius(\sample) = 0$ \pcomment{approx.\ cell radius of $\sample$}

    \FOR{$i=1$ \textbf{to} $\hammerlimit$}
    \STATE $\spoke \leftarrow \funct{RandomSpoke}(\sample, 0, |\domainsym|)$ \label{alg:meshing_random_line_spoke}
%  \pcomment{\Cref{sec:operation}}
    \FOR{each sample $\sampleprime \in \neighborcandidates$}
    \STATE $\pi(\sample, \sampleprime) \leftarrow$ hyperplane between $\sample$ and $\sampleprime$
    \STATE trim $\spoke$ with $\pi(\sample, \sampleprime)$
%    \STATE $\ell_{old} \leftarrow \funct{length}(\spoke)$
%    \STATE $\ell_{new} \leftarrow \funct{length}(\spoke)$
    \IF{$\ell$ got shorter}
    \STATE $\samplestar \leftarrow \sampleprime$
    \ENDIF
    \ENDFOR
    \STATE $\dgraphast \leftarrow \dgraphast \bigcup \left\{ \overline{\sample \samplestar} \right\}$ \pcomment{without duplication}
    \STATE $\neighborset \leftarrow \neighborset \bigcup \{\samplestar\}$
    \STATE $\cellradius(\sample) = \max\left(\cellradius(\sample), \funct{length}(\ell)\right)$
    \ENDFOR
    
    \IF{$R=\textrm{true}$}
    \STATE \pcomment{update edges of neighbors, removing some}
    \FOR{each sample $\sampleprime \in \neighborset$}
      \STATE $\neighborcandidates \leftarrow \funct{Neighbors}(\sampleprime) \bigcup \{\sample\}$ 
      \STATE $\dgraphast \leftarrow \dgraphast \setminus \funct{Edges}(\sampleprime)$ \pcomment{remove all edges}
      \STATE $\funct{Recurse}(\sampleprime,\dgraphast,\neighborcandidates,\textrm{false})$ \pcomment{restore some}
    \ENDFOR
  \ENDIF

    \RETURN{} $\dgraphast$
  \end{algorithmic}
  
  \Caption{Add a vertex to the approximate Delaunay graph.}{}
% Each spoke $\spoke$ has $\rnumber = 0$ in line~\ref{alg:meshing_random_line_spoke}.
  \label{alg:meshing}
\end{algorithm}
%
%
%\begin{algorithm}
%  \begin{algorithmic}[1]
%   
%    \REQUIRE new vertex $\sample$ and the existing Delaunay graph $\dgraphast$
%    \ENSURE $\dgraphast$ with $\sample$ added
%
%    \FOR{$i=1$ \textbf{to} $\hammerlimit$}
%    \STATE $\spoke \leftarrow \funct{RandomLineSpoke}(\sample, 0, \spokesize)$ \pcomment{\Cref{sec:operation}} \label{alg:meshing_random_line_spoke}
%    \FOR{each sample $\sampleprime$ near $\sample$}
%    \STATE $\pi(\sample, \sampleprime) \leftarrow$ hyperplane between $\sample$ and $\sampleprime$
%    \STATE $\ell_{old} \leftarrow \funct{length}(\spoke)$
%    \STATE trim $\spoke$ with $\pi(\sample, \sampleprime)$
%    \STATE $\ell_{new} \leftarrow \funct{length}(\spoke)$
%    \IF{$\ell_{new} < \ell_{old}$}
%    \STATE $\samplestar \leftarrow \sampleprime$
%    \ENDIF
%    \ENDFOR
%    \STATE $\dgraphast \leftarrow \dgraphast \bigcup \left\{ \overline{\sample \samplestar} \right\}$ \pcomment{set union without duplication}
%    \ENDFOR
%
%    \RETURN{} $\dgraphast$
%  \end{algorithmic}
%  \Caption{Approximate Delaunay graph via our method.}
%{Note that each spoke $\spoke$ has $\rnumber = 0$ in line~\ref{alg:meshing_random_line_spoke}.
%\scott{If you want to describe $w$, then say ``Some distributions have a bound on the maximum possible value of $r_c.$ We can set $w = \max{r_c},$ or $\infty$ if there is no bound.}
%\liyi{Here we assume each spoke $\spoke$ is sufficiently long to begin with.}
%}
%  \label{alg:meshing}
%\end{algorithm}

%
Our basic idea is to throw random \linespokestext{} to tease out the significant Delaunay edges from a set of spatial neighbors.
This is a very simple method that scales well across different dimensions.
It is summarized in \Cref{alg:meshing} with details as follows.
We construct the \nothing{approximate }graph $\dgraphast$ for each vertex $\sample$ in turn.
We initialize its edge pool with all vertices that are close enough to possibly share a Delaunay edge with $\sample$.
We next identify vertices from this pool who are actual Delaunay neighbors of $\sample$ with the following probabilistic method.
%\liyi{what probabilistic method? this is entirely non-reproducible.} \scott{The method we describe next.}
Using \spokedarts{}, we throw $\hammerlimit$ \linespokestext{}.
We trim each spoke $\spoke$ using the separating hyperplane between $\sample$ and each vertex $\sampleprime$ in the pool.  
There is one pool vertex $\samplestar$ whose hyperplane trims $\spoke$ the most.
(In degenerate cases where multiple vertices trim the spoke the most and equally, we can pick an arbitrary one for $\samplestar$.)
The far end of the trimmed spoke $\witness$ is equidistant from $\sample$ and $\samplestar$, and no other vertex is closer.
Hence $\witness$ is the witness that $\sample$ and $\samplestar$ share a Voronoi-face (Delaunay-edge), and $\overline{\sample \samplestar}$ is added to $\dgraphast$.

\nothing{
\liyi{(January 19, 2014)
This algorithm is about Delaunay graph, connecting edges among vertices.
However, our optimization requires Voronoi regions.
How do we compute these? It is not all that trivial in high dimensions, right?
I guess this is done as I described under \Cref{alg:gopt} caption.
}%liyi
\scott{(January 25, 2014) A Voronoi face is dual to a Delaunay edge, meaning they are always defined by the same pair of vertices.}
}%nothing

\muhammad{(January 19, 2014) To increase efficiency of capturing Delaunay neighbors using \spokedart{} I had to extend the spoke across the sphere center. So, instead of getting just one Delaunay neighbor per spoke we get two neighbors.}%muhamamd

% samitch: for some reason, using width=0.46 breaks each figure into its own line on at least some latex implementations.
\begin{figure}[tbh] 
 \centering
   \subfloat[computation time]
   {
     \label{fig:spoke_delaunay_speed}
     \includegraphics[width=0.46\linewidth]{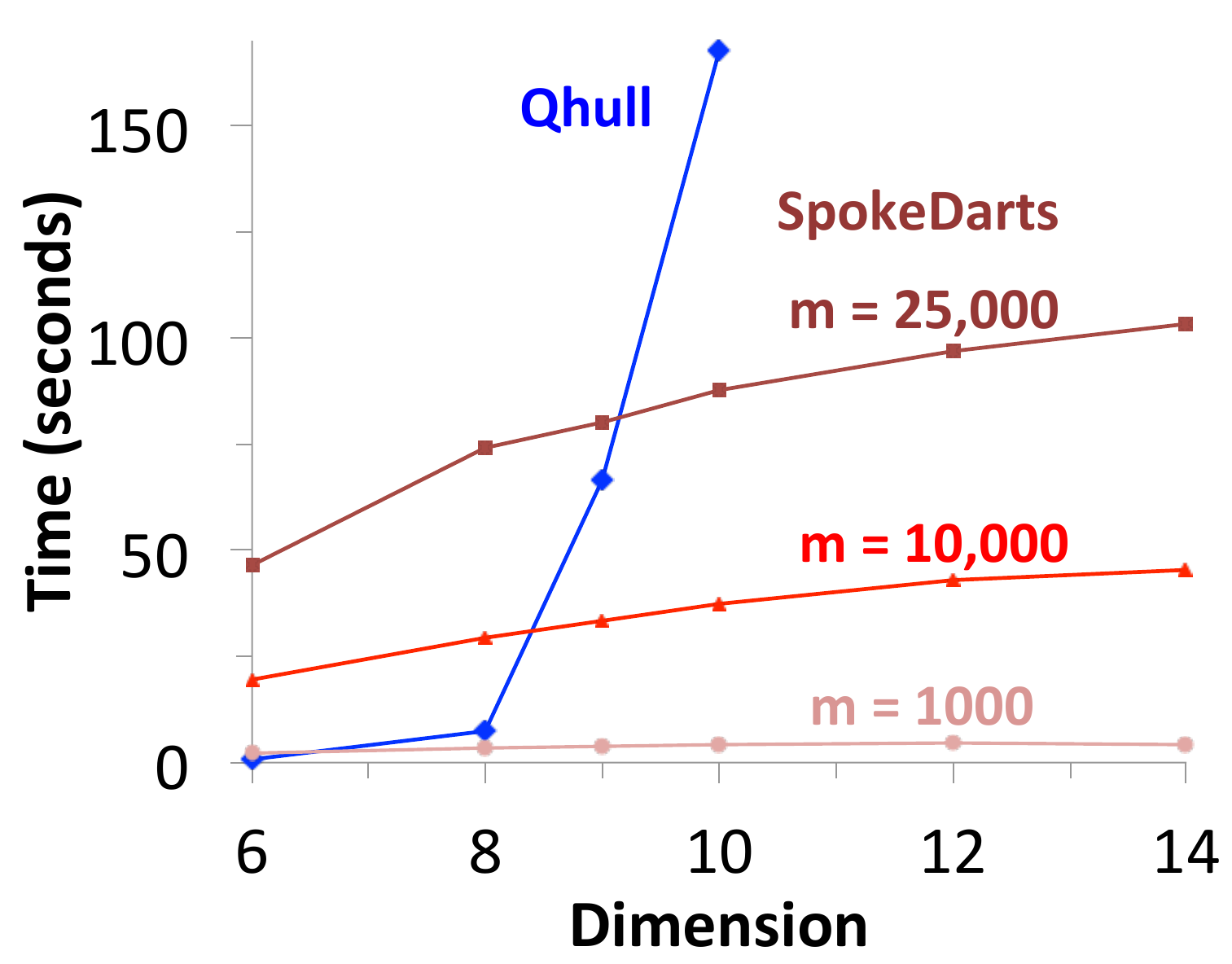}
   }
   \subfloat[memory requirement]
   {
     \label{fig:spoke_delaunay_memory}
     \includegraphics[width=0.46\linewidth]{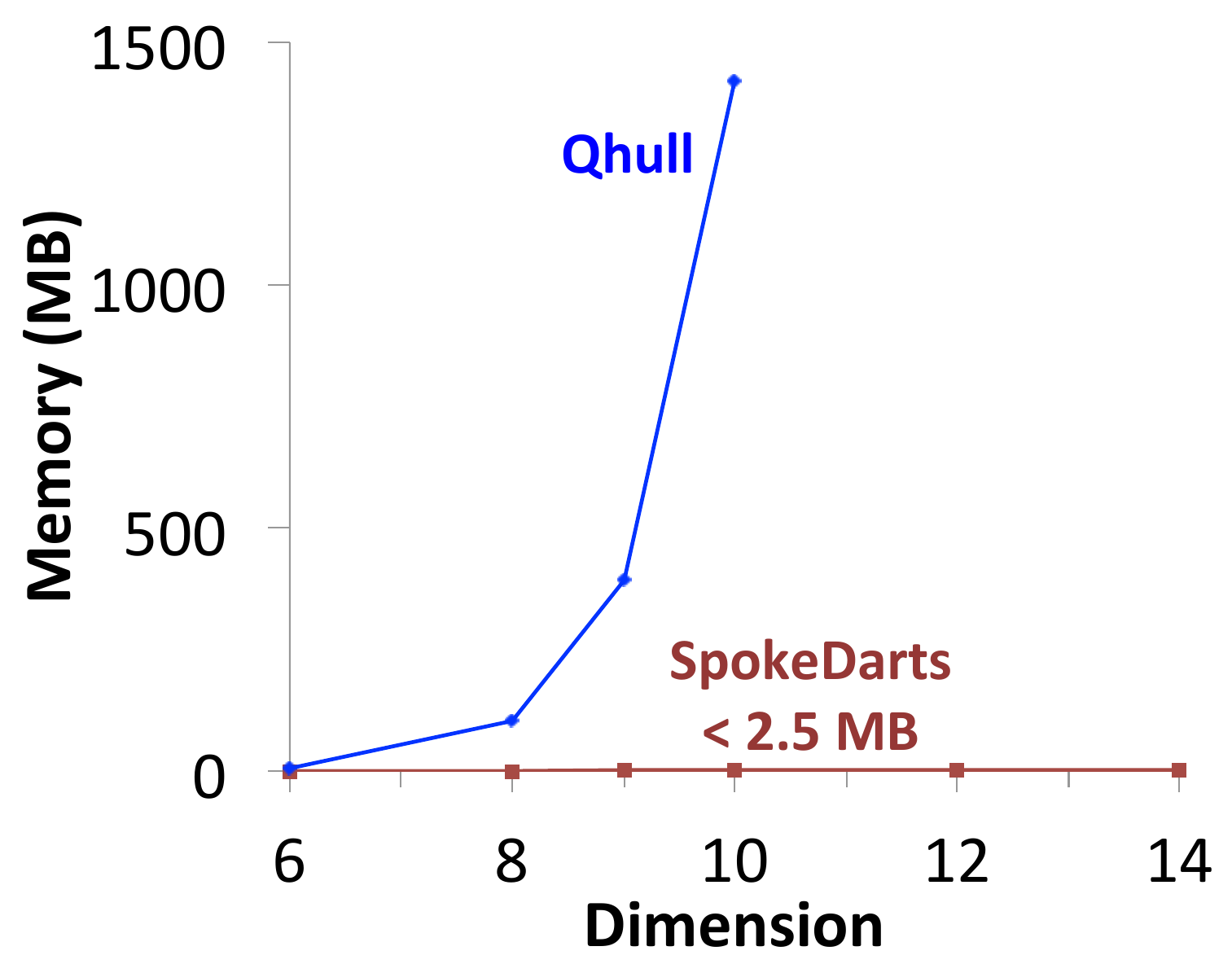}
   } 
   
  \Caption{Comparison of speed \protect\subref{fig:spoke_delaunay_speed} and memory \protect\subref{fig:spoke_delaunay_memory} between Qhull and \hidmethod{} for an approximate Delaunay graph.}
{Qhull becomes infeasible beyond $\dimnumber = 10$ whereas our method scales well.%
% in both memory and speed across different dimensions $\dimnumber$ and $\hammerlimit$ values. 
}
\label{fig:spoke_delaunay_vs_qhull}
\end{figure}

% samitch: for some reason, using width=0.46 breaks each figure into its own line on at least some latex implementations.
\begin{figure}[tbh] 
  \centering
   \subfloat[effects of $\hammerlimit$ on \% of missing \nothing{delaunay }edges]
   {
     \label{fig:m_on_missing_ratio}
       \includegraphics[width=0.46\linewidth]{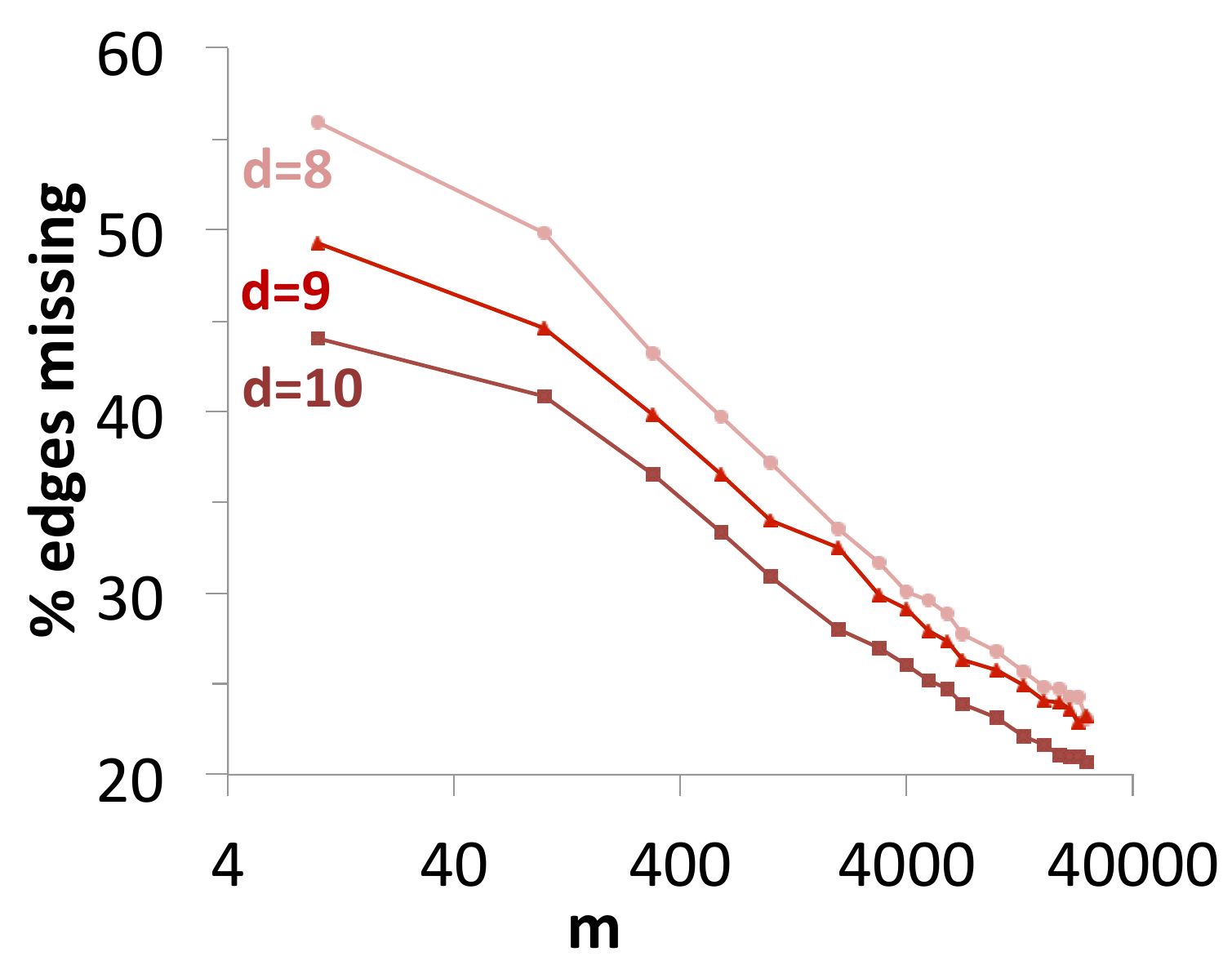}
    } 
   \subfloat[effects of $\hammerlimit$ on time]
   {
     \label{fig:m_on_time}
       \includegraphics[width=0.46\linewidth]{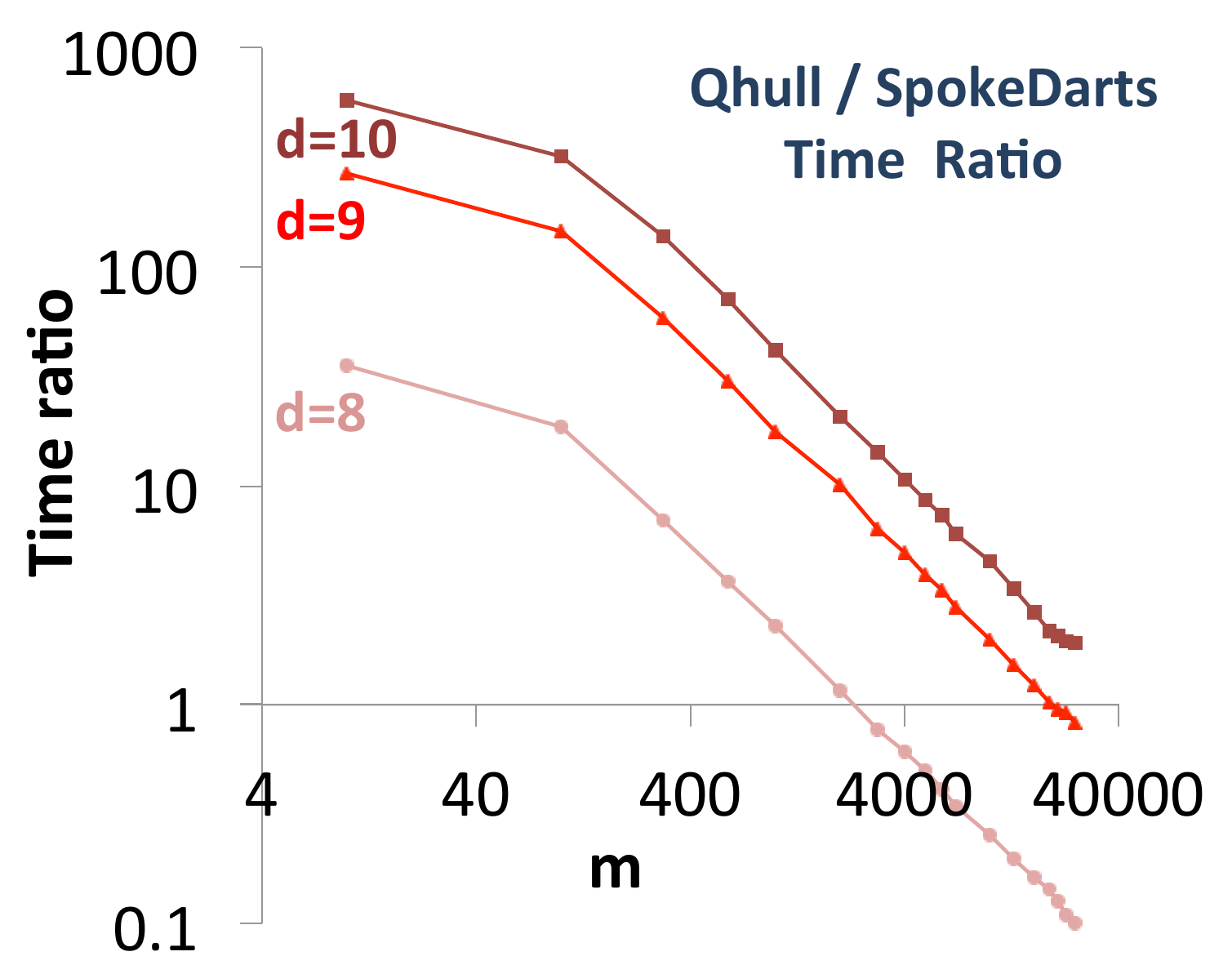}
    }
   \Caption{Effects of $\hammerlimit$ on the approximate Delaunay graph.}
  {As $\hammerlimit$ increases, fewer Delaunay edges are missed \subref{fig:m_on_missing_ratio} but run-time increases \subref{fig:m_on_time}.%
\scott{These graphs were redone so that the line for d=10 is the same color in both plots, etc. Also the x-axis range and labels are be the same.}%
\liyi{(June 23, 2016) It is interesting to note that in \subref{fig:m_on_missing_ratio}, higher dimensions tend to have lower miss rates. Should we elaborate and explain a bit more on this?}%liyi
}
  \label{fig:m_effects}
\end{figure}

\nothing{
\liyi{6D and 8D seem to be flipped in  \subref{fig:m_on_missing_ratio}. Why?}
}%nothing

The reason we tend to find the significant neighbors with high probability is obvious from the above algorithm description.
Spokes sample the solid angle around each vertex $\sample$ uniformly, so the probability that a given spoke hits a given Voronoi face is proportional to the solid angle the face subtends at $\sample$. 
As the number of spokes $\hammerlimit$ increases we are more likely to also find less significant neighbors, and $\dgraphast \rightarrow \dgraph$. 
%(This is analogous to our blue noise sampling algorithm in \Cref{alg:spoke}, where larger $\hammerlimit$ increases our chance of finding even the small voids.)

\subsection{Experiments}
We demonstrate the efficiency of our approach against Qhull \cite{barber1996quickhull}.
It is a commonly-used code for \emph{full} convex hulls and Delaunay triangulations, 
and hence suffers from the curse of dimensionality.
We know of no tools for approximate Delaunay graphs to compare against.
\scott{(January 24, 2014) Something that doesn't make this comparison very fair is that I think QHull is generating the faces of all dimensions, whereas we are just generating the edges. I think we should look into the algorithms in CGAL, and perhaps the options within QHull, to see if there is some closer thing we should compare against. If we are ``lucky'' a reviewer will suggest them for us ;)}%
\scott{I think some CGAL algorithms may have superseded Qhull.}%
\nothing{
\liyi{(December 24, 2013) Citation for Qhull, and a short description on why we compare against it, e.g. state of art for high dimensional Delaunay graph?}
}%nothing
As test input, we used Poisson-disk point sets over the unit-box domain in various dimensions.
\nothing{
\liyi{(December 24, 2013) Need more descriptions about these domains and rationale on why we choose them.}%liyi
}%nothing
For each case, we used Qhull to generate the exact solution $\dgraph$ and our method for the approximate solution $\dgraphast$.
\nothing{
\liyi{(January 14, 2014)
I am not sure why this is a fair comparison at all.
Is there any approximate Delaunay graph method that is more efficient in high dimensions?
}%liyi
}%nothing
\nothing{\liyi{January 14, 2014) sentence not seem to make any sense or purpose.}
For this analysis, we chose $m$ large enough to capture $80\%$ of the Delaunay edges\note{; in practice $m$ cannot be selected this way}.
\liyi{(December 24, 2013) ???? So exactly which $m$ values should be used?}
}%nothing
As \Cref{fig:spoke_delaunay_vs_qhull} shows, the memory and time requirements of Qhull grows significantly as $d$ increases. Qhull required memory that might not be practical for $d \geq 11$. On the other hand, our method shows a linear growth for time and memory with $d$. We see that our method became competitive for $d \ge 9$. 
\Cref{fig:m_effects} shows the effect of $\hammerlimit$ on the time and number of missed edges.
\nothing{
\mohamed{Here we explain how to utilize RCS together with MPS to retrieve the approximate Delaunay Graph.}
}%nothing

% \usepackage{algorithm}
% \usepackage[noend]{algpseudocode}
%\usepackage{amsfonts,epsfig,graphics}

% \input{symbols2}

%%%%%%%%%%%%%%%%%%%%%%%%%%%%%%%%%%%%%%%%%%%%%%%%%%%%%%%%%%

\section{Rethinking Global Optimization using Voronoi Decompositions} \label{sec:optimization}

%%%%%%%%%%%%%%%%%%%%%%%%%%%%%%%%%%%%%%%%%%%%%%%%%%%%%%%%%%

A variety of disciplines, --- science, engineering or even economics, --- seek the \emacsquote{best} answer for a question under study. This usually requires solving a global optimization problem, where we have to explore the parameter space of a function $f$ in order to find the optimum point $\hat{f}$ of some objective function, possibly under a set of feasibility constraints. For many problems, local optimality is not sufficient and a global optimal point is desired. For simple analytical functions, some algorithms are guaranteed to find the global minimum. However, no method is guaranteed to find the global minimum for all functions, or even come close in finite time. For example, no method is guaranteed to find the minimum of a function resembling white noise. In practice, heuristic stochastic techniques are usually the best, and sometimes the only option~\cite{horst2002handbook}. Lipschitzian optimization is an important category of global optimization methods. Shubert~\shortcite{shubert1972sequential} explores the parameter space and provides convergence based on the Lipschitz constant $K$ of the objective function, where a function $f$ is \emph{Lipschitz continuous} with constant $K > 0$ if
\begin{equation}
|f(x_i)-f(x_j)| \leq K |x_i-x_j| 
\end{equation}

for all $x_i \neq x_j$ in the domain of $f$. The DIRECT algorithm~\cite{jones1993lipschitzian} extends Shubert's work to higher dimensions and does not require knowledge of $K$, decomposing a domain into nested hyperrectangle partitions.

In this section, we demonstrate how \hidmethod{} can further generalize DIRECT by sampling points in random directions, not necessarily aligned with grid lines and replacing the nested hyperrectangle partitions with \textit{random approximations} to Voronoi cells. Using classical test functions, we briefly illustrate how \hidmethod{} has a significantly improved optimization performance. We believe this opens the door to new solutions of optimization problems.  Below, we outline our optimization approach, called ``Opt-darts'' and provide a careful comparison between Opt-darts and DIRECT.  

To our knowledge, our method is the first stochastic Lipschitzian optimization technique.  Our use of the phrase ``stochastic'' refers to the randomness in the Voronoi cell seed locations within our algorithm; we are not referring to optimization of a stochastic objective function. %that combines the benefits of guaranteed convergence in Jones~\cite{jones1993lipschitzian} and high dimensional efficiency in Spall~\cite{spall2005introduction}. 
Computing and refining random Voronoi cells has been intractable in high dimensions due to the exponential growth of Voronoi vertices as the dimension increases, and this is probably why this direction has not been explored before. Our spoke darts algorithm enables the size estimation of the Voronoi Cells without explicitly calculating and storing these Voronoi vertices. This allows tractable cell refinement needed in solving global optimization problems.  

%%%%%%%%%%%%%%%%%%%%%%%%%%%%%%%%%%%%%%%%%%%%%%%%%%%%%%%%%%

\subsection{DIRECT algorithm}

%%%%%%%%%%%%%%%%%%%%%%%%%%%%%%%%%%%%%%%%%%%%%%%%%%%%%%%%%%

The DIRECT (DIviding RECTangles) algorithm~\cite{jones1993lipschitzian} was developed for optimization of ``black-box'' functions (often expensive engineering simulations) which may be nonlinear, non-convex, and multi-modal.  
DIRECT is a global optimization approach that combines 
global exploration of the space with local search around the best solution and does not require gradient information.  
DIRECT partitions the domain into hyperrectangles. 
It refines those rectangles, typically by trisecting each rectangle along one of its long sides.  
The refinement process creates nested hyperrectangles that could contain a point whose function value is smaller than the smallest $f^\ast$ found so far. 
This refinement recurses until reaching the maximum number of iterations, or the remaining possible improvement is small.
An important aspect of DIRECT is that it does not just pick one hyperrectangle for refinement.  Instead, 
several hyperrectangles are selected based on relative weightings of local versus global search.  

Each rectangle $i$ is associated with two quantities: 1) a function evaluation $f_i$ at its center and 2) a size estimate $h_i$ given by the distance from the rectangle center $c_i$ to any of its corners. 
The lower convex hull $\mathcal{H}$ of the 2D data points $\{h_i,f_i\}$ lists the cells to be refined next. This convex hull is a Pareto curve that represents the tradeoff between local search (search around the best values of $f_i$) and global search (search around points with large $h_i$ because they have not been refined much yet).  

To avoid overrefining the cell with the current best solution $f^\ast$, an artificial data point is added with $\{h_0 = 0, f_0 = f^\ast - \epsilon|f^\ast|\}$, where $\epsilon$ (typically set to $10^{-4}$) is a parameter to balance global and local searches. A cell is refined by choosing an axis-aligned direction, and splitting the cell into three equal-sized cells in that direction. $\mathcal{H}$ is updated every time its cells are refined. This refinement recurses until the sample budget is exhausted. Note that limiting the cell refinement to those in $\mathcal{H}$ explores the most probable locations for a new best solution without any assumptions about the Lipschitz constant $K$ of the underlying function.

%%%%%%%%%%%%%%%%%%%%%%%%%%%%%%%%%%%%%%%%%%%%%%%%%%%%%%%%%%

\subsection{Opt-darts algorithm: our method}

%%%%%%%%%%%%%%%%%%%%%%%%%%%%%%%%%%%%%%%%%%%%%%%%%%%%%%%%%%

In this section, we first highlight limitations of DIRECT and how they are addressed in Opt-Darts. We then present the details of how Opt-darts chooses the first sample, estimates a cell size, and refines a cell. We then summarize the algorithm in a pseudocode.

\subsubsection{Motivation} 
DIRECT had a number of algorithmic limitations: 1) a cell can only be a hyperrectangle, 2) nested refinement, where a new cell can not extend beyond the boundaries of the refined cell, and 3) new sample points can only be on an axis-aligned direction with the refined cell's center. To mitigate these limitations, our algorithm (Opt-darts) uses Voronoi cells rather than hyperrectangles. From a Lipschitzian perspective, the cell size $h_i$ should be the distance from its sample point $c_i$ (cell seed) and its furthest Voronoi vertex $v_i$. This offers a much more accurate neighborhood representation. On the other hand, nested refinement may result in false convergence (a phenomenon often reported by DIRECT users). This happens when DIRECT persistently refines a cell that is close to $\hat{f}$ but does not actually contain it; see Figure~\ref{fig:opt} for an illustration. Voronoi cells do not follow the nested refinement approach; each new sample includes the domain points closest to it, with no boundary constraints. In DIRECT, axis-aligned sampling results in a stair-pattern marching towards the global solution. Alleviating this constraint increases the possibility of sampling points closer to $\hat{f}$. Moreover, while DIRECT adds new samples in the interior of the refined cell, Opt-dart has the flexibility of adding points on the refined cell's boundaries. This is more efficient for space filling.

\subsubsection{Algorithm} The Opt-darts algorithm is summarized in \Cref{alg:optdarts}.

\subsubsection{Sampling first point} We pick the first point uniformly randomly from the middle 1/3 of the domain.  

\subsubsection{Cell size estimation} Starting at a cell seed $c_i$, we throw two sets of spokes: $d$ spokes in axis-aligned directions (mimicking DIRECT), and $2d$ more spokes along random directions, for a total of $3d$ spokes. (One may use more spokes to more accurately estimate the cell size, at the price of higher computational cost.) Each spoke starts infinite, with anchor point $c_i$, and we trim each end by separating hyperplanes until its end points are on the cell's boundary. If an end point is too close to the domain boundary, we reduce its length so its distance to the nearest boundary plane is at least 1/3 the distance from the center to that plane. We label the end points $p_r$ and $p_l$. We say the length of the spoke is $\min(\|c_i p_l)\|_2, \|c_i p_r\|_2)$. We approximate the cell size by the longest such length, with spoke with end points $l_i$ and $r_i$. 
%We label that associated end points $l_i$ and $r_i$ and store them as the refinement candidates of that cell.

\subsubsection{Cell refinement} When cell $i$ is chosen for refinement,  $l_i$ and $r_i$ are added as new samples, implicitly creating two new Voronoi cells and modifying nearby cells. 

%% ------------------------------------------------------------------------
\begin{algorithm}
	\Caption{Opt-darts}{}
        \label{alg:optdarts}
	\begin{algorithmic}[1]
		\REQUIRE sample budget $N$, function to optimize $f$
   	        \ENSURE global optimum estimate $f^\ast \approx \hat{f}$   
		\STATE Sample first point $x_1$, evaluate $f_1=f(x_1)$
		\STATE Estimate cell size $h_1$
		\STATE $n \gets 1$
		\WHILE{$n \leq N$}
		\STATE Construct 2D lower convex hull $\mathcal{H}$ of $\{h_i,f_i\}$
		\FOR {each $c_i \in \mathcal{H}$}
		\STATE Refine cell $c_i$, evaluate $f$ at new points
		\STATE Update cell sizes of new and refined cells
		\STATE $n \gets n + 2$
		\STATE \algorithmicif { $n \geq N$} \algorithmicthen { \algorithmicend}
 		\ENDFOR
		\ENDWHILE
		\RETURN{} $f^\ast=\min\{f_i\}$ \pcomment{best solution found}
	\end{algorithmic}
\end{algorithm}
%% ------------------------------------------------------------------------

%%%%%%%%%%%%%%%%%%%%%%%%%%%%%%%%%%%%%%%%%%%%%%%%%%%%%%%%%%

\subsection{Analytical experiments}

%%%%%%%%%%%%%%%%%%%%%%%%%%%%%%%%%%%%%%%%%%%%%%%%%%%%%%%%%%

In this section, we used two standard test functions, {\tt Easom} and {\tt Bohachevsky}~\cite{jamil2013}, over a variable number of dimensions to illustrate the difference between Opt-darts and DIRECT. 
Two of the test suites which list these functions~\cite{jamil2013,yang2010} have been collectively cited more than 350 times.  They are also available in many online tools and from test function libraries in Matlab~\cite{burkardt,leong}, and R~\cite{bossek}.  We chose these two functions to represent two extreme behaviors in the neighborhood of the global minimum $\hat{f}$. The {\tt Easom} function approaches the global minimum via very high gradient.  It is almost flat everywhere and has a deep ``pinhole'' region where the optimum lies.  In contrast, the {\tt Bohachevsky} function approaches $\hat{f}$ via an almost flat region that looks like a shallow bowl. Both functions are noisy, and have many local minima. The global minimum $\hat{f} = 0$ for both functions, and is located at the origin, $\forall d$. Note that finding $\hat{f}$ for the class of functions like {\tt Easom} gets significantly harder as dimension increases. This problem is not as significant for the class of functions like {\tt Bohachevsky}. Figure~\ref{fig:opt} illustrates an informative comparison of DIRECT and Opt-darts in terms of point placement using evaluations of the {\tt Easom} function. In Table~\ref{tab:gopt_performance} in \Cref{sec:gopt}, we compare the number of function evaluations needed to be within $10^{-4}$ of the true global minimum $\hat{f}$. As shown in the table, opt-darts was able to achieve orders of speedup over DIRECT.

% ------------------------------
\begin{figure}[htb]
	\centering
	\begin{tabular}{lcccc}
	 	\rotatebox[origin=l]{90}{\small{ DIRECT }}
		& \includegraphics[width=0.19\linewidth]{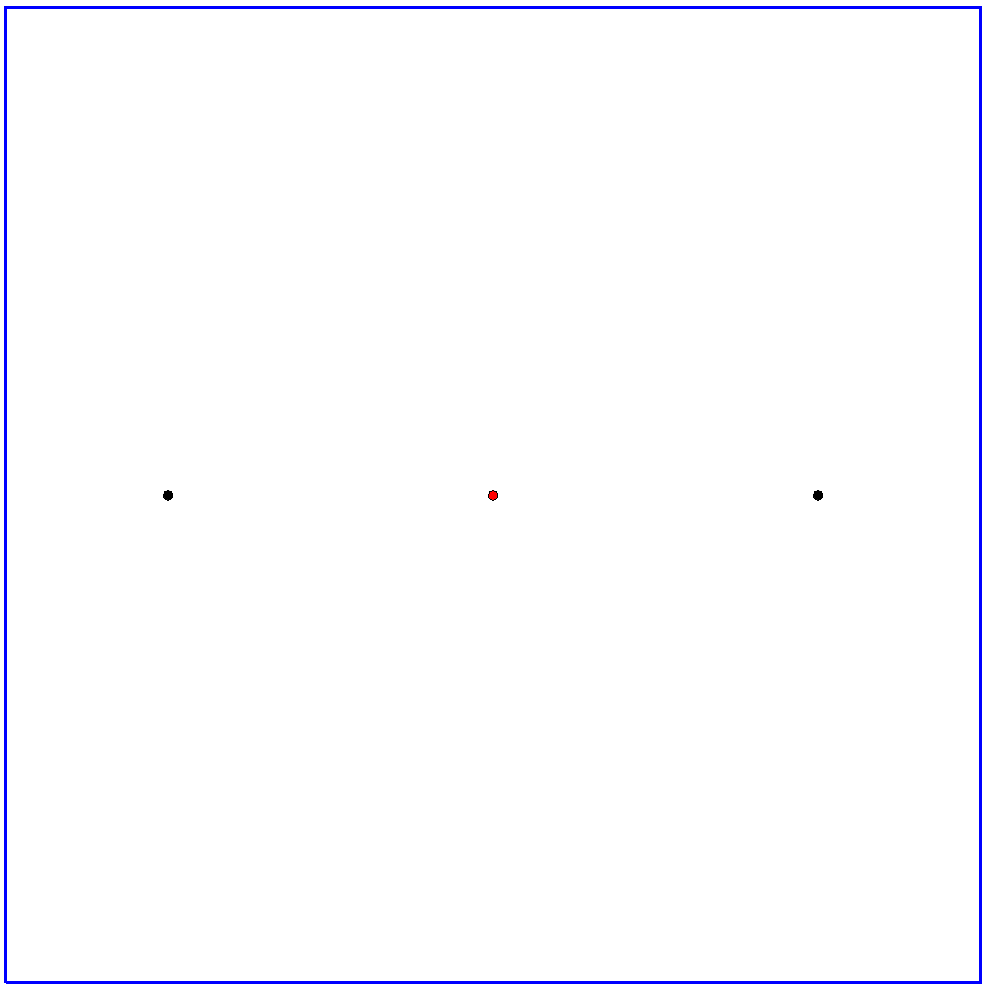}
		& \includegraphics[width=0.19\linewidth]{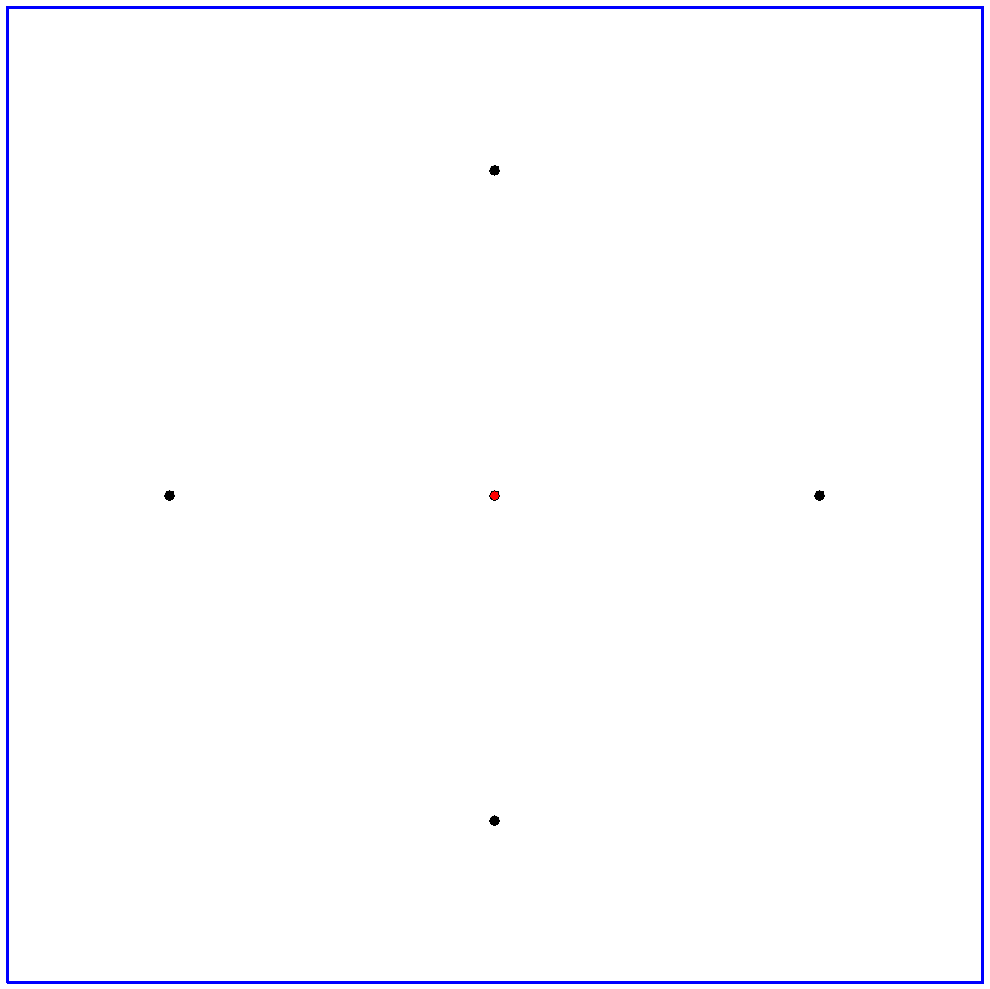}
		& \includegraphics[width=0.19\linewidth]{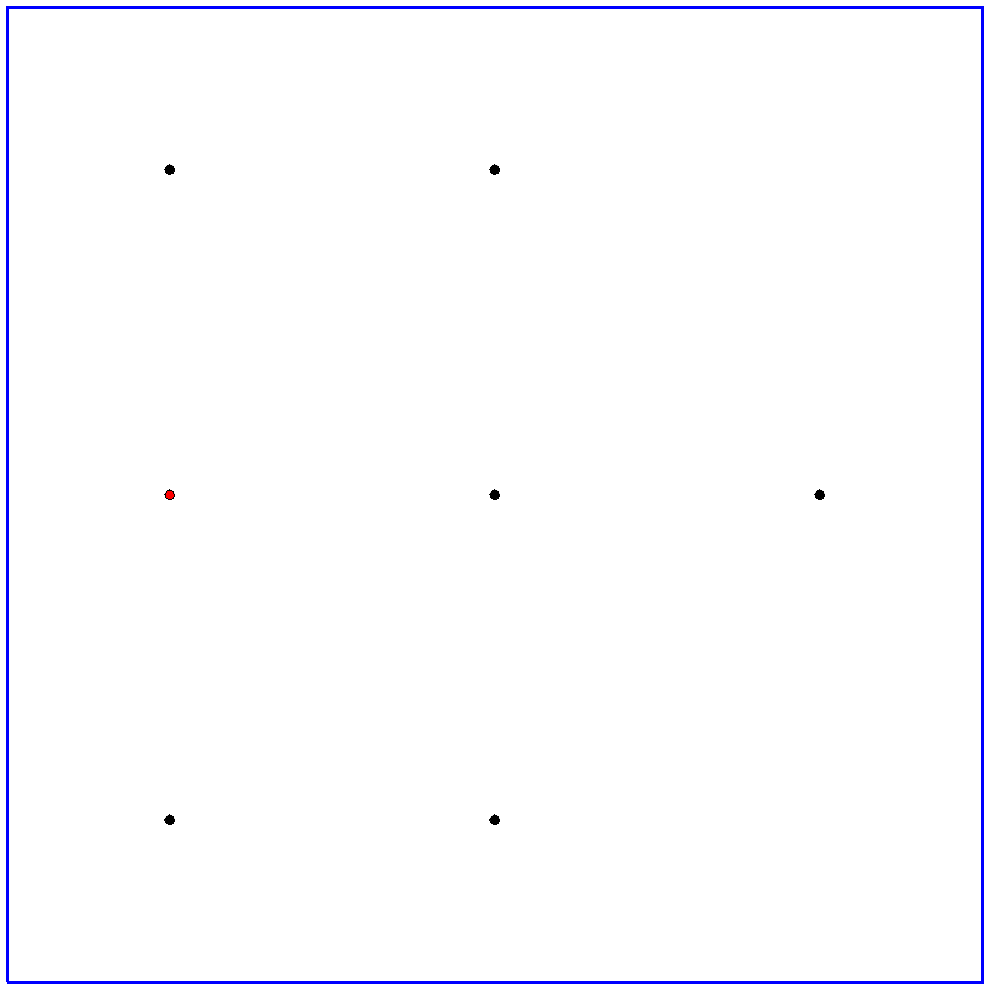}
		& \includegraphics[width=0.19\linewidth]{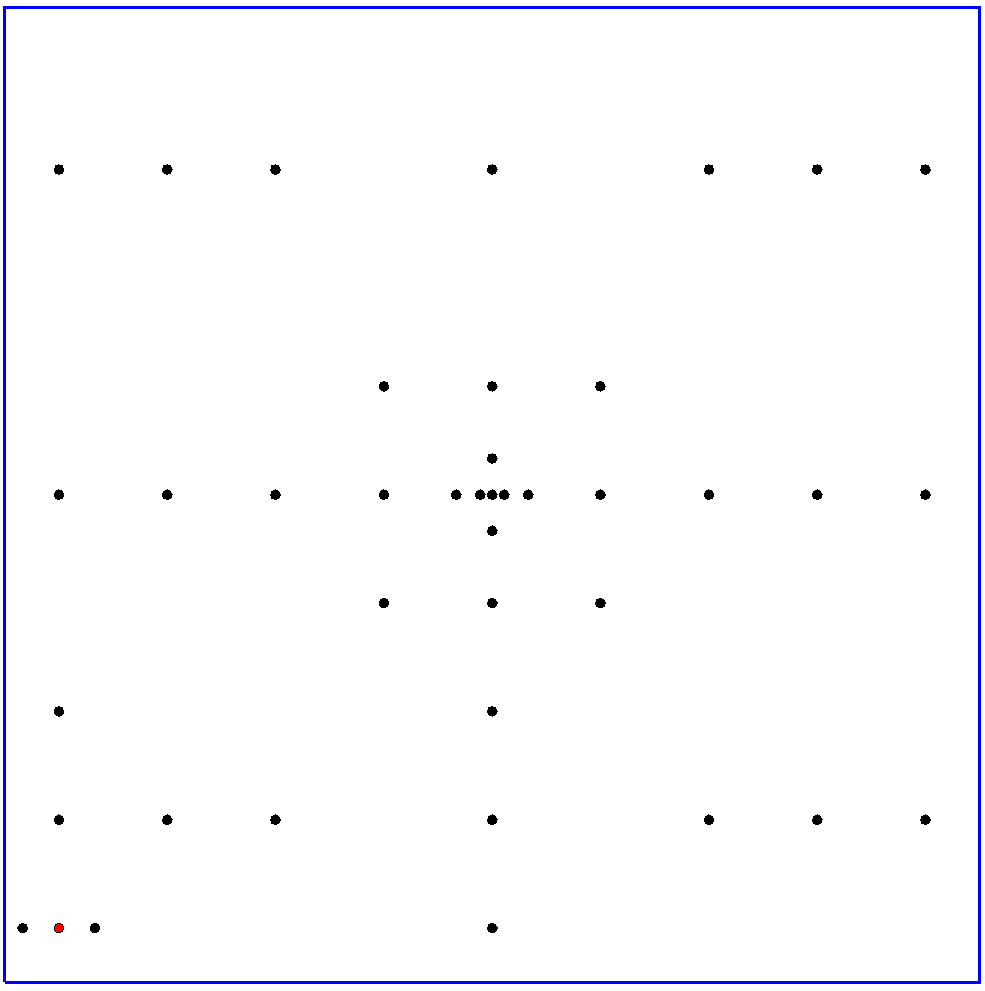} 
		\\
	 	\rotatebox[origin=l]{90}{\small{ Opt-darts }}
		& \includegraphics[width=0.19\linewidth]{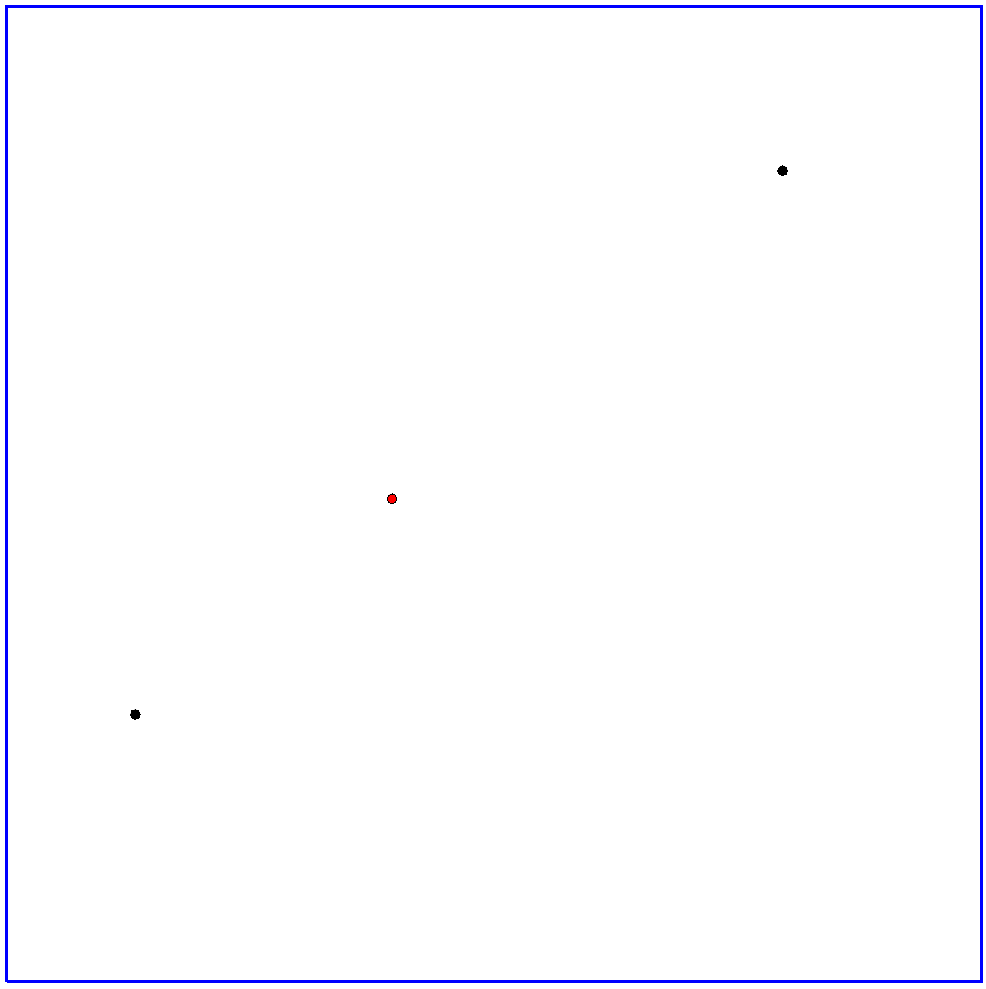}
		& \includegraphics[width=0.19\linewidth]{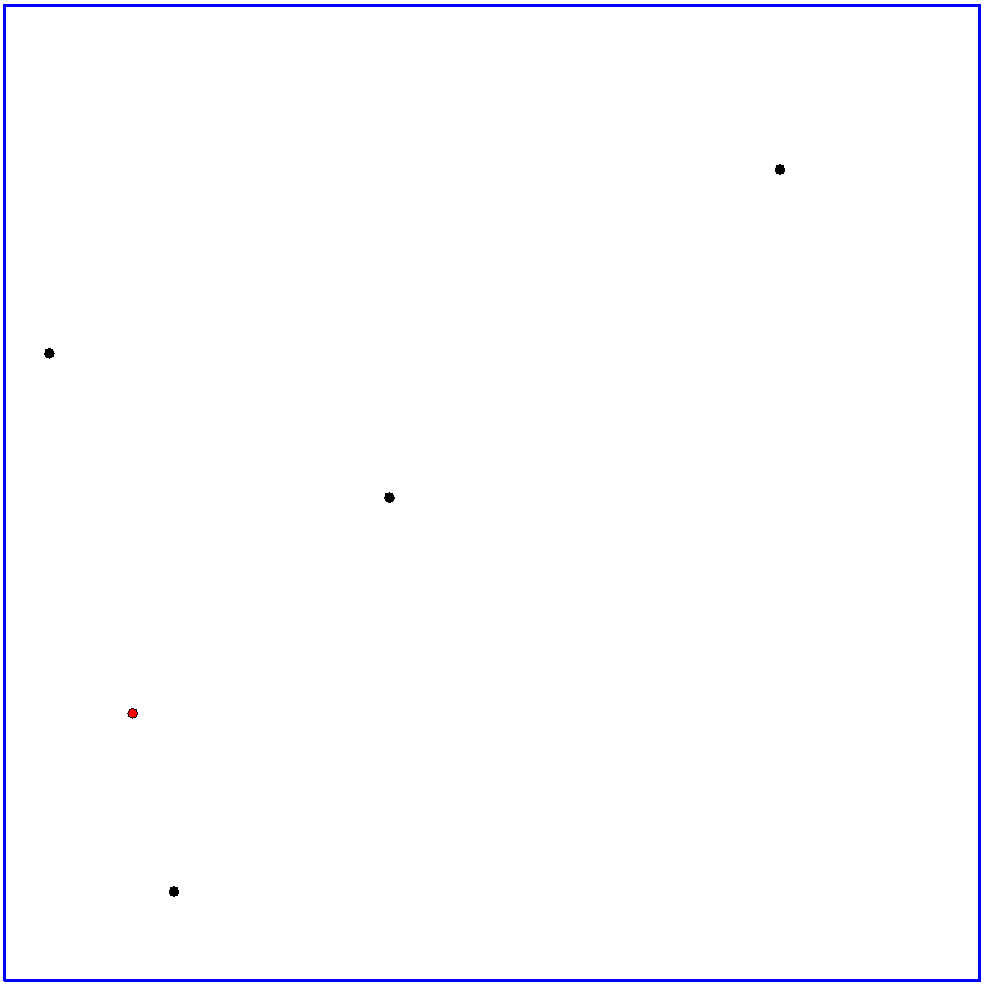}
		& \includegraphics[width=0.19\linewidth]{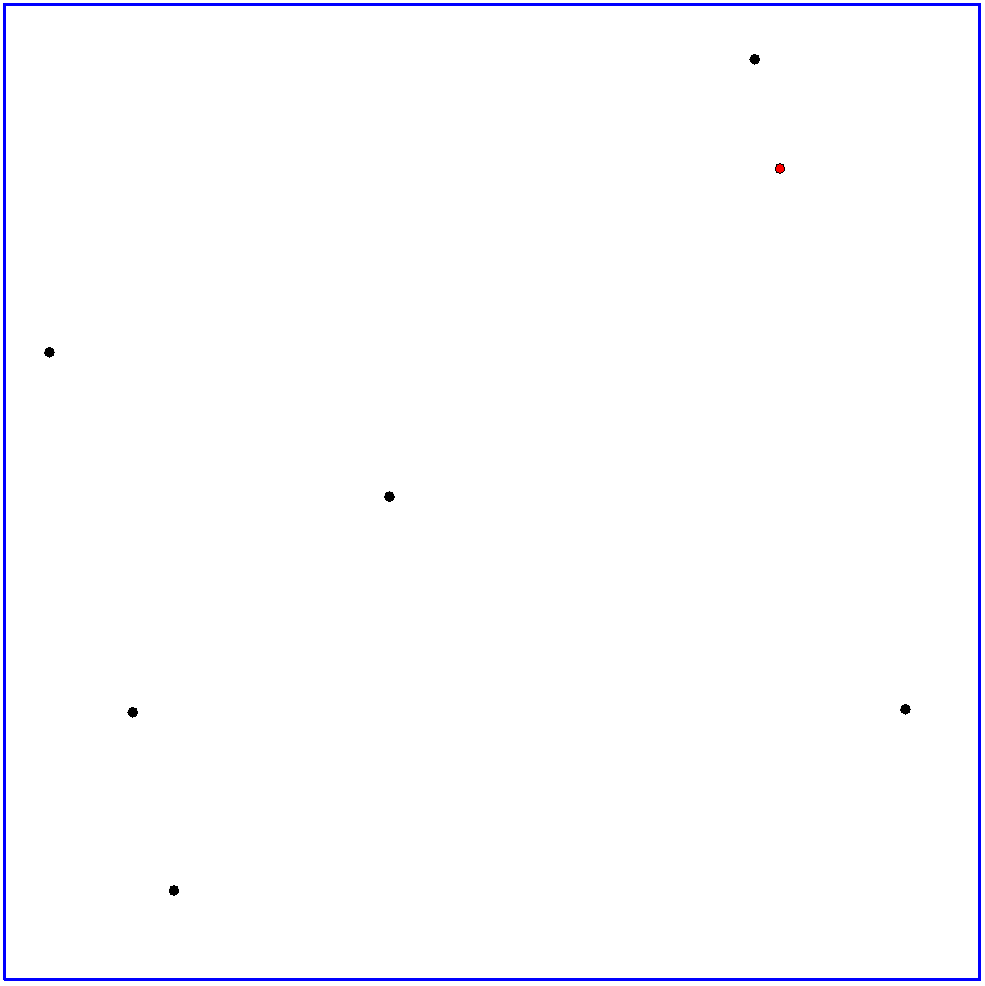}
		& \includegraphics[width=0.19\linewidth]{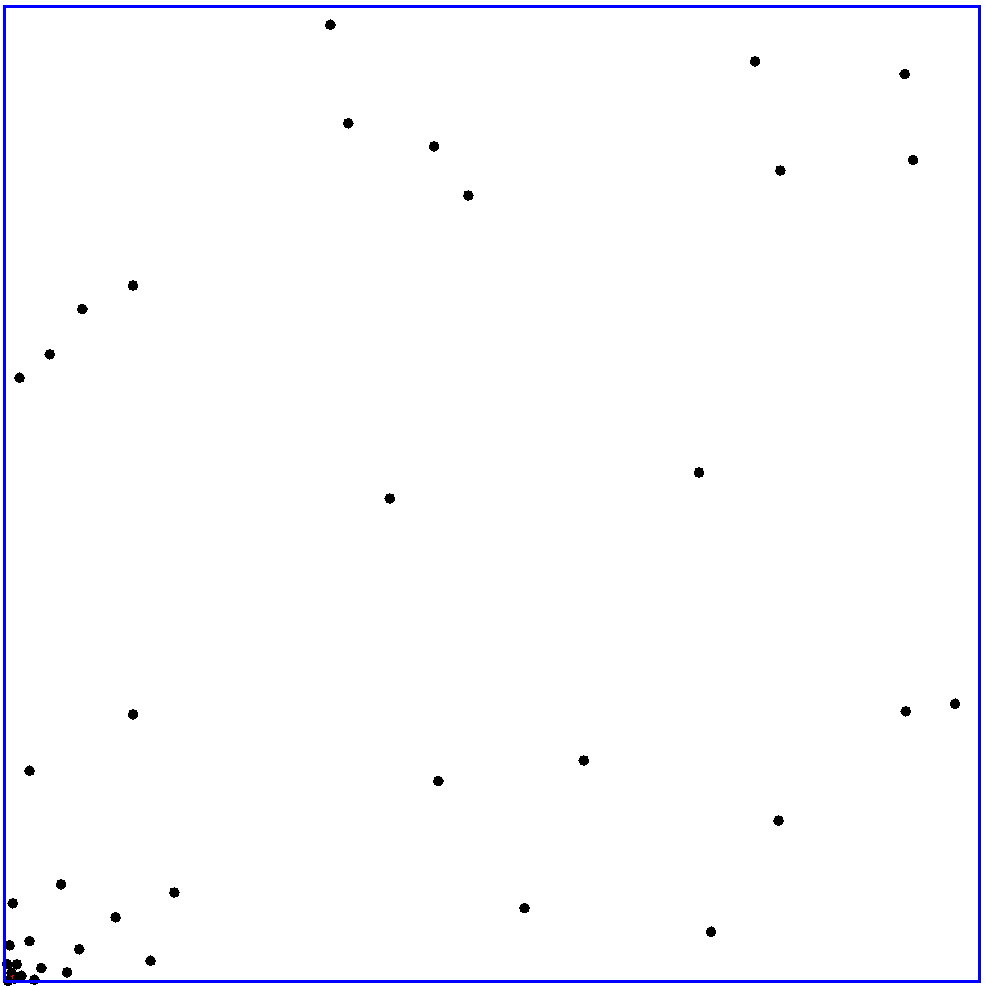} 
		\\
	 	\rotatebox[origin=l]{90}{\tiny{}}
		& $n = 3$ & $n = 5$ & $n = 7$ & $n = 41$
	\end{tabular} 
	\Caption{Contrasting sample placements, over the 2D Easom test function.}
        {The global minimum $\hat{f}$ is located at the lower left corner of the domain. Opt-darts approached it much faster than DIRECT.}
	\label{fig:opt}
\end{figure}
% ------------------------------------------------

\section{Application: Rendering}
\label{sec:rendering}

\begin{figure}[tbh]

\setlength{\tempwidth}{.3\linewidth}
\settoheight{\tempheight}{\includegraphics[width=\tempwidth]{example-image-a}}%
\centering
\hspace{\baselineskip}
\columnname{Stratified}\hfil
\columnname{Low-discrepancy}\hfil
\columnname{Spoke-darts}\\
\rowname{\hspace{36pt} 4D, 16-spp}
\ifthenelse{\equal{\isarxiv}{1}}
{
  \subfloat{\includegraphics[width=\tempwidth]{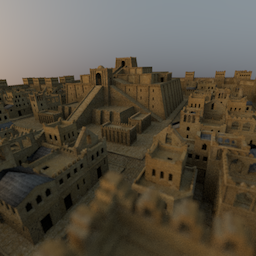}}\label{a}\hfil
  \subfloat{\includegraphics[width=\tempwidth]{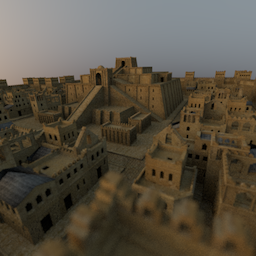}}\label{b}\hfil
  \subfloat{\includegraphics[width=\tempwidth]{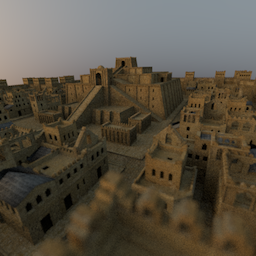}}\label{c}\\
}
{
  \subfloat{\includegraphics[width=\tempwidth]{figs/mitsuba/babylon/16spp/stratified.png}}\label{a}\hfil
  \subfloat{\includegraphics[width=\tempwidth]{figs/mitsuba/babylon/16spp/ldsampler.png}}\label{b}\hfil
  \subfloat{\includegraphics[width=\tempwidth]{figs/mitsuba/babylon/16spp/spoke11_4D.png}}\label{c}\\
}
\rowname{\hspace{36pt} 8D, 256-spp}
\ifthenelse{\equal{\isarxiv}{1}}
{
  \subfloat{\includegraphics[width=\tempwidth]{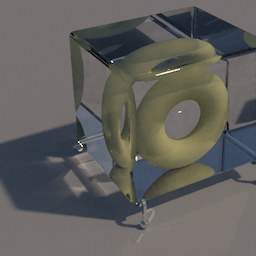}}\label{d}\hfil
  \subfloat{\includegraphics[width=\tempwidth]{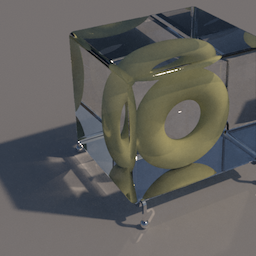}}\label{e}\hfil
  \subfloat{\includegraphics[width=\tempwidth]{figs/mitsuba/torus-in-glass/256spp/spoke_8D-small.png}}\label{f}\\
}
{
  \subfloat{\includegraphics[width=\tempwidth]{figs/mitsuba/torus-in-glass/256spp/stratified.png}}\label{d}\hfil
  \subfloat{\includegraphics[width=\tempwidth]{figs/mitsuba/torus-in-glass/256spp/ldsampler.png}}\label{e}\hfil
  \subfloat{\includegraphics[width=\tempwidth]{figs/mitsuba/torus-in-glass/256spp/spoke_8D.png}}\label{f}\\
}
 \Caption{Rendering by Mitsuba using two of the default Mitsuba samplers and Spoke-darts.}
{%
 The Babylon scene (top) renders antialiased depth-of-field using 16 samples-per-pixel (spp) in 4 dimensions, and the torus-in-glass scene (bottom) renders antialiased using 256-spp in 8D.
} 
 \label{fig:mitsuba}
\end{figure}

For a traditional rendering demonstration, we have integrated spoke-darts into the Mitsuba physically-based renderer~\cite{Mitsuba}.
Most rendering algorithms in Mitsuba use point samplers to generate multidimensional point sets, providing a good base for applying and comparing different sampling methods.

\paragraph{Scenes}

We have chosen two scenes for this rendering experiment: Babylon and torus-in-glass, as shown in \Cref{fig:mitsuba}.
The Babylon case uses 4d samples corresponding to the 2D camera screen space + the 2D lens space to generate defocus blur.
The torus-in-glass cases demonstrates a bidirectional path tracer using 8d samples corresponding to 2D for the sky emitter, 2D for the camera screen, and 2D for each bounce along each camera and light path.

\nothing{
\liyi{(June 23, 2016)
What are the 4 and 8 dimensions sampled for \Cref{fig:mitsuba}?
For example, I guess for the Babylon case 4d is for 2D screen space + 2D lens space for defocus blur, and for the torus-in-glass scene 8D is caused by refraction and caustics?
In general, we need more descriptions for these rendered scenes.
}%liyi 

\anjul{(June 23, 2016)
It's a bidirectional path tracer and the dimensions correspond to 2 for the
emitter (sky in this scene), 2 for the camera, and 2 for each bounce along the
way. The renderer actually requests 10 dimensional samples for this scene but
for performance I stop at 8 and deliver a random 2D point for the next 2
dimensions. Both stratified and ldsamplers also stop at 8 as a mitsuba
default.}
}%nothing

\paragraph{Results}
In \Cref{fig:mitsuba}, we compare the rendering quality of our method against the high-quality samplers within Mitsuba: multidimensional stratified sampling, and low-discrepancy sampling based on \cite{Kollig:2003:EIH}.
These samplers all seem to be well-spaced only along pairs of dimensions, such as the x-y camera samples and the u-v lens samples, but not the joint domains in higher dimensions such as the 4D camera + lens domain.
In contrast, spoke-darts samples are well-spaced along the joint domains.
In \Cref{fig:mitsuba}, the rendering quality using spoke-darts is comparable to that of using Mitsuba's samplers.
As analyzed in \cite{Reinert:2016:PBS}, the rendering quality depends on the projected sample distributions, which might explain the comparable quality of Mitsuba samplers and our method.
Nevertheless, our method guarantees good sample distributions in any dimensions and performs at least as well in projected dimensions, even without explicit consideration of projected distributions as in \cite{Reinert:2016:PBS}.

%Extending our method for projective blue noise is a potential future work.
%The following three applications benefit from our method more directly and clearly.

\nothing{
\liyi{(June 22, 2016)
So no clear quality improvement over alternative methods?
}%liyi 

\anjul{(June 23, 2016)
Unfortunately I haven't seen any noticeable improvement in image quality for
path tracing and depth-of-field. I can probably try other algorithms if we can
think of ones that may benefit from high-dimensional blue noise. To be
completely honest I am not sure if this result is contributing much to the
paper.}

\liyi{(June 23, 2016)
What does this mean? We are not doing some kind of projective blue noise \cite{Reinert:2016:PBS}.
}%liyi

\anjul{(June 23, 2016)
Default mitsuba samplers all seem to only be well-spaced along pairs of
dimensions. E.g. for depth of field, the x,y domain (pixel samples) is
low-discrepancy and the u,v domain (lens) is low discrepancy, but the x,y,u,v
4D domain isn't low-discrepancy.

BTW I feel that this property effectively gives mitsuba samplers somewhat
projective blue noise properties and is probably the reason that it sometimes
has better quality.}
}%nothing

\nothing{
\liyi{(June 24, 2016)
High dimensional sampling for rendering remains a separate SIGGRAPH paper (if not direction) all by itself, so is probably beyond the scope of this paper to explore.
I have spun the story as above.
}%liyi

\anjul{(June 26, 2016)
The story looks good to me. Thanks for paraphrasing! I added a citation for the low-discrepancy sampler.
}
}%nothing

\section{Application: Motion Planning}
\label{sec:motion_planning}

\nothing{
\liyi{(May 9, 2013)
Given \cite{Park:2013:RT} is already in public domain, we need to be clear on the main novelty here: high dimensional sampling.
My understanding is that \cite{Park:2013:RT} can handle only sufficiently low dimensional scenarios? And if so, what is the maximal dimensionality handled so far?
The RRT + MPS + parallelization stuff should not be presented as our contribution here as it has been published.
I also think it is very dangerous to reuse graphs from prior publications.

Cite \cite{Park:2013:PRRT} somewhere as an anonymous material containing implementation details.
}%liyi
\scott{Chonhyon tells me we should cite park2014poisson not Park:2013:RT}
}%nothing

Motion planning algorithms are frequently used in robotics, gaming, CAD/CAM, and animation~\cite{yamane2004synthesizing,overmars2005path,pan2010hybrid}. 
The main goal is to compute a collision-free path for real or virtual robots among obstacles. Furthermore, the resulting path may need to satisfy additional constraints, such as path smoothness, dynamics constraints, and plausible motions.
This problem has been extensively studied for more than three decades.
Two main challenges are:
\begin{description}
\item[Speed.]
The computation needs to be fast enough for interactive applications and dynamic environments.
\item[High dimensionality.] 
High Degrees-Of-Freedom (DOF) robots are very common.
For example, the simplest models for humans (or humanoid robots) have tens of DOF, capable of motions like walking, sitting, bending, or picking objects.
\end{description}

Some of the most popular algorithms for high-DOF robots use sample-based planning \cite{lavalle2001randomized}.
The main idea is to generate random collision-free sample points in the high-dimensional configuration space, and join the nearby points using local collision-free paths.
Connected paths provide a roadmap or tree for path computation or navigation.
In particular, RRT (Rapidly-exploring Random Tree) \cite{Kuffner:2000:RRT} incrementally builds a tree from the initial point towards the goal configuration.
RRT is relatively simple to implement and widely used in many applications.
\nothing{
However, current implementations are mainly limited to static environments, and are unable to provide real-time performance in dynamic scenes.
}%nothing

\nothing{
\liyi{bitmap quality unacceptable. Need vector graphics.}
\begin{figure}  
 \centering
   
 \subfloat[White-noise RRT]
 {\label{fig:treeprrt}
   \includegraphics[width=0.45\linewidth]{figs/motionplanning/prrt_zoom.png}
 }
 \subfloat[Poisson-disk RRT]
 {\label{fig:treepoisson}
   \includegraphics[width=0.45\linewidth]{figs/motionplanning/poisson_zoom.png}
 }
  
 \Caption{Comparison of RRT trees generated using different planning approaches.}
{\subref{fig:treeprrt}: tree generated via white noise sampling tends to have many redundant nodes that are close to other nodes in the tree (e.g., the new nodes $\mathbf y_2$, $\mathbf y_3$, and $\mathbf y_4$ are close to $\mathbf y$).
\subref{fig:treepoisson}: tree generated using Poisson-disk sampling has fewer redundant nodes due to the empty disk property of samples.
Image courtesy of \cite{Park:2013:RT}.
}
 
 \label{fig:rrt_tree}
\end{figure}

\nothing{
 \subfloat[RRT]
 {\label{fig:treerrt}
   \includegraphics[width=0.33\linewidth]{figs/motionplanning/rrt_zoom.png}
 } 
\subref{fig:treerrt}: The tree corresponding to the original RRT algorithm is generated according to the Voronoi bias of the sequential algorithm.
}%nothing

}%nothing

However, prior RRT methods generate samples via white noise (a.k.a.\ Poisson process).
These samples are not uniformly spaced in the configuration space, leading to suboptimal computation.
%See \Cref{fig:rrt_tree}.
Park et al.~\shortcite{park2016ieeetro} demonstrated that using Poisson-disk sampling instead can lead to more efficient exploration of the configuration space. We summarize \cite{park2016ieeetro} in \Cref{alg:rrt}.
\nothing{
\mps{} samples have low-dispersion and low-discrepancy, which are desired sampling properties for motion planning.
Dispersion is defined as the largest empty \emacsquote{ball} of unoccupied space and low-dispersion prevents the clustering of samples, and thereby results in a distribution through the configuration space.
}%nothing
However, the algorithm described in \cite{park2016ieeetro} assumes availability of precomputed Poisson-disk samples that can guide the selection of new points which are not too close to prior points. It starts with uniform sampling in the high dimensional space, and generates more adaptive samples in tight space or narrow passages. 
% can be used to generate more samples in tight spaces. 
%The performance of RRT planning can be further improved by multi-core GPUs.
Furthermore, the Poisson-disk sampling can be used to design a parallel version of RRT algorithm that can map well to current commodity processors, including multi-core CPUs and many-core GPUs.
Using a precomputed sampling that is shared by all threads allows efficient detection when a tree branch reaches an area that is already explored, and avoids redundant exploration.

\begin{algorithm}
  \begin{algorithmic} [1]
    \REQUIRE configurations $\mathbf{x}_{init}$ and $\mathbf x_{goal}$ in domain $\domainsym$
    \REQUIRE Poisson-disk sample set $\mathbf P$ precomputed via \Cref{alg:spoke}
    \ENSURE RRT Tree $\mathbf{T}$
    \STATE $\mathbf{T}.\funct{add}(\mathbf{x}_{init})$
    \STATE $\mathbf{P}.\funct{add}(\mathbf{x}_{goal})$
    %\STATE \pcomment{can handle multiple threads easily}
    \FORP{$i =1$ \textbf{to} $m$} \pcomment{multiple threads}
    \WHILE{$\mathbf x_{goal} \notin \mathbf{T}$}
    %\STATE \pcomment{pick a random point from the domain}
    \STATE $\mathbf{y} \leftarrow \funct{RandomSample}(\domainsym)$
    \STATE $\mathbf{T} \leftarrow \funct{Extend}(\mathbf{T}, \mathbf{y}, \mathbf P)$
    \ENDWHILE
    \ENDFP
    \RETURN{} $\mathbf{T}$
  \end{algorithmic}
  \Caption{Parallel Poisson-RRT with precomputed samples.}{}
  \label{alg:rrt}
\end{algorithm}

% moved to teaser \input{hrp4_fig} 

We use the novel spoke-darts algorithm to precompute the  high-dimension sample set via \hidmethod{} and also adaptively refine this set to compute collision-free paths through narrow passages. This high-dimensional set is used by the resulting Poisson-RRT based motion planning algorithm~\cite{park2016ieeetro}. Furthermore, it is used to design a practical parallel RRT in  high-dimensional configurations space, e.g., for a $23$ DOF robot.
%we extend Park et al.~\shortcite{park2014poisson} to 23 DOF robots, and 
%Previously Park et al.\ was restricted to relatively low dimensional spaces, $d \leq 6$.
\nothing{
High-dimensional maximal Poisson-disk sampling can be used in the motion planning of high-DOF or human-like characters.
}%nothing
We highlight the performance of this novel parallel Poisson RRT planner on three well-known motion planning benchmark scenarios from OMPL~\cite{sucan2012the-open-motion-planning-library}.
% to evaluate the performance of the planning algorithm. 
These scenarios all have $6$ DOF, and vary in their level of difficulty.
We also compute the motion of the HRP-4 robot with $23$ DOF; see \figref{fig:teaser_motion_planning}.
The total times taken by the \emph{planner} are shown in \Cref{tab:rrt_performance}.
%This includes only planner time; 
\nothing{
For \emph{sampling} time, the only competition comes from line darts, and we have demonstrated in \Cref{fig:mps_maximality} that our \hidmethod{} is more efficient. 
We observe up to $24$X speedup and can obtain real-time performance due to maximal Poisson sampling (even for $23$ dimensional spaces).
}%nothing
\nothing{
\liyi{(May 10, 2013)
These computation times are for motion planning, not sampling, whose performance ought to have been reported somewhere else (I hope). 
}%liyi
\mohamed{(May 13, 2013)
In this high dimensions, the only competition comes from line darts and we have demonstrated in figure~\ref{fig:mps_maximality} that our \spokedart{} method is more efficient toward achieving maximality. 
}%mohamed
\liyi{(May 13, 2013) Good. Updated above, so that people would not be confused.}
}%nothing

\begin{table}
\tabtextsize
\centering
\begin{center}
\resizebox{\columnwidth}{!}{%
\renewcommand{\arraystretch}{1.2}
\begin{tabular}{|c| r| r| r|r|}
\hline
Benchmark & DOF & RRT (1 CPU core) & GPU Poisson-RRT & Speed-up \\
\hline
Easy			&	6 	&	0.34		&	0.03 	& 12.14$\times$	\\					
AlphaPuzzle	&	6 	&	32.76	&	1.31 	& 24.93$\times$	\\			
Apartment	&	6	&	191.79	&	11.88 	& 16.15$\times$	\\	
HRP-4             & 	23	&	6.17 	& 	0.32 	& 19.28$\times$ \\
\hline			

\end{tabular}%
}
\end{center}
\Caption{Comparison of the performances of our GPU-based Poisson-RRT planning algorithms and a reference single-core CPU algorithm.}
{We compared the planning time for different benchmarks using 100 trials.}
\label{tab:rrt_performance}
\end{table}

}
{}
% ----------------------------------

\ifthenelse{\equal{\final}{0}}
{
\clearpage
\pagenumbering{roman}
\section{Meta}
\label{sec:meta}

\subsection{Roster}

\begin{description}

\item[Chonhyon]

RRT \Cref{sec:motion_planning}

Improving the efficiency of spoke darts in higher dimensions and show favorable comparisons against \kddarts\ \Cref{sec:analysis}

\item[Muhammad]

Meshing applications for approximate Delaunay graph \Cref{sec:high_d_meshing}

\item[Mohamed]

Lipschitzian optimization \Cref{sec:optimization}

\item[Mohamed/Scott]

Initial algorithm write up in \Cref{sec:algorithm}

\item[Laura]
Run experiments and compare against DIRECT as well as other techniques.

\item[Li-Yi]
Paper writing and strategic planning
\end{description}

\liyi{(December 2, 2013)
Please help fill in missing information.
}%liyi

\subsection{TODO}
\label{sec:todo}

\liyi{(January 16, 2014)
To help keep track of things I have created this list, sorted in roughly order of priority (high to low).
}%liyi

\begin{description}
\nothing{
\item[SIS submission creation]
Create a sis submission, and add me \email{liyiwei@stanfordalumni.org} or at least tell me the paper number.
Note that the creation deadline is *earlier* than paper deadline!

\mohamed{done. Please confirm you have access.}
\liyi{Yup, I can access, and I have made the necessary changes so that our submission is now marked as complete.}
\mohamed{great. Thanks.}
}%nothing

\item[\Cref{sec:optimization}]
\nothing{
We absolutely need to compare our method against state of art for a benchmark of high dimensional functions.
Otherwise, we better remove this part entirely.

\mohamed{That's what I am working on.}
\liyi{great!
On a second thought, this is actually the only known application for our high dimensional approximate meshing in \Cref{sec:meshing}.
So it is kind of a domino effect; we better keep this application.
=D
}
\mohamed{Opt-dart is quite significant from theoretical aspect. It is the first alternative to DIRECT since 20 years. I am working hard to achieve good results in the practical zone not only the asymptotic regime that the theory suggest. That's my main priority now, however, It's consuming a lot of energy. Hopefully will finish in time.}
\liyi{If it is not consuming a lot of energy we are not doing it right. =D I have been whacking my brain off just to rewrite (almost the entire) paper.}
\mohamed{I understand and I am not complaining. I am actually enjoying that. I just wanted to notify you where I am putting my gun powder.}
}%nothing

Finish \Cref{tab:gopt_performance}.

\item[\Cref{sec:implementation}]

This implementation part is not clear at all.
We need to describe both the data structures and how we deal with neighbors, and how exactly we avoid curse of dimensionality.

\item[\Cref{sec:parameters}]

Related to the reproducibility issue above, we need to tell people how to choose parameters and variations of our method.

\nothing{
\item[\Cref{sec:meshing}]

Need someone to initiate \Cref{alg:meshing}, as the current algorithm description is not human understandable.
\mohamed{Ok :D. Scott's task with me (Friday).}

\liyi{Done a major pass. Lose ends to be covered for the detailed writing item below.}
}%nothing
 
\item[\Cref{sec:analysis}]
Resolve remaining issues in \Cref{fig:coverage} and \Cref{fig:pairwise_dist}.

\item[\Cref{sec:meshing}]
Resolve remaining issues in \Cref{fig:m_effects}.

\nothing{
\item[\Cref{fig:teaser}]
Try to produce images for all applications. Worst case, we keep only \Cref{fig:teaser_motion_planning} and move it to \Cref{sec:motion_planning}.

\mohamed{Worst case can be using a figure of one the functions that we will use for global optimization.}
\liyi{that would be for \Cref{fig:teaser_optimization}.
How about \Cref{fig:teaser_delaunay_graph}? Mohamed, can you suggest a better figure for meshing?}
}%nothing

\nothing{
\item[\Cref{fig:delaunay_spoke_darts}]
See figure caption for suggestions.
\liyi{(January 20, 2014) Done.}
}%nothing

\item[Writing]
Need more writing passes.
There are still many loose ends.

\nothing{
\item[Video]
I think we can just submit the video segment for \Cref{fig:teaser_motion_planning}; we already have this.
\mohamed{OK}
\liyi{(January 18, 2014) I have uploaded the robotics video from our siga13 submission to sis.}
}%nothing

\nothing{
\item[Math proof]
Add them to the appendix as supplementary materials if Scott got time.
}%nothing

\item[\Cref{sec:optimization}]
Find a high dimensional function for graphics applications.

\mohamed{Can we say that shape optimization of various models would require solving a global optimization problem without reaaly doing shape optimization (it's just too much for application of application of a tool we developed.)}
\liyi{Not sure which shape optimization you were referring to. Can you cite a few (SIGGRAPH) papers?}

\end{description}

\subsection{Schedule}

\pargrph{SIGGRAPH 2014 deadline}

\deadline{Tuesday, 21 January 2014, 22:00 UTC/GMT}

\url{http://s2014.qltdclient.com/submitters/technical-papers}

\subsection{SIS}
\label{sec:sis}

\liyi{Please make sure this part is consistent with what we have under SIS!}%liyi

\mohamed{ok, will do.}

\note{(50 word summary)
We present spoke darts for efficient high dimensional blue noise sampling with applications in meshing, optimization, and motion planning.
}%note

\note{
\begin{description}
\item[Topic areas]
\begin{itemize}
\item
Primary topic area: Modeling/Geometry

\item
Seconary topic area: Animation/Simulation
%Rendering (realistic or non-phorealistic rendering)
\end{itemize}

\item[Sorting topics]
\begin{itemize}
\item
Primary sorting topic: Rendering - Stochastic Sampling

\item
Modeling - Mesh Generation

\item
Methods and Applications - Optimization

\item
Animation - Motion Planning

\end{itemize}
\end{description}
}%note

% included in introduction instead, as a sort of teaser
\subsection{Graphics}
\label{sec:graphics}

\liyi{
\pargrph{Visualization via parallel coordinates}

Continuous parallel coordinates \cite{Heinrich:2009:CPC} can be used to visualize high dimensional functions.
In particular, each $N$ dimensional function is visualized by $N$ vertical coordinate lines, with each point $(x_i, x_j)$ within a pair of adjacent coordinate lines represent the density $-$ amount of intersection of that $N-2$ dimensional hyperplane $(x_i, x_j)$ with the input function.
This can be computed via either gathering (as stated above), or scattering $N$-dimensional samples inside the function to the output parallel coordinates.

The gathering is of no use to us, it is boils down to sampling on 2D output space (regular grid will do) followed by complicated intersection computation in $N$ dimensional space.

Scattering is relevant to high dimensional blue noise sampling.
The basic idea is to sample the original function with a bunch of $N$-dimensional points, and scatter them into the 2D output parallel coordinates.
The choice of the sampling scheme matters a lot, as non-uniform sampling can introduce noise or bias.
\cite{Yan:2014:VSH} demonstrated that blue noise sampling can be a good strategy.

I think our method can be directly applied, not just for 2D surfaces embedded in $N$ dimensional spaces (as in \cite{Yan:2014:VSH}) but also arbitrary $M$ dimensional functions embedded in $N$ (as long as $M \leq N$).
At the core, this is just Monte Carlo integration using blue noise sampling, and parallel coordinates is just a particular form of such integration.

We can actual do general Monte Carlo integration, but a particular application in parallel coordinates should suffice.

\pargrph{Visualization via interval arithmetic}

Inspired by \cite{Tupper:2001:RTG}, I wonder if it makes sense to apply our method to visualize very high dimensional functions, such as finding the roots (not just points but also higher dimensional primitives like curves and surfaces).
Naturally, we want to use as few samples as possible to achieve as much accuracy as possible.
If the function is Lipchitizian continuous, we can do it analogous to optimization in \Cref{sec:optimization}, right?

If so, I would recommend this application, as it is lower hanging than rendering and fits the collective expertise of Sandia folks better.

Let me know.

\pargrph{Texturing}

Log here due to a random comment by John Owens.

Some procedural texturing, e.g. cellular texture \cite{Worley:1996:CTB}, uses Voronoi diagrams.
But it is not clear to me why anyone wants to do this in high dimensional though.

\pargrph{Rendering}
We can perform 2D screen space $\times$ 2D aperture space (defocus/depth blur) $\times$ 1D time dimension (motion blur) $\times$ 2D light space (soft shadow) = 7D sampling.
Additional options include 2D surface material (e.g. matte BRDF), 2/3D geometry, and many-d light bounces (reflection, refraction, volumetric scattering).

Blue noise is most beneficial when each sample stands out (e.g. stippling), so integration, which lies at the core of rendering, inherently benefits less from blue noise, especially at higher dimensions.
One possibility is to design the scene so that the kernels get narrowed down in various places so that the problem becomes essentially lower dimension in various places.

In particular, we can have both low and high frequency stuff for all scene set ups, such as:
\begin{description}
\item[Light]
Smooth area light, sharp spot/point light, high frequency area light source with zone-plate pattern, or a certain combination;

\item[Geometry] 
Objects with different sizes and shapes (e.g. grass blades can benefit from proper anti-aliasing);

\item[Motion]
Objects moving in different speeds and directions;

\item[Material]
Both sharp and smooth BRDFs or even BSSRDF (subsurface scattering), probably smoothly varying on the same objects;

\item[Aperture]
Ordinary depth of field will naturally lead to varying degree of blur across the image plane according to scene depth.
\end{description}

Another concern: my understanding is that kd-dart \cite{kddarts_arxiv} is inherently more suitable for integration, especially those that can be analytically computed over a dart plane.
So we need to make sure none of the scene functions are suitable for analytical integration.

}%liyi

\scott{BSSRDF sounds interesting. It's pretty new to me but it looks like we can get 9D: wavelength, 2D incident angle, 2D reflected angle, 2D spatial over the material, 2D position offset from internal reflectances. Here is a related problem Sandia cares about: put special paint on something as a tamper-detection seal, take a picture; come back in several years and take another picture and tell the difference between natural aging and the seal being broken. 
See 

\url{http://www.cs.sandia.gov/~samitch/papers/CATreport.pdf} pages 33--35.} % tilde ~ = %7E

\dmanocha{(March 25, 2014)

Here is an old paper in SIGGRAPH On 'use of interval arithmetic in computer graphics':
\url{http://research.microsoft.com/en-us/um/people/johnsny/caltech/snyder0_siggraph92.pdf}

They showed that many problems in graphics were solved using interval arithmetic (a sort of optimization). Can you make a strong case that k-darts can be used for these?

For example, if you are given two parametric surfaces in 3D (Bezier Patches or Algebraic Patches) and you want to find the closest points between them. You can pose them as solving 4 or 6 dimensional optimization problems (for global minima).

}%dmanocha
\liyi{(April 8, 2014)
Function visualization \cite{Tupper:2001:RTG} is another potential application for interval arithmetics.
}%liyi

\section{Blog}
\begin{description}
\item[November 11, 2013]
{
\mohamed{
Li-Yi,

After Nov. 19, I will be mostly working on our SIGGRAPH projects.

For the Intersection Search and Center Search algorithms, I see you have split these into 2 papers (low-d and hi-d). Can you please clarify your vision here (using email) so that Dinesh can participate? ... We can then archive the email thread in note.tex

We need to make sure that we are all on the same page with regard to that project.

FYI, I am planning to have another submission for Mesh optimization via resampling to achieve the following targets:

1. Mesh Simplification (removing of Steiner points that don't contribute to desired quality)

2. Non-obtuse triangulation, Voronoi Meshes with no short edges

3. Untangling of high-order elements 

The common theme their is to transfer our objective into geometric constraints and identifying the solution domain. We have tested this approach for provably good mesh simplification and our initial results are quite promising.

I don't think this would overlap with the lo-d approach since we don't rely on Poisson-disks there at all.

Thanks,
Mohamed
}%mohamed

\mohamed{
I now personally prefer to keep Intersection\_Serach and Center\_Search together cause the recursive Spoke-darts approach I am gonna introduce to enable maximal sampling in higher dimensions rely on both algorithms.
}%mohamed

\liyi{
The separation is about application domain instead of algorithms.
See our discussions below for more details.

In a nutshell, one paper is about high dimensional sampling and another is about sampling from unknown sizing functions.
More details can be found in the current drafts under \foldername{lod} and \foldername{hid}.
Since algorithms are not our main innovation, we can distribute them whatever we see fit as long as they are different enough to avoid the impression of self-scooping.

(You can just email this entire compiled pdf to Dinesh if necessary. We need to get him into this svn thing as otherwise he will not be effective for communication. If necessary, I can get a student to show him the ropes via remote desktop. I will enforce this as a pre-condition for participation for whoever I might bring in to the project in the future. =D)
}%liyi

\mohamed{
Ok. I will leave this for both of you to decide. 

with regard to splitting, sampling from unknown sizing functions can be in high dimensions as well  (e.g. UQ, Motion Planning, Global Optimization) so the classification is still ambiguous.

}%mohamed

\liyi{
That is exactly the two papers should be separated: the 2 concepts, high dimensionality and sampling unknown sizing functions, are entirely orthogonal.

Each paper should have only one key point. Mixing multiple up is only going to confuse the presentation and make a bloated paper.

Hope this is clear.

PS
I am thinking about making a script so that it can automatically svn check-in email comments into the blog section.
But there is no guarantee I can pull that off.
=D
}%liyi
}
\item[August 22, 2013]
{
\liyi{
[Continue discussion for previous post to avoid deep nesting]

From what I can see from Scott's comments, we just need more work beyond just writing for each low-d and high-d paper.
Merging two smaller weak units will just produce a gigantic weak one, so I say we bite the bullet, split, and make each one a strong submission.

I would love to hear if anyone has better suggestions.

Cheers.
}%

\mohamed{
After reading your comments, I totally agree with you on this and I think the extended spoke darts would strengthen the high dimensional paper and 3d meshes and improved robustness of the sizing function detection would strengthen the low-d one. Let's go for it!
}

}
\item[August 20, 2013: plan]
{
\liyi{
After looking at the paper again, here is my suggestion, and why.

We split the paper into two parts: the low dimensional part, and the high dimensional part.

I have several reasons against having a unified paper:
\begin{itemize}
\item
Advancing front is not our idea.
Thus, using it as our central point can only hurt, not help.

\item
The low and high dimensional algorithms are sufficiently quite different, and can actually be made non-overlap (see below).

\item
The applications are totally disjoint.

\item
It is simply too big a paper to put both parts together.
\end{itemize}

\begin{description}
\item[Low dimensional part]

The key selling point for the low dimensional part would be implicit sizing function, which, as far as I can see, is the real novelty from the perspective of sampling.
The known sizing function part can be either removed or mentioned as bonus/minor contributions.

For algorithm, just have \addinball{} with user tunable radius, and treat \addatpoint{} as a special case (zero radius ball).
\addonray{} is basically useless from what I can see.
This gives us a single, clean, algorithm.

Note: I cannot understand the sizing function section \nothing{Section~\ref{sec:algorithm_sizing_function}} so I might have missed something, but the results in the no sizing function mesh figure \nothing{Figure~\ref{fig:no_sizing_function_mesh}} indicated that our approach can be a gem.
(From what I can see, it is our best result over the entire paper.)

For applications, in addition to 2D meshing, can we also do 3D volumetric meshing (e.g. tet- or hex-meshing)?
I am aware that this is a sufficiently important application; see e.g. \cite{Jacobson:2013:RIS}.

\item[High dimensional part]

To my best knowledge, very high dimensional blue noise sampling is still an open problem. So we have enough novelty here.

We have two applications already; we probably need another one, likely a more graphics one.

For algorithm, just have \addonray{}.
Maybe I missed something, but \addinball{} is not really used or useful.

\end{description}

Please let me know what you thought, in particular if I have missed anything.
(Given the length and writing issue of the paper, I might.)

Cheers.
}%liyi

\scott{

Two papers is a fine idea.

I totally agree with you about advancing front not being novel enough on its own to carry a paper; the applications being disjoint; and the paper being too big already.

The algorithms could be put together, or apart, so I think we should decide one-paper-or-two based on other criteria.

\begin{description}
\item[Low dimensional part]

I think the low-d paper is actually not as close to being ready, because there is more past work in the literature and the standards are higher there.

I have comments previously embedded about the sizing function discovery not being mature yet.

I think you are right about just going with \addinball{}.

For low-d, doing tet meshing would be nice, but we do not know how to avoid slivers yet. Slivers = tets with large Delaunay ball to edge length ratio, but whose points are close to coplanar so there is no lower bound on the dihedral angles in the tet. Control volume methods are OK with slivers, but finite elements are not. So it is OK but not great to have slivers.

For low-d, we also have to show runtime on par with Delaunay refinement for the community to take us seriously as a practical alternative.

--

I actually am not thrilled about the results in no sizing function mesh figure\nothing{Figure~\ref{fig:no_sizing_function_mesh}}.
The figures on the right and left have the same minimum angle, but our mesh (left) has many more points than Delaunay refinement produces.

One thing we need to investigate, is that DR can get lucky and have no small angles even if the Delaunay disk sizes are large and the edges are small, as long as they don't occur together in a particular way. Our analysis and algorithm is based on preventing the conditions for a worst case to arise anywhere.
Something to work on.

\item[High dimensional part]

Yes, I think any practical algorithm for blue-noise or well-spaced is novel and good enough to get in, and we're doing both.

Right, just do \addonray{}.

As far as a high-d graphics application, that was a tricky thing and one of the reasons we thought a low+high paper was the way to go before. 
I don't know what applications make sense. 
Is there some high dimensional integration problem that graphics cares about?
\end{description}
}
}
\item[August 13, 2013]
{
\mohamed{
In addition to fixing the writeup, I will be working on increasing the efficiency of the Spoke darts algorithm for solving the MPS problem in high dimensions. I have a nice idea for modifying a spoke dart to include a random circle sample from each existing sphere. In 2d this will result in a maximal solution with exactly ONE miss per disk. In higher dimensions this sampled circle should increase the efficiency of void retrieval. Another technical improvement lies in the approximate Delaunay meshing algorithm in higher dimension. I think we can convert that algorithm to an exact one as well regardless of the dimension. If succeed this will become the meshing algorithm that does not suffer from the curse-of-dimensionality.  
}
}

\item[July 26, 2013: SIGGRAPH meeting note]
{
\liyi{
So Mohamed, Scott, Dinesh, and I met during SIGGRAPH 2013 to discuss the next step.
Basically, we agree to take another shot at SIGGRAPH 2014, with the major task in writing.
There are several possibilities to restructure the paper (not all mutually exclusive and some could be combined):
\begin{description}
\item[Shorten] the paper into 10 pages. Make it more like a paper for SIGGRAPH than some math/geometry journal.

\item[Reposition] the paper so that the high dimensional part plays the main role, with the low dimensional part being bonus features.
For this, we will have to find additional high dimensional (graphics) applications beyond motion planning.

\item[Split] the high dimensional and low dimensional algorithms into two papers.

\item[Spin-off] the automatic sizing function part into a separate paper.
\end{description}

I will take a full pass of the paper to refresh my memory and comment more later.
Mohamed and Scott, please help fill in if I have missed anything.

Cheers.
}%liyi
}
\end{description}

\section*{Cemetery}

\section{Background - Old Version}
\label{sec:background}

\liyi{(August 20, 2013)
This part is too long; materials not very relevant to our contributions can be trimmed or removed.
A good sanity check (as well as writing style) is to describe briefly, at the end of each group of prior methods, their relevant to our method.
If this cannot be done strongly, it indicates either our method is not novel enough, or that group of prior methods are not very relevant.
}%liyi
\scott{(August 20, 2013)
Li-Yi's suggestion above seems very reasonable.
To me, all of what we have currently is relevant to either the low or high dimensional case, or both, but I agree that the connection isn't clear and we should describe, it otherwise the reader's eyes will glaze over.
Which parts are relevant will be split if we split the paper into two.

General note: for many of your other comments, I've included a detailed answer about ``why is this relevant.'' 
My hope and intent is that this provides information in order to write a better description and motivation for the paper.  A lot times I thought, "wow, if the reason we need this isn't clear, the writing of the paper must have been really bad,
'' so I wanted to try to succinctly say what I was trying to say, to make the next version more clear.
}%scott

\nothing{
\note{Why people should care about blue noise sampling in high dimensions. Why existing methods are not enough.}
%\scott{Suggest Li-Yi write this part.}
}%nothing

\subsection{Blue Noise and High Dimensions}
Maximal Poisson-disk Sampling (MPS) is a random process for generating well-spaced blue noise.
In two dimensions, we have a provably-correct $O(n \log n)$ algorithm for MPS~\cite{Ebeida:2011:EMP}. In higher dimensions ``Simple MPS''  provably achieves an MPS up to roundoff, with run-time that is empirically $O(n \log n)$ for fixed $d$~\cite{Ebeida:2012:SAM}. Because the method uses and refines a background grid, the dependence on $d$ is exponential, $O(2^d),$ and the method is impractical for $d>6$.

``Fast Well-Spaced''~\cite{Miller:2013:FAW} is a new theoretical algorithm  to generate non-random well-spaced points in high dimensions.
%It enriches a set of $p$ points to a well-spaced set.
%Voronoi neighbors are maintained.
%It includes a fast neighbor search datastructure.
%The output is deterministically well-spaced. %??
It runs in expected time $2^{O(d)}(p \log p + n)$ where $p$ is the input size.
\liyi{(August 20, 2013) previously ``This is excellent theoretical complexity.'' Why is an exponential term considered to be \emacsquote{excellent theoretical complexity}? This seems self contradicting the curse-of-dimensionality part right below.}%liyi
\scott{(August 20, 2013) ``excellent'' because it is the best known, and doesn't have any $n^d$ or $d!$ terms. But it isn't perfect. I think it would be hard to get something that doesn't involve involving the number of neighbors, which is roughly $1.3^d = 2^O(d)$. I can explain better; attempt at below.}
This complexity is the best known. The $2^{O(d)}$ term is not ideal, but since the number of neighbors can be roughly $1.3^d$, it would be difficult to avoid at least some exponential dependence on $d$.
As for Simple MPS, we speculate that the practicality of Fast Well-Spaced (if it is implemented) will depend on the constants, particularly in the $2^{O(d)}$ term.
\todo{say something about the cathedral of theory results that fast well-spaced relies on, vs. our relatively simple and straightforward and self-containted result.}

\subsection{Challenges in High Dimensions}
%\{Curse of Dimensionality}
The difficulty of obtaining efficient datastructures and algorithms in high dimensions is called the ``curse-of-dimensionality.'' 
We recap some fundamental issues. The goal is to avoid time and memory complexities that have non-constant exponents: in order of increasing complexity one often encounters $2^d, d^d,$ and $n^d.$ In Graphics we often expect linear algorithms, but $O(nd)$ is needed  to simply list the $d$ coordinates of $n$ points. 

\pargrph{Neighbors}
The most fundamental issue is the neighborhood of a point or disk.
Examining sample neighborhoods is a key step among all known blue noise methods.
In a packing of non-overlapping unit balls, the number of balls which can touch a single ball, the ``kissing number'' $K,$ grows exponentially with dimension.  
A general upper bound is $K < (1.32042. . .)^{d(1+o(1))}$~\cite{kissing}. 
That is, $K = 2^{O(d)}$.
This is compounded if the spheres are of different radii.
For non-unit balls, whose radius function follows a Lipschitz constant $L$, 
we speculate that $K$ is also $2^{O(d)}$, but with worse constants.
This incurs exponential-in-$d$ growth in computation speed, memory usage, or both \cite{Samet:2005:FMM}.

\mohamed{(Dec. 27, 2013) This is important: The good news here is $K$ is bounded by $n$. Hence classical dart throwing is $O(n^2)$; we throw $m = O(n)$ darts each one check its conflict in $O(n)$ time. However, we terminate without any guarantee on how close we are to maximality. Line-spokes require that for each sample point you throw $O(n)$ line-spokes each require $O(n)$ to check conflict and hence we end up with $O(n^3)$ time complexity, however we have now a bound on the aspect ratio of the associated sample points. a $k$-d tree may reduce that to $O(n^2 \log n)$, however that require  $K << n$, which is harder to achieve as the number of dimensions grow due to memory constraints. $n$ would be too large to store!
}

Herein we consider at least one $K$ term to be unavoidable. Since in practice constants matter, we seek to avoid any complexity in $K$ worse than $O(K \log K)$ and avoid general $2^{O(d)}$ terms.
Constructing intersections of disks may be $O(n^{d/2})$, and we avoid this.

\todo{... mention failures of existing techniques to extend to high d? that is in background?}%todo
\liyi{(May 1, 2013) Yes, need to elaborate the preceding part a bit more. In particular, we should spell out speed and memory complexity of prior methods.}%liyi
\note{
Compounding this is a further combinatorial blow up of how neighborhoods around these neighbors may overlap.
\liyi{(May 1, 2013) Someone please elaborate the preceding sentence a bit.}%liyi
}%note

\nothing{
\liyi{(May 10, 2013)
I did a quick high level structuring of this section.
The goal here is to provide a solid knowledge of how prior sampling methods deal with dimensionality issues.
Reviewers should be able to just read this part and have a solid grasp of related prior works and more importantly buy into the contribution of our paper.
We highly recommend summarizing key information into \Cref{tab:complexity}.

Mohamed/Scott, I will let you guys make a first draft, as you ought to know all these algorithms/data-structures/complexities stuff very well (many of which came from your algorithms).
}%liyi
}%nothing

\pargrph{Proximity Queries}
In very high dimensions, $K$ grows so quickly that for practical $n$ \emph{most} of the samples tend to be near one another.
In this way the very definition of a \emacsquote{local neighbor} begins to break down, and there is less to be gained by attempting to find them. 
In any particular distribution, the actual number of neighbors is at most all the points, 
so any of our dependence on $K$ is limited by $n$.
%(Not so with general $2^{O(d)}$ terms.)

Regardless of the number of actual neighbors, proximity queries are difficult to do efficiently in high dimensions. This is an active research topic in Computational Geometry~\cite{proximitysurvey,Miller:2013:FAW}. Just computing the distance between two points is $O(d)$, because one must consider all coordinates. Many query datastructures rely on some sorting for each dimension, or nested sub-structures for each dimension.  Most speedups over brute force require complicated datastructures. On the other hand, generating the neighbors of a single ball  by brute force is only $O(dn)$ time and $O(d+n)$ space, which may be efficient enough.
\scott{Need lots of catching up to do here. Get help from Manocha? cite what?}

\pargrph{Boxes}
``Simple MPS''~\cite{Ebeida:2012:SAM} used a box-tree (quad-tree) for proximity. Because of the well-spaced nature of the point set, the size of the box-tree was limited to $O(n 2^d)$. In theory this seems reasonable, but in practice the $2^d$ term limited the method to $d \le 5$ because of memory issues.
The curse-of-dimensionality for MPS methods that rely on a background grid was recently summarized~\cite{kddarts_arxiv}.

The main problem is that the ratio of the volume of a sphere to that of its bounding box decreases rapidly with dimension. The volume ratio for a sphere to its inscribed box \emph{increases} rapidly.
Either box becomes a worse proxy for a sphere as the dimension increases.
Either there are a lot of empty boxes, or they contain a lot of sample points. 
Memory usage is inefficient. 

%Refining boxes is an inefficient way to represent uncovered areas, or other bounded shapes, because a box contains $2^d$ subboxes, and only a few are likely to cross the boundary.
%
\subsection{Non-point Samples}
\Darts~\cite{kddarts_arxiv} uses axis-aligned lines and higher dimensional hyperplane searches to find regions of interest. This is especially efficient for exploring arrangements of spheres, because the intersection of a sphere with an axis-aligned hyperplane is a simple analytic sphere, in the lower dimensional space of the hyperplane. Our method is essentially a local, advancing front version of the Relaxed Maximal Poisson-disk Sampling method in \darts~\cite{kddarts_arxiv}.

Sun et al.~\shortcite{Sun:2013:LSS} samples lines and line-segments for rendering applications including 3D motion blur, 4D lens blur, and 5D temporal light fields.
The method relies on prior algorithms for determining sample positions and will share similar limitations in higher dimensional spaces.

\todo{ subsection{Low discrepancy sequences}: 
add some mention in final draft}

\liyi{(May 10, 2013)
they clearly have fast speed and low memory consumption.
But I am not sure if they can work well in high dimensions.
In any case, they should be mentioned.
}%liyi

\subsection{Advancing Front}
\label{sec:background_advancing_front}
Many advancing front methods have been proposed for meshing. 
To our knowledge, the first one based on sphere-packing appeared in 1991~\cite{firstspheremesh}. It was described for general dimensions. A more recent version, ``TangentPack,'' is very fast for 2 and 3 dimensions~\cite{fastspherepacking}.
TangentPack places non-overlapping half-radius disks. A disk is tangent to two prior ones in 2D, and three prior ones in 3D.
We were unaware of TangentPack when we developed our method, but have since noticed that it is equivalent to 
our \lodmethod-\addatpoint\ variation for low dimensions.

``Biting'' is another advancing front disk packing algorithm for 2D~\cite{Li99biting:advancing}.
It places full-radius Poisson-like disks. Disk centers  are chosen randomly, anywhere on the boundary of a prior disk.
Both TangentPack and Biting produce many exactly-$r$ distances between disks, resulting in different spectra than the output of MPS.
They maintain the geometry and combinatorial structure of the boundary of the disks surrounding the uncovered region. 
The concepts readily extend to higher dimensions, but in practice the $O(d^d)$ combinatorial complexity of intersections may become unaffordable. 

The graphics literature contains an MPS-like advancing front algorithm for blue noise in 2D~\cite{Dunbar:2006:PDS}. 
The disk distribution is not maximal. The main drawback of extending it to higher dimensions is that it again requires an expensive construction of the front.

\nothing{
\todo{MPS-like advancing front algorithm review}
}%nothing

\subsection{Controlled Randomness and Uniformity}
``Variable Radii Poisson-Disk Sampling''~\cite{Mitchell:2012:VRPD} defined two useful concepts for us. The first is ``two-radii MPS'', which gives disks two radius, a large one $\rcoverage$ that determines domain coverage and a smaller one $\rconflict$ defining the distance between points. Their ratio $\beta = \frac{\rcoverage}{\rconflict} \ge1$ defines a measure of maximality, affects the randomness (spectrum) of the distribution, and provides provable bounds on the quality of a triangulation of the points. We use $\beta$ as an input control parameter. The second concept is ``spatial MPS'' where the disk radius varies throughout the domain according to a sizing function. This concept predates that paper, but that paper quantified the limits and effect of the sizing function's Lipschitz constant $L$.
``Sifted Disks''~\cite{SiftedDisks} is a post-processing algorithm that can improve blue noise and reduce the number of sample points, while preserving $\beta=1.$

\nothing{\liyi{(August 20, 2013) Doesn't seem very relevant.}
\subsection{Voronoi and Power Diagrams}
The Voronoi cell for a sample point is the domain points closer to that sample than any other. 
The perpendicular bisector of the line between two samples defines a separator between their cells.
The power cell generalizes this to disks. The separator is again perpendicular to the line between the samples, but offset by the square of the disk radii. For overlapping disks, the separator passes through the $d-2$ dimensional-sphere defined by the intersection of the $d-1$ spherical surfaces of the disk.
}%nothing

\nothing{
\subsection{Well-spaced Blue-noise}
\todo{define well-spaced and blue noise. they are related so do this together.}
Blue noise distributions plays an important role in many graphics applications (rendering , ... etc).

In meshing, ``well-spaced'' is the traditional moniker for points that are locally separated from one another yet globally dense enough to cover the entire domain. Well-spaced leads to mesh elements with 
desirable shape~\cite{miller1999radius}.

\subsection{Poisson-disk Radii}
In this paper, Poisson-disk does not refer to the Poisson process for generating disk centers, the classical definition of dart throwing.
Instead we use ``Poisson-disk'' to remind of us two key properties related to the power diagram.
The sample points are centers of disks that inhibits the insertion of new points within it, of radius $\rx$.
We also consider the maximum distance from a sample point to its farthest power vertex, $\rvor$.
The traditional definition of a Maximal Poisson-disk Sampling MPS has $\rvor \le \rx$.
Otherwise, if $\rvor > \rx$, then the power vertex is outside all disks, 
an uncovered domain point. 
This is also used in two-radius Poisson-disk sampling~\cite{Mitchell:2012:VRPD}
and the definition of coverage $\beta$~\cite{SiftedDisks}.
}

\subsection{Neighbor collection}
\label{sec:implementation}

%nothing

For a sample on the front, for each spoke we trim it by iterating over the nearby samples.
When generating blue noise, a sample is a neighbor if its distance is less than $3\rnumber.$
It saves time to collect a sample's neighbors once before throwing any of its spokes.
Those wishing to reproduce our output may simply iterate over all the samples and gather the indices of the neighbors in an array.  (This brute-force approach is sufficient for our run-time proofs.)

In our implementation, we have found that using a $k$-d tree saves time in moderate dimensions.
We maintain a $k$-d tree of the entire point set.
We collect the subtree of neighbors.
We update the tree and subtree as we successfully add new disks.
If the neighbor list is huge, $\neighbors \rightarrow \samplenumber,$ as can happen when $\dimnumber$ is very large, then these trees do not save any time over an array, but they are not significantly more costly either.

\nothing{
\mohamed{
We use a $k$-d tree to store the entire point set and two sub-trees to collect neighbors for current sample $\sample$ whose spokes are being trimmed.
The first (second) sub-tree gathers the neighbors within distance $2\rnumber$ ($3\rnumber$) from $\sample$ to invalidate its degenerate/full spokes, respectively.
Note that these trees are just pointers to the coordinates of the sample points which are stored only once.
}

\liyi{(January 21, 2014) Unless we can say how we switch from trees to linear arrays I recommend not mentioning the stuff below.}
\note{
However, these $k$-d trees only help when $\neighbors \ll \samplenumber$, where $\neighbors$ is the number of neighbors and $\samplenumber$ is the total number of samples.
When $\neighbors \rightarrow \samplenumber$ these trees start to behave no better than the set of all current samples.
}%note
}%nothing

\nothing{
\mohamed{Li-Yi, please check if this quick description is clear enough. I just added it so that we are all on the same page with regard to the new spokes I have introduced recently.}

\liyi{(December 26, 2013) please see my updates and comments above.}

\mohamed{(Dec 27, 2013) Done. I have also written a description of the opt-dart algorithm that I am currently implementing.}
}%nothing

\nothing{
\liyi{
\nothing{
(January 19, 2014)
I am not sure if this belongs to here, and if so, we should elevate it to the implementation part \Cref{sec:implementation} about an important general issue, on how to keep track of neighbors.
}%nothing
(January 20, 2014)
I transferred this from \Cref{sec:meshing} because if this is an important issue we should mention it here about finding nearby neighbors.
}%liyi
\scott{
{\bf Pool.} The initial pool of candidate neighbors may be all the points in the distribution. 
However, if we know that the points are well-spaced, e.g.\ the Poisson-disk radius function obeys a Lipschitz constant $L$, then we may reduce this pool.
The $L$ provides a limit to the ratio of the distance of the farthest Delaunay neighbor to the nearest neighbor.
Using standard libraries we can compute an approximate nearest-neighbor distance, which is an upper bound on the true nearest neighbor distance.
% We can also use our algorithm to dynamically update the nearest-neighbor lower bound, periodically discarding far pool vertices.
Combining these gives an upper bound on the distance to the true farthest neighbor, and hence we can eject all farther candidates from the pool.
As the dimension becomes extreme, approximate nearest neighbors are more expensive to calculate, and fewer candidates can be discarded, so just initializing the pool to all the points is most efficient.
}%scott
}%nothing

% samitch: for some reason, using width=0.46 breaks each figure into its own line on at least some latex implementations.
\begin{figure}[tbh] 
 \centering
   \subfloat[performance comparison]
   {
     \label{fig:spoke_delaunay_vs_qhull_old}
     \includegraphics[width=0.46\linewidth]{figs/spoke_delaunay_vs_qhull.pdf}
  }
  \subfloat[effects of $\hammerlimit$]
  {
    \label{fig:m_vs_missingratio_old}
      \includegraphics[width=0.46\linewidth]{figs/m_vs_missingratio.pdf}
   } 
\nothing{
  \Caption{Comparisons between Qhull and our \hidmethod{} method.}
{\subref{fig:spoke_delaunay_vs_qhull_old}: speed and memory for getting the exact Delaunay triangulation using Qhull, and an approximate Delaunay graph using \spokedarts{} with different $\hammerlimit$.
\subref{fig:m_vs_missingratio_old}: as $\hammerlimit$ increases, the percentage of missing Delaunay edges decreases and the computation increases.
}
}%nothing
   \subfloat[consumed time]
   {
     \label{fig:spoke_delaunay_a}
     \includegraphics[width=0.46\linewidth]{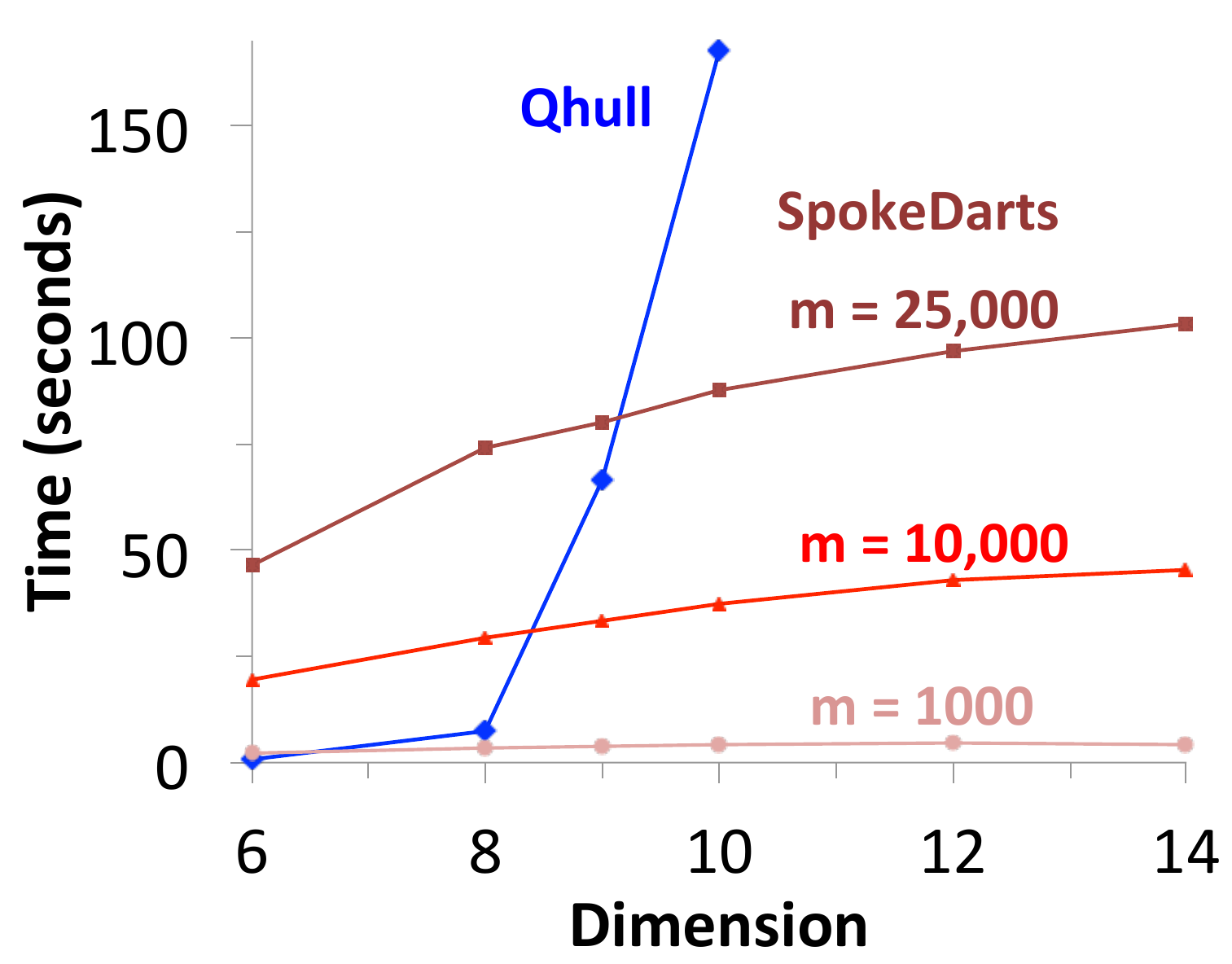}
   }
   \subfloat[memory requirements]
   {
     \label{fig:spoke_delaunay_b}
     \includegraphics[width=0.46\linewidth]{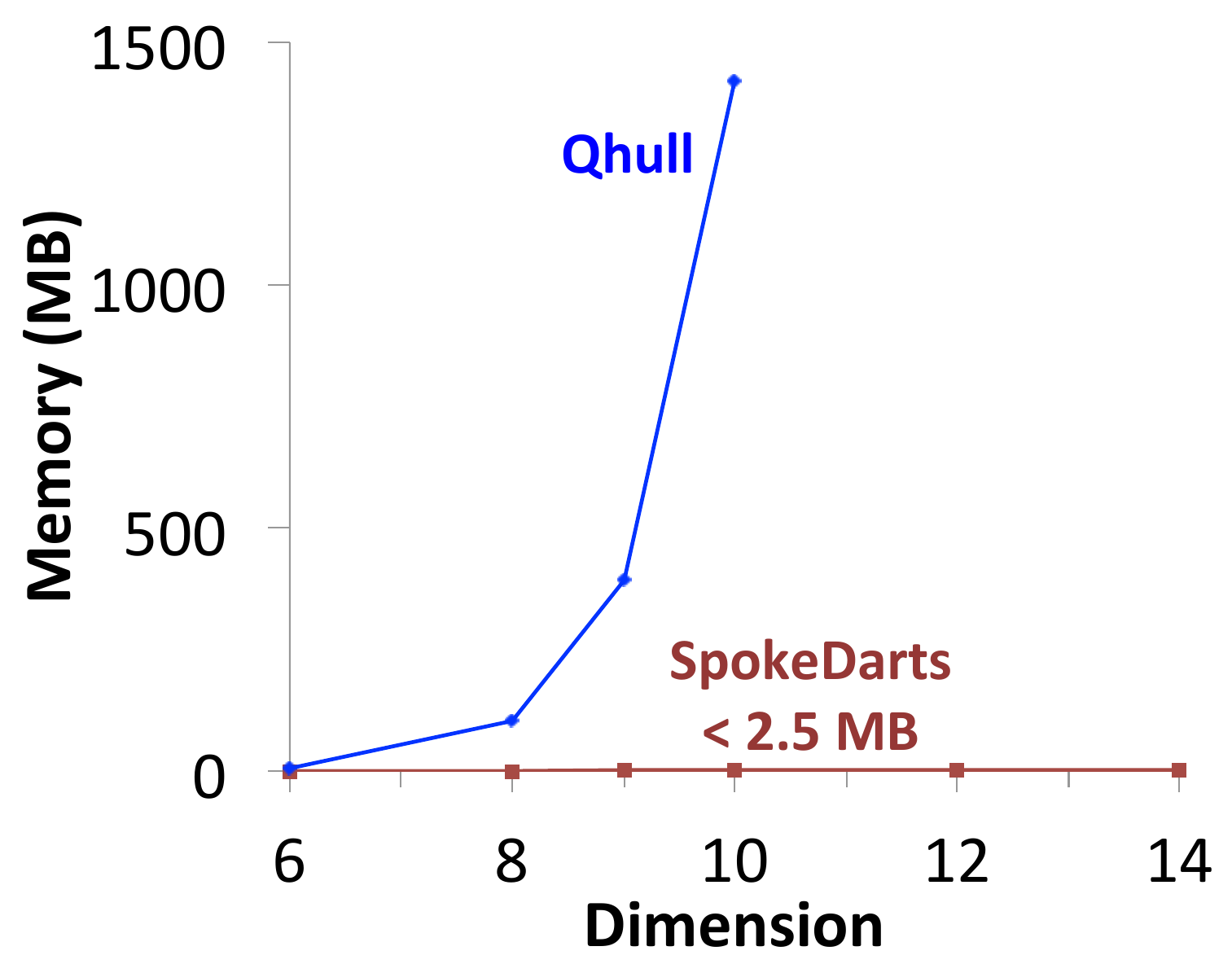}
   } 
   
   \subfloat[effects of $\hammerlimit$ on \% of missing \nothing{delaunay }edges]
   {
     \label{fig:spoke_delaunay_c}
       \includegraphics[width=0.46\linewidth]{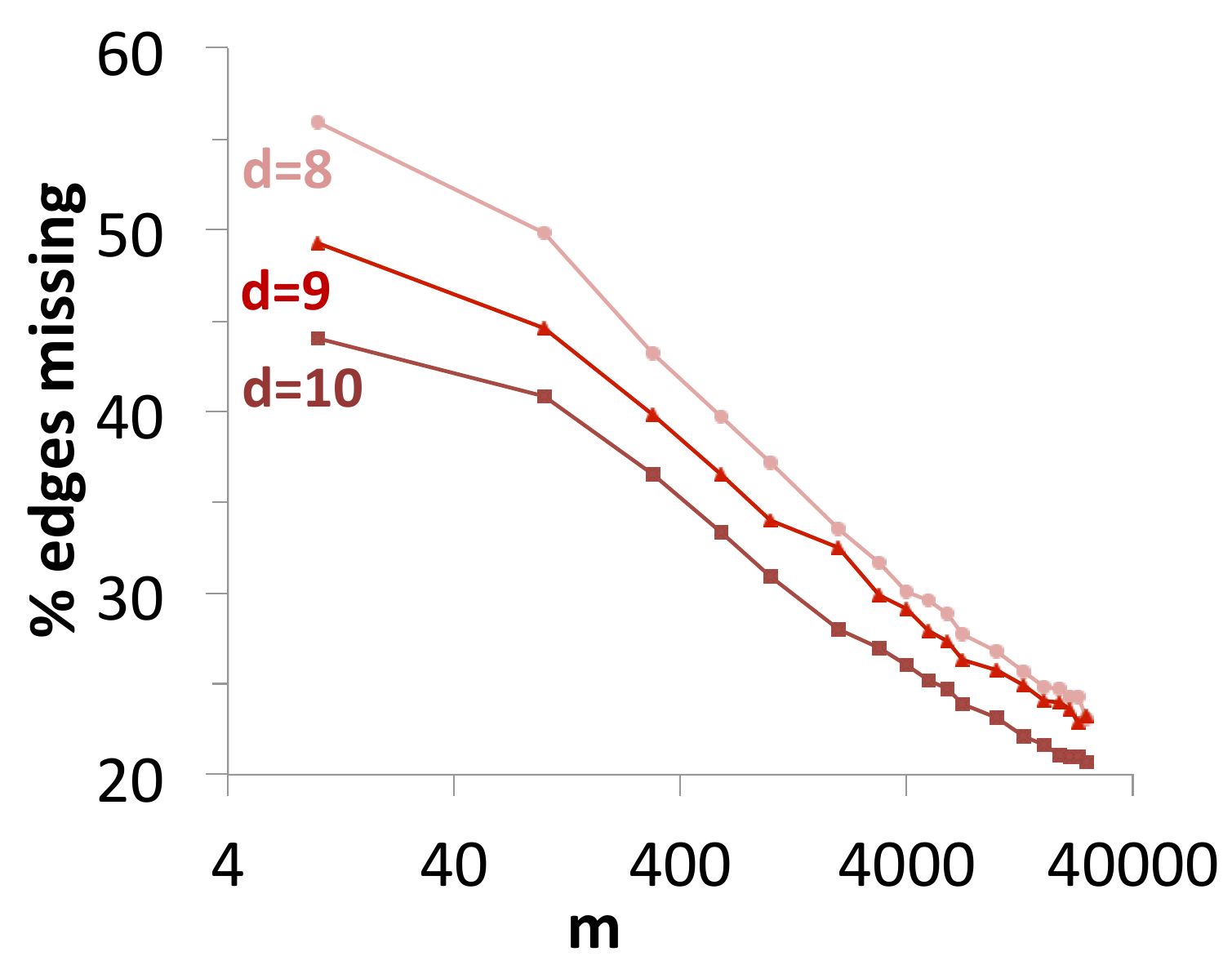}
    } 
   \subfloat[effects of $\hammerlimit$ on time]
   {
     \label{fig:spoke_delaunay_d}
       \includegraphics[width=0.46\linewidth]{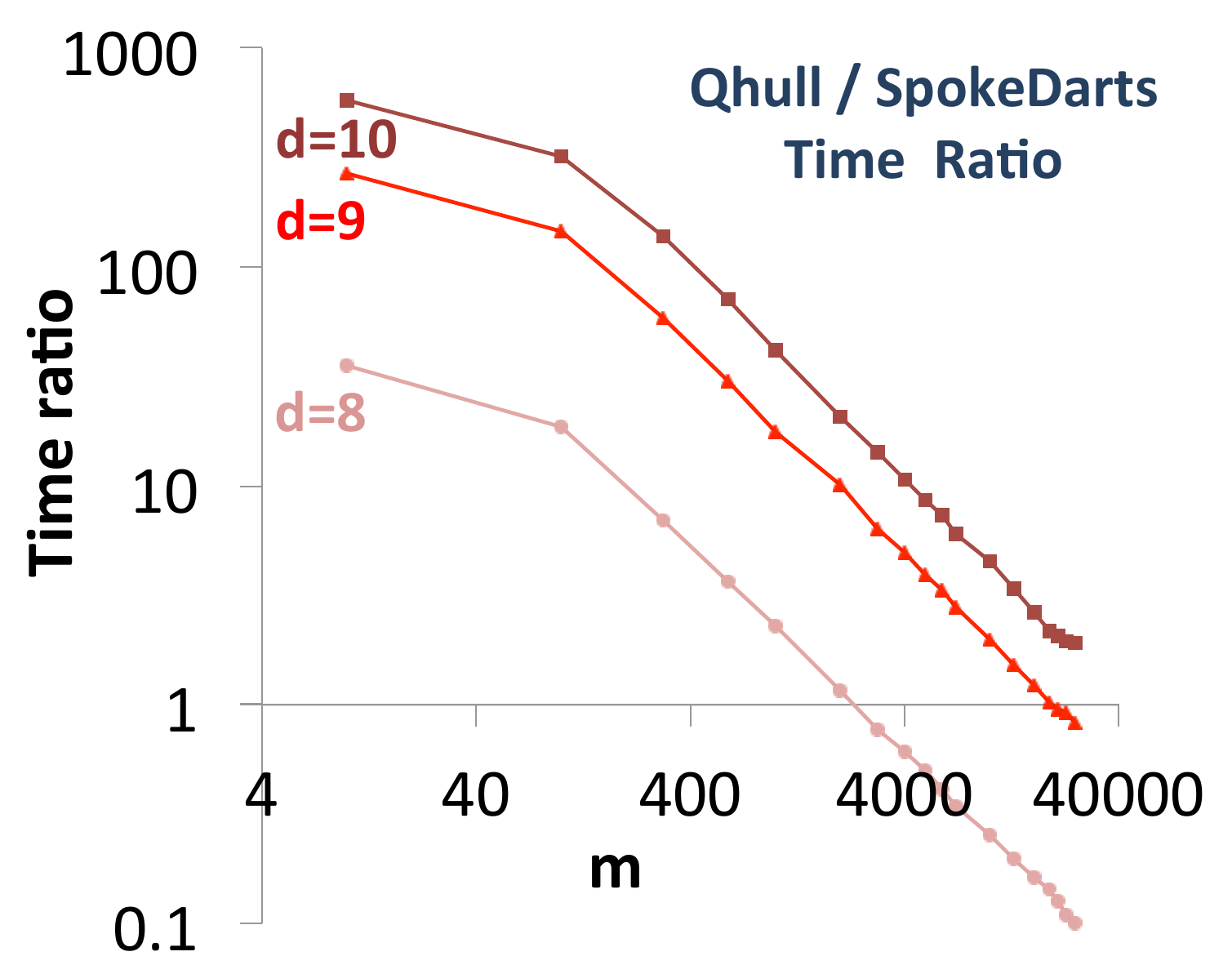}
    } 

  \Caption{Comparisons between Qhull and our \hidmethod{} method.}
{\subref{fig:spoke_delaunay_a}: speed comparison for getting the exact Delaunay triangulation using Qhull, and an approximate Delaunay graph using \spokedarts{} with different values of $\hammerlimit$.
\subref{fig:spoke_delaunay_b}: Our \hidmethod{} method generally consumes memory less than 2.5 MBs in different dimensions, and using different values of $\hammerlimit$.
\subref{fig:spoke_delaunay_c}: as $\hammerlimit$ increases, percentage of missing Delaunay edges decreases but computation also increases for our method \subref{fig:spoke_delaunay_d}.

\nothing{
\muhammad{
It's clear from the Qhull performance graph that there is a jump in the memory/ time consumption in $d > 9$. On the other hand, our algorithm gives better performance with only 20\%  of missing edges.
}}%nothing

\liyi{(January 15, 2014)
I see a few issues for this comparion.
(1) For \subref{fig:spoke_delaunay_vs_qhull_old}, we cannot claim trend with just 3 data points.
Can we plot more dimensions, say from 2 to 20 or even larger?
It is OK (actually, good) if competing techniques cannot handle; we just cut short their curves and say they have to capitulate.
(2) For \subref{fig:m_vs_missingratio_old}, what we mean by \emacsquote{\% of missing Delaunay edges}, number of edges, or sum of solid angles?
In the main text our argument is about the latter (missing small solid angles), so it pays off to show both.
\nothing{
(3) A more serious issue is why Qhull should be our target of comparison at all; it is exact while our method is approximate.
We need to justify, or find a better method to compare against.
We are not the first high dimensional approximate meshing method, right?
If time is running short, at least add \kddart\ to comparison here.
}%nothing
}%liyi
\mohamed{
Qhull required memory that we could't afford for d > 10. Up to my knowledge we ARE the first approximate meshing (I haven't seen this term before). In high dimensions people (e.g. in manifold learning, computational topology ... etc.) use nearest neighbors for connecting near-by points  ... I believe, spoke darts will shift many of these communities to use approximate meshing instead since it accounts for directionality not only distance.
}
\liyi{OK this is important. In nowhere of our paper have we stated this crucial fact that we are the first to compute Approximate Delaunay, or even the first one to propose this notion. I have updated the main text. So, we are safe on this comparison side. But the question will remain on our novel proposal of approximate Delaunay; can be a dangerous thing if the reviewers think it is useless and just our fancy imagination. Hopefully our second paragraph is convincing enough.}

\muhammad{(January 19, 2014) I've updated \Cref{fig:delaunay_spoke_darts} to include more dimensions. If the figures are too crowded we can split them into 4 figures. To extend Fig. \subref{fig:spoke_delaunay_a} beyond $d = 10$, $\hammerlimit$ is fixed instead of fixing the \% missing delaunay edges.}
\muhammad{(January 19, 2014) The graphs are now split. The old ones are commented}
}
    \label{fig:delaunay_spoke_darts}
\end{figure}  

\nothing{
\liyi{(May 17, 2013) plot fonts are too small to be legible, so I enlarged them for now.}
\muhammad{(January 14, 2014) Looks better now ?}
\liyi{(January 14, 2014) Yup.}
}%nothing

\begin{figure}[tbh]
  \centering

  \subfloat[different spoke types]
  { \label{fig:histogram_dart_type}
     \includegraphics[width=0.48\linewidth]{figs/histogram_type2wide.pdf} %0.02 bin width, adding neighboring bins
  }  
  \subfloat[different $\hammerlimit$]
% fixed now! thanks Chonhyon \scott{there is something wrong here. with increasing misses the peak decreases and the flat region decreases, but it is supposed to be normalized to the total number of points produced, so the area under the curve should be invariant.}
  { \label{fig:histogram_m}
     \includegraphics[width=0.48\linewidth]{figs/histogram_misses2zoom.pdf}  % distance 1-1.25 showing just peaks
  }

 \Caption{Radial profiles, from differential domain distributions \protect\cite{Wei:2011:DDA}.}
 {\subref{fig:histogram_dart_type} Different sampling types. Spokes all use the same $\hammerlimit = 1000$.
 \Linespokestext{}, line darts (\kddarts{} with lines), and Simple MPS produce nearly identical peaks, but line darts is  farthest from saturation.
\subref{fig:histogram_m} \Linespokestext{} with different $\hammerlimit.$
The peak becomes more pronounced as saturation is approached.
All distributions are quite flat for larger distances.
\scott{figs are fixed IMO}
\scott{todo: start the figures at exactly $r=1,$ not $0.95<1.$}
\chpark{Li-Yi, if I reduce the a-axis max to 1.25, it is better to see the peaks. Is it what you thinking? }
\liyi{Yes, we don't need to show a lot of the boring flat region on the right.}
\scott{I do think we need to convey that out to $r=2$ it is indeed flat and there are not any surprising secondary peaks. One graph like that, or some text somewhere, might suffice to convey that thought and allow us to zoom in on the peaks in these figures.}

}
 \label{fig:pairwise_dist}
 \label{fig:histograms}
 %BND_AdvFront\Drafts\hid\figs\pairwise_dist.xlsx
\end{figure}

  \nothing{
  \subfloat[Number of inserted points for different dart types]
  { \label{fig:histogram1_1}
     \includegraphics[width=0.48\linewidth]{figs/histogram_type1.pdf}
  }    
  \subfloat[Number of points for different $\hammerlimit$]
  { \label{fig:histogram2_1}
     \includegraphics[width=0.48\linewidth]{figs/histogram_misses1.pdf} 
  }
  }%nothing

\nothing{
\scott{I agree with Li-Yi. I suggest someone (Mohamed?) brings up the excel sheets on a mac and right-click saves the picture as a pdf. I'm not on a mac right now.}
\liyi{fonts too small in this scale and there are strange boundaries in the plots. These pdf file appear to be rasterized rather than vector graphics; you can see jpeg artifacts upon zoom-in.}
\scott{Chonhyun, please add MPS to both right-column plots so I can compare to it.}
\scott{Top right, the peaks are not clear. Need to draw a horizontal line or something to show where the peak is for each method.} 
\scott{I'm confused. (c) and (d) should be the same spokes. But (c) appears to be normal line-spokes and (d) is degenerate ones? I suggest normal line-spokes for each, or perhaps two rows, one showing degenerate, and one normal.}
\liyi{I suggest zoom-in to the interesting regions near the peak, especially for \subref{fig:histogram_dart_type}, so that we can see more clearly the differences around the peaks. Right now, all curves are cluttered together.}
\liyi{Consistency: say $\hammerlimit$ instead of \emacsquote{miss}?}
\liyi{(January 21, 2014) Font are too small; try to be consistent with \Cref{fig:mps_maximality}.}
}%nothing

\input{rebuttal}

}
{}
\fi

\end{document}